\documentclass{aa}

\usepackage{graphicx}
\usepackage{txfonts}
\usepackage{longtable}
\usepackage{multirow}
\usepackage{textcomp}

\usepackage{morefloats}
\newcommand\nobr{\mbox{-}}
\newcommand\kms{km~s$^{-1}$}

\newcommand\micron{~\hbox{\textmu}m}
\usepackage{color}

\begin{document}

   \title{Probing the effects of external irradiation on low-mass protostars through unbiased line surveys\thanks{Based on observations with the Atacama Pathfinder EXperiment (APEX) telescope. APEX is a collaboration between the Max Planck Institute for Radio Astronomy, the European Southern Observatory, and the Onsala Space Observatory.}}

   \author{J. E. Lindberg\inst{1,2}\thanks{NASA Postdoctoral Program Fellow.}
          \and
          J. K. J{\o}rgensen\inst{1}
          \and
          Y. Watanabe\inst{3}
          \and
          S. E. Bisschop\inst{1}
          \and
          N. Sakai\inst{3}
          \and
          S. Yamamoto\inst{3}
          }

   \institute{
          {Centre for Star and Planet Formation, Niels Bohr Institute and Natural History Museum of Denmark, University of Copenhagen, {\O}ster Voldgade 5-7, DK\nobr1350 Copenhagen K, Denmark}
                \and
      {NASA Goddard Space Flight Center, Astrochemistry Laboratory, Mail Code 691, 8800 Greenbelt Road, Greenbelt, MD 20771, USA}\\
      \email{johan.lindberg@nasa.gov}
                    \and
      {Department of Physics, The University of Tokyo, 7-3-1 Hongo, Bunkyo-ku, Tokyo, 113-0033, Japan}\\
         }

   \date{Received March 30, 2015; accepted September 7, 2015}


  \abstract
   {The envelopes of molecular gas around embedded low-mass protostars show different chemistries, which can be used to trace their formation history and physical conditions. The excitation conditions of some molecular species can also be used to trace these physical conditions, making it possible to constrain for instance sources of heating and excitation.}
   {We study the range of influence of an intermediate-mass Herbig~Be protostar. We also study the effect of feedback from the environment on the chemical and physical properties of embedded protostars.}
   {We followed up on an earlier line survey of the Class~0/I source R~CrA IRS7B in the 0.8~mm window with an unbiased line survey of the same source in the 1.3~mm window using the Atacama Pathfinder Experiment (APEX) telescope. We also studied the excitation of the key species H$_2$CO, CH$_3$OH, and \textit{c}\nobr C$_3$H$_2$ in a complete sample of the 18 embedded protostars in the Corona Australis star-forming region. Radiative transfer models were employed to establish abundances of the molecular species.}
   {We detect line emission from 20 molecular species (32 including isotopologues) in the two surveys. The most complex species detected are CH$_3$OH, CH$_3$CCH, CH$_3$CHO, and CH$_3$CN (the latter two are only tentatively detected). CH$_3$CN and several other complex organic molecules are significantly under-abundant in comparison with what is found towards hot corino protostars. The H$_2$CO rotational temperatures of the sources in the region decrease with the distance to the Herbig~Be star R~CrA, whereas the \textit{c}\nobr C$_3$H$_2$ temperatures remain constant across the star-forming region.}
   {The high H$_2$CO temperatures observed towards objects close to R~CrA suggest that this star has a sphere of influence of several 10\,000~AU in which it increases the temperature of the molecular gas to 30--50~K through irradiation. The chemistry in the IRS7B envelope differs significantly from many other embedded protostars, which could be an effect of the external irradiation from R~CrA.}

   \maketitle
%

\section{Introduction}

The molecular composition of the envelopes of deeply embedded low-mass protostars can be used to trace their history and physical conditions. As a result of observations of these sources as well
as of chemical modelling, the relevance and origin of many of these molecules are now relatively well understood \citep[see e.g.][]{herbst09}. The studied molecules include organic, silicon- and sulphur-bearing, and deuterated species \citep[e.g.][]{blake94,vandishoeck95,schoier02}. \citet{jorgensen04b} correlated the presence and abundance of nine molecular species, discovering strong correlations between some species and anti-correlations between others. This enabled  constructing an empirical chemical network of molecular species.

Data from spectroscopy observations of molecular emission lines can also be used to trace the physical properties of protostellar envelopes. The rotational temperature of the molecular gas, which can be used as a proxy for the kinetic temperature of the gas given certain assumptions, can be measured by excitation diagrams of the observed molecules \citep[see e.g.][]{goldsmith99,jorgensen05b}. By comparing the rotational temperatures of different molecular species, insights about their distribution and formation paths can be gained. Isotopical ratios can also be used as tracers of formation history.

During the past decade, the existence of an inner ($R\lesssim100$~AU) hot region in the envelopes of deeply embedded low-mass protostars with a chemistry similar to that of high-mass hot cores has been proposed. These dense ($n>10^6$~cm$^{-3}$) regions are referred to as hot corinos \citep{ceccarelli04}, and they are characterised by the presence and relatively high abundance of complex organic molecules such as CH$_3$OH, CH$_3$OCH$_3$, and CH$_3$OCHO \citep{bottinelli04a,bottinelli07,jorgensen05a}. High angular resolution interferometric observations show that complex organics are often concentrated in the inner $\lesssim 1\arcsec$ ($R \lesssim 50$--$100$~AU) of low-mass protostars  \citep[e.g.][]{bottinelli04b,kuan04,jorgensen05a,bisschop08,maury14} -- most likely reflecting that they evaporate from the icy dust grains at high temperatures ($T \gtrsim 100$~K). IRAS~16293-2422 is considered as the prototypical hot corino, but at least three more such sources have been discovered.

\citet{sakai08,sakai09b} found that the chemistry towards two deeply embedded low-mass protostellar sources was distinctly different from that of the hot corinos. The two sources L1527 and IRAS 15398-3359 are dubbed warm carbon-chain chemistry (WCCC) sources. They are characterised by comparably high abundances of long carbon-chain molecules such as HC$_3$N, C$_4$H, and C$_6$H, and at the same time by low or moderate abundances of complex organic species. The difference between the chemical properties observed towards hot corinos and WCCC sources could be an effect of different collapse timescales \citep{sakai09b,sakai13}. This
mechanism has been verified by chemical modelling \citep{hassel08}.

The presence of complex organic molecules in low-mass star formation is, however, not unique to the hot inner regions of deeply embedded protostars. The spectral signatures of several complex organic molecules have recently been detected towards the prestellar core L1689B, with temperatures $\sim 10$~K \citep{bacmann12}. Complex organic molecules have also been found towards protostellar outflows such as that of L1157, with similarly low temperatures \citep{arce08}.

The transitional Class~0/I protostar R~CrA IRS7B is located in the R~CrA cloud (NGC~6729) within the Corona Australis (CrA) star-forming region. At a distance of 130~pc \citep{neuhauser08}, this is one of the nearest low-mass star-forming regions, making it ideal for the study of the envelope and core chemistry of the individual sources by single-dish (sub)mm observations. The population of young stellar objects (YSOs) in the region was inventoried through multi-wavelength studies by \citet{peterson11}, who identified 116 YSOs and classified them according to their spectral slope following the definitions of \citet{greene94}. Out of these sources, 14 were found to be Class~I sources or younger, and 5 were found to be flat-spectrum sources (transitional sources between Class~I and Class~II). Most of these deeply embedded sources are concentrated in the R~CrA cloud, within a few 1000~AU from the Herbig~Be star R~CrA. \citet{lindberg12} found large-scale H$_2$CO emission in the region, exhibiting gas temperatures $>30$~K. The high temperatures, which cannot be explained by internal radiation from the low-mass stars, were suggested to be a result of external irradiation from R~CrA.

\citet{watanabe12} presented an unbiased line survey of R~CrA IRS7B in the 0.8~mm window and a few scans in the 0.7~mm window, using the Atacama Submillimeter Telescope Experiment (ASTE) 10~m telescope. These observations showed enhancements of the radicals CN and C$_2$H, which trace photon-dominated regions (PDRs). Furthermore, neither any complex organic molecules characteristic for hot corino chemistry (except the ubiquitous CH$_3$OH) nor any long carbon chains characteristic for WCCC were detected. High-resolution Atacama Large Millimeter/submillimeter Array (ALMA) observations towards R~CrA IRS7B also indicate an absence or unusually low abundances of the complex organic species CH$_3$CN and CH$_3$OCH$_3$ \citep{lindberg14_alma}.

This paper complements the work by \citet{watanabe12} by adding an unbiased line survey in the 1.3~mm window, which increases the number of detected molecular species as well as the number of molecular transitions in many of the excitation diagrams, and thus also the range of $E_{\mathrm{u}}$ in those diagrams. We also investigate other similar protostellar sources in Corona Australis to study the spatial variations of the chemical and physical properties in the star-forming region. This paper finally uses radiative transfer modelling of the emission from several of the species to establish molecular abundances for comparison with other sources.

Section~\ref{sec:obs} describes the observational methods and data reduction strategy. Section~\ref{sec:results} describes the spectra resulting from the observations, and Sect.~\ref{sec:analysis} discusses rotational diagrams, radiative transfer models, and isotopologue ratios of many of the observed molecules. Section~\ref{sec:discussion} discusses the large-scale heating in the CrA region and the chemistry of the envelopes. Section~\ref{sec:conclusions} lists our conclusions.

\section{Observations}
\label{sec:obs}

\subsection{APEX unbiased line survey of R CrA IRS7B}
\label{sec:obs_irs7b}

The observations were carried out with the Atacama Pathfinder Experiment 12~m telescope \citep[APEX;][]{gusten06} in position-switching mode in August and October 2010. The frequency range between 217.9~GHz and 245.5~GHz was covered with 17 spectral setups on the SHeFI (Swedish Heterodyne Facility Instrument) receiver APEX-1 \citep{vassilev08}. This survey was unbiased in the sense that it covered a wide spectral range without targeting any particular molecular species. In addition, we used parts of the APEX-1 data that were observed by \citet{lindberg12} in April 2010 to cover the range between 217.2~GHz and 217.9~GHz. All observations were centred at the sub-mm/IR point source (Class 0/I YSO) IRS7B in the R~CrA cloud (see Table~\ref{tab:sourcelist} for coordinates). The two sets of observations have an overlap around 218~GHz, which we used to investigate the calibration accuracy. The small differences found in the line intensities can all be explained with a calibration uncertainty $\lesssim10\%$.

\begin{table*}
        \centering
        \caption[]{Observed sources and the rms of the observations in the APEX surveys of CrA. See Table~\ref{tab:obsparam} for spectral window parameters.}
        \label{tab:sourcelist}
        \begin{tabular}{l l l c l l l}
                \noalign{\smallskip}
                \hline
                \hline
                \noalign{\smallskip}
                Source name\tablefootmark{a} & RA & Dec & YSO Class\tablefootmark{b} & \multicolumn{3}{c}{rms [mK~(km~s$^{-1}$)$^{-1}$]}\\ 
                & (J2000.0) & (J2000.0) & & 1.4 mm & 0.9 mm & 0.8 mm \\ 
                \noalign{\smallskip}
                \hline
                \noalign{\smallskip}
                IRS7B & 19 01 56.40 & $-$36 57 28.1 & I & 11\tablefootmark{c} & 13 & 40 \\
                CrA-46 & 18 55 56.32 & $-$37 00 07.1 & Flat & 19 & ... & ... \\
                CrA-3 & 18 59 43.92 & $-$37 04 01.1 & Flat & 19 & 32 & ... \\
                CrA-5 & 19 00 15.55 & $-$36 57 57.7 & I & 10 & 21 & 27 \\
                LS-RCrA1 & 19 01 33.56 & $-$37 00 30.3 & Flat & 20 & 22 & ... \\
                Haas4 & 19 01 40.67 & $-$36 56 04.9 & Flat & 16 & 20 & ... \\
                IRS2 & 19 01 41.56 & $-$36 58 31.2 & I & \phantom{0}7 & 14 & ... \\
                IRS5A & 19 01 48.03 & $-$36 57 22.2 & I & 11 & 18 & 36 \\
                IRS5N & 19 01 48.46 & $-$36 57 14.7 & I & 11 & 15 & 38 \\
                IRS1 & 19 01 50.68 & $-$36 58 09.7 & I & 12 & 20 & 34 \\
                IRS7A & 19 01 55.32 & $-$36 57 21.9 & I & ... & 14 & 38 \\
                CrA-24 & 19 01 55.60 & $-$36 56 51.1 & I & 20 & 20 & 36 \\
                SMM~2 & 19 01 58.54 & $-$36 57 08.5 & I & 20 & 15 & 36 \\
                CXO42 & 19 02 01.96 & $-$36 54 00.0 & I & \phantom{0}8 & 17 & ... \\
                CrA-44\tablefootmark{d} & 19 02 58.67 & $-$37 07 35.9 & I & 12 & 19 & ... \\
                CrA-33 & 19 03 01.03 & $-$37 07 53.4 & Flat & 19 & 16 & ... \\
                VV~CrA & 19 03 06.80 & $-$37 12 49.1 & I & \phantom{0}9 & 12 & ... \\
                CrA-37 & 19 03 55.24 & $-$37 09 35.9 & I & 20 & 15 & ... \\
                \noalign{\smallskip}
                \hline
        \end{tabular}
        \tablefoot{
                \tablefoottext{a}{We use the notation of \citet{peterson11} for all sources except IRS5A, IRS5N, and SMM~2, for which we use the notation of \citet{lindberg12} and \citet{lindberg14_herschel}. These are referred to as CrA-19, CrA-20, and CrA-43, respectively, by \citet{peterson11}.}
                \tablefoottext{b}{Assigned from the spectral index $\alpha$ \citep{peterson11}. Class~I sources are Class~I or younger.}
                \tablefoottext{c}{Only partially covered by the IRS7B line survey. The observations do not cover 216.1--217.2~GHz, and the 217.2~GHz--217.9~GHz portion is covered only at a lower S/N level (37~mK~(km~s$^{-1}$)$^{-1}$).}
                \tablefoottext{d}{Also known as IRAS~32.}
                \\
        }
\end{table*}

For all new scans reported in this work, rms levels between 21 and 35~mK~channel$^{-1}$ were reached with velocity resolutions ranging between 0.17~\kms\ and 0.15~\kms\ (122~kHz), corresponding to rms levels of 8--14~mK~(km~s$^{-1}$)$^{-1}$. The scan from \citet{lindberg12}, which here was used to cover 217.2~GHz--217.9~GHz, is not as deep as the new data presented here. It has an rms level of 84~mK~channel$^{-1}$, or 35~mK~(km~s$^{-1}$)$^{-1}$.

Because of unstable quasi-sinusoidal baselines in the observed spectra \citep[see][]{vassilev08}, we had to perform relatively advanced baseline fitting after carefully identifying channels containing any spectral line emission. The exact method used for this baseline fitting is described in Appendix~\ref{app:baseline}. The calculation and subtraction of baselines were performed using our own scripts as described in the appendix, but Gaussian line fitting and calculations of line strengths were afterwards performed using the GILDAS CLASS package\footnote{GILDAS CLASS (Continuum and Line Analysis Single-dish Software) is developed by the IRAM institute, Grenoble, France, and can be downloaded from \url{http://www.iram.fr/IRAMFR/GILDAS}.}. To account for some smaller features left after the baseline subtractions, a zeroth-order baseline was subtracted locally around all fitted lines before the extraction of line parameters.

All line intensities are given in the $T_{\mathrm{mb}}$ (main beam brightness temperature) scale, which means that the antenna temperatures $T_A^*$ are corrected with the APEX main beam efficiency $\eta_{\mathrm{mb}} = 0.75$. The beam sizes of the observations range between 26\arcsec\ and 29\arcsec\ depending on the observed frequency. All errors given in this work are at the $1\sigma$ level if not otherwise stated.

As a complement to the APEX observations, we use the spectral lines detected in the ASTE 0.8~mm and 0.7~mm observations of \citet{watanabe12}. This enables a more accurate excitation analysis of the various molecular species thanks to the larger range of $E_{\mathrm{u}}$ in the used transitions. The beam size of these ASTE observations ranges between 21\arcsec\ and 23\arcsec\ for the 0.8~mm window, and between 16\arcsec\ and 17\arcsec\ for the few lines in the 0.7~mm window. For the 0.8~mm window in
particular, the beam size is quite similar to that of the APEX observations, meaning that the two sets of observations cover roughly the same spatial region. From previous SMA observations we know that the spatial extent of the molecular line emission is similar to or larger than these beams \citep{lindberg12}. For this reason, and to be consistent with \citet{watanabe12}, we thus calculated beam-averaged column densities. When combining data points with different beam sizes in excitation diagrams, this means that we assumed that the beam-averaged column densities are similar in both beams, or in other words, that the emission is extended and uniform.

\subsection{Unbiased survey of H$_2$CO, CH$_3$OH, and \textit{c}\nobr C$_3$H$_2$ in CrA}

\begin{figure*}[!htb]
    \centering  
    \includegraphics{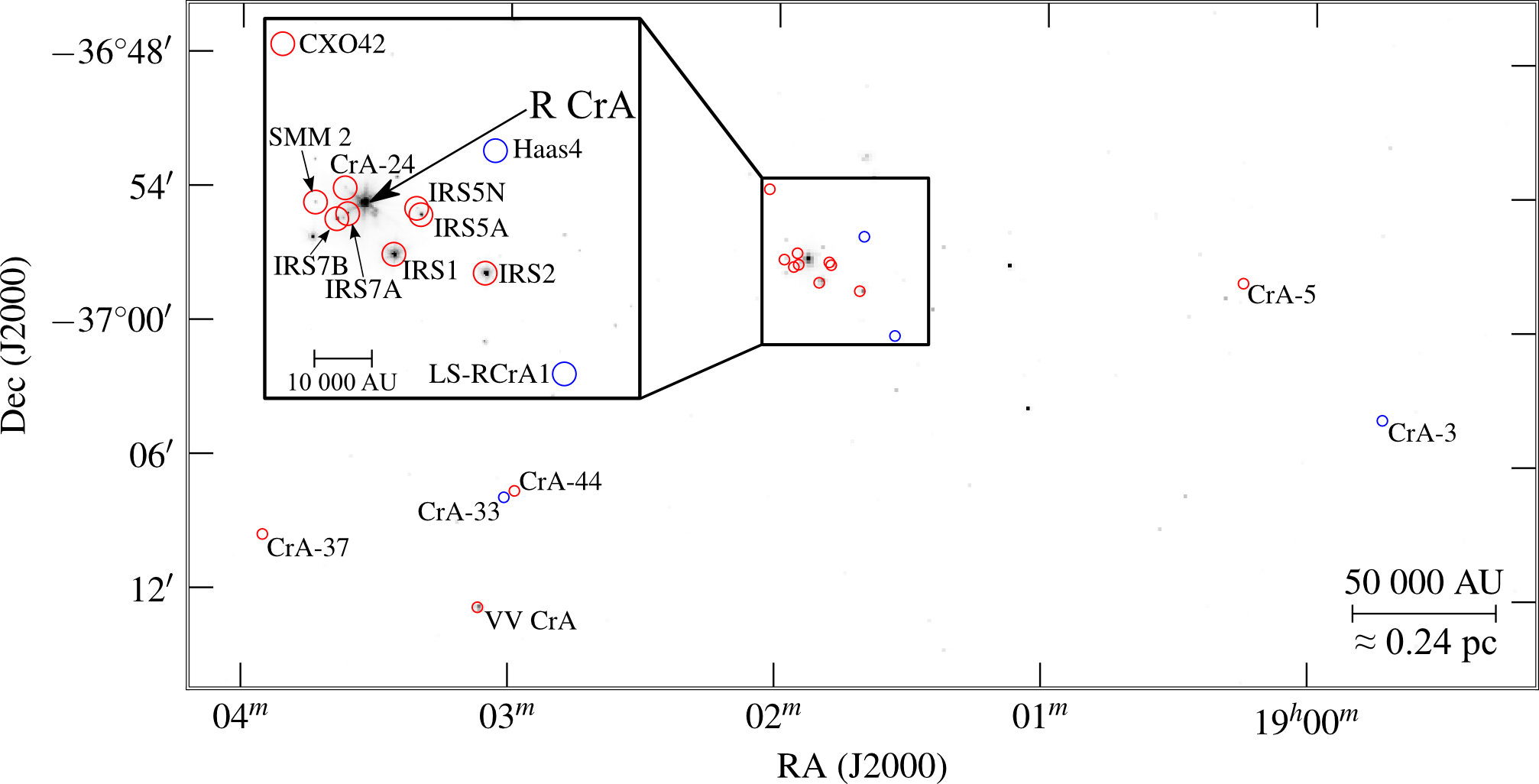}
    \caption{Overview of the observed sources in the H$_2$CO, CH$_3$OH, and \textit{c}\nobr C$_3$H$_2$ survey. The circles show the APEX beam size in the 1.4~mm window. Red circles denote Class~0/I objects and blue circles are flat-spectrum sources. CrA-46 is located $\sim40$\arcmin\ ($\sim1.5$~pc) west of the right edge of the plot. The inset shows a zoom-in of the region around R~CrA enlarged by a factor $\sim2$. The greyscale background is a \textit{Spitzer} 4.5\micron\ image. (A colour version of this plot is available in the online journal.)}\label{fig:overview}
\end{figure*}

\begin{table}
        \centering
        \caption[]{Observational parameters of the APEX observations (this work; IRS7B line survey and CrA source survey) and the ASTE line survey of IRS7B \citep{watanabe12}.}
        \label{tab:obsparam}
        \begin{tabular}{l l c c}
                \noalign{\smallskip}
                \hline
                \hline
                \noalign{\smallskip}
                Window & Frequency range & \multicolumn{2}{c}{Beam size} \\
                & [GHz] & [\arcsec] & [AU] \\
                \noalign{\smallskip}
                \hline
                \noalign{\smallskip}
                IRS7B survey & 217.2--245.5\tablefootmark{a} & 26--29 & 3300--3800 \\
                1.4 mm & 216.1--220.1 & 29 & 3700 \\
                0.9 mm & 336.9--340.9 & 19 & 2400 \\
                0.8 mm & 361.8--365.8\tablefootmark{b} & 17 & 2200 \\
                \noalign{\smallskip}
                \hline
                \noalign{\smallskip}
                ASTE 0.8 mm & 332.0--364.0 & 21--23 & 2700--3000 \\
                ASTE 0.7 mm & 435.0--437.0 & 17 & 2300 \\
                ASTE 0.7 mm & 459.2--461.2 & 16 & 2100 \\
                \noalign{\smallskip}
                \hline
        \end{tabular}
        \tablefoot{
                \tablefoottext{a}{The 217.2~GHz--217.9~GHz portion is only covered at a low S/N level.}
                \tablefoottext{b}{In most sources only 361.8--364.3~GHz, due to high noise levels and baseline issues in the upper sideband.}
        }
\end{table}

In addition to the unbiased line survey of IRS7B discussed above, IRS7B and 17 other protostellar sources within the Corona Australis star-forming region were observed in more narrow frequency bands using the APEX telescope. The selected sample consists of all sources defined as Class~I sources (or younger) or flat-spectrum sources in the survey of YSOs in CrA conducted by \citet{peterson11}. Only the faint Class~I source CXO34 was excluded because of its close proximity to the two much stronger sources IRS7A and IRS7B, which would cause confusion. The observed protostellar sources are shown on a map in Fig.~\ref{fig:overview}, and their positional data are listed in Table~\ref{tab:sourcelist}. The observations were performed in up to three frequency setups: 216.1--220.1~GHz (1.4~mm), 336.9--340.9~GHz (0.9~mm), and 361.8--365.8~GHz (0.8~mm), see also Table~\ref{tab:obsparam}. The frequency setups were chosen to cover a wide $E_{\mathrm{u}}$ range of H$_2$CO, CH$_3$OH, and \textit{c}\nobr C$_3$H$_2$ lines. The 1.4~mm window was observed with the SHeFI APEX-1 receiver and the other two windows with the SHeFI APEX-2 receiver \citep{vassilev08}. Observations in the higher-frequency windows were only executed for sources with a significant number of detected lines (see Table~\ref{tab:sourcelist} for details on observed spectral windows). The observations were carried out in September and November 2011. This survey is unbiased in the sense that it covers (nearly) all embedded sources in CrA, albeit only in a limited number of spectral setups, targeting certain molecular species.

All line intensities are given on the $T_{\mathrm{mb}}$ scale. The APEX main beam efficiency $\eta_{\mathrm{mb}} = 0.75$ for the 1.4~mm window and $\eta_{\mathrm{mb}} = 0.73$ for the 0.9~mm and 0.8~mm windows. The unsmoothed spectra have 76~kHz channel widths (corresponding to 0.10~\kms\ for the 1.4~mm window, 0.067~\kms\ for the 0.9~mm window, and 0.063~\kms\ for the 0.8~mm window). The rms levels when smoothed to 1~\kms\ channels are listed in Table~\ref{tab:sourcelist}. In the 1.4~mm window, the rms level is between 7~mK~(\kms)$^{-1}$ and 20~mK~(\kms)$^{-1}$, which can be compared to a noise level of 11~mK~(\kms)$^{-1}$ for the 1.4~mm window in the IRS7B survey (above).

Like in the observations described in Sect.~\ref{sec:obs_irs7b}, the spectra in this survey also have unstable baselines, and the same baseline corrections as were used for the IRS7B data were applied (see Appendix~\ref{app:baseline}). The 0.8~mm band was particularly affected by unstable baselines with both wide and narrow features. In the lower sideband of the 0.8~mm observations these effects could mostly be accounted for, but in the upper sideband the features were too strong to make proper use of the spectra except for in smaller stable chunks. Thus, only a few spectral lines were identified in this sideband, and then only in the sources where the strongest line emission was found.

\section{Results}
\label{sec:results}

\subsection{IRS7B line survey}
\label{sec:results_irs7b}

We detect 102 spectral lines in the IRS7B line survey between 217.2~GHz and 245.5~GHz. We successfully identify 87 of the detected lines, and the identifications correspond to 16 molecular species (25 if counting isotopologues separately). The online database Splatalogue\footnote{\url{http://splatalogue.net/}} was used for line identification. The detected spectral lines and their parameters are listed in Table~\ref{tab:irs7b_survey} in Appendix~\ref{app:survey_tables}. A compressed version of the APEX spectrum of the IRS7B observations can be found in Fig.~\ref{fig:fullspec}, and a full spectrum is available in Appendix~\ref{app:spectrum_irs7b} (Figs.~\ref{fig:largespec1}--\ref{fig:largespec8}).

\begin{figure*}[!htb]
        \centering  
        \includegraphics{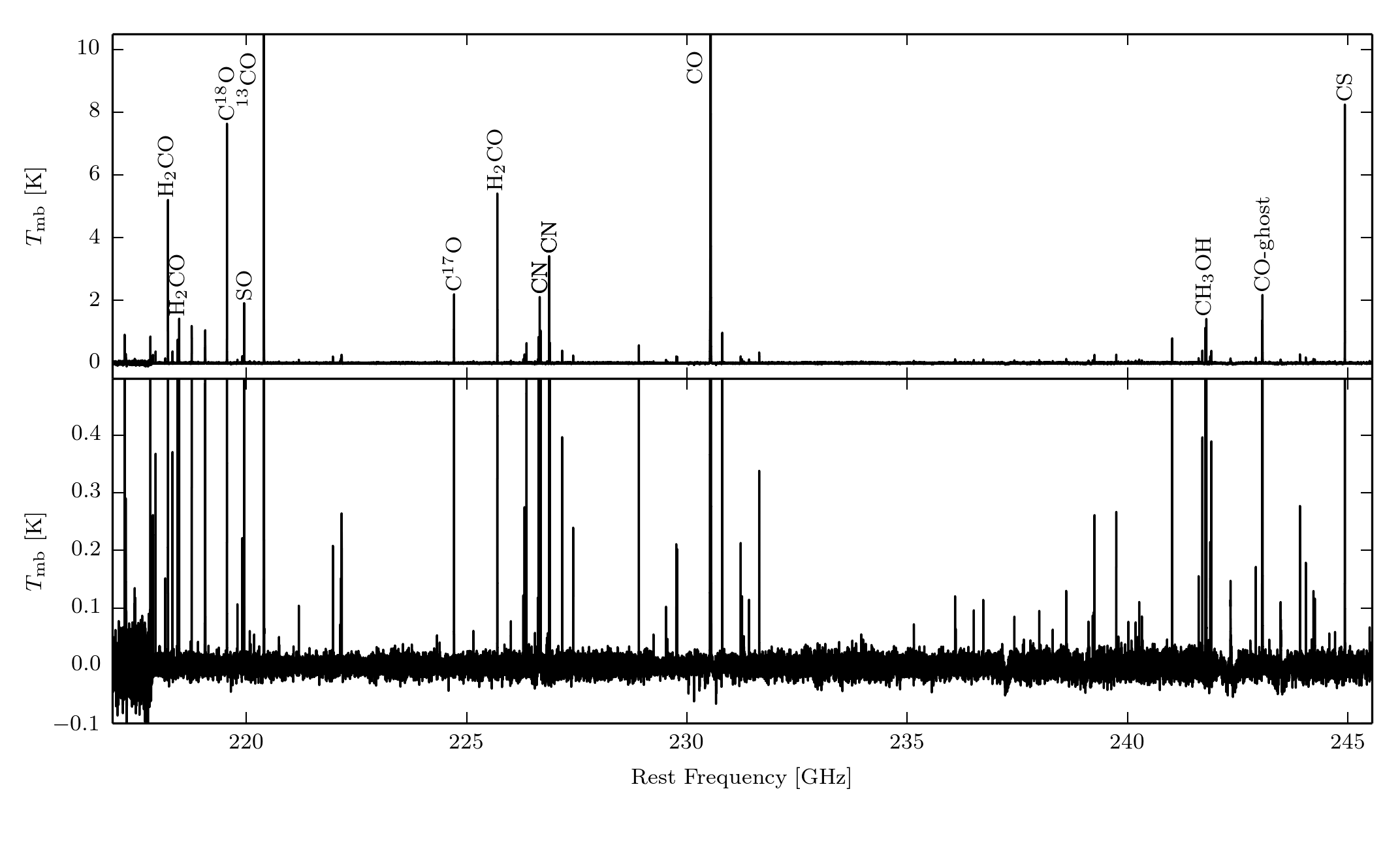}
        \caption{1.3~mm spectrum of IRS7B, cut at 10.5~K (top panel) and 0.5~K (bottom panel). A high-resolution spectrum can be found in Appendix~\ref{app:spectrum_irs7b}.}\label{fig:fullspec}
\end{figure*}

\begin{figure}[!tb]
        \centering  
        \includegraphics[width=1.0\linewidth]{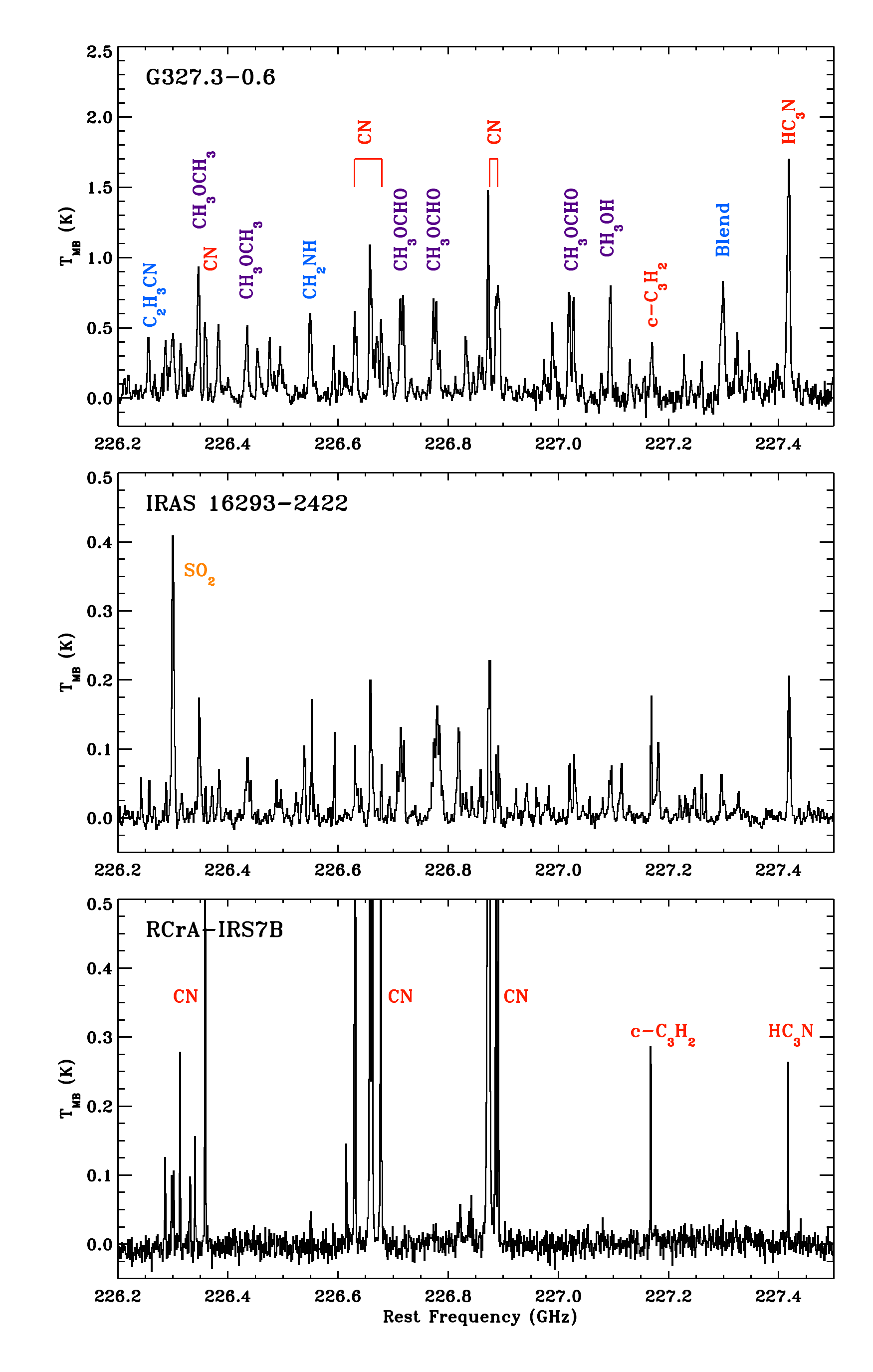}
        \caption{Single-dish spectra around 227~GHz of the hot core G327.3\nobr 0.6 (top; observed with APEX; \citealt{bisschop13}; Bisschop et~al., in prep.), the hot corino IRAS~16293\nobr 2422 \citep[middle; observed with IRAM 30~m;][]{caux11}, and R~CrA IRS7B (bottom; this work). The labels in the upper plot correspond well to the lines also seen in the middle plot. The colours denote oxygen-bearing (purple), nitrogen-bearing (blue), sulphur-bearing (yellow), and typcial WCCC and PDR (red) species. (A colour version of this plot is available in the online journal.)}\label{fig:comparison}
\end{figure}

In Fig.~\ref{fig:comparison}, we compare the line emission observed towards IRS7B in a window around 227~GHz with the emission detected towards a typical high-mass hot core and a typical low-mass hot corino source. This simplistic comparison shows that the hot corino bears more resemblance to the hot core than to IRS7B.

Many of the species detected towards IRS7B were previously detected in the ASTE survey \citep{watanabe12}, but we detect six molecular species not detected in that survey: HNCO, H$_2$CCO, CH$_3$CCH, CH$_3$CHO, CH$_3$CN, and HC$_3$N. HC$_3$N was previously detected in SMA/APEX observations \citep{lindberg12}, but the five other species have, to the best of our knowledge, not previously been detected towards IRS7B. The CH$_3$CN and CH$_3$CHO detections are only tentative: For CH$_3$CN, only two faint spectral lines are seen, and the two lines at 239.133~GHz and 239.138~GHz, expected to be of similar intensity, are not detected. For CH$_3$CHO, three lines are detected at 3--$6\sigma$ levels, but three lines at 219.820~GHz, 232.691~GHz, and 235.684~GHz, expected to be at least as strong, are non-detected. We also detect some isotopologues not seen in the ASTE survey: $^{13}$CN, C$^{15}$N, $^{13}$CO, C$^{18}$O, $^{13}$CS, C$^{33}$S, and DNC.

We find that all identified spectral lines have LSR velocities in the range $5.0-6.6$~km~s$^{-1}$ (median 5.7~km~s$^{-1}$). The line widths show a somewhat larger variation, with the lowest values around 1~km~s$^{-1}$ for species like $^{13}$CN, \textit{c}\nobr C$_3$H$_2$ and CH$_3$CCH, and the highest values above 6~km~s$^{-1}$ for CS and CO. The median line width for the non-blended lines to which a Gaussian could be fitted is 2.0~km~s$^{-1}$.

Of the 102 detected lines, 15 are unidentified. A handful of these are on the 3--$5\sigma$ level, and could thus be flukes, but a few U lines are strong. In Appendix~\ref{app:ulines} we demonstrate our attempts to identify these U lines.

\subsection{H$_2$CO, CH$_3$OH, and \textit{c}\nobr C$_3$H$_2$ survey in CrA}
\label{sec:results_sourcesurvey}

Of the 18 targeted sources in this survey, only CrA-46 shows no spectral line emission at all. Towards an additional four sources, only CO isotopologues were detected, but towards the remaining 13 sources several different molecular species were detected. The species that were detected towards at least one of the sources in this source survey are H$_2$CO, H$_2^{13}$CO, CH$_3$OH, \textit{c}\nobr C$_3$H$_2$, C$^{18}$O, C$^{17}$O, DCO$^+$, HC$^{18}$O$^+$, SiO, SO, $^{34}$SO, CN, $^{13}$CN, C$^{15}$N, C$^{34}$S, C$_2$D, DCN, HNC, and H$_2$CS. The detected spectral lines and their parameters are listed in Tables~\ref{tab:cra46}--\ref{tab:cra37} in Appendix~\ref{app:survey_tables}, except for the IRS7B parameters, which are given together with the results of the IRS7B line survey in Table~\ref{tab:irs7b_survey}. The spectra can be found in Figs.~\ref{fig:irs7b_2}--\ref{fig:cra37_2} in Appendix~\ref{app:spectrum_sources}.

With the exception of SiO, which traces shocked gas at high absolute velocities relative to the protostar, all spectral lines have LSR velocities in the range 4.6--6.6~km~s$^{-1}$ towards all sources where spectral lines are detected (CrA-24 and SMM~2 have two velocity components, however, which is discussed below). This confirms the conclusion of \citet{peterson11} that all sources where we detect line emission are situated in the same star-forming region, none of them representing background sources.

In CrA-24 and SMM~2, the H$_2$CO, CH$_3$OH, SO, and SiO spectral lines show two velocity components: one low-velocity component with LSR velocities in the range 1.2--3.1~km~s$^{-1}$ (outflow component) and one component in agreement with the other sources, with LSR velocities in the range 4.6--6.4~km~s$^{-1}$ (on-source component). The SiO emission is strongest in the low-velocity component, while the other species are stronger in the on-source component. The low-velocity component is probably an effect of a spatial overlap with a large-scale ($\sim 10\,000$~AU) outflow originating in IRS7B, extending in a north-eastward direction. This outflow has been detected in SEST mapping observations of the CH$_3$OH $2_{-2}\rightarrow1_{-1}$,~E and $2_{0}\rightarrow1_{0}$,~A+ transitions at 96.7~GHz (Miettinen et al., in prep.). Traces of this outflow can also be seen in ALMA observations of CH$_3$OH lines at 340~GHz \citep{lindberg14_alma}. For the CrA-24 and SMM~2 data, we separated the line emission of the two velocity components by fitting two Gaussians to each spectral line. The total line intensities, the on-source components, and the outflow components are reported in three separate tables for each of these two sources in Appendix~\ref{app:survey_tables}. The lines of all species except H$_2$CO, CH$_3$OH, SO, and SiO show only one velocity component, which is consistent with the on-source component. These lines are only listed in the tables with total line intensities.

We note that most sources with a large number of line detections are found in the R~CrA cloud. CrA-44 and VV~CrA also show large numbers of line detections, while the remaining sources far from R~CrA show relatively few line detections.

\section{Analysis}
\label{sec:analysis}

\subsection{Rotational diagram analysis -- IRS7B}
\label{sec:rotdiag_irs7b}

We performed a rotational diagram analysis \citep[see e.g.][]{goldsmith99} of several of the detected molecular species. This analysis requires the assumptions that the gas is in local thermal equilibrium (LTE) and that all lines are optically thin. In addition to the APEX data, we included ASTE data previously reported by \citet{watanabe12} in the fits. We performed this analysis for all the molecular species detected towards IRS7B for which at least two transitions with significantly different $E_{\mathrm{u}}$ were detected (a minimum $E_{\mathrm{u}}$ difference of 15~K was chosen to avoid making rotational diagram fits with large errors). This includes H$_2$CO, D$_2$CO, H$_2^{13}$CO, CH$_3$OH, CH$_3$CCH, H$_2$CS, SO, SO$_2$, CS, C$^{34}$S, CN, DCN, HC$_3$N, C$_2$H, \textit{c}-C$_3$H$_2$, and CH$_3$CHO. To establish approximate column densities, we also performed fits for H$_2$CCO, HNCO, and CH$_3$CN assuming a fixed temperature of $30\pm5$~K. We also fixed the rotational temperature to $30\pm5$~K for CH$_3$CHO because the three detected lines are relatively faint. The temperature 30~K was chosen to match the rotational temperatures measured for other organic species. We excluded CO and its isotopologues because their lines are probably optically thick. We used the CDMS \citep{cdms} and JPL \citep{jpl} databases to retrieve molecular spectroscopy data.

From previous interferometric observations we know that the line emission in the region is very extended in comparison with the APEX and ASTE beams \citep{lindberg12}, and we therefore calculated beam-averaged column densities. This requires the assumption that the column densities are relatively uniform across the range of beam sizes. Since the 0.9~mm and 0.8~mm band APEX observations of IRS7B partially coincide with the ASTE survey, we tested this assumption by comparing line strengths of lines covered in both surveys. We find that lines of H$_2$CO, H$_2$CS, HNC, DCN, and most CN lines are within errors between the two measurement sets, but lines of C$^{17}$O, C$^{34}$S, and CH$_3$OH are 1.5--2.3 times stronger in the ASTE data. This might be explained by the different spatial distribution of these molecules in combination with the difference in the APEX and ASTE beam sizes -- the interferometric observations of \citet{lindberg12} show that the CH$_3$OH emission peaks $\sim15\arcsec$ south of IRS7B, and this emission probably contributes more to the signal in the larger ASTE beam than in the APEX beam. When lines were available from both datasets, the ASTE data were used instead of the APEX data because the S/N ratio is higher in the ASTE data and because the APEX 1.3~mm beam size is more similar to the ASTE 0.8--0.9~mm beam sizes than the APEX 0.8--0.9~mm beam sizes. Three H$_2$CO lines detected in the APEX 0.8~mm observations with frequencies just above the upper boundary of the ASTE survey (364~GHz) were included, however. To account for the calibration uncertainty of the observations, we introduced a $10\%$ calibration error in addition to the rms noise.

For the H$_2$CO and H$_2^{13}$CO lines we assumed an ortho-to-para ratio of 1.6 \citep[see][]{dickens99,jorgensen05b}, for D$_2$CO we assumed the statistical value of 2, and for \textit{c}\nobr C$_3$H$_2$, H$_2$CS, and H$_2$CCO we assumed ortho-to-para ratios of 3 \citep{lucas00,minh91,ohishi91}. For the molecules with A and E states (CH$_3$OH, CH$_3$CCH, CH$_3$CN, and CH$_3$CHO), we used the statistical A/E ratios as provided by the CDMS database.

\citet{mangum93} showed that line ratios of H$_2$CO transitions with the same $J_\mathrm{u}$-level are excellent probes of the kinetic temperature of the molecular gas. However, line ratios of transitions involving different $J_\mathrm{u}$-levels also
strongly depend on the density $n(\mathrm{H}_2)$. This is further discussed in Appendix~\ref{app:h2co}. To account for this effect, we calculated rotational temperatures for each $J_\mathrm{u}$-level separately. We then used the weighted average of these two as the H$_2$CO temperature, and found the column density $N(\mathrm{H}_2\mathrm{CO})$ and the molecular density $n(\mathrm{H}_2)$ by comparing the measured $3_{03}\rightarrow2_{02}$ and $5_{05}\rightarrow4_{04}$ line strengths and RADEX estimates of these transitions (see Appendix~\ref{app:h2co} for a thorough description of the method used). The H$_2$ number density towards IRS7B is found to be $(8.9\pm1.0)\times10^5$~cm$^{-3}$.

The same problem probably applies to H$_2^{13}$CO and D$_2$CO, but the low number of detected lines, their low S/N levels, and the lack of collisional data for these species make it difficult to perform fits on the separate $J_\mathrm{u}$-level transitions. The fits to these two isotopologues are thus highly uncertain.

\begin{table}[!tb]
	\centering
	\caption[]{Rotational diagram parameters for IRS7B.}
	\label{tab:rotdiag_params}
	\begin{tabular}{l l l}
		\noalign{\smallskip}
		\hline
		\hline
		\noalign{\smallskip}
		Molecule & $T_{\mathrm{rot}}$ & $N_{\mathrm{rot}}$ \\ 
		& [K] & [cm$^{-2}$] \\ 
		\noalign{\smallskip}
		\hline
		\noalign{\smallskip}
		H$_2$CO\tablefootmark{a} & $40.3\pm2.9$ & $(1.0\pm0.1)\times 10^{14}$ \\ 
		D$_2$CO & $31.2\pm4.2$ & $(1.8\pm0.3)\times 10^{12}$ \\ 
		H$_2^{13}$CO & $13.6\pm1.8$ & $(3.4\pm1.3)\times 10^{12}$ \\ 
		CH$_3$OH & $27.6\pm0.9$ & $(1.6\pm0.1)\times 10^{14}$ \\ 
		CH$_3$CCH & $32.9\pm2.1$ & $(1.1\pm0.2)\times 10^{14}$ \\ 
		H$_2$CS & $23.2\pm2.1$ & $(8.4\pm2.1)\times 10^{12}$ \\ 
		SO & $28.9\pm1.9$ & $(7.1\pm1.1)\times 10^{13}$ \\ 
		SO$_2$ & $38.0\pm2.5$ & $(1.8\pm0.2)\times 10^{13}$ \\ 
		CS & $23.8\pm2.6$ & $(7.3\pm1.8)\times 10^{13}$ \\ 
		C$^{34}$S & $18.5\pm1.9$ & $(6.4\pm1.7)\times 10^{12}$ \\ 
		CN & $\phantom{0}9.9\pm0.3$ & $(1.8\pm0.1)\times 10^{14}$ \\ 
		DCN & $10.9\pm0.9$ & $(2.6\pm0.6)\times 10^{12}$ \\ 
		HC$_3$N & $15.6\pm1.9$ & $(6.1\pm6.6)\times 10^{14}$ \\ 
		C$_2$H & $16.9\pm2.8$ & $(2.8\pm1.2)\times 10^{14}$ \\ 
		$c$-C$_3$H$_2$ & $16.9\pm0.7$ & $(8.5\pm0.9)\times 10^{12}$ \\ 
		HNCO & $30$\tablefootmark{b} & $(5.1\pm1.9)\times 10^{12}$ \\ 
		CH$_3$CN & $30$\tablefootmark{b} & $(8.7\pm3.9)\times 10^{11}$ \\ 
		CH$_3$CHO & $30$\tablefootmark{b} & $(1.6\pm0.7)\times 10^{13}$ \\ 
		\noalign{\smallskip}
		\hline
	\end{tabular}
	\tablefoot{
		\tablefoottext{a}{The H$_2$CO fit is a weighted average of the $J_\mathrm{u}=3$ and $J_\mathrm{u}=5$ fits. The two fits were also used to estimate the H$_2$ number density, $n(\mathrm{H}_2) = (8.9\pm1.0)\times10^5$~cm$^{-3}$.}
		\tablefoottext{b}{For these species, the $E_{\mathrm{u}}$ of the observed lines are in too narrow a range for the rotational temperature to be accurately calculated, or all detected lines are of a very low S/N level. We assumed a temperature of $30\pm5$~K (based on the rotational temperatures of similar species) to be able to estimate the column densities of these molecules.}
	}
\end{table}

\begin{figure*}[!htb]
        \centering
        $\begin{array}{ccc}
        \includegraphics{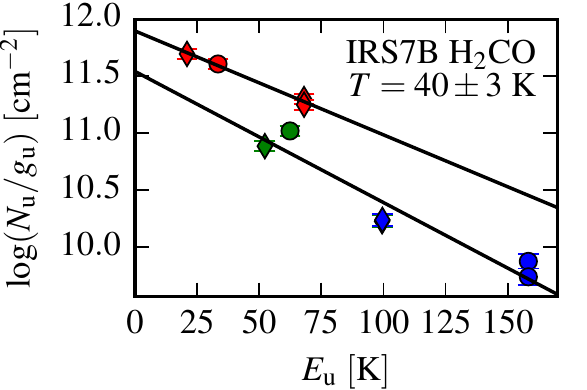} &
        \includegraphics{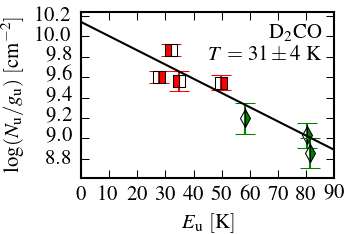} &
        \includegraphics{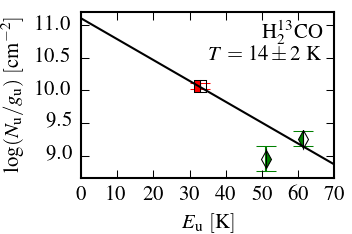} \\
        \includegraphics{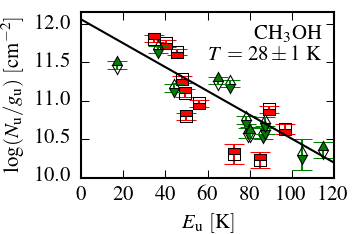} &
        \includegraphics{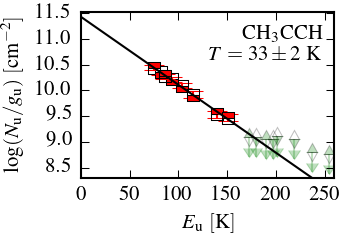} &
        \includegraphics{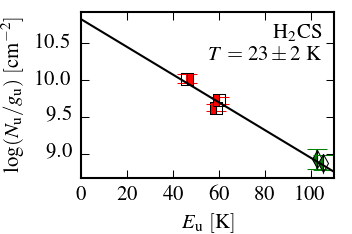} \\
        \includegraphics{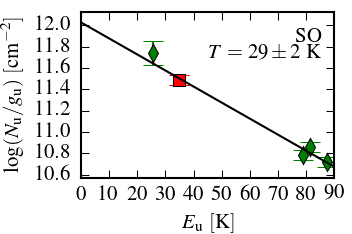} &
        \includegraphics{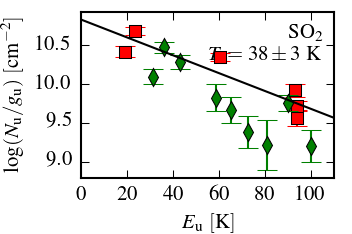} &
        \includegraphics{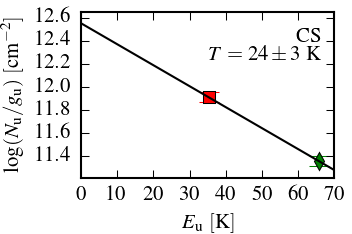} \\
        \includegraphics{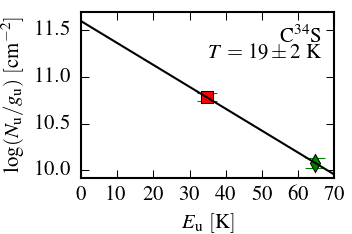} &
        \includegraphics{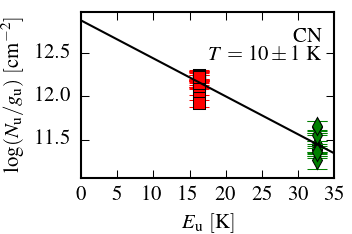} &
        \includegraphics{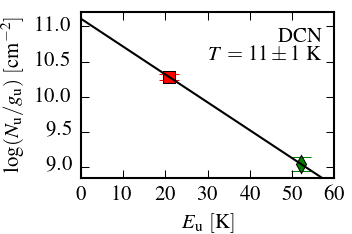} \\
        \includegraphics{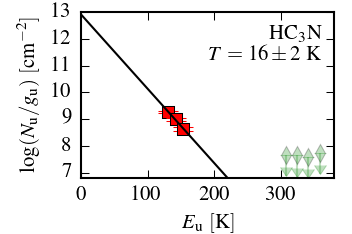} &
        \includegraphics{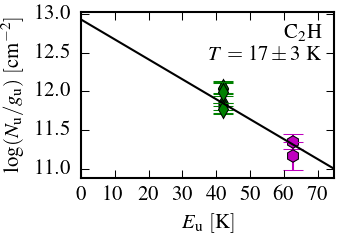} &
        \includegraphics{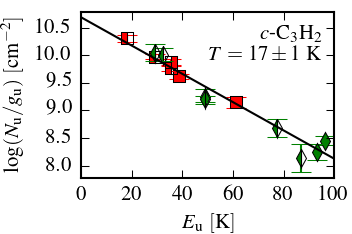} \\
        \end{array}$
        \caption{Rotational diagrams of molecular species detected towards IRS7B. Red square data points are 1.3~mm lines observed with APEX, blue circle data points are 0.8~mm lines observed with APEX (this work); green diamond data points are 0.8~mm lines observed with ASTE, and magenta hexagon data points are 0.7~mm lines observed with ASTE \citep{watanabe12}. Upper limit data points are semi-transparent. For H$_2$CO, H$_2^{13}$CO, D$_2$CO, \textit{c}\nobr C$_3$H$_2$, and H$_2$CS, the ortho lines are filled in the left half, the para lines are filled in the right half, and lines that are blends of ortho transitions and para transitions are completely filled. For CH$_3$OH and CH$_3$CCH, the A lines are filled in the top half and the E lines are filled in the bottom half. (A colour version of this plot is available in the online journal.)}\label{fig:irs7b_rotdiags}
\end{figure*}

The results of all the IRS7B rotational diagram fits are listed in Table~\ref{tab:rotdiag_params}, and the rotational diagrams are shown in Fig.~\ref{fig:irs7b_rotdiags}. The rotational temperatures of the different molecules are found to vary between 10~K and 40~K. From their rotational temperatures, the molecules can be divided into two groups: on one hand the nitrogen-bearing species (i.e. cyanides) and unsaturated hydrocarbon molecules at 10--17~K, and on the other hand, the oxygen-bearing organic molecules and sulphur-bearing species at 19--40~K. H$_2^{13}$CO, however, has a lower rotational temperature, but the fit is made with only three lines, involving transitions with different $J_{\mathrm{u}}$.

The rotational diagram of CH$_3$OH shows a considerable scatter (see Fig.~\ref{fig:irs7b_rotdiags}), which like in the case of H$_2$CO suggests non-LTE conditions. To be able to accurately evaluate the temperature and column density of this species, we employed non-LTE RADEX models. From interferometric observations, we know that H$_2$CO and CH$_3$OH are present in the same spatial regions around IRS7B \citep{lindberg12}, and we therefore assumed the H$_2$ density calculated from the H$_2$CO excitation analysis ($n(\mathrm{H}_2)=8.9\times10^5$~cm$^{-3}$).

Our CH$_3$OH models show that with the expected physical properties, a scatter in the CH$_3$OH rotational diagram similar to what we observe should appear due to non-LTE excitation of the molecule. We performed a least-$\chi^2$ fit with the temperature and column density as free parameters. The errors used in the $\chi^2$ calculations are a combination of the rms error and a $10\%$ calibration uncertainty. We find that the lowest $\chi^2$ is achieved for $T=28$~K and $N(\mathrm{CH}_3\mathrm{OH}) = 2.1\times10^{14}$~cm$^{-2}$. The best-fit temperature and column density are similar to the rotational diagram solution (see below). Figure~\ref{fig:radex_ch3oh} shows a contour plot of the reduced-$\chi^2$ values of fit.

\begin{figure}[!htb]
        \centering
        \includegraphics{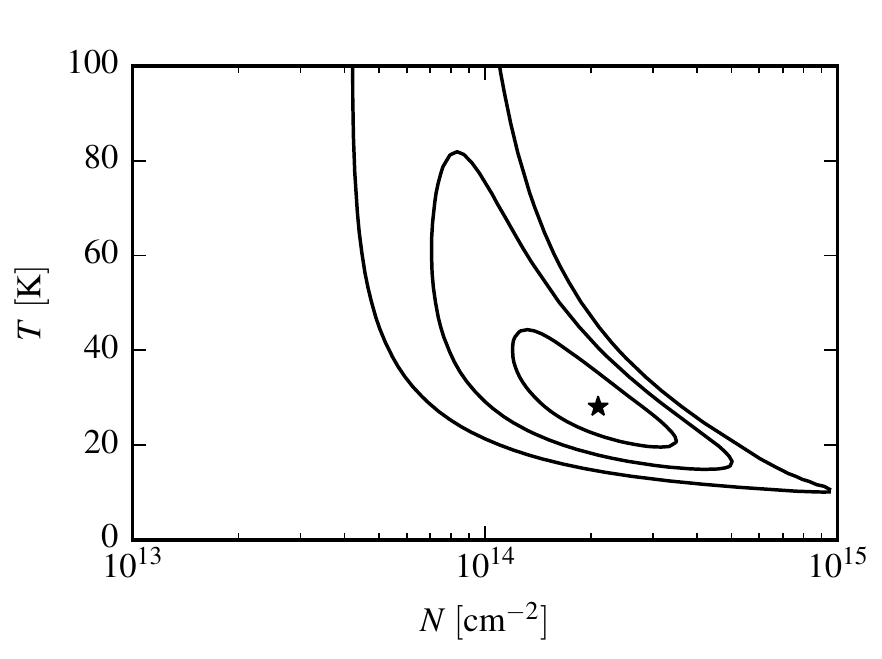}
        \caption{Reduced-$\chi^2$ fit to the CH$_3$OH lines observed towards IRS7B assuming $n(\mathrm{H}_2)=8.9\times10^5$~cm$^{-3}$. The contours indicate $1\sigma$, $2\sigma$, and $3\sigma$ certainty. The star symbol shows the best fit.}\label{fig:radex_ch3oh}
\end{figure}

The median line width of the nitrogen-bearing and hydrocarbon species (1.6~\kms) is slightly lower than the median line width of the organic and sulphur-bearing species (2.1~\kms). The median LSR velocity is not significantly different between the two groups of molecules. The differences in temperature and velocity dispersion suggest that the two groups of molecules have different spatial origins. Interestingly, molecules from the two groups were identified as having two different spatial distributions in interferometric spectral line mapping of the region \citep{lindberg12}: the warmer molecules H$_2$CO and CH$_3$OH were found in two ridges north and south of the YSOs in the R~CrA cloud, whereas the cooler molecules DCN, CN, \textit{c}\nobr C$_3$H$_2$, and HC$_3$N were found to peak in an east-west band around the protostars and/or in a faint outflow extending northwards from IRS7B. CH$_3$CCH does not exhibit the expected properties -- despite being an unsaturated hydrocarbon and having a small velocity dispersion, it does have a rotational temperature similar to the second group of molecules, so its nature in relation to the other species remains unclear.

\subsection{Rotational diagram analysis -- other sources}
\label{sec:rot_sosu}

Rotational diagrams of H$_2$CO, CH$_3$OH, and \textit{c}\nobr C$_3$H$_2$ were produced for the sources included in the CrA survey where at least two lines of the respective molecule were detected. For sources with only one detected line of a given species, upper limits of the rotational temperature and lower limits of the column density were calculated. As above, we assumed ortho-to-para ratios of 1.6 for H$_2$CO and 3 for \textit{c}\nobr C$_3$H$_2$, and an A/E ratio of 1 for CH$_3$OH. For the IRS7B rotational diagrams in this section we used only the APEX observations, and only those parts of the IRS7B survey that are also covered by the source survey, to avoid any bias between IRS7B and the other sources, even though more lines are available for IRS7B in the full line survey and the ASTE data. Thus, the rotational diagram parameters for IRS7B reported in this section are not identical to those in the previous section. We used the same calibration uncertainty of $10\%$ as in the rotational diagrams for IRS7B.

Since no mm line interferometry data are available for most of the sources in the sample, we cannot determine the spatial extent of the molecular line emission. We therefore used the same approach as for IRS7B, calculating the beam-averaged column densities.

As discussed in Sect.~\ref{sec:rotdiag_irs7b}, H$_2$CO rotational diagrams must treat transitions involving different $J_\mathrm{u}$-levels separately. In the sources where lines from both the $J_\mathrm{u}=3$ and $J_\mathrm{u}=5$ levels are detected, we thus calculated two rotational temperatures and used the weighted average of these. We then estimated the column density and the number density with the method described in Appendix~\ref{app:h2co}. We find that towards the sources where we could measure the H$_2$ density, it is between $4\times10^5$~cm$^{-3}$ and $10^6$~cm$^{-3}$, except for towards the outflow component of SMM~2, where the density is lower. It is difficult to constrain the densities and column densities for the outflow components of SMM~2 and CrA-24, since the densities are low and the column densities are high, which enhances both the optical depth effects and the non-LTE effects.

\begin{figure*}[!htb]
        \centering
    $\begin{array}{ccc}
    \includegraphics{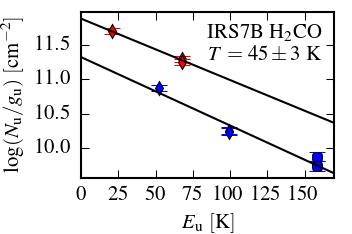} &
    \includegraphics{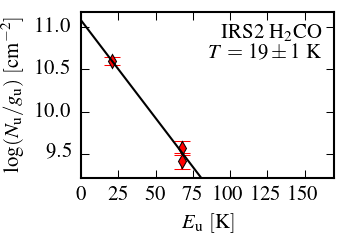} &
    \includegraphics{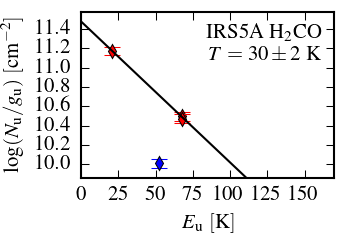} \\
    \includegraphics{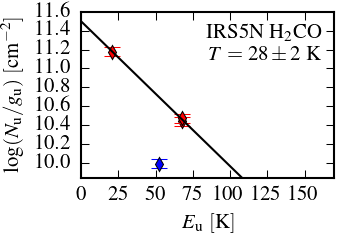} &
    \includegraphics{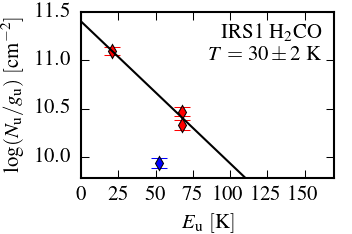} &
    \includegraphics{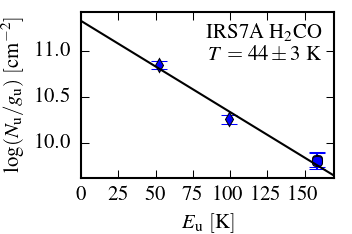} \\
    \includegraphics{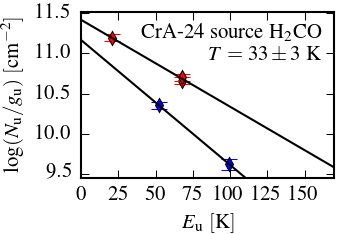} &
    \includegraphics{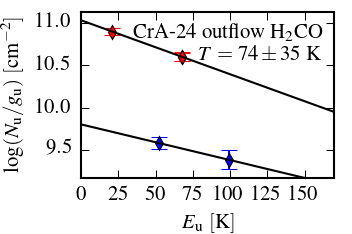} &
    \includegraphics{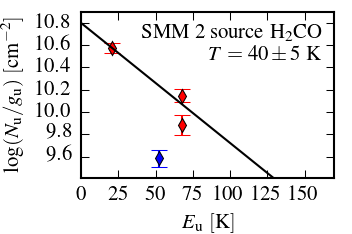} \\
    \includegraphics{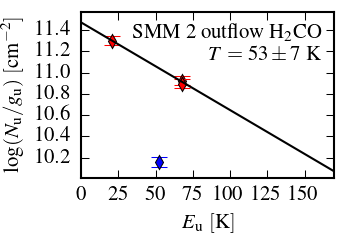} &
    \includegraphics{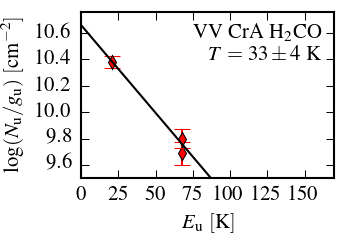} \\
    \end{array}$
    \caption{H$_2$CO rotational diagrams of the sources in the CrA survey where at least two H$_2$CO lines were detected. The ortho lines are filled in the left half and the para lines are filled in the right half. Red square data points are 1.3~mm lines and blue circle data points are 0.8~mm lines, all observed with APEX. (A colour version of this plot is available in the online journal.)}\label{fig:h2co_rotdiags}
\end{figure*}

\begin{figure*}[!htb]
        \centering
    $\begin{array}{ccc}
    \includegraphics{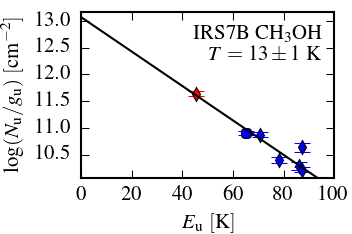} &
    \includegraphics{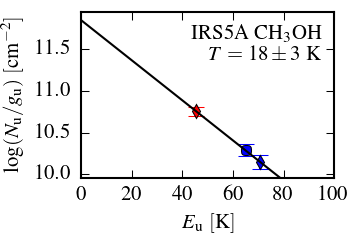} &
    \includegraphics{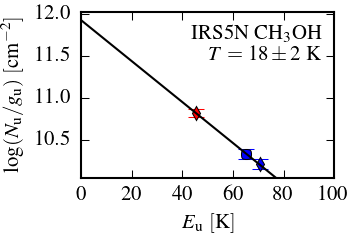} \\
    \includegraphics{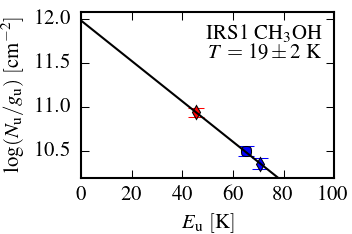} &
    \includegraphics{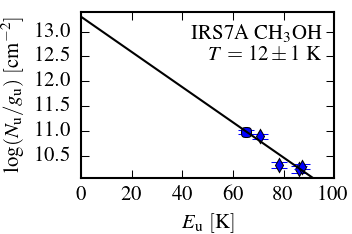} &
    \includegraphics{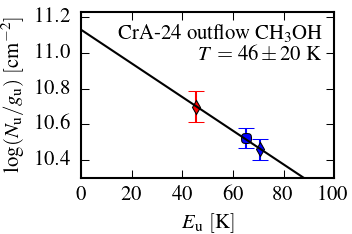} \\
    \includegraphics{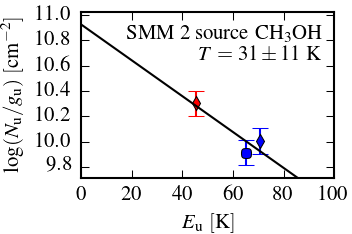} &
    \includegraphics{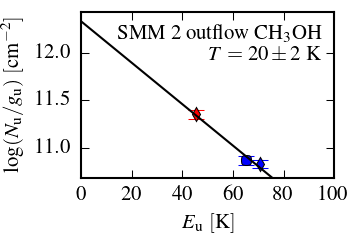} \\
    \end{array}$
    \caption{CH$_3$OH rotational diagrams of the sources in the CrA survey where at least two CH$_3$OH lines were detected and a positive rotational temperature could be fitted. The A lines are filled in the top half and the E lines are filled in the bottom half. Red square data points are 1.3~mm lines and blue circle data points are 0.9~mm lines, all observed with APEX. (A colour version of this plot is available in the online journal.)}\label{fig:ch3oh_rotdiags}
\end{figure*}

\begin{figure*}[!htb]
        \centering
    $\begin{array}{ccc}
    \includegraphics{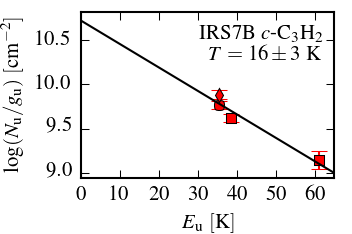} &
    \includegraphics{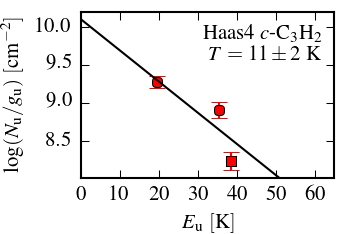} &
    \includegraphics{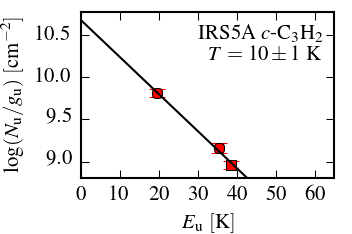} \\
    \includegraphics{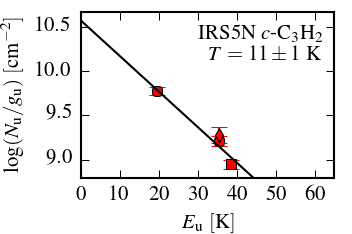} &
    \includegraphics{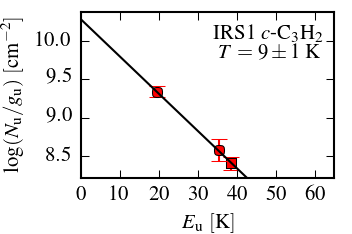} &
    \includegraphics{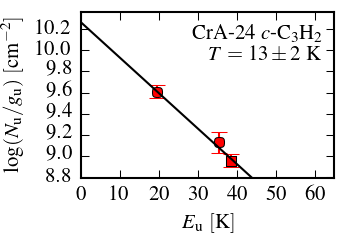} \\
    \includegraphics{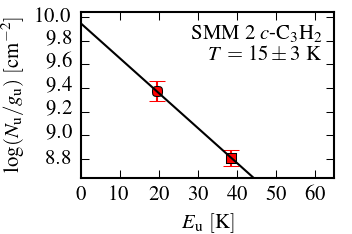} &
    \includegraphics{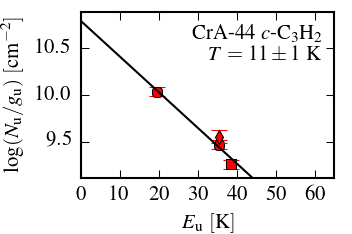} &
    \includegraphics{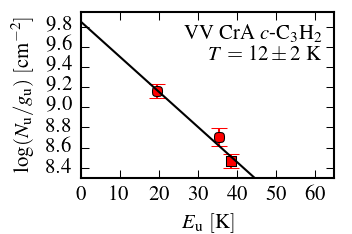} \\
    \end{array}$
    \caption{\textit{c}\nobr C$_3$H$_2$ rotational diagrams of the sources in the CrA survey where at least two \textit{c}\nobr C$_3$H$_2$ lines were detected. The ortho lines are filled in the left half, the para lines are filled in the right half, and the lines that are blends of ortho transitions and para transitions are completely filled. All data points are 1.3~mm lines observed with APEX. (A colour version of this plot is available in the online journal.)}\label{fig:c-c3h2_rotdiags}
\end{figure*}

\begin{table*}
\centering
\caption[]{Rotational diagram parameters for H$_2$CO in the survey of CrA sources. Sources where the fit could not be performed are excluded. The H$_2$ densities $n$ and column densities $N$ were measured by the method described in Appendix~\ref{app:h2co}, which can only be made for the sources where both the $3_{03}\rightarrow2_{02}$ and the $5_{05}\rightarrow4_{04}$ lines have been detected. To facilitate comparison between IRS7B and the other sources, only the lines covered by the source survey are included, which explains why IRS7B shows different values here and in Table~\ref{tab:rotdiag_params}.}
\label{tab:rotdiag_h2co}
\begin{tabular}{l l l l l l}
\noalign{\smallskip}
\hline
\hline
\noalign{\smallskip}
Source name & $T_{J_\mathrm{u}=3}$\tablefootmark{a} & $T_{J_\mathrm{u}=5}$\tablefootmark{b} & $T_{\mathrm{av}}$\tablefootmark{c} & $n$ & $N$\tablefootmark{d} \\
& [K] & [K] & [K] & [cm$^{-3}$] & [cm$^{-2}$] \\
\noalign{\smallskip}
\hline
\noalign{\smallskip}
IRS7B & \phantom{< }$48.8\pm\phantom{0}6.2$ & $\phantom{0}43.8\pm\phantom{0}2.7$ & $45.3\pm\phantom{0}3.4$ &  \phantom{< }$7.7\times10^5$ &  \phantom{< }$1.0\times10^{14}$ \\ 
Haas4 & $< 24.3$ & \phantom{0}... & ... &  \phantom{< }... &  \phantom{< }... \\ 
IRS2 & \phantom{< }$18.9\pm\phantom{0}1.3$ & \phantom{0}... & $18.9\pm\phantom{0}1.3$ &  \phantom{< }... &  \phantom{< }... \\ 
IRS5A & \phantom{< }$29.8\pm\phantom{0}2.4$ & \phantom{0}... & $29.8\pm\phantom{0}2.4$ &  \phantom{< }$4.0\times10^5$ &  \phantom{< }$3.7\times10^{13}$ \\ 
IRS5N & \phantom{< }$28.4\pm\phantom{0}2.1$ & \phantom{0}... & $28.4\pm\phantom{0}2.1$ &  \phantom{< }$3.9\times10^5$ &  \phantom{< }$3.9\times10^{13}$ \\ 
IRS1 & \phantom{< }$29.5\pm\phantom{0}2.4$ & \phantom{0}... & $29.5\pm\phantom{0}2.4$ &  \phantom{< }$4.6\times10^5$ &  \phantom{< }$2.8\times10^{13}$ \\ 
IRS7A & \phantom{< }... & $\phantom{0}43.6\pm\phantom{0}3.1$ & $43.6\pm\phantom{0}3.1$ &  \phantom{< }... &  \phantom{< }... \\ 
CrA-24 source & \phantom{< }$40.3\pm\phantom{0}4.3$ & $\phantom{0}28.0\pm\phantom{0}3.1$ & $33.2\pm\phantom{0}2.7$ &  \phantom{< }$1.3\times10^6$ &  \phantom{< }$2.6\times10^{13}$ \\ 
CrA-24 outflow & \phantom{< }$68.6\pm12.9$ & $103.7\pm68.1$ & $74.2\pm34.7$ & \phantom{< }...\tablefootmark{e} & \phantom{< }...\tablefootmark{e} \\ 
SMM 2 source & \phantom{< }$40.4\pm\phantom{0}5.3$ & \phantom{0}... & $40.4\pm\phantom{0}5.3$ &  \phantom{< }$6.3\times10^5$ &  \phantom{< }$6.6\times10^{12}$ \\ 
SMM 2 outflow & \phantom{< }$52.6\pm\phantom{0}7.3$ & \phantom{0}... & $52.6\pm\phantom{0}7.3$ &  $<2\phantom{.0}\times10^5$\tablefootmark{e} & $>6\phantom{.0}\times10^{13}$\tablefootmark{e} \\ 
CXO42 & $< 87.5$ & \phantom{0}... & ... & \phantom{< }... & \phantom{< }... \\ 
CrA-44 & $< 21.2$ & \phantom{0}... & ... & \phantom{< }... & \phantom{< }... \\ 
CrA-33 & $< 54.5$ & \phantom{0}... & ... & \phantom{< }... & \phantom{< }... \\ 
VV~CrA & \phantom{< }$32.7\pm\phantom{0}3.7$ & \phantom{0}... & $32.7\pm\phantom{0}3.7$ & \phantom{< }... & \phantom{< }... \\ 
\noalign{\smallskip}
\hline
\end{tabular}
\tablefoot{
        \tablefoottext{a}{Rotational temperatures of the transitions with $J_\mathrm{u}=3$.}
        \tablefoottext{b}{Rotational temperatures of the transitions with $J_\mathrm{u}=5$.}
        \tablefoottext{c}{Weighted average of the two previous, if they are both measured, otherwise the same as the only one measured.}
        \tablefoottext{d}{The total H$_2$CO column density assuming an ortho-to-para ratio of 1.6.}
        \tablefoottext{e}{For these sources, the column density and H$_2$ density could not be constrained from the $3_{03}\rightarrow2_{02}$/$5_{05}\rightarrow4_{04}$ ratio.}
}
\end{table*}

\begin{table*}
        \centering
        \caption[]{Rotational diagram parameters for CH$_3$OH and \textit{c}\nobr C$_3$H$_2$ in the survey of CrA sources. Sources where neither of the fits could be performed are excluded. To facilitate comparison between IRS7B and the other sources, only the lines covered by the source survey are included, which explains why IRS7B shows different values here and in Table~\ref{tab:rotdiag_params}.}
        \label{tab:rotdiag_cra}
        \begin{tabular}{l l l l l}
                \noalign{\smallskip}
                \hline
                \hline
                \noalign{\smallskip}
                Source name & \multicolumn{2}{c}{CH$_3$OH} & \multicolumn{2}{c}{\textit{c}\nobr C$_3$H$_2$} \\
                & $T$ & $N$\tablefootmark{a} & $T$ & $N$\tablefootmark{b} \\
                & [K] & [cm$^{-2}$] & [K] & [cm$^{-2}$] \\
                \noalign{\smallskip}
                \hline
                \noalign{\smallskip}
                IRS7B &  \phantom{< }$13.5\pm\phantom{0}0.6$ & \phantom{<}$(4.5\pm1.0)\times 10^{14}$ & \phantom{< }$16.5\pm2.6$ & \phantom{<}$(8.6\pm3.2)\times 10^{12}$ \\ 
                Haas4 &  \phantom{< }... & \phantom{<}... & \phantom{< }$10.6\pm1.7$ & \phantom{<}$(1.1\pm0.5)\times 10^{12}$ \\ 
                IRS2 & $< 16.2$ & $>1.2\phantom{\pm0.00)}\times 10^{13}$ & \phantom{< }... & \phantom{<}... \\ 
                IRS5A &  \phantom{< }$18.1\pm\phantom{0}2.6$ & \phantom{<}$(4.5\pm2.1)\times 10^{13}$ & \phantom{< }$\phantom{0}9.9\pm0.8$ & \phantom{<}$(3.7\pm0.9)\times 10^{12}$ \\ 
                IRS5N &  \phantom{< }$17.7\pm\phantom{0}2.2$ & \phantom{<}$(5.3\pm2.2)\times 10^{13}$ & \phantom{< }$10.9\pm0.9$ & \phantom{<}$(3.3\pm0.8)\times 10^{12}$ \\ 
                IRS1 &  \phantom{< }$18.9\pm\phantom{0}2.4$ & \phantom{<}$(6.7\pm2.6)\times 10^{13}$ & \phantom{< }$\phantom{0}8.9\pm1.0$ & \phantom{<}$(1.3\pm0.5)\times 10^{12}$ \\ 
                IRS7A &  \phantom{< }$12.2\pm\phantom{0}0.9$ & \phantom{<}$(6.4\pm3.0)\times 10^{14}$ & \phantom{< }... & \phantom{<}... \\ 
                CrA-24 &  \phantom{< }... & \phantom{<}... & \phantom{< }$13.0\pm1.7$ & \phantom{<}$(2.1\pm0.7)\times 10^{12}$ \\ 
                CrA-24 source &  \phantom{< }... & \phantom{<}... & \phantom{< }... & \phantom{<}... \\ 
                CrA-24 outflow &  \phantom{< }$46.3\pm20.2$ & \phantom{<}$(4.4\pm2.7)\times 10^{13}$ & \phantom{< }... & \phantom{<}... \\ 
                SMM 2 &  \phantom{< }... & \phantom{<}... & \phantom{< }$14.7\pm2.9$ & \phantom{<}$(1.2\pm0.5)\times 10^{12}$ \\ 
                SMM 2 source &  \phantom{< }$30.6\pm11.4$ & \phantom{<}$(1.4\pm1.0)\times 10^{13}$ & \phantom{< }... & \phantom{<}... \\ 
                SMM 2 outflow &  \phantom{< }$19.9\pm\phantom{0}2.3$ & \phantom{<}$(1.6\pm0.6)\times 10^{14}$ & \phantom{< }... & \phantom{<}... \\ 
                CrA-44 &  \phantom{< }... & \phantom{<}... & \phantom{< }$11.4\pm0.9$ & \phantom{<}$(6.0\pm1.4)\times 10^{12}$ \\ 
                CrA-33 & $< 14.4$ & $>2.2\phantom{\pm0.00)}\times 10^{13}$ & $< 10.7$ & $>5.8\phantom{\pm0.00)}\times 10^{11}$ \\ 
                VV~CrA &  \phantom{< }... & \phantom{<}... & \phantom{< }$12.5\pm1.8$ & \phantom{<}$(7.8\pm2.9)\times 10^{11}$ \\ 
                \noalign{\smallskip}
                \hline
        \end{tabular}
\tablefoot{
        \tablefoottext{a}{The CH$_3$OH column density assuming an A/E ratio of 1.}
        \tablefoottext{b}{The \textit{c}\nobr C$_3$H$_2$ column density assuming an ortho-to-para ratio of 3.}}

\end{table*}

\begin{figure}[!tb]
        \centering
    $\begin{array}{c}
    \includegraphics{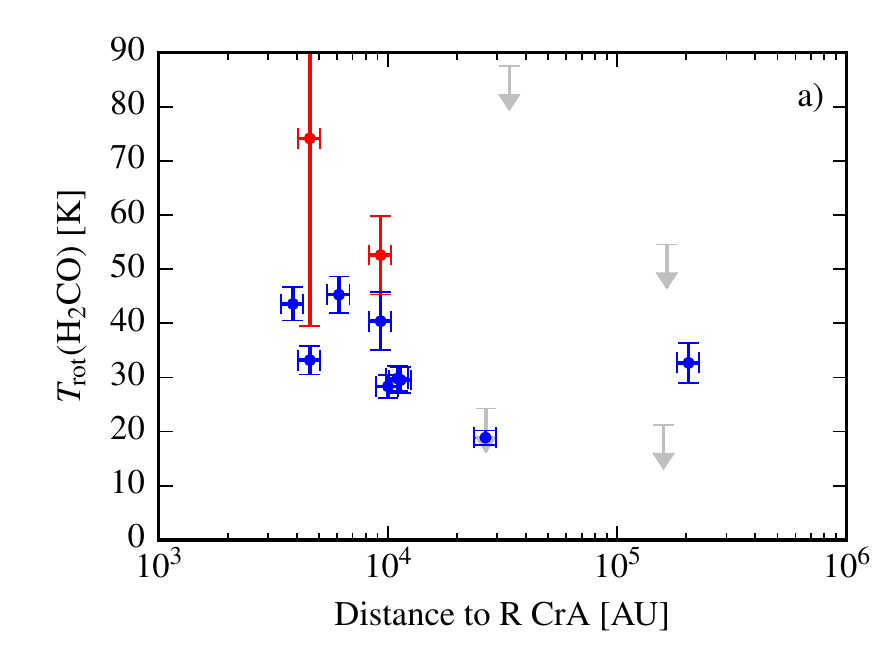} \\
    \includegraphics{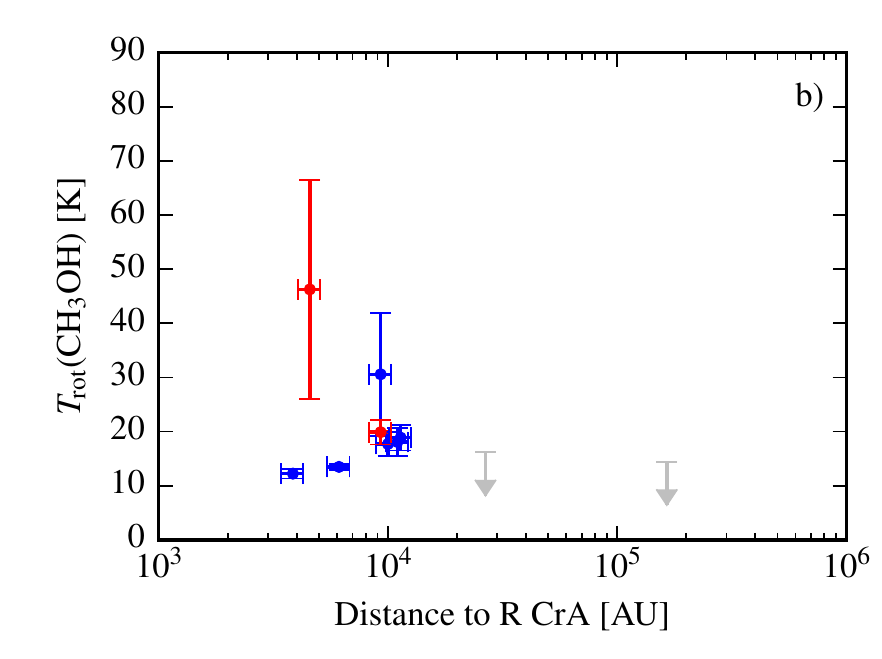} \\
    \includegraphics{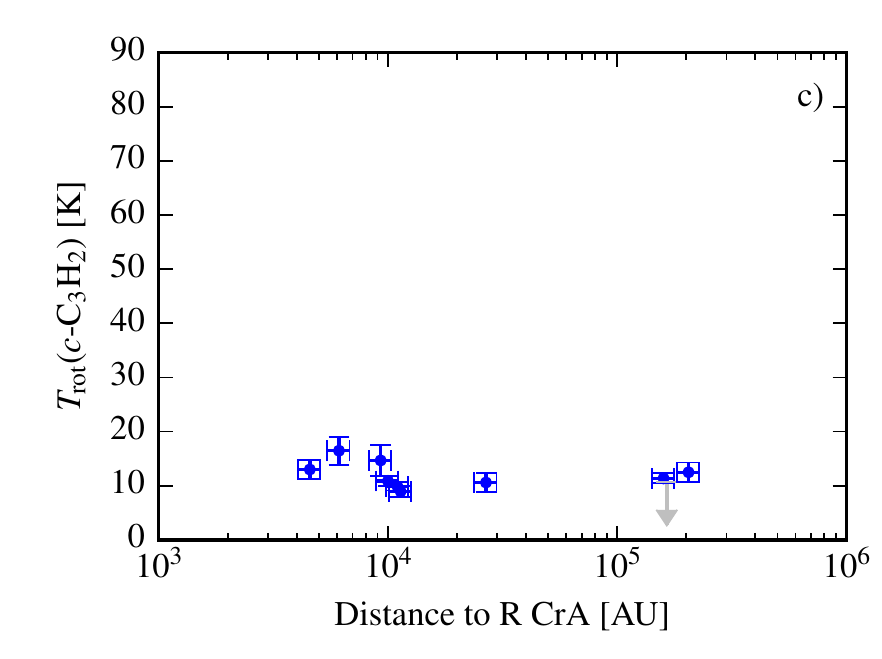} \\
    
    \end{array}$
    \caption{Rotational temperatures for a)~H$_2$CO, b)~CH$_3$OH, and c)~\textit{c}\nobr C$_3$H$_2$ as functions of distance to R~CrA. Upper limits are shown in grey. In the H$_2$CO plot, CrA-24 and SMM~2 have two data points each, one for the on-source component (blue) and one for the outflow component (red), and the same applies to the CH$_3$OH plot when the rotational temperature could be measured. The CH$_3$OH rotational temperature most likely does not reflect a physical temperature due to non-LTE excitation in combination with few data points. (A colour version of this plot is available in the online journal.)}\label{fig:dist_relations}
\end{figure}

The rotational diagrams are shown in Figs.~\ref{fig:h2co_rotdiags}--\ref{fig:c-c3h2_rotdiags} and the fitted parameters are listed in Tables~\ref{tab:rotdiag_h2co}--\ref{tab:rotdiag_cra}. In Fig.~\ref{fig:dist_relations}, the derived H$_2$CO, CH$_3$OH, and \textit{c}\nobr C$_3$H$_2$ rotational temperatures are plotted versus the estimated distance to R~CrA (the statistically most likely distance assuming a uniform three-dimensional spherical cloud with appropriate error bars are used to illustrate the uncertainty of projected distances).
We find that the H$_2$CO temperatures show a large spread throughout the cloud (19--74~K), the CH$_3$OH temperatures show a smaller spread (12--46~K), while the \textit{c}\nobr C$_3$H$_2$ temperatures are uniform around $9$--$16$~K. In the sources CrA-24 and SMM~2, the two velocity components of the H$_2$CO and CH$_3$OH spectral lines (corresponding to on-source and outflow emission, see Sect.~\ref{sec:results_sourcesurvey}) are treated separately. The H$_2$CO temperatures of the outflow components are significantly higher than the H$_2$CO temperatures of the on-source components. This is different for CH$_3$OH. Furthermore, the CH$_3$OH temperatures do not show the same wide range of temperatures as the H$_2$CO, even though these molecules have been shown to co-exist on large scales towards IRS7B \citep{lindberg12}. However, CH$_3$OH observed in low-density outflow regions has been found to be extremely sub-thermally excited \citep{bachiller95}, and the CH$_3$OH rotational temperature is not very sensitive to local increases in kinetic temperature \citep{bachiller98}. As seen in the CH$_3$OH rotational diagram of the IRS7B survey (Fig.~\ref{fig:irs7b_rotdiags}), this molecule shows a large scatter in the rotational diagram due to non-LTE effects. This shows that it is difficult to interpret CH$_3$OH rotational diagrams containing as few as three data points. We therefore choose not to discuss the CH$_3$OH rotational temperatures any further.

The large spread in the H$_2$CO temperatures might have been caused by optical depth effects, in particular for the H$_2$CO $3_{03}\rightarrow2_{02}$ line. We investigated this using the H$_2^{13}$CO $3_{12}\rightarrow2_{11}$ transition, which is detected towards seven of the studied sources. We assumed the ISM value for the $^{12}$C/$^{13}$C ratio ($69\pm6$; \citealt{wilson99}), an ortho-to-para ratio of 1.6, and excitation temperatures derived from the respective H$_2$CO rotational diagrams. For a majority of the sources, we find that this line ratio corresponds to optical depths $\tau$ of the H$_2$CO $3_{03}\rightarrow2_{02}$ line between 0.3 and 0.8. For IRS2 and the SMM~2 on-source component the numbers are higher (1.1 and 2.2), but it should be noted that these numbers are uncertain since the H$_2^{13}$CO $3_{12}\rightarrow2_{11}$ line is at a low S/N-level in these sources. The SMM~2 on-source value may also have errors originating in the separation between the on-source and outflow components. We conclude that the estimated H$_2$CO optical depths are so low that the detected spread in temperatures must be real.

The \textit{c}\nobr C$_3$H$_2$ rotational temperatures are found to be constant at around 10--15~K for all sources. Non-LTE analysis using RADEX \citep{vandertak07} shows that the observed line strengths are consistent with LTE and low optical depths, and we therefore expect that the measured rotational temperatures reflect the kinetic temperature of the \textit{c}\nobr C$_3$H$_2$ gas. The H$_2$CO temperatures are found to be strongly elevated compared to the \textit{c}\nobr C$_3$H$_2$ temperatures ($T(\mathrm{H}_2\mathrm{CO})\sim30$--50~K) in the sources close to R~CrA (see Fig.~\ref{fig:dist_relations}). This is further discussed in Sect.~\ref{sec:discussion}.

\subsection{Molecular abundances in IRS7B and other protostars}
\label{sec:molabund}

\subsubsection{Rotational diagrams}

The only complex molecules detected towards IRS7B are CH$_3$OH, CH$_3$CCH, CH$_3$CN, and CH$_3$CHO, the latter two only tentatively detected. In typical hot corino sources, complex organic molecules are present at comparably high abundance levels. To investigate whether the relatively low number of complex organic molecules detected towards IRS7B is significant, we compared relative abundances (and upper limits for non-detections) of organic molecules in IRS7B with the typical hot corino source IRAS~16293-2422, the WCCC source L1527, the quiescent source B1-b, and the L1157 outflow. To account for the uncertainty and difference in the sizes of the emission regions of the different sources, we normalised the abundances with that of CH$_3$OH. We estimated the relative abundances by calculating the ratio of column densities as calculated from the rotational diagrams (see Table~\ref{tab:rotdiag_params}), thus assuming LTE and optically thin lines. The large uncertainties on some of the column density measurements will, however, introduce large uncertainties in these ratios.

To estimate upper limits on abundances of molecules not detected in IRS7B, we used Splatalogue and the line survey of IRAS~16293-2422 \citep{caux11} to identify the strongest lines in the covered spectral bands. Assuming a temperature of 30~K and a line width of 2~km~s$^{-1}$, we calculated the abundance that would generate a line that would just reach the $3\sigma$ detection limit in the APEX or ASTE observations of the band covering that certain line. In some ambiguous cases, several lines with similar line characteristics were tested, and the one generating the lowest upper limit was used. The lines used were for CH$_3$OCHO the 227.560--227.564~GHz octuplet and for HCOOH the 220.038~GHz line. For CH$_3$OCH$_3$ we found that the 225.599~GHz quadruplet produced the best upper limit in the APEX survey, but the 358.447--358.457~GHz octuplet covered by the ASTE survey \citep{watanabe12} provided an upper limit three times lower, so this line was used instead. It is difficult to calculate upper limits of the abundances of complex organic molecules relative to CH$_3$OH in the CrA source survey because the CH$_3$OH excitation is poorly understood in these sources (see Sect.~\ref{sec:rot_sosu}).

\begin{figure*}[!tb]
        \centering  
        \includegraphics{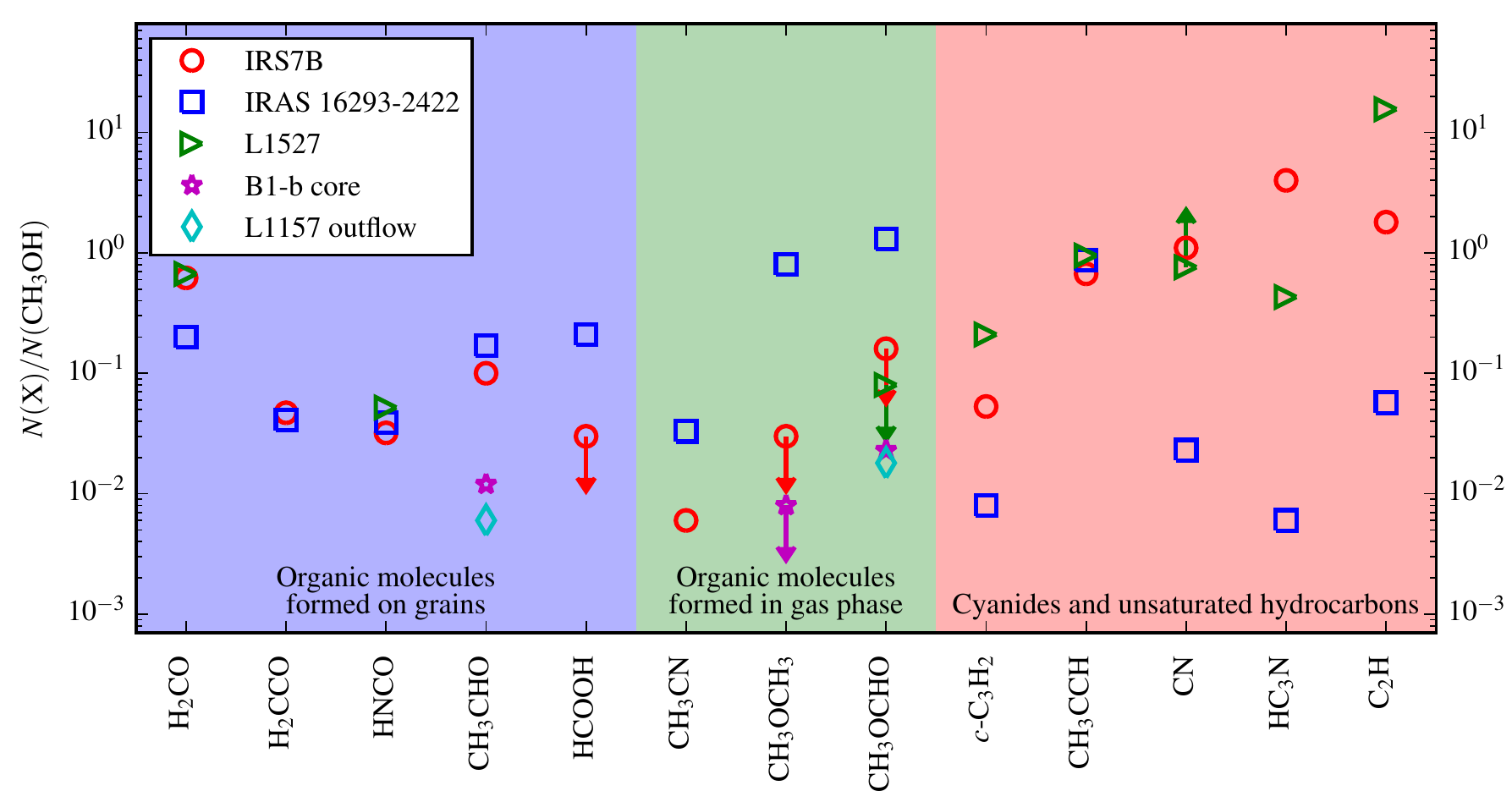}
        \caption{Graphical representation of molecular abundances relative to the CH$_3$OH abundance as given in Table~\ref{tab:abundances_rotdiag} for IRS7B, the hot corino source IRAS~16293-2422, the WCCC source L1527, the quiescent protostar B1-b, and the L1157 outflow. See the notes of Table~\ref{tab:abundances_rotdiag} for references. For a discussion on the classification of gas and grain formation of the organic species, see \citet{charnley97,charnley01}. (A colour version of this plot is available in the online journal.)}\label{fig:abundances}
\end{figure*}

The resulting abundance ratios or upper limits of a selection of molecules measured towards IRS7B, the hot corino IRAS~16293-2422, the WCCC source L1527, the quiescent protostar B1-b, and the L1157 outflow are shown in Table~\ref{tab:abundances_rotdiag} and in Fig.~\ref{fig:abundances}.

\begin{table*}
\centering
\caption[]{Molecular abundances relative to CH$_3$OH ($N$(X)/$N$(CH$_3$OH)) towards IRS7B and other protostellar sources as established from rotational diagrams. See also Fig.~\ref{fig:abundances}.}
\label{tab:abundances_rotdiag}
\begin{tabular}{l r r r r r}
\noalign{\smallskip}
\hline
\hline
\noalign{\smallskip}
Molecule & IRS7B & I16293\tablefootmark{a} & L1527\tablefootmark{b} &  B1-b core\tablefootmark{c} & L1157 outflow\tablefootmark{d} \\ 
\noalign{\smallskip}
\hline
\noalign{\smallskip}
CH$_3$OH & $\equiv1$\phantom{.000} & $\equiv1$\phantom{.000} & $\equiv1$\phantom{.000} & $\equiv1$\phantom{.000} & $\equiv1$\phantom{.000} \\ 
H$_2$CO & 0.62\phantom{0} & 0.20\phantom{0} & 0.67\phantom{0} & ...\phantom{000}  & ...\phantom{000} \\ 
H$_2$CCO & 0.047 & 0.041 & ...\phantom{000} & ...\phantom{000}  & ...\phantom{000} \\
HNCO & 0.032 & 0.039 & 0.052 & ...\phantom{000}  & ...\phantom{000} \\
CH$_3$CHO & 0.10\phantom{0} & 0.17\phantom{0} & ...\phantom{000} & 0.012 & 0.006 \\
HCOOH & $<0.03$\phantom{0} & 0.21\phantom{0} & ...\phantom{000} & ...\phantom{000}  & ...\phantom{000} \\
CH$_3$CN & 0.006 & 0.033 & ...\phantom{000} & ...\phantom{000}  & ...\phantom{000} \\
CH$_3$OCH$_3$ & $<0.03$\phantom{0} & 0.80\phantom{0} & ...\phantom{000} & $<0.008$ & ...\phantom{000} \\
CH$_3$OCHO & $<0.16$\phantom{0} & 1.3\phantom{00} & $<0.08$\phantom{0} & 0.023 & 0.018 \\
\textit{c}\nobr C$_3$H$_2$ & 0.053 & 0.008 & 0.21\phantom{0} & ...\phantom{000}  & ...\phantom{000} \\
CH$_3$CCH & 0.67\phantom{0} & 0.87\phantom{0} & 0.95\phantom{0} & ...\phantom{000}  & ...\phantom{000} \\
CN & 1.1\phantom{00} & 0.023 & $>0.76$\tablefootmark{e} & ...\phantom{000}  & ...\phantom{000} \\
HC$_3$N & $\sim4$\phantom{.000} & 0.006 & 0.43\phantom{0} & ...\phantom{000}  & ...\phantom{000} \\
C$_2$H & 1.8\phantom{00} & 0.057 & 15.6\phantom{00} & ...\phantom{000}  & ...\phantom{000} \\
\noalign{\smallskip}
\hline
\end{tabular}
\tablefoot{
        \tablefoottext{a}{IRAS~16293-2422 \citep{vandishoeck95,schoier02,cazaux03}.}
        \tablefoottext{b}{\citet{sakai13} and references therein. Observations were performed with several different single-dish telescopes with beam sizes between 10\arcsec\ and 30\arcsec. Since the source is extended \citep{sakai10}, we used the beam-averaged column densities.}
        \tablefoottext{c}{\citet{oberg10}.}
        \tablefoottext{d}{\citet{bachiller97,arce08}.}
        \tablefoottext{e}{\citet{agundez08}.}
        }
\end{table*}

We note that several of the organic species are significantly less abundant towards IRS7B compared to IRAS~16293-2422, including CH$_3$CN, CH$_3$OCH$_3$, CH$_3$OCHO, and HCOOH, which makes the chemical parameters of IRS7B more similar to the L1157 molecular outflow and the quiescent source B1-b, although the CH$_3$OH gas is much warmer towards IRS7B ($28\pm1$~K towards IRS7B, $10\pm5$~K towards B1-b, and $12\pm2$~K towards the L1157 outflow). Simpler organic species like H$_2$CO, H$_2$CCO, and HNCO have similar abundances towards IRS7B and IRAS~16293-2422, whereas unsaturated hydrocarbons and cyanides (\textit{c}\nobr C$_3$H$_2$, CN, HC$_3$N, and C$_2$H) are over-abundant towards IRS7B compared to IRAS~16293-2422. The unsaturated hydrocarbons \textit{c}\nobr C$_3$H$_2$ and C$_2$H are, however, under-abundant towards IRS7B compared to the WCCC source L1527, and C$_4$H is non-detected towards IRS7B \citep{sakai09a}. This shows that IRS7B is not a WCCC source. The given upper limits, as well as the values calculated for H$_2$CCO, HNCO, CH$_3$CN, and CH$_3$CHO, were calculated under the assumption that the temperature of these species is 30~K. A lower temperature would increase the ratios for these species. 

Interferometric observations of CH$_3$OH show that this emission is present on large scales around IRS7B, in particular in two extended ridges north and south of IRS7B \citep{lindberg12}, which are partially covered by the APEX and ASTE beams. These beam-averaged abundance ratios therefore probably reflect the composition of that large-scale molecular gas heated by R~CrA \citep{lindberg12}, rather than the compact inner envelope of IRS7B.

\subsubsection{Radiative transfer modelling}

We used the radiative transfer code RATRAN \citep{ratran} to model the line strengths of some of the molecular species detected towards IRS7B. The one-dimensional envelope and temperature model made by \citet{lindberg12} and refined by \citet{lindberg14_alma} was employed. This is an envelope with a power-law density profile ($\rho \sim r^{-1.5}$) heated both by a central object and from the outside by irradiation corresponding to that of the nearby Herbig~Be star R~CrA. We used the RATRAN models to calculate abundances for 13 important species for which most or all lines are expected to be optically thin (see \citealt{jorgensen04b}). The model abundances were kept constant as a function of radius for all species except CN, since the modelled species are not expected to be heavily affected by hot corino emission. On the other hand, we let the abundance of the photo-destruction product CN take the shape of a step function, allowing for a higher CN abundance in the outer envelope, where the gas is more affected by photo-destruction from UV radiation originating in R~CrA.

In addition to the APEX line data of IRS7B reported in this work, we used 
\textit{c}\nobr C$_3$H$_2$, C$^{17}$O, C$^{34}$S, SO, CN, HC$^{18}$O$^+$, and DCN line data from the ASTE survey of IRS7B \citep{watanabe12}.

Molecular data files from the LAMDA database \citep{lamda} were employed when available. In addition, the same molecular datafiles as were used by \citet{jorgensen04b} were used for CN and HC$_3$N, since data for hyperfine lines of CN and high-$J$ transitions of HC$_3$N are not available in the LAMDA database. For isotopologue species where no LAMDA data files are available, the collisional data of the main isotopologue were used, but the line parameters of the isotopologue were acquired from the CDMS database \citep{cdms}. This was done for $^{13}$CS, C$^{34}$S, C$^{33}$S, DCN, and DNC.

The line widths and LSR velocities were estimated from Gaussian fits to the spectral lines. The abundance was left as a free parameter, and the best value was found through a least-$\chi^2$ fit. The error used for the $\chi^2$ calculations is the combination of the rms error and a calibration uncertainty of $10\%$ for both the APEX and the ASTE observations.

Table~\ref{tab:abundances_other} lists the resulting modelled abundances together with data for other low-mass embedded protostellar sources in the literature. The abundance of the optically thick species CO, CS, and HCO$^+$ were calculated from optically thin isotopologues using isotope ratios from the literature: C, O, and $^{34}$S isotope ratios were taken from \citet{wilson99} and $^{33}$S isotope ratios from \citet{chin96}.

\begin{figure}[!htb]
        \centering
        \includegraphics{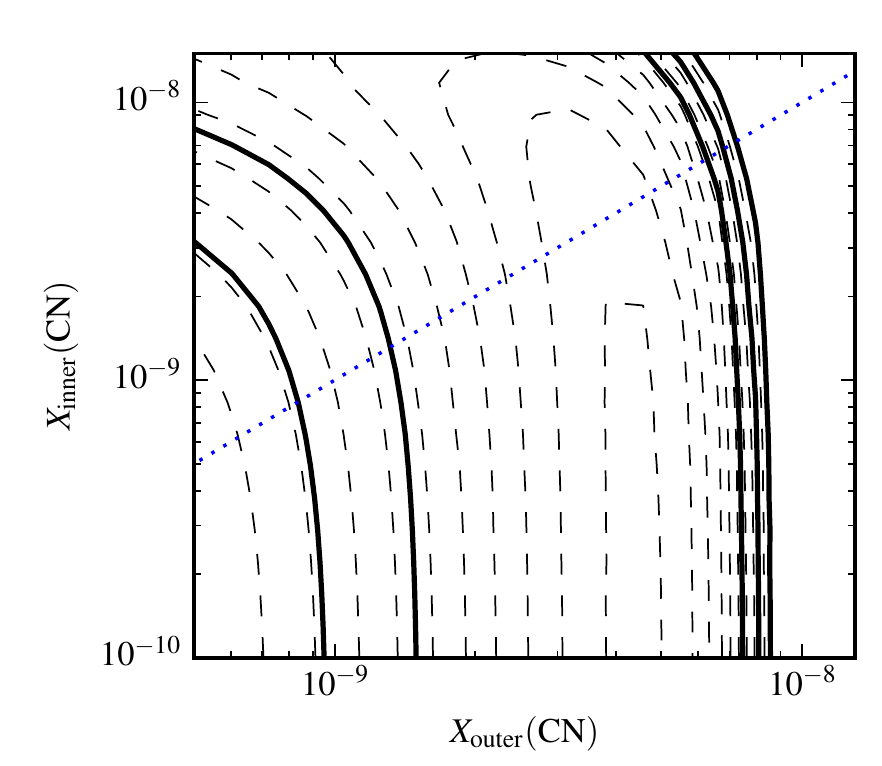}
        \caption{Reduced-$\chi^2$ fit to the CN abundance. The solid lines show the $1\sigma$, $2\sigma$, and $3\sigma$ confidence levels. The dashed lines show the reduced-$\chi^2$ values in steps of 5. The border between inner and outer abundance lies where the temperature falls below 30~K, which occurs at 370~AU. The blue dotted line shows the values for equal inner and outer abundances. (A colour version of this plot is available in the online journal.)}\label{fig:chi2_cn}
\end{figure}

A contour plot of the reduced-$\chi^2$ values as function of the CN abundances is shown in Fig.~\ref{fig:chi2_cn}. The fit clearly is relatively insensitive to the inner abundance ($X_{\mathrm{inner}}\lesssim 10^{-8}$), but fixes the outer abundance to $X_{\mathrm{outer}}\sim \mathrm{few}\times10^{-9}$.

While the IRS7B abundances of many of the more common molecules (CO, HCO$^+$, SO, and CS) are comparable to those in other sources, we find that the CN-bearing species (CN, DCN, DNC, and HC$_3$N) and  \textit{c}\nobr C$_3$H$_2$ are strongly enhanced in IRS7B compared to the other protostellar envelopes. This, in combination with the low upper limits of complex organic molecules discussed above, demonstrates that the chemical properties of the gas around IRS7B strongly differs from those of most other protostars studied before. Previous studies have shown that small hydrocarbon molecules such as \textit{c}\nobr C$_3$H$_2$ and C$_2$H are enhanced in PDR regions. In high-UV flux regions this is a result of gas-phase chemistry \citep{cuadrado15}, but in regions with more moderate UV flux levels it is a consequence of photodestruction of PAHs \citep{montillaud13}. For further references and discussion, see also \citet{guzman15}.

\begin{table*}
        \centering
        \caption[]{Results of RATRAN models of fractional abundances of certain molecules in R~CrA IRS7B and similar modelling results from the literature for other protostellar sources. The reduced-$\chi^2$ values for IRS7B assume a $10\%$ calibration error and can only be computed for species with at least two spectral lines.}
        \label{tab:abundances_other}
        \begin{tabular}{l l r l l l l}
                \noalign{\smallskip}
                \hline
                \hline
                \noalign{\smallskip}
                 & IRS7B & & I16293\tablefootmark{a} & Class 0\tablefootmark{b} & Class I\tablefootmark{b} & Prestellar\tablefootmark{b} \\ 
                Molecule& Abundance & $\chi^2_{\mathrm{red}}$ & Abundance & Abundance & Abundance & Abundance \\ 
                \noalign{\smallskip}
                \hline
                \noalign{\smallskip}
                CO\tablefootmark{c} & $1.4\times10^{-4}$ & ...\phantom{000} & $3.3\times10^{-5}$ & $2.1\times10^{-5}$ & $1.1\times10^{-4}$ & $1.4\times10^{-5}$ \\
                C$^{17}$O & $5.7\times10^{-8}$ & 6.9\phantom{00} & $1.6\times10^{-8}$ & ... & ... & ... \\
                C$^{18}$O & $3.2\times10^{-7}$ & ...\phantom{000} & $6.2\times10^{-8}$ & ... & ... & ... \\
                HCO$^+$\tablefootmark{c} & $3.8\times10^{-9}$ & ...\phantom{000} & $1.4\times10^{-9}$ & $2.2\times10^{-9}$ & $1.1\times10^{-8}$ & $8.0\times10^{-10}$ \\
                HC$^{18}$O$^+$ & $6.8\times10^{-12}$ & 3.6\phantom{00} & $6.4\times10^{-12}$ & ... & ... & ... \\
                SO & $1.9\times10^{-9}$ & 33.6\phantom{00} & $4.4\times10^{-9}$ & $2.0\times10^{-9}$ & $2.3\times10^{-9}$ & $1.5\times10^{-9}$ \\ 
                CS\tablefootmark{c} & $6.7\times10^{-9}$ & ...\phantom{000} & $3.0\times10^{-9}$ & $1.5\times10^{-9}$ & $4.3\times10^{-9}$ & $1.8\times10^{-9}$ \\
                $^{13}$CS & $8.8\times10^{-11}$ & ...\phantom{000} & ... & ... & ... & ... \\ 
                C$^{34}$S & $2.1\times10^{-10}$ & 16.2\phantom{00} & $1.2\times10^{-10}$ & ... & ... & ... \\ 
                C$^{33}$S & $6.8\times10^{-11}$ & ...\phantom{000} & ...  & ... & ... & ...\\ 
                CN outer\tablefootmark{d} & $4.5\times10^{-9}$ & \multirow{2}{*}{8.5\phantom{00}} & \multirow{2}{*}{$8.0\times10^{-11}$} & \multirow{2}{*}{$6.9\times10^{-10}$} & \multirow{2}{*}{$3.0\times10^{-9}$} & \multirow{2}{*}{$4.8\times10^{-9}$} \\ 
                CN inner\tablefootmark{d} & $<1\times10^{-9}$ \\
                DCN & $7.2\times10^{-11}$ & 14.1\phantom{00} & $1.3\times10^{-11}$ & $1.4\times10^{-11}$ & ... & $7.7\times10^{-12}$ \\ 
                DNC & $6.8\times10^{-11}$ & ...\phantom{000} & $4.2\times10^{-12}$ & ... & ... & ... \\
                HNCO & $1.2\times10^{-10}$ & 0.002 & $1.3\times10^{-10}$ & ... & ... & ... \\
                HC$_3$N & $3.3\times10^{-9}$ & 11.2\phantom{00} & $1.5\times10^{-10}$ & $3.5\times10^{-10}$ & $3.1\times10^{-9}$ & $8.2\times10^{-10}$ \\
                \textit{c}\nobr C$_3$H$_2$ & $3.0\times10^{-10}$ & 12.9\phantom{00} & $1.6\times10^{-11}$ & ... & ... & ... \\
                \noalign{\smallskip}
                \hline
        \end{tabular}
        \tablefoot{
                \tablefoottext{a}{IRAS 16293-2422, \citet{schoier02}.}
                \tablefoottext{b}{Average of sources in \citet{jorgensen04b}.}
                \tablefoottext{c}{The abundances for these optically thick species have been calculated from the optically thin isotopologues assuming local ISM isotope ratios \citep{wilson99,chin96}.}
                \tablefoottext{d}{The CN fit was performed with a step in the abundance profile at 370~AU (where $T\approx30$~K).}
        }
\end{table*}

\subsection{Isotopic fractionation in CrA}

The [C$^{18}$O]/[C$^{17}$O] abundance ratio in IRS7B is found to be 5.6, considerably higher than the expected value in the local ISM, 3.6 \citep{wilson99}. It should, however, be noted that the C$^{17}$O fit includes both the 3--2 and 2--1 lines, whereas the C$^{18}$O fit only includes the 2--1 line. The uncertainty of the large-scale CO distribution and the differing beam sizes could contribute to the unusual isotope ratio, and if the ratio is calculated only with the 2--1 lines, a [C$^{18}$O]/[C$^{17}$O] ratio of 4.0 is reached. The [$^{34}$CS]/[C$^{13}$S] ratio 2.4 is slightly lower than the local ISM value 2.9, and the [$^{34}$CS]/[$^{33}$CS] ratio 3.1 is lower than the local ISM value $6.3\pm1.0$.

We also investigated the D/H ratio of H$_2$CO towards IRS7B. As mentioned in Sect.~\ref{sec:rotdiag_irs7b}, the excitation conditions of the H$_2$CO isotopologues is somewhat uncertain. Since the D$_2$CO rotational diagram fit is much better than the H$_2^{13}$CO fit, we applied the D$_2$CO rotational temperature to the H$_2^{13}$CO excitation diagram to acquire a better estimate on the H$_2^{13}$CO column density. Assuming the local ISM value for the $^{12}$C/$^{13}$C ratio ($69\pm6$, \citealt{wilson99}), we estimated the [D$_2$CO]/[H$_2$CO] ratio towards IRS7B to $0.018\pm0.003$. The calculated ratio is within the errors of the value of \citet{watanabe12}, who calculated the ratio for temperatures fixed at 15~K, 20~K, and 25~K. Their [D$_2$CO]/[H$_2$CO] ratio for 25~K is $0.021\pm0.003$. They also observed HDCO, and calculated the [HDCO]/[H$_2$CO] ratio to $0.052\pm0.008$ assuming the same temperature. We compared these values with [D$_2$CO]/[H$_2$CO] and [HDCO]/[H$_2$CO] ratios in other sources. Although the [D$_2$CO]/[H$_2$CO] value is elevated in comparison with what was found in Orion~KL ([D$_2$CO]/[H$_2$CO] = 0.003; \citealt{turner90}), the level of deuteration is lower than what is found in typical hot corino sources and other Class~0 protostars (\citealt{parise06}; e.g. [D$_2$CO]/[H$_2$CO] = $0.05\pm0.008$ in IRAS~16293-2422). This could suggest that grain-surface chemistry is less important for the formation of H$_2$CO in IRS7B than in other low-mass protostellar sources, but the lower level of deuteration could also be an effect of the higher large-scale temperature due to the external irradiation. However, while the [D$_2$CO]/[H$_2$CO] and [HDCO]/[H$_2$CO] ratios are biased by the D/H ratio in the precursor molecules, the incremental fractionation ratio [HDCO]/[H$_2$CO]:[D$_2$CO]/[HDCO] only depends on the formation mechanisms for H$_2$CO \citep{turner90,rodgers02}. The incremental fractionation ratio of H$_2$CO in IRS7B is $0.15\pm0.04$, 
assuming the [D$_2$CO]/[H$_2$CO] from this work and the [HDCO]/[H$_2$CO] ratio of \citet{watanabe12} at 25~K. This is a low value, but still within the errors of the hot corino values \citep{parise06}, so we cannot draw any clear conclusions as to whether there is a difference in the formation path of H$_2$CO in IRS7B versus hot corinos or not.

\citet{watanabe12} also measured the DCN/HCN ratio, which was found to be $0.010\pm0.001$,
similar to the IRAS~16293-2422 value \citep{vandishoeck95}.

We detect $^{13}$CN and C$^{15}$N towards several of the sources in CrA. In most sources, the $^{12}$CN lines are only observed in the $3\rightarrow2$ rotational transitions, while the isotopologue species are observed in the $2\rightarrow1$ rotational transitions. The uncertainty in the excitation conditions makes the estimation of isotope ratios difficult. For IRS7B, we have $2\rightarrow1$ observations also for the $^{12}$CN lines, and (excluding the strongest $^{12}$CN lines, which are expected to be optically thick) we find a CN/C$^{15}$N value of $229\pm50$, which is within the errors of the $^{14}$N/$^{15}$N ratios calculated from IRS7B observations using  H$^{13}$CN/HC$^{15}$N ratios ($287\pm36$) and HN$^{13}$C/H$^{15}$NC ratios ($259\pm34$) by \citet{wampfler14}.

Faint $^{34}$SO lines are detected towards IRS7A (but not IRS7B). The resulting [SO]/[$^{34}$SO] ratio is found to be $11\pm10$, which is almost within the errors of the ISM value $\sim22$ \citep{wilson99}.

\section{Discussion}
\label{sec:discussion}

\begin{figure}[!tb]
        \centering  
        \includegraphics{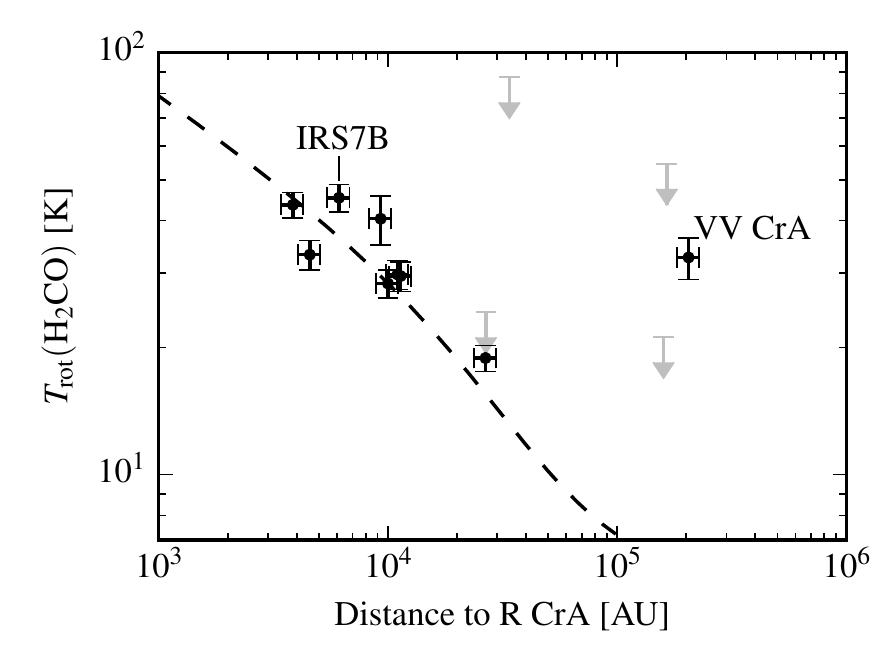}
        \caption{1-D radiative transfer model of heating from R~CrA (dashed line) plotted together with the H$_2$CO rotational temperatures measured in the CrA sources as a function of distance to R~CrA. For CrA-24 and SMM~2, only the on-source components are shown. Upper limits are shown in grey. See also Fig.~\ref{fig:dist_relations}a.}\label{fig:h2co_model}
\end{figure}

In Sect.~\ref{sec:rot_sosu}, the H$_2$CO rotational temperatures in the survey of CrA sources were found to be highest close to the Herbig~Be star R~CrA (Fig.~\ref{fig:dist_relations}a). This suggests that the sphere of thermal influence of R~CrA \citep[see][]{adams85} stretches out to radii greater than 10\,000~AU. To study if this is plausible, radiative transfer modelling was employed. Using the dust radiative transfer code \textit{Transphere} \citep{transphere}, we constructed a one-dimensional model of the molecular cloud around R~CrA. The density was assumed to be constant at $n = 10^4$~cm$^{-3}$ throughout the cloud, starting at a distance of 200~AU from R~CrA. The luminosity of R~CrA was set to $144~L_{\odot}$ (see \citealt{lindberg14_herschel} for a thorough discussion of the luminosity of R~CrA). For this configuration, \textit{Transphere} calculates the temperature in the region self-consistently as a function of radius. The model is shown compared to the data in Fig.~\ref{fig:h2co_model}, and we conclude that it provides a good fit to the H$_2$CO rotational temperature data points within a radius of $\sim30\,000$~AU from R~CrA. We attribute the deviations outside the error bars to irregularities in the distribution of gas in the molecular cloud, with the exception for the source VV~CrA.

The protostar VV~CrA has a projected distance of more than 0.5~pc (100\,000~AU) to R~CrA, but still has a comparably high H$_2$CO rotational temperature ($33\pm4$~K). This cannot be explained by the irradiation from R~CrA. However, VV~CrA is a fairly luminous protostar, with a higher flux at 5.8\micron\ than R~CrA \citep{peterson11}. Using 2MASS, Spitzer, IRAS, PACS, and SCUBA data of VV~CrA \citep[][and references therein]{sicilia13}, we estimate the bolometric luminosity of the source to $13 L_{\odot}$, which is not sufficient to explain the measured H$_2$CO temperature. However, by the use of the Robitaille database of SED models for YSOs \citep{robitaille06} and their
online\footnote{\url{http://caravan.astro.wisc.edu/protostars/}} fitting tool
\citep{robitaille07}, we find that the best matches to the SED are protostellar models with luminosities $\gtrsim40L_{\odot}$. Such a luminosity can heat the gas to 30~K on scales of several 1000~AU, and the elevated temperature in the envelope of VV~CrA could thus be attributed to internal irradiation from VV~CrA itself. It should, however, be noted that the SED models of \citet{robitaille06} work best with protostars at later stages of evolution.

The \textit{c}\nobr C$_3$H$_2$ rotational temperatures measured in the source survey are much lower than the H$_2$CO temperatures and independent of the distance to R~CrA. This can be explained by the \textit{c}\nobr C$_3$H$_2$ gas being more compact than the H$_2$CO gas, and thus more shielded against external irradiation. SMA interferometry of the region around IRS7B shows \textit{c}\nobr C$_3$H$_2$ emission that is more concentrated towards the low-mass YSOs than the H$_2$CO emission is, which supports this hypothesis \citep{lindberg12}, but H$_2$CO and \textit{c}\nobr C$_3$H$_2$ mapping of additional sources is required to confirm this explanation.

We also investigated the chemical composition of the molecular gas detected towards IRS7B. A simple comparison of spectra from three different star-forming regions -- a hot core, a hot corino, and IRS7B (Fig.~\ref{fig:comparison}) -- suggests that the chemistry of the low-mass hot corino source IRAS~16293\nobr 2422 has more in common with the high-mass hot core G327.3-0.6 than with the low-mass protostar IRS7B. Towards IRS7B, the PDR-tracing CN lines dominate the spectrum completely, whereas the two other sources have spectra dominated by various complex organic molecules. The only complex molecules detected in IRS7B are CH$_3$OH, CH$_3$CCH, CH$_3$CN, and CH$_3$CHO (the latter two only tentatively detected). CH$_3$OCH$_3$ and CH$_3$CN are non-detected also in high-resolution ALMA observations of the inner envelope of IRS7B \citep{lindberg14_alma}.
We also find that the relative abundances of several complex organic molecules in IRS7B are significantly lower than what is found in typical hot corino sources. However, simple cyanides (CN and HCN) and the unsaturated hydrocarbons \textit{c}\nobr C$_3$H$_2$ and HC$_3$N are over-abundant in IRS7B. This makes IRS7B similar to the warm carbon-chain chemistry sources, but considering that the C$_2$H abundance is an order of magnitude higher in L1527 than in IRS7B and that a very low upper limit of the C$_4$H abundance in IRS7B was measured by \citet{sakai09a}, IRS7B is not likely to be a typical WCCC source. Rather, the detected molecular abundances reflect a PDR-like chemistry resulting from the strong irradiation onto the molecular envelope from R~CrA.

The CH$_3$OH abundance observed towards IRS7B is only around 3--$6\times10^{-9}$ (we calculated this indirectly using the rotational diagram column density of CH$_3$OH and the rotational diagram column densities and RATRAN abundances of HNCO and \textit{c}\nobr C$_3$H$_2$). This is very low when compared to hot corino abundances ($3\times10^{-7}$ in IRAS~16293-2422; \citealt{cazaux03}) and could be explained by the elevated temperature in the region, which inhibits grain formation of CH$_3$OH due to evaporation of both CO and H atoms. The low abundances of CH$_3$OCHO and CH$_3$OCH$_3$ relative to CH$_3$OH can be explained if we assume that these molecules are generally formed through gas-phase reactions (see, e.g., \citealt{charnley97,peeters06}). In this scenario, a CH$_3$OH abundance $\lesssim10^{-8}$ is too low for an efficient formation of these molecules (Taquet et~al., in prep.). H$_2$CCO and CH$_3$CHO are, on the other hand, more likely to form on grains \citep{charnley01} and should form at a rate similar to CH$_3$OH, which could explain why the H$_2$CCO/CH$_3$OH and CH$_3$CHO/CH$_3$OH ratios are similar to what is found in IRAS~16293-2422. Another possibility is that the tentatively detected complex molecules could exist in a more shielded inner region, at temperatures at the same levels as those of the hydrocarbon molecules (10--15~K), similar to the properties of the complex organic molecules detected in the prestellar core L1689B \citep{bacmann12}. If, on the other hand, we assume a grain-formation scenario for all complex organic molecules, the elevated temperatures should enhance the formation of complex organic molecules as long as precursor molecules such as H$_2$CO exist on the icy grains \citep{garrod06}. The CH$_3$CN found towards protostellar envelopes is believed to be formed from gas-phase reactions between HCN and CH$_3^+$ \citep{rodgers01,wang10}. The low CH$_3$CN abundance and high CN and DCN abundances observed towards IRS7B are thus puzzling, but might be due to a relative lack of CH$_4$, the precursor species of CH$_3^+$. The exact reason for a low CH$_3$CN abundance needs to be further investigated.

\begin{figure}[!tb]
        \centering  
        \includegraphics{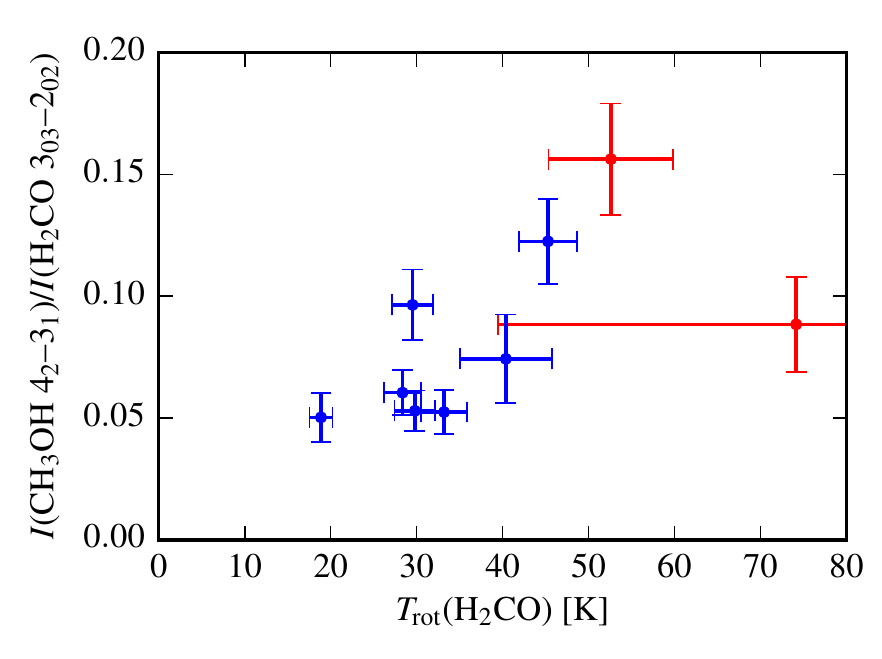}
        \caption{As a proxy for the CH$_3$OH/H$_2$CO abundance ratio, we here show the line ratio between the CH$_3$OH $4_2\rightarrow3_1$,~E line at 218.440~GHz and the H$_2$CO $3_{03}\rightarrow2_{02}$ line at 218.222~GHz in the CrA sources as a function of the H$_2$CO rotational temperature. CrA-24 and SMM~2 are shown with two data points each, one for the on-source component (blue) and one for the outflow component (red). (A colour version of this plot is available in the online journal.)}\label{fig:ch3oh_h2co}
\end{figure}

Laboratory experiments \citep{watanabe04,fuchs09} and numerical simulations \citep{cuppen09} have suggested that the hydrogenation reactions of CO forming H$_2$CO and CH$_3$OH on CO-H$_2$O ices are strongly temperature dependent, with the CH$_3$OH/H$_2$CO ratio increasing significantly at relatively small increases in temperature. Since the CH$_3$OH rotational diagrams of the source survey have too few data points to be reliable (see Sect.~\ref{sec:rot_sosu}), we did not use the CH$_3$OH/H$_2$CO abundance ratio, but instead used the ratio between the CH$_3$OH $4_2\rightarrow3_1$,~E line at 218.440~GHz and the H$_2$CO $3_{03}\rightarrow2_{02}$ line at 218.222~GHz as a more useful proxy for the CH$_3$OH/H$_2$CO abundance ratio. We compare this ratio with the H$_2$CO rotational temperature in Fig.~\ref{fig:ch3oh_h2co} and see a positive trend between the CH$_3$OH/H$_2$CO ratio and the H$_2$CO temperature. We suggest that the high level of irradiation has caused most of the CO to evaporate from the ices in the whole region around R~CrA, which has caused the relatively low CH$_3$OH abundances (see above). However, the low amounts of CO that are still left on the grain surfaces will form CH$_3$OH more efficiently in the warmest protostellar envelopes, causing the trend in Fig.~\ref{fig:ch3oh_h2co}. This effect should be strongest at $T<20$~K \citep{watanabe03}, but the trend observed above 20~K could be explained if the CH$_3$OH now observed in the gas phase was formed on the icy grains at a time when R~CrA was less luminous than today.

\begin{figure*}[!htb]
        \centering  
        \includegraphics{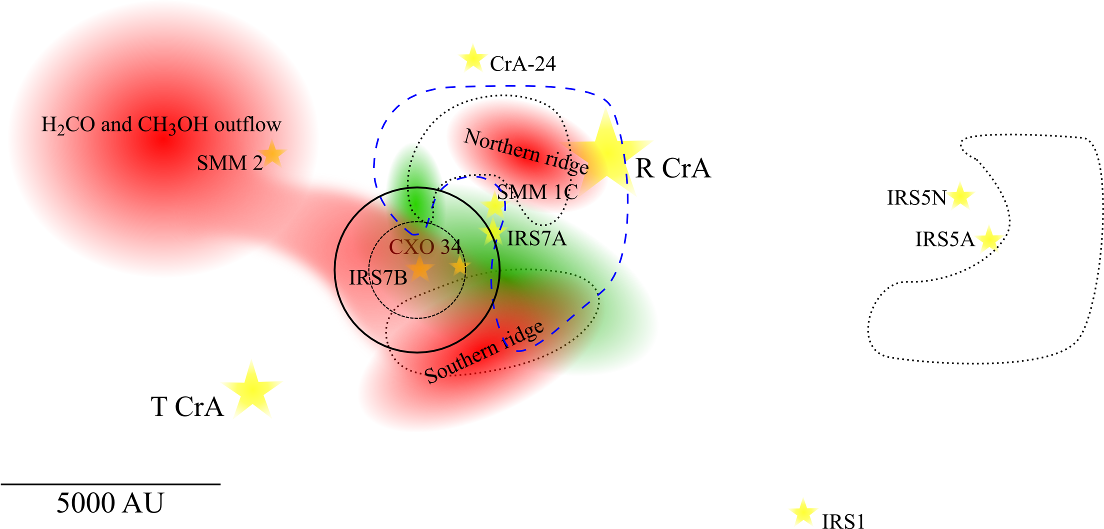}
        \caption{Schematic drawing of the R~CrA cloud and its surroundings. Protostars are shown in yellow. Red structures indicate extended H$_2$CO and CH$_3$OH emission, and blue striped structures show extended emission from hydrocarbon and nitrogen-bearing species (Miettinen et al., in prep.; \citealt{lindberg12}). The brown dashed contour shows extended high-$J$ ($J\sim20$) CO emission, and the black dotted contours show extended 110\micron\ dust continuum emission not associated with the point sources \citep[from \textit{Herschel} PACS observations;][]{lindberg14_herschel}. The solid black circle shows the APEX beam at 218~GHz centred at IRS7B, and the dashed black circle shows the APEX beam at 363~GHz. Large parts of the region have not been covered by all observation modes, and thus the map is probably incomplete. This map covers only a small part of the sources in the CrA source survey (cf. Fig.~\ref{fig:overview}). (A colour version of this plot is available in the online journal.)}\label{fig:drawing}
\end{figure*}

\citet{lindberg12} found that the embedded protostellar sources in the R~CrA cloud are encompassed by dense molecular gas traced by ridges of H$_2$CO on scales of at least 5\,000~AU. A large part of the emission observed in the APEX beam centred at IRS7B is expected to originate in these ridges. Towards CrA-24 and SMM~2, we find signs of large-scale H$_2$CO and CH$_3$OH gas at velocities somewhat lower than the typical $v_{\mathrm{LSR}}$ of the sources in the CrA region. This is consistent with CH$_3$OH emission extending towards the northeast from the R~CrA cloud observed in SEST CH$_3$OH maps (Miettinen et~al., in prep.). A schematic drawing of the distribution of dense molecular gas, dust emission, and YSOs in the vicinity of R~CrA is shown in Fig.~\ref{fig:drawing}. A direct comparison with the typical hot corino sources might therefore be argued against, because in such sources the emission from complex organic molecules originates in relatively compact regions close to the central source \citep[e.g.][]{kuan04,bottinelli04b,bottinelli08,bisschop08}. However, \citet{lindberg14_alma} separated the faint CH$_3$OH emission originating in the IRS7B central source from the surrounding emission using high-resolution ALMA interferometry data. The faint on-source CH$_3$OH lines can either be explained by a very low CH$_3$OH abundance in the inner envelope ($X(\mathrm{CH}_3\mathrm{OH})\sim10^{-10}$; an order of magnitude lower than the already low large-scale CH$_3$OH abundance), or by a more typical CH$_3$OH abundance in the inner envelope ($X(\mathrm{CH}_3\mathrm{OH})\sim10^{-8}$) in combination with a flattened density profile in the inner 100~AU caused by the presence of a Keplerian disc. This latter case would still allow for a higher CH$_3$OH abundance and the presence of more complex organic molecules in the inner envelope of IRS7B at similar abundances as found in hot corinos. Chemical modelling of the irradiated envelope and ALMA observations with higher sensitivity and resolution will help clarify the causes of the chemistry observed towards IRS7B and the inner structure of the source.

\section{Conclusions}
\label{sec:conclusions}

We have performed an unbiased line survey of the Class~0/I source R~CrA IRS7B to investigate its chemical and physical properties. We also presented a survey of the H$_2$CO, CH$_3$OH, and \textit{c}\nobr C$_3$H$_2$ emission for the full sample of embedded YSOs in the Corona Australis region. The observations were carried out with the (sub)millimetre single-dish APEX telescope, but we also used supplementary data from the ASTE telescope. These are our main conclusions:

   \begin{enumerate}
      \item In comparison with other well-studied Class~0/I sources, the deeply embedded YSO IRS7B shows a different chemistry. In particular, very strong CN emission is seen, which indicates a PDR-like chemistry. The only signs of complex organic molecules or large carbon-chain species detected towards IRS7B are CH$_3$OH, CH$_3$CCH, and faint CH$_3$CN and CH$_3$CHO emission. The CH$_3$OH abundance is also very low (3--$6\times10^{-9}$) in comparison with to other embedded protostars. The values or estimated upper limits of the abundances of several complex organic species relative to CH$_3$OH are much lower than the relative abundances detected towards typical hot corino sources, while species such as CN, HC$_3$N, and \textit{c}\nobr C$_3$H$_2$ show strongly elevated relative abundances. High-resolution ALMA observations have also failed to detect complex organics on smaller scales \citep{lindberg14_alma}.
      \item The molecules detected towards R~CrA IRS7B can be grouped into two categories by their rotational temperatures: the warm carbon-chain molecules and nitrogen-bearing species show temperatures around 10--17~K, whereas the organic mole\-cules and sulphur-bearing species are warmer at 19--40~K. Molecules from the two groups have previously been found to have distinct spatial distributions in SMA/APEX interferometry observations of the region \citep{lindberg12}. 
      \item The H$_2$CO rotational temperatures of the sources in the CrA survey near the Herbig~Be star R~CrA are elevated to 30--50~K, and the H$_2$CO temperature is also found to decrease with the distance to R~CrA. This temperature gradient is consistent with a 1-D radiative transfer model with R~CrA as the heating source. The H$_2$CO temperatures of the outflow components of the sources with two velocity components are higher than the corresponding on-source component temperatures. The \textit{c}\nobr C$_3$H$_2$ rotational temperatures are on the other hand found to be fairly constant around 10--15~K for all sources where it could be measured. This could be a shielding effect as a result of the \textit{c}\nobr C$_3$H$_2$ being present at much smaller spatial scales than the H$_2$CO.
      \item The CH$_3$OH/H$_2$CO is found to increase with the H$_2$CO rotational temperature. This can be explained by the higher temperature increasing the efficiency of hydrogenation reactions of CO, as suggested by laboratory studies.
   \end{enumerate}

We have shown that the chemistry of protostellar envelopes can be strongly influenced by external irradiation. Follow-up studies with interferometric observations are needed to reveal the different origins of various organic molecules and unsaturated hydrocarbons in protostellar envelopes. To understand why the abundance levels of some but not all organic molecules are lowered, their exact formation paths need to be studied using astrochemical modelling, in particular assuming elevated temperatures on large scales.

\begin{acknowledgements}

We thank Steven Charnley for helpful suggestions and discussions. We also thank the anonymous referee for helpful comments and suggestions that have improved the manuscript. Research at Centre for Star and Planet Formation is funded by the Danish National Research Foundation and the University of Copenhagen's programme of excellence. This research was also supported by an appointment to the NASA Postdoctoral Program at the NASA Goddard Space Flight Center to J.E.L., administered by Oak Ridge Associated Universities through a contract with NASA, and by a Lundbeck Foundation Group Leader Fellowship to J.K.J. Y.W, N.S, and S.Y. acknowledge financial support from Grant-in-Aid from the Ministry of Education, Culture, Sports, Science, and Technologies of Japan (25108005).

\end{acknowledgements}

\bibliographystyle{aa}
\bibliography{lindberg_arxiv2}

\Online

\clearpage
\begin{appendix}

        \onecolumn
        
        \section{Baseline subtraction}
        \label{app:baseline}
        
        As noted in Sect.~\ref{sec:obs}, the spectra observed by APEX exhibit unstable quasi-sinusoidal baselines \citep[see][]{vassilev08}. When investigating the spectra, we found that polynomial and/or sinusoidal baselines were not giving desirable results. To account for this, we developed a relatively advanced baseline-fitting algorithm, computing running-mean baselines, which is described here.
        
\begin{figure}[!htb]
                \centering
                $\begin{array}{cc}
                \includegraphics{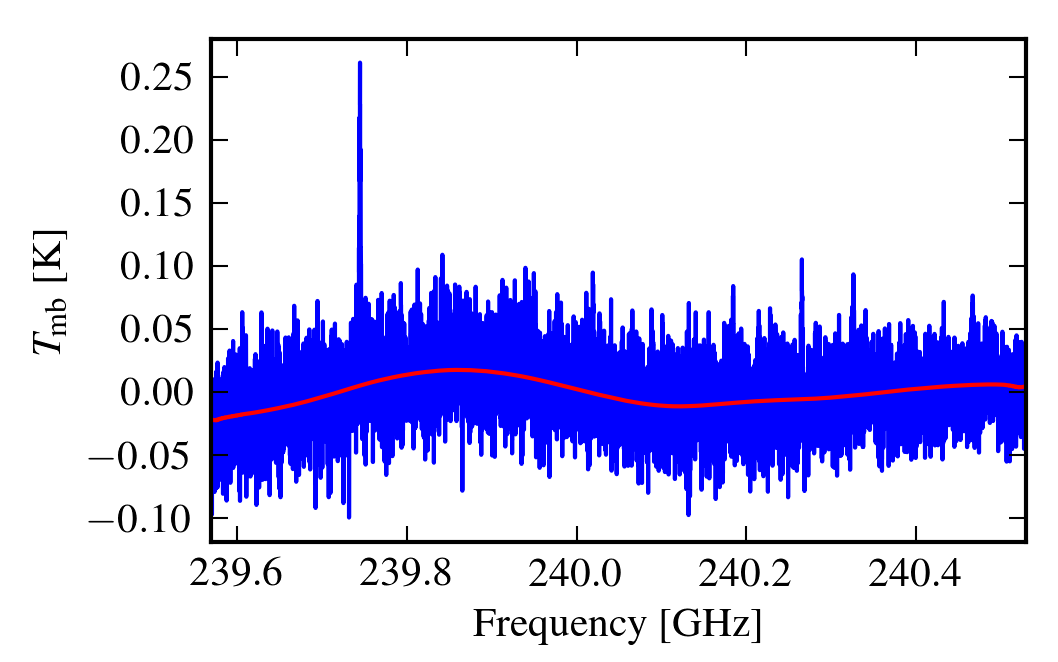} &
                \includegraphics{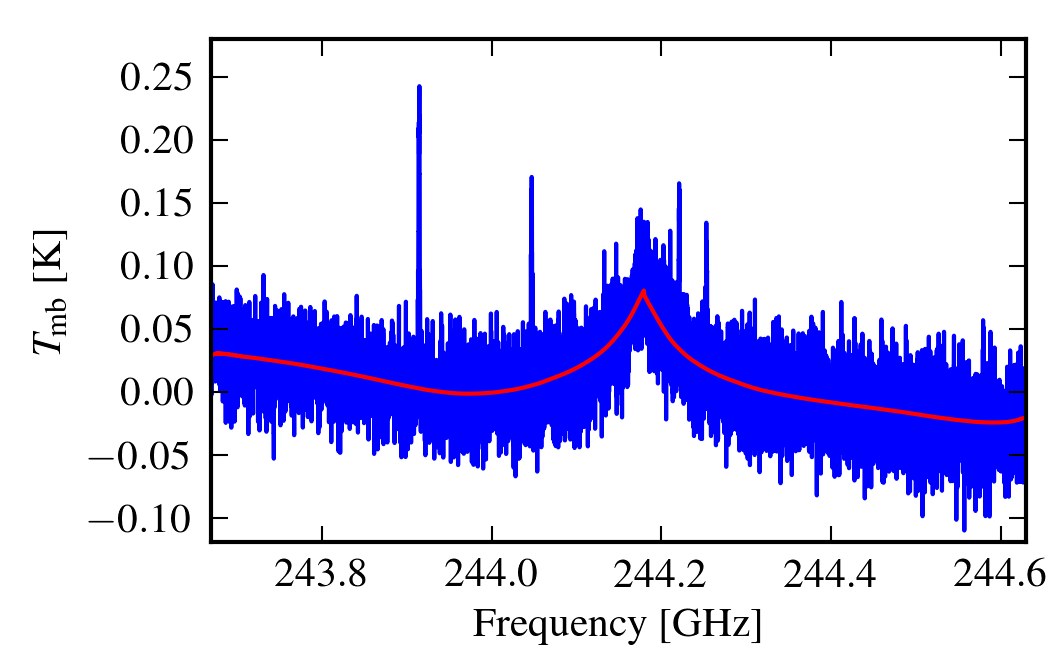}\\
                \end{array}$
                
                \caption{Examples of spectral data before they
were treated by the running-mean baseline-fitting routine (blue) and the fitted baseline that was subsequently subtracted (red). The right spectrum has a discontinuity in the fitted baseline that is due to an atmospheric line. (A colour version of this plot is available in the online journal.)}\label{fig:baseline}
        \end{figure}

        In principle, the fitted baseline is a boxcar smoothing of the spectrum (where line channels have been removed) with weights increasing towards the centre of the box: For each channel in the spectrum, the algorithm calculates the weighted mean of the line-free channels in a box around the channel. The weights have a Gaussian distribution around the central channel of the box. The width of this Gaussian must be adjusted so that all irregularities in the baseline are removed but no real spectral features are removed. In addition to the quasi-sinusoidal baselines, strong atmospheric lines arising from the different elevations of the position-switching on- and off-positions made it necessary to introduce discontinuities in the fitted baseline. By this method, a piecewise smooth but in general non-analytical baseline can be fitted to the data (see examples in Fig.~\ref{fig:baseline}). Some artefacts of the strongest of these atmospheric lines still remain after the baseline subtraction and appear as broad spectral line features in the resulting spectra. In this dataset, they are easy to distinguish from the source lines due to the large difference in line widths.
        
        \section{Observed spectral line parameters}
        \label{app:survey_tables}
        
        The integrated intensities and other parameters of the detected spectral lines are listed in the tables of this appendix. The values of $v_{\mathrm{LSR}}$, $\Delta v$, and $T_{\mathrm{peak}}$ are results of Gaussian fits. In the cases where no values are given, Gaussian fits could not be performed to that line because
of irregular line shapes.
        
        \begin{footnotesize}
                \longtab{}
\begin{longtable}{l l c c c c c}
                        \caption[]{Observed line strengths towards IRS7B, both for the 1.3~mm unbiased line survey and for the IRS7B part of the source survey, which also covered the 0.9~mm and 0.8~mm windows.}\\
                        \hline
                        \hline
                        \noalign{\smallskip}
                        Frequency\tablefootmark{a} & Molecule & Transition & $v_{\mathrm{LSR}}$ & $\Delta v$ & $T_{\mathrm{mb}}$ & $\int T_{\mathrm{mb}} dv$ \\
                        $[$MHz$]$ & & & $[$km~s$^{-1}]$ & $[$km~s$^{-1}]$ & $[$K$]$ & $[$K km~s$^{-1}]$ \\
                        \noalign{\smallskip}
                        \hline
                        \noalign{\smallskip}
                        217238.538 & DCN & $J=3\rightarrow2$ & 6.1 & 1.8 & \phantom{0}1.04 & $\phantom{00}2.01 \pm 0.07$ \\ 
                        217467.150 & $^{13}$CN & $N=2\rightarrow1,J=5/2\rightarrow3/2,F_1=3\rightarrow2,F=4\rightarrow3$ & 5.8 & 0.9 & \phantom{0}0.24 & $\phantom{00}0.24 \pm 0.05$ \\ 
                        217467.150 & $^{13}$CN & $N=2\rightarrow1,J=5/2\rightarrow3/2,F_1=3\rightarrow2,F=3\rightarrow2$ & ... & ... & ...  & ... \\ 
                        217822.148 & \textit{c}\nobr C$_3$H$_2$ & $6_{06}\rightarrow5_{15}$ & 5.8 & 1.6 & \phantom{0}0.95 & $\phantom{00}1.68 \pm 0.07$ \\ 
                        217822.148 & \textit{c}\nobr C$_3$H$_2$ & $6_{16}\rightarrow5_{05}$ & ... & ... & ...  & ... \\
\end{longtable}
                \tablefoot{
                        \tablefoottext{a}{Rest frequency. For the unidentified lines, this is calculated assuming a source velocity of 5.7~km~s$^{-1}$.} \\
                }
                
        \end{footnotesize}
        
        \begin{table}[!htb]
                \centering
                \caption[]{Observed line strengths towards CrA-46.}
                \label{tab:cra46}
                \begin{footnotesize}
                        \begin{tabular}{l l c c c c c c}
                                \noalign{\smallskip}
                                \hline
                                \noalign{\smallskip}
                                Frequency & Molecule & Transition & $v_{\mathrm{LSR}}$ & $\Delta v$ & $T_{\mathrm{mb}}$ & $\int T_{\mathrm{mb}} dv$ \\
                                $[$MHz$]$ & & & $[$km~s$^{-1}]$ & $[$km~s$^{-1}]$ & $[$K$]$ & $[$K km~s$^{-1}]$ \\
                                \noalign{\smallskip}
                                \hline
                                \noalign{\smallskip}
                                No detections \\
                                \noalign{\smallskip}
                                \hline
                        \end{tabular}
                \end{footnotesize}
        \end{table}
        
        \begin{table}[!htb]
                \centering
                \caption[]{Observed line strengths towards CrA-3.}
                \label{tab:cra3}
                \begin{footnotesize}
                        \begin{tabular}{l l c c c c c c}
                                \noalign{\smallskip}
                                \hline
                                \noalign{\smallskip}
                                Frequency & Molecule & Transition & $v_{\mathrm{LSR}}$ & $\Delta v$ & $T_{\mathrm{mb}}$ & $\int T_{\mathrm{mb}} dv$ \\
                                $[$MHz$]$ & & & $[$km~s$^{-1}]$ & $[$km~s$^{-1}]$ & $[$K$]$ & $[$K km~s$^{-1}]$ \\
                                \noalign{\smallskip}
                                \hline
                                \noalign{\smallskip}
                                219560.354 & C$^{18}$O & $J=2\rightarrow1$ & 6.0 & 1.1 & \phantom{0}0.63 & $\phantom{00}0.69 \pm 0.03$ \\ 
                                \noalign{\smallskip}
                                \hline 
                                \noalign{\smallskip}
                                337061.050 & C$^{17}$O & $J=3\rightarrow2$ & 5.8 & 0.3 & \phantom{0}0.26 & $\phantom{00}0.06 \pm 0.03$ \\ 
                                \noalign{\smallskip}
                                \hline
                        \end{tabular}
                \end{footnotesize}
        \end{table}
        
        \begin{table}[!htb]
                \centering
                \caption[]{Observed line strengths towards CrA-5.}
                \label{tab:cra5}
                \begin{footnotesize}
                        \begin{tabular}{l l c c c c c c}
                                \noalign{\smallskip}
                                \hline
                                \noalign{\smallskip}
                                Frequency & Molecule & Transition & $v_{\mathrm{LSR}}$ & $\Delta v$ & $T_{\mathrm{mb}}$ & $\int T_{\mathrm{mb}} dv$ \\
                                $[$MHz$]$ & & & $[$km~s$^{-1}]$ & $[$km~s$^{-1}]$ & $[$K$]$ & $[$K km~s$^{-1}]$ \\
                                \noalign{\smallskip}
                                \hline
                                \noalign{\smallskip}
                                219560.354 & C$^{18}$O & $J=2\rightarrow1$ & 6.2 & 0.7 & \phantom{0}1.96 & $\phantom{00}1.40 \pm 0.02$ \\ 
                                \noalign{\smallskip}
                                \hline
                        \end{tabular}
                \end{footnotesize}
        \end{table}
        
        \begin{table}[!htb]
                \centering
                \caption[]{Observed line strengths towards LS-RCrA1.}
                \label{tab:lsrcra1}
                \begin{footnotesize}
                        \begin{tabular}{l l c c c c c c}
                                \noalign{\smallskip}
                                \hline
                                \noalign{\smallskip}
                                Frequency & Molecule & Transition & $v_{\mathrm{LSR}}$ & $\Delta v$ & $T_{\mathrm{mb}}$ & $\int T_{\mathrm{mb}} dv$ \\
                                $[$MHz$]$ & & & $[$km~s$^{-1}]$ & $[$km~s$^{-1}]$ & $[$K$]$ & $[$K km~s$^{-1}]$ \\
                                \noalign{\smallskip}
                                \hline
                                \noalign{\smallskip}
                                219560.354 & C$^{18}$O & $J=2\rightarrow1$ & 6.0 & 1.4 & \phantom{0}0.86 & $\phantom{00}1.22 \pm 0.04$ \\ 
                                \noalign{\smallskip}
                                \hline
                        \end{tabular}
                \end{footnotesize}
        \end{table}
        
        \begin{table}[!htb]
                \centering
                \caption[]{Observed line strengths towards Haas4.}
                \label{tab:haas4}
                \begin{footnotesize}
                        \begin{tabular}{l l c c c c c c}
                                \noalign{\smallskip}
                                \hline
                                \noalign{\smallskip}
                                Frequency & Molecule & Transition & $v_{\mathrm{LSR}}$ & $\Delta v$ & $T_{\mathrm{mb}}$ & $\int T_{\mathrm{mb}} dv$ \\
                                $[$MHz$]$ & & & $[$km~s$^{-1}]$ & $[$km~s$^{-1}]$ & $[$K$]$ & $[$K km~s$^{-1}]$ \\
                                \noalign{\smallskip}
                                \hline
                                \noalign{\smallskip}
                                216112.582 & DCO$^+$ & $J=3\rightarrow2$ & \phantom{0}5.6 & 0.9 & \phantom{0}0.61 & $\phantom{00}0.55 \pm 0.02$ \\ 
                                216278.756 & \textit{c}\nobr C$_3$H$_2$ & $3_{03}\rightarrow2_{21}$ & \phantom{0}5.4 & 0.5 & \phantom{0}0.24 & $\phantom{00}0.11 \pm 0.02$ \\                              217104.919 & SiO & $J=5\rightarrow4$ & 10.4 & 5.6 & \phantom{0}0.15 & $\phantom{00}1.08 \pm 0.06$ \\ 
                                217822.148 & \textit{c}\nobr C$_3$H$_2$ & $6_{06}\rightarrow5_{15}$ & \phantom{0}5.4 & 0.6 & \phantom{0}0.12 & $\phantom{00}0.07 \pm 0.02$ \\                              217822.148 & \textit{c}\nobr C$_3$H$_2$ & $6_{16}\rightarrow5_{05}$ & ... & ... & ... & ... \\ 
                                217940.046 & \textit{c}\nobr C$_3$H$_2$ & $5_{14}\rightarrow4_{23}$ & \phantom{0}5.7 & 0.9 & \phantom{0}0.11 & $\phantom{00}0.12 \pm 0.03$ \\
\hline
                        \end{tabular}
                \end{footnotesize}
        \end{table}
        
        \begin{table}[!htb]
                \centering
                \caption[]{Observed line strengths towards IRS2.}
                \label{tab:irs2}
                \begin{footnotesize}
                        \begin{tabular}{l l c c c c c c}
                                \noalign{\smallskip}
                                \hline
                                \noalign{\smallskip}
                                Frequency & Molecule & Transition & $v_{\mathrm{LSR}}$ & $\Delta v$ & $T_{\mathrm{mb}}$ & $\int T_{\mathrm{mb}} dv$ \\
                                $[$MHz$]$ & & & $[$km~s$^{-1}]$ & $[$km~s$^{-1}]$ & $[$K$]$ & $[$K km~s$^{-1}]$ \\
                                \noalign{\smallskip}
                                \hline
                                \noalign{\smallskip}
                                216112.582 & DCO$^+$ & $J=3\rightarrow2$ & 6.0 & 0.4 & \phantom{0}0.87 & $\phantom{00}0.37 \pm 0.01$ \\ 
                                218222.192 & H$_2$CO & $3_{03}\rightarrow2_{02}$ & 6.0 & 0.6 & \phantom{0}1.97 & $\phantom{00}1.29 \pm 0.01$ \\ 
                                218440.063 & CH$_3$OH & $4_{2}\rightarrow3_{1}$,~E & 6.0 & 0.4 & \phantom{0}0.16 & $\phantom{00}0.06 \pm 0.01$ \\ 
                                218475.632 & H$_2$CO & $3_{22}\rightarrow2_{21}$ & 6.0 & 0.6 & \phantom{0}0.07 & $\phantom{00}0.05 \pm 0.01$ \\ 
                                218760.066 & H$_2$CO & $3_{21}\rightarrow2_{20}$ & 6.2 & 0.9 & \phantom{0}0.07 & $\phantom{00}0.07 \pm 0.01$ \\ 
                                219560.354 & C$^{18}$O & $J=2\rightarrow1$ & 6.2 & 1.0 & \phantom{0}7.72 & $\phantom{00}8.47 \pm 0.02$ \\
\hline
                        \end{tabular}
                \end{footnotesize}
        \end{table}
        
        \begin{table}[!htb]
                \centering
                \caption[]{Observed line strengths towards IRS5A.}
                \label{tab:irs5}
                \begin{footnotesize}
                        \begin{tabular}{l l c c c c c c}
                                \noalign{\smallskip}
                                \hline
                                \noalign{\smallskip}
                                Frequency & Molecule & Transition & $v_{\mathrm{LSR}}$ & $\Delta v$ & $T_{\mathrm{mb}}$ & $\int T_{\mathrm{mb}} dv$ \\
                                $[$MHz$]$ & & & $[$km~s$^{-1}]$ & $[$km~s$^{-1}]$ & $[$K$]$ & $[$K km~s$^{-1}]$ \\
                                \noalign{\smallskip}
                                \hline
                                \noalign{\smallskip}
                                216112.582 & DCO$^+$ & $J=3\rightarrow2$ & 5.7 & 0.7 & \phantom{0}0.99 & $\phantom{00}0.81 \pm 0.02$ \\ 
                                216278.756 & \textit{c}\nobr C$_3$H$_2$ & $3_{03}\rightarrow2_{21}$ & 5.6 & 0.9 & \phantom{0}0.38 & $\phantom{00}0.39 \pm 0.02$ \\ 
                                216373.320 & C$_2$D & $N=3\rightarrow2,J=7/2\rightarrow5/2,F=5/2\rightarrow3/2$ & 5.9 & 1.5 & \phantom{0}0.16 & $\phantom{00}0.26 \pm 0.02$ \\ 
                                216373.320 & C$_2$D & $N=3\rightarrow2,J=7/2\rightarrow5/2,F=7/2\rightarrow5/2$ & ... & ... & ... & ... \\ 
                                216428.320 & C$_2$D & $N=3\rightarrow2,J=5/2\rightarrow3/2,F=7/2\rightarrow5/2$ & 5.5 & 1.0 & \phantom{0}0.12 & $\phantom{00}0.12 \pm 0.01$ \\ 
                                216428.320 & C$_2$D & $N=3\rightarrow2,J=5/2\rightarrow3/2,F=5/2\rightarrow3/2$ & ... & ... & ... & ... \\
\hline
                        \end{tabular}
                \end{footnotesize}
        \end{table}
        
        \begin{table}[!htb]
                \centering
                \caption[]{Observed line strengths towards IRS5N.}
                \label{tab:irs5n}
                \begin{footnotesize}
                        \begin{tabular}{l l c c c c c c}
                                \noalign{\smallskip}
                                \hline
                                \noalign{\smallskip}
                                Frequency & Molecule & Transition & $v_{\mathrm{LSR}}$ & $\Delta v$ & $T_{\mathrm{mb}}$ & $\int T_{\mathrm{mb}} dv$ \\
                                $[$MHz$]$ & & & $[$km~s$^{-1}]$ & $[$km~s$^{-1}]$ & $[$K$]$ & $[$K km~s$^{-1}]$ \\
                                \noalign{\smallskip}
                                \hline
                                \noalign{\smallskip}
                                216112.582 & DCO$^+$ & $J=3\rightarrow2$ & 5.7 & 0.9 & \phantom{0}0.80 & $\phantom{00}0.78 \pm 0.02$ \\ 
                                216278.756 & \textit{c}\nobr C$_3$H$_2$ & $3_{03}\rightarrow2_{21}$ & 5.7 & 1.3 & \phantom{0}0.26 & $\phantom{00}0.35 \pm 0.02$ \\ 
                                216373.320 & C$_2$D & $N=3\rightarrow2,J=7/2\rightarrow5/2,F=5/2\rightarrow3/2$ & 5.9 & 1.6 & \phantom{0}0.17 & $\phantom{00}0.29 \pm 0.02$ \\ 
                                216373.320 & C$_2$D & $N=3\rightarrow2,J=7/2\rightarrow5/2,F=7/2\rightarrow5/2$ & ... & ... & ... & ... \\ 
                                216428.320 & C$_2$D & $N=3\rightarrow2,J=5/2\rightarrow3/2,F=7/2\rightarrow5/2$ & 5.5 & 1.6 & \phantom{0}0.10 & $\phantom{00}0.15 \pm 0.02$ \\ 
                                216428.320 & C$_2$D & $N=3\rightarrow2,J=5/2\rightarrow3/2,F=5/2\rightarrow3/2$ & ... & ... & ... & ... \\
\hline
                        \end{tabular}
                \end{footnotesize}
        \end{table}
        
        \begin{table}[!htb]
                \centering
                \caption[]{Observed line strengths towards IRS1.}
                \label{tab:irs1}
                \begin{footnotesize}
                        \begin{tabular}{l l c c c c c c}
                                \noalign{\smallskip}
                                \hline
                                \noalign{\smallskip}
                                Frequency & Molecule & Transition & $v_{\mathrm{LSR}}$ & $\Delta v$ & $T_{\mathrm{mb}}$ & $\int T_{\mathrm{mb}} dv$ \\
                                $[$MHz$]$ & & & $[$km~s$^{-1}]$ & $[$km~s$^{-1}]$ & $[$K$]$ & $[$K km~s$^{-1}]$ \\
                                \noalign{\smallskip}
                                \hline
                                \noalign{\smallskip}
                                216112.582 & DCO$^+$ & $J=3\rightarrow2$ & 5.8 & 0.8 & \phantom{0}0.44 & $\phantom{00}0.38 \pm 0.02$ \\ 
                                216278.756 & \textit{c}\nobr C$_3$H$_2$ & $3_{03}\rightarrow2_{21}$ & 5.8 & 0.8 & \phantom{0}0.15 & $\phantom{00}0.13 \pm 0.02$ \\ 
                                216373.320 & C$_2$D & $N=3\rightarrow2,J=7/2\rightarrow5/2,F=5/2\rightarrow3/2$ & 6.0 & 1.2 & \phantom{0}0.10 & $\phantom{00}0.13 \pm 0.02$ \\ 
                                216373.320 & C$_2$D & $N=3\rightarrow2,J=7/2\rightarrow5/2,F=7/2\rightarrow5/2$ & ... & ... & ... & ... \\ 
                                216428.320 & C$_2$D & $N=3\rightarrow2,J=5/2\rightarrow3/2,F=7/2\rightarrow5/2$ & 5.8 & 0.8 & \phantom{0}0.09 & $\phantom{00}0.06 \pm 0.02$ \\ 
                                216428.320 & C$_2$D & $N=3\rightarrow2,J=5/2\rightarrow3/2,F=5/2\rightarrow3/2$ & ... & ... & ... & ... \\
\hline
                        \end{tabular}
                \end{footnotesize}
        \end{table}
        
        \begin{table}[!htb]
                \centering
                \caption[]{Observed line strengths towards IRS7A.}
                \label{tab:irs7a}
                \begin{footnotesize}
                        \begin{tabular}{l l c c c c c c}
                                \noalign{\smallskip}
                                \hline
                                \noalign{\smallskip}
                                Frequency & Molecule & Transition & $v_{\mathrm{LSR}}$ & $\Delta v$ & $T_{\mathrm{mb}}$ & $\int T_{\mathrm{mb}} dv$ \\
                                $[$MHz$]$ & & & $[$km~s$^{-1}]$ & $[$km~s$^{-1}]$ & $[$K$]$ & $[$K km~s$^{-1}]$ \\
                                \noalign{\smallskip}
                                \hline
                                \noalign{\smallskip}
                                337061.050 & C$^{17}$O & $J=3\rightarrow2$ & 5.1 & 2.0 & \phantom{0}2.39 & $\phantom{00}5.63 \pm 0.04$ \\ 
                                337396.459 & C$^{34}$S & $J=7\rightarrow6$ & 5.0 & 2.4 & \phantom{0}0.21 & $\phantom{00}0.54 \pm 0.03$ \\ 
                                337580.147 & $^{34}$SO & $8_8\rightarrow7_7$ & 4.6 & 3.8 & \phantom{0}0.07 & $\phantom{00}0.34 \pm 0.04$ \\ 
                                338083.195 & H$_2$CS & $10_{1,10}\rightarrow9_{1,9}$ & 5.2 & 1.4 & \phantom{0}0.08 & $\phantom{00}0.11 \pm 0.02$ \\ 
                                338124.488 & CH$_3$OH & $7_{0}\rightarrow6_{0}$,~E & 5.0 & 2.2 & \phantom{0}0.12 & $\phantom{00}0.24 \pm 0.02$ \\ 
                                338344.588 & CH$_3$OH & $7_{-1}\rightarrow6_{-1}$,~E & 5.1 & 2.4 & \phantom{0}0.36 & $\phantom{00}0.87 \pm 0.03$ \\
\hline
                        \end{tabular}
                \end{footnotesize}
        \end{table}
        
        \begin{table}[!htb]
                \centering
                \caption[]{Observed line strengths towards CrA-24, total flux (both source and outflow components).}
                \label{tab:cra24}
                \begin{footnotesize}
                        \begin{tabular}{l l c c c c c c}
                                \noalign{\smallskip}
                                \hline
                                \noalign{\smallskip}
                                Frequency & Molecule & Transition & $v_{\mathrm{LSR}}$ & $\Delta v$ & $T_{\mathrm{mb}}$ & $\int T_{\mathrm{mb}} dv$ \\
                                $[$MHz$]$ & & & $[$km~s$^{-1}]$ & $[$km~s$^{-1}]$ & $[$K$]$ & $[$K km~s$^{-1}]$ \\
                                \noalign{\smallskip}
                                \hline
                                \noalign{\smallskip}
                                216112.582 & DCO$^+$ & $J=3\rightarrow2$ & 4.9 & 1.2 & \phantom{0}0.26 & $\phantom{00}0.31 \pm 0.03$ \\ 
                                216278.756 & \textit{c}\nobr C$_3$H$_2$ & $3_{03}\rightarrow2_{21}$ & 4.8 & 0.9 & \phantom{0}0.27 & $\phantom{00}0.24 \pm 0.02$ \\ 
                                216373.320 & C$_2$D & $N=3\rightarrow2,J=7/2\rightarrow5/2,F=5/2\rightarrow3/2$ & 5.2 & 1.5 & \phantom{0}0.10 & $\phantom{00}0.19 \pm 0.04$ \\ 
                                216373.320 & C$_2$D & $N=3\rightarrow2,J=7/2\rightarrow5/2,F=7/2\rightarrow5/2$ & ... & ... & ... & ... \\ 
                                217104.919 & SiO & $J=5\rightarrow4$ & 1.9 & 4.5 & \phantom{0}0.10 & $\phantom{00}0.49 \pm 0.05$ \\ 
                                217238.538 & DCN & $J=3\rightarrow2$ & 4.7 & 1.1 & \phantom{0}0.24 & $\phantom{00}0.27 \pm 0.03$ \\
\hline
                        \end{tabular}
                \end{footnotesize}
        \end{table}
        
        \begin{table}[!htb]
                \centering
                \caption[]{Observed line strengths towards CrA-24, on-source velocity component. The integrated intensities are calculated from Gaussian fits.}
                \label{tab:cra24source}
                \begin{footnotesize}
                        \begin{tabular}{l l c c c c c c}
                                \noalign{\smallskip}
                                \hline
                                \noalign{\smallskip}
                                Frequency & Molecule & Transition & $v_{\mathrm{LSR}}$ & $\Delta v$ & $T_{\mathrm{mb}}$ & $\int T_{\mathrm{mb}} dv$ \\
                                $[$MHz$]$ & & & $[$km~s$^{-1}]$ & $[$km~s$^{-1}]$ & $[$K$]$ & $[$K km~s$^{-1}]$ \\
                                \noalign{\smallskip}
                                \hline
                                \noalign{\smallskip}
                                217104.919 & SiO & $J=5\rightarrow4$ & 4.6 & 2.4 & \phantom{0}0.06 & $\phantom{00}0.17 \pm 0.03$ \\ 
                                218222.192 & H$_2$CO & $3_{03}\rightarrow2_{02}$ & 5.0 & 1.8 & \phantom{0}2.61 & $\phantom{00}5.09 \pm 0.03$ \\ 
                                218440.063 & CH$_3$OH & $4_{2}\rightarrow3_{1}$,~E & 5.1 & 1.5 & \phantom{0}0.17 & $\phantom{00}0.27 \pm 0.03$ \\ 
                                218475.632 & H$_2$CO & $3_{22}\rightarrow2_{21}$ & 5.0 & 1.9 & \phantom{0}0.42 & $\phantom{00}0.84 \pm 0.03$ \\ 
                                218760.066 & H$_2$CO & $3_{21}\rightarrow2_{20}$ & 4.9 & 1.9 & \phantom{0}0.46 & $\phantom{00}0.92 \pm 0.03$ \\ 
                                219908.525 & H$_2^{13}$CO & $3_{12}\rightarrow2_{11}$ & 4.6 & 1.6 & \phantom{0}0.10 & $\phantom{00}0.16 \pm 0.03$ \\
\hline
                        \end{tabular}
                \end{footnotesize}
        \end{table}
        
        \begin{table}[!htb]
                \centering
                \caption[]{Observed line strengths towards CrA-24, outflow velocity component. The integrated intensities are calculated from Gaussian fits.}
                \label{tab:cra24outflow}
                \begin{footnotesize}
                        \begin{tabular}{l l c c c c c c}
                                \noalign{\smallskip}
                                \hline
                                \noalign{\smallskip}
                                Frequency & Molecule & Transition & $v_{\mathrm{LSR}}$ & $\Delta v$ & $T_{\mathrm{mb}}$ & $\int T_{\mathrm{mb}} dv$ \\
                                $[$MHz$]$ & & & $[$km~s$^{-1}]$ & $[$km~s$^{-1}]$ & $[$K$]$ & $[$K km~s$^{-1}]$ \\
                                \noalign{\smallskip}
                                \hline
                                \noalign{\smallskip}
                                217104.919 & SiO & $J=5\rightarrow4$ & 1.2 & 2.2 & \phantom{0}0.13 & $\phantom{00}0.30 \pm 0.03$ \\ 
                                218222.192 & H$_2$CO & $3_{03}\rightarrow2_{02}$ & 2.0 & 3.2 & \phantom{0}0.75 & $\phantom{00}2.57 \pm 0.04$ \\ 
                                218440.063 & CH$_3$OH & $4_{2}\rightarrow3_{1}$,~E & 2.7 & 3.5 & \phantom{0}0.06 & $\phantom{00}0.23 \pm 0.04$ \\ 
                                218475.632 & H$_2$CO & $3_{22}\rightarrow2_{21}$ & 2.2 & 3.7 & \phantom{0}0.18 & $\phantom{00}0.71 \pm 0.04$ \\ 
                                218760.066 & H$_2$CO & $3_{21}\rightarrow2_{20}$ & 1.6 & 3.7 & \phantom{0}0.18 & $\phantom{00}0.73 \pm 0.04$ \\ 
                                219908.525 & H$_2^{13}$CO & $3_{12}\rightarrow2_{11}$ & 1.3 & 1.1 & \phantom{0}0.05 & $\phantom{00}0.06 \pm 0.02$ \\
\hline
                        \end{tabular}
                \end{footnotesize}
        \end{table}
        
        \begin{table}[!htb]
                \centering
                \caption[]{Observed line strengths towards SMM 2, total flux (both source and outflow components).}
                \label{tab:smm2}
                \begin{footnotesize}
                        \begin{tabular}{l l c c c c c c}
                                \noalign{\smallskip}
                                \hline
                                \noalign{\smallskip}
                                Frequency & Molecule & Transition & $v_{\mathrm{LSR}}$ & $\Delta v$ & $T_{\mathrm{mb}}$ & $\int T_{\mathrm{mb}} dv$ \\
                                $[$MHz$]$ & & & $[$km~s$^{-1}]$ & $[$km~s$^{-1}]$ & $[$K$]$ & $[$K km~s$^{-1}]$ \\
                                \noalign{\smallskip}
                                \hline
                                \noalign{\smallskip}
                                216112.582 & DCO$^+$ & $J=3\rightarrow2$ & 4.8 & 0.5 & \phantom{0}0.36 & $\phantom{00}0.21 \pm 0.03$ \\ 
                                216278.756 & \textit{c}\nobr C$_3$H$_2$ & $3_{03}\rightarrow2_{21}$ & 5.0 & 0.5 & \phantom{0}0.26 & $\phantom{00}0.14 \pm 0.02$ \\ 
                                216373.320 & C$_2$D & $N=3\rightarrow2,J=7/2\rightarrow5/2,F=5/2\rightarrow3/2$ & 5.4 & 0.8 & \phantom{0}0.09 & $\phantom{00}0.08 \pm 0.03$ \\ 
                                216373.320 & C$_2$D & $N=3\rightarrow2,J=7/2\rightarrow5/2,F=7/2\rightarrow5/2$ & ... & ... & ... & ... \\ 
                                217104.919 & SiO & $J=5\rightarrow4$ & 1.3 & 6.0 & \phantom{0}0.14 & $\phantom{00}1.08 \pm 0.08$ \\ 
                                217238.538 & DCN & $J=3\rightarrow2$ & 4.3 & 2.7 & \phantom{0}0.13 & $\phantom{00}0.28 \pm 0.03$ \\
\hline
                        \end{tabular}
                \end{footnotesize}
        \end{table}
        
        \begin{table}[!htb]
                \centering
                \caption[]{Observed line strengths towards SMM 2, on-source velocity component. The integrated intensities are calculated from Gaussian fits.}
                \label{tab:smm2source}
                \begin{footnotesize}
                        \begin{tabular}{l l c c c c c c}
                                \noalign{\smallskip}
                                \hline
                                \noalign{\smallskip}
                                Frequency & Molecule & Transition & $v_{\mathrm{LSR}}$ & $\Delta v$ & $T_{\mathrm{mb}}$ & $\int T_{\mathrm{mb}} dv$ \\
                                $[$MHz$]$ & & & $[$km~s$^{-1}]$ & $[$km~s$^{-1}]$ & $[$K$]$ & $[$K km~s$^{-1}]$ \\
                                \noalign{\smallskip}
                                \hline
                                \noalign{\smallskip}
                                218222.192 & H$_2$CO & $3_{03}\rightarrow2_{02}$ & 5.1 & 1.3 & \phantom{0}0.87 & $\phantom{00}1.23 \pm 0.02$ \\ 
                                218440.063 & CH$_3$OH & $4_{2}\rightarrow3_{1}$,~E & 5.2 & 0.8 & \phantom{0}0.11 & $\phantom{00}0.09 \pm 0.02$ \\ 
                                218475.632 & H$_2$CO & $3_{22}\rightarrow2_{21}$ & 5.1 & 1.6 & \phantom{0}0.08 & $\phantom{00}0.14 \pm 0.03$ \\ 
                                218760.066 & H$_2$CO & $3_{21}\rightarrow2_{20}$ & 4.6 & 1.5 & \phantom{0}0.16 & $\phantom{00}0.26 \pm 0.02$ \\ 
                                219908.525 & H$_2^{13}$CO & $3_{12}\rightarrow2_{11}$ & 4.6 & 2.2 & \phantom{0}0.07 & $\phantom{00}0.16 \pm 0.03$ \\ 
                                219949.442 & SO & $6_5\rightarrow5_4$ & 5.7 & 0.5 & \phantom{0}0.23 & $\phantom{00}0.13 \pm 0.01$ \\
\hline
                        \end{tabular}
                \end{footnotesize}
        \end{table}
        
        \begin{table}[!htb]
                \centering
                \caption[]{Observed line strengths towards SMM 2, outflow velocity component. The integrated intensities are calculated from Gaussian fits.}
                \label{tab:smm2outflow}
                \begin{footnotesize}
                        \begin{tabular}{l l c c c c c c}
                                \noalign{\smallskip}
                                \hline
                                \noalign{\smallskip}
                                Frequency & Molecule & Transition & $v_{\mathrm{LSR}}$ & $\Delta v$ & $T_{\mathrm{mb}}$ & $\int T_{\mathrm{mb}} dv$ \\
                                $[$MHz$]$ & & & $[$km~s$^{-1}]$ & $[$km~s$^{-1}]$ & $[$K$]$ & $[$K km~s$^{-1}]$ \\
                                \noalign{\smallskip}
                                \hline
                                \noalign{\smallskip}
                                217104.919 & SiO & $J=5\rightarrow4$ & 1.3 & 6.6 & \phantom{0}0.32 & $\phantom{00}1.08 \pm 0.08$ \\ 
                                218222.192 & H$_2$CO & $3_{03}\rightarrow2_{02}$ & 2.7 & 2.9 & \phantom{0}2.11 & $\phantom{00}6.55 \pm 0.03$ \\ 
                                218440.063 & CH$_3$OH & $4_{2}\rightarrow3_{1}$,~E & 2.6 & 3.3 & \phantom{0}0.29 & $\phantom{00}1.02 \pm 0.04$ \\ 
                                218475.632 & H$_2$CO & $3_{22}\rightarrow2_{21}$ & 2.5 & 2.8 & \phantom{0}0.51 & $\phantom{00}1.54 \pm 0.03$ \\ 
                                218760.066 & H$_2$CO & $3_{21}\rightarrow2_{20}$ & 2.4 & 2.4 & \phantom{0}0.56 & $\phantom{00}1.44 \pm 0.03$ \\ 
                                219908.525 & H$_2^{13}$CO & $3_{12}\rightarrow2_{11}$ & 2.0 & 1.8 & \phantom{0}0.09 & $\phantom{00}0.17 \pm 0.03$ \\
\hline
                        \end{tabular}
                \end{footnotesize}
        \end{table}
        
        \begin{table}[!htb]
                \centering
                \caption[]{Observed line strengths towards CXO42.}
                \label{tab:cxo42}
                \begin{footnotesize}
                        \begin{tabular}{l l c c c c c c}
                                \noalign{\smallskip}
                                \hline
                                \noalign{\smallskip}
                                Frequency & Molecule & Transition & $v_{\mathrm{LSR}}$ & $\Delta v$ & $T_{\mathrm{mb}}$ & $\int T_{\mathrm{mb}} dv$ \\
                                $[$MHz$]$ & & & $[$km~s$^{-1}]$ & $[$km~s$^{-1}]$ & $[$K$]$ & $[$K km~s$^{-1}]$ \\
                                \noalign{\smallskip}
                                \hline
                                \noalign{\smallskip}
                                218222.192 & H$_2$CO & $3_{03}\rightarrow2_{02}$ & 5.8 & 1.9 & \phantom{0}0.07 & $\phantom{00}0.14 \pm 0.02$ \\ 
                                219560.354 & C$^{18}$O & $J=2\rightarrow1$ & 5.4 & 1.2 & \phantom{0}2.40 & $\phantom{00}3.34 \pm 0.02$ \\ 
                                \noalign{\smallskip}
                                \hline 
                                \noalign{\smallskip}
                                337061.050 & C$^{17}$O & $J=3\rightarrow2$ & 5.4 & 0.8 & \phantom{0}0.20 & $\phantom{00}0.16 \pm 0.02$ \\ 
                                \noalign{\smallskip}
                                \hline
                        \end{tabular}
                \end{footnotesize}
        \end{table}
        
        \begin{table}[!htb]
                \centering
                \caption[]{Observed line strengths towards CrA-44.}
                \label{tab:cra44}
                \begin{footnotesize}
                        \begin{tabular}{l l c c c c c c}
                                \noalign{\smallskip}
                                \hline
                                \noalign{\smallskip}
                                Frequency & Molecule & Transition & $v_{\mathrm{LSR}}$ & $\Delta v$ & $T_{\mathrm{mb}}$ & $\int T_{\mathrm{mb}} dv$ \\
                                $[$MHz$]$ & & & $[$km~s$^{-1}]$ & $[$km~s$^{-1}]$ & $[$K$]$ & $[$K km~s$^{-1}]$ \\
                                \noalign{\smallskip}
                                \hline
                                \noalign{\smallskip}
                                216112.582 & DCO$^+$ & $J=3\rightarrow2$ & 5.7 & 0.6 & \phantom{0}0.83 & $\phantom{00}0.51 \pm 0.02$ \\ 
                                216278.756 & \textit{c}\nobr C$_3$H$_2$ & $3_{03}\rightarrow2_{21}$ & 5.7 & 0.8 & \phantom{0}0.71 & $\phantom{00}0.64 \pm 0.02$ \\ 
                                216373.320 & C$_2$D & $N=3\rightarrow2,J=7/2\rightarrow5/2,F=5/2\rightarrow3/2$ & 6.0 & 1.4 & \phantom{0}0.20 & $\phantom{00}0.28 \pm 0.02$ \\ 
                                216373.320 & C$_2$D & $N=3\rightarrow2,J=7/2\rightarrow5/2,F=7/2\rightarrow5/2$ & ... & ... & ... & ... \\ 
                                216428.320 & C$_2$D & $N=3\rightarrow2,J=5/2\rightarrow3/2,F=7/2\rightarrow5/2$ & 5.5 & 1.0 & \phantom{0}0.18 & $\phantom{00}0.19 \pm 0.02$ \\ 
                                216428.320 & C$_2$D & $N=3\rightarrow2,J=5/2\rightarrow3/2,F=5/2\rightarrow3/2$ & ... & ... & ... & ... \\
\hline
                        \end{tabular}
                \end{footnotesize}
        \end{table}
        
        \begin{table}[!htb]
                \centering
                \caption[]{Observed line strengths towards CrA-33.}
                \label{tab:cra33}
                \begin{footnotesize}
                        \begin{tabular}{l l c c c c c c}
                                \noalign{\smallskip}
                                \hline
                                \noalign{\smallskip}
                                Frequency & Molecule & Transition & $v_{\mathrm{LSR}}$ & $\Delta v$ & $T_{\mathrm{mb}}$ & $\int T_{\mathrm{mb}} dv$ \\
                                $[$MHz$]$ & & & $[$km~s$^{-1}]$ & $[$km~s$^{-1}]$ & $[$K$]$ & $[$K km~s$^{-1}]$ \\
                                \noalign{\smallskip}
                                \hline
                                \noalign{\smallskip}
                                216112.582 & DCO$^+$ & $J=3\rightarrow2$ & 5.7 & 0.4 & \phantom{0}0.55 & $\phantom{00}0.26 \pm 0.02$ \\ 
                                216278.756 & \textit{c}\nobr C$_3$H$_2$ & $3_{03}\rightarrow2_{21}$ & 5.7 & 0.1 & \phantom{0}0.17 & $\phantom{00}0.06 \pm 0.02$ \\ 
                                218222.192 & H$_2$CO & $3_{03}\rightarrow2_{02}$ & 5.7 & 0.4 & \phantom{0}0.76 & $\phantom{00}0.31 \pm 0.02$ \\ 
                                218440.063 & CH$_3$OH & $4_{2}\rightarrow3_{1}$,~E & 5.7 & 0.5 & \phantom{0}0.17 & $\phantom{00}0.10 \pm 0.03$ \\ 
                                219560.354 & C$^{18}$O & $J=2\rightarrow1$ & 5.7 & 0.8 & \phantom{0}4.51 & $\phantom{00}3.92 \pm 0.03$ \\ 
                                219949.442 & SO & $6_5\rightarrow5_4$ & 5.7 & 0.3 & \phantom{0}0.45 & $\phantom{00}0.15 \pm 0.02$ \\ 
                                \noalign{\smallskip}
                                \hline 
                                \noalign{\smallskip}
                                337061.050 & C$^{17}$O & $J=3\rightarrow2$ & 5.6 & 0.7 & \phantom{0}0.24 & $\phantom{00}0.18 \pm 0.02$ \\ 
                                \noalign{\smallskip}
                                \hline
                        \end{tabular}
                \end{footnotesize}
        \end{table}
        
        \begin{table}[!htb]
                \centering
                \caption[]{Observed line strengths towards VV~CrA.}
                \label{tab:vvcra}
                \begin{footnotesize}
                        \begin{tabular}{l l c c c c c c}
                                \noalign{\smallskip}
                                \hline
                                \noalign{\smallskip}
                                Frequency & Molecule & Transition & $v_{\mathrm{LSR}}$ & $\Delta v$ & $T_{\mathrm{mb}}$ & $\int T_{\mathrm{mb}} dv$ \\
                                $[$MHz$]$ & & & $[$km~s$^{-1}]$ & $[$km~s$^{-1}]$ & $[$K$]$ & $[$K km~s$^{-1}]$ \\
                                \noalign{\smallskip}
                                \hline
                                \noalign{\smallskip}
                                216112.582 & DCO$^+$ & $J=3\rightarrow2$ & 4.9 & 0.7 & \phantom{0}0.15 & $\phantom{00}0.12 \pm 0.01$ \\ 
                                216278.756 & \textit{c}\nobr C$_3$H$_2$ & $3_{03}\rightarrow2_{21}$ & 5.0 & 0.7 & \phantom{0}0.13 & $\phantom{00}0.09 \pm 0.01$ \\ 
                                217238.538 & DCN & $J=3\rightarrow2$ & 4.8 & 0.8 & \phantom{0}0.05 & $\phantom{00}0.04 \pm 0.01$ \\ 
                                217822.148 & \textit{c}\nobr C$_3$H$_2$ & $6_{06}\rightarrow5_{15}$ & 5.0 & 0.8 & \phantom{0}0.13 & $\phantom{00}0.12 \pm 0.01$ \\ 
                                217822.148 & \textit{c}\nobr C$_3$H$_2$ & $6_{16}\rightarrow5_{05}$ & ... & ... & ... & ... \\ 
                                217940.046 & \textit{c}\nobr C$_3$H$_2$ & $5_{14}\rightarrow4_{23}$ & 5.1 & 0.8 & \phantom{0}0.08 & $\phantom{00}0.07 \pm 0.01$ \\
\hline
                        \end{tabular}
                \end{footnotesize}
        \end{table}
        
        \begin{table}[!htb]
                \centering
                \caption[]{Observed line strengths towards CrA-37.}
                \label{tab:cra37}
                \begin{footnotesize}
                        \begin{tabular}{l l c c c c c c}
                                \noalign{\smallskip}
                                \hline
                                \noalign{\smallskip}
                                Frequency & Molecule & Transition & $v_{\mathrm{LSR}}$ & $\Delta v$ & $T_{\mathrm{mb}}$ & $\int T_{\mathrm{mb}} dv$ \\
                                $[$MHz$]$ & & & $[$km~s$^{-1}]$ & $[$km~s$^{-1}]$ & $[$K$]$ & $[$K km~s$^{-1}]$ \\
                                \noalign{\smallskip}
                                \hline
                                \noalign{\smallskip}
                                219560.354 & C$^{18}$O & $J=2\rightarrow1$ & 5.7 & 0.6 & \phantom{0}1.12 & $\phantom{00}0.72 \pm 0.03$ \\ 
                                \noalign{\smallskip}
                                \hline
                        \end{tabular}
                \end{footnotesize}
        \end{table}

        \clearpage
        \section{Complete IRS7B spectrum at 217--245.5~GHz}
        \label{app:spectrum_irs7b}
        
        All spectra in this and the next section are cut off at 0.5~K to also show the weaker spectral lines.
        
        \begin{figure*}[!htb]
                \centering  
                $\begin{array}{c}
                \includegraphics{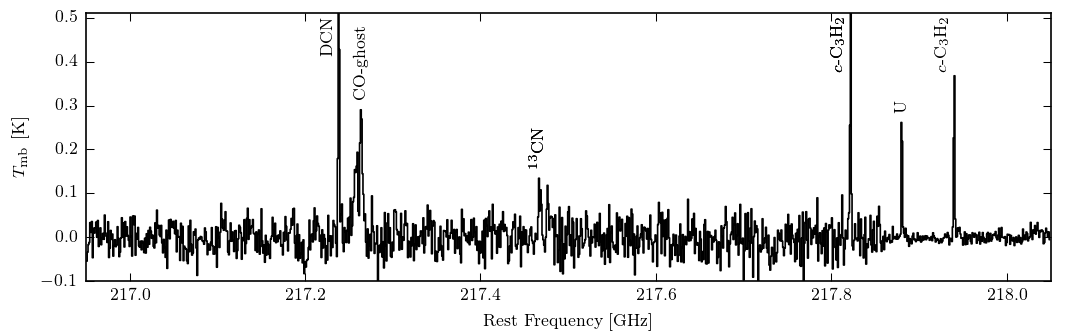} \\
                \includegraphics{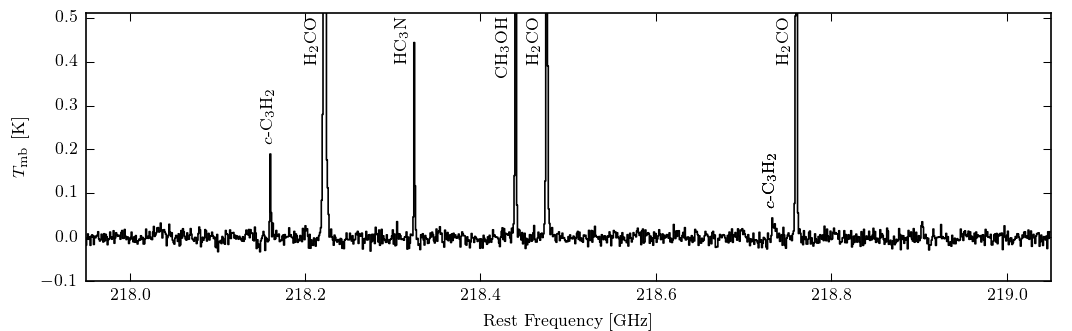} \\
                \includegraphics{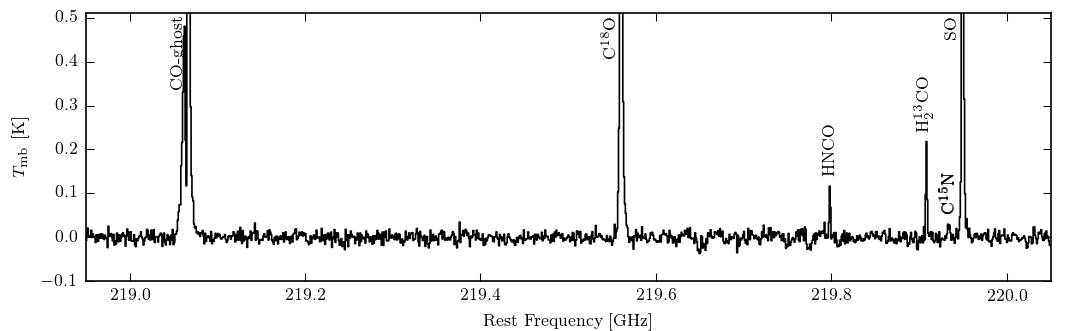} \\
                \end{array}$
                \caption{Full spectrum of IRS7B, smoothed by a factor 8, corresponding to a channel width of 1.2--1.4~\kms.}\label{fig:largespec1}
        \end{figure*}
        
        \begin{figure*}[!htb]
                \centering  
                $\begin{array}{c}
                \includegraphics{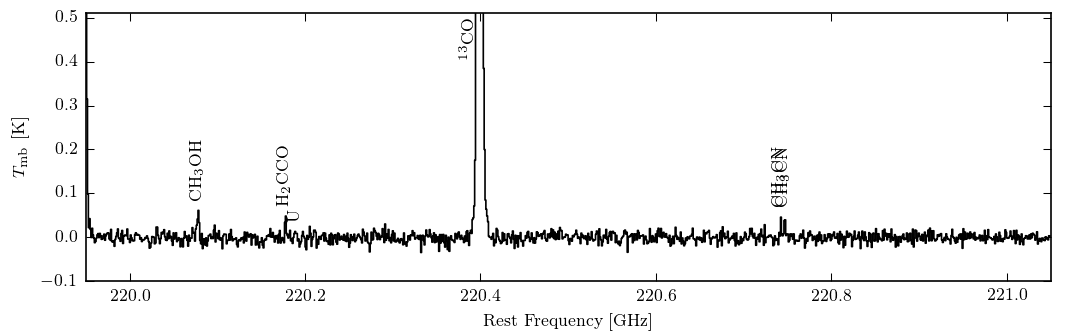} \\
                \includegraphics{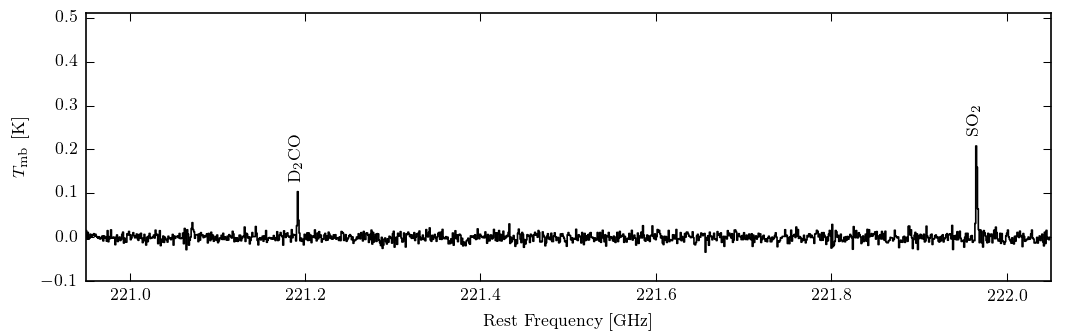} \\
                \includegraphics{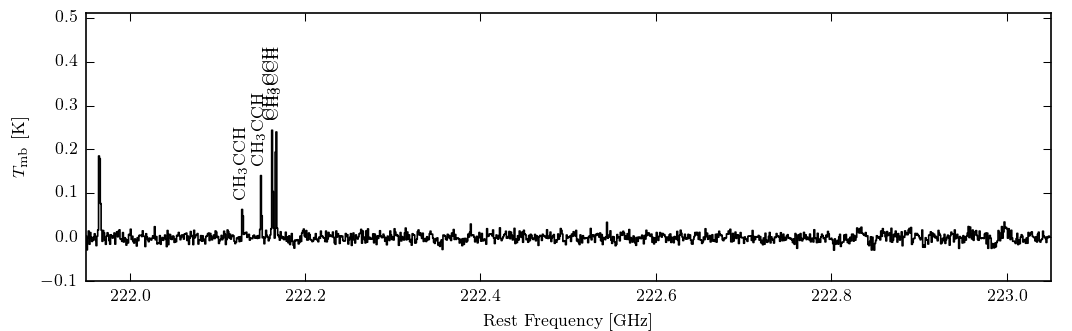} \\
                \includegraphics{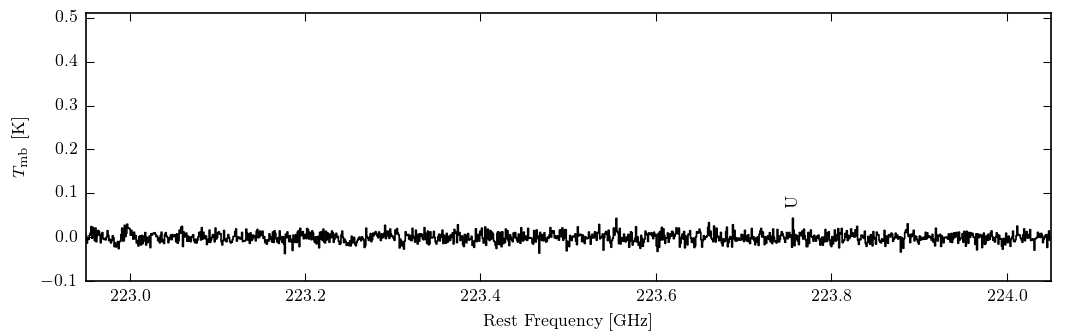} \\
                \end{array}$
                \caption{See Fig.~\ref{fig:largespec1}.}\label{fig:largespec2}
        \end{figure*}
        
        \begin{figure*}[!htb]
                \centering  
                $\begin{array}{c}
                \includegraphics{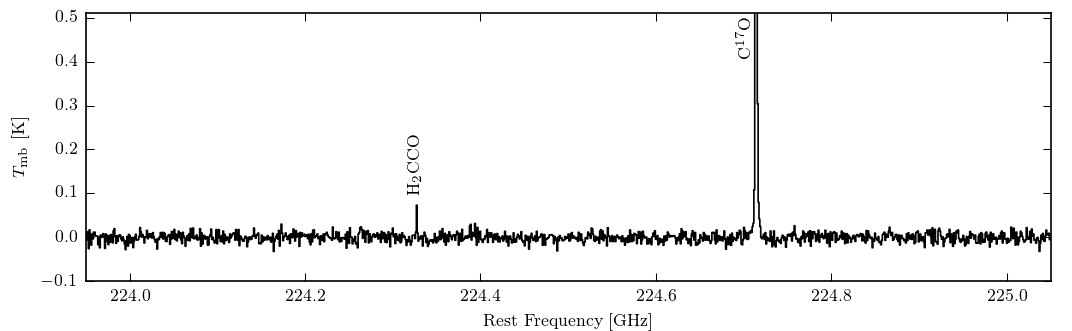} \\
                \includegraphics{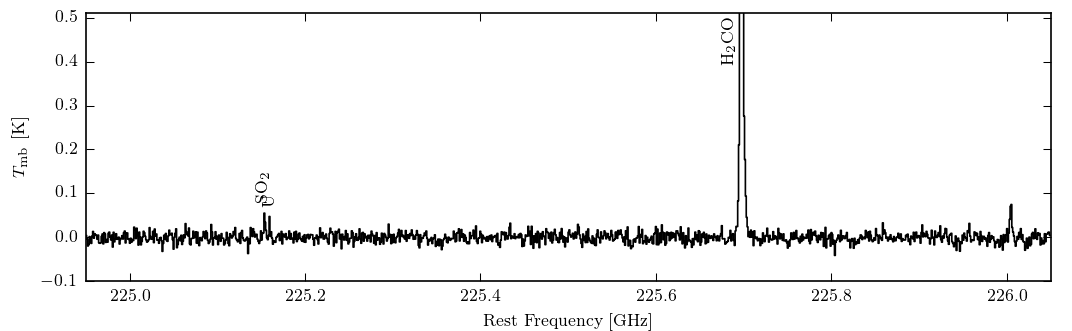} \\
                \includegraphics{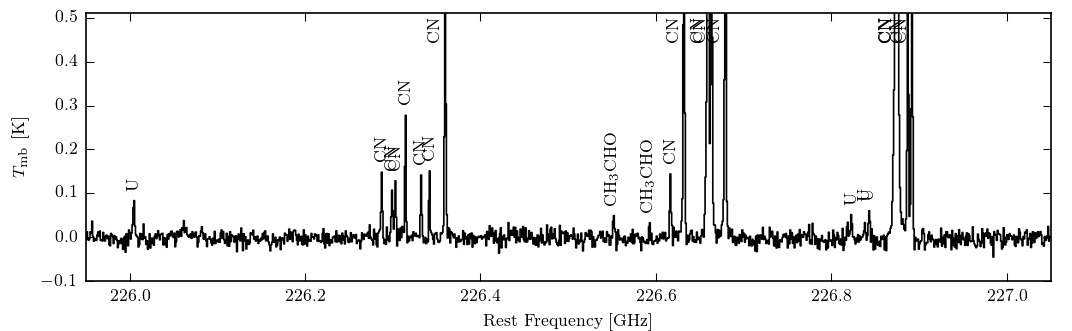} \\
                \includegraphics{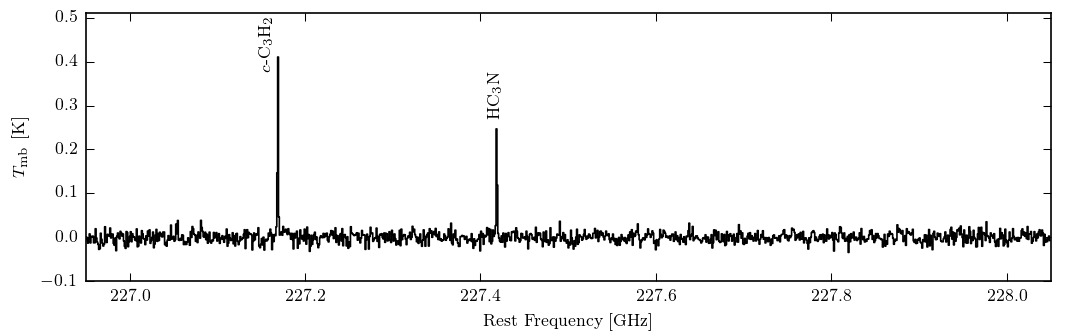} \\
                \end{array}$
                \caption{See Fig.~\ref{fig:largespec1}.}\label{fig:largespec3}
        \end{figure*}
        
        \begin{figure*}[!htb]
                \centering  
                $\begin{array}{c}
                \includegraphics{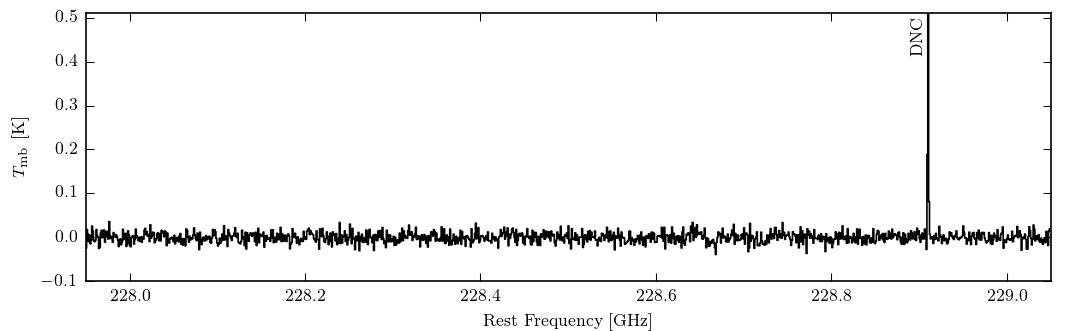} \\
                \includegraphics{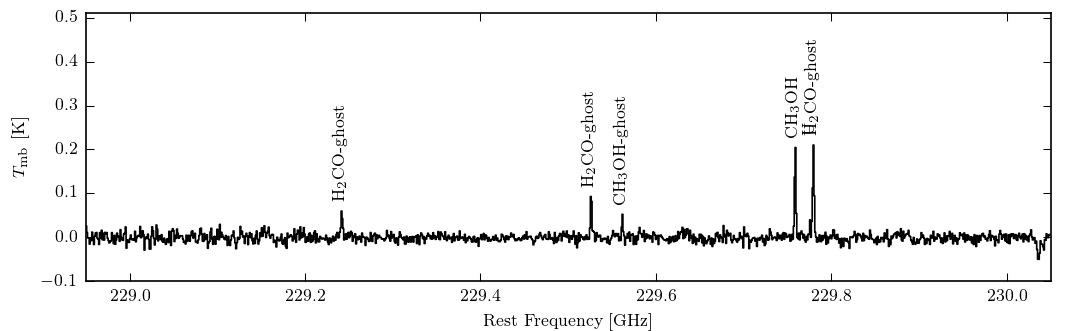} \\
                \includegraphics{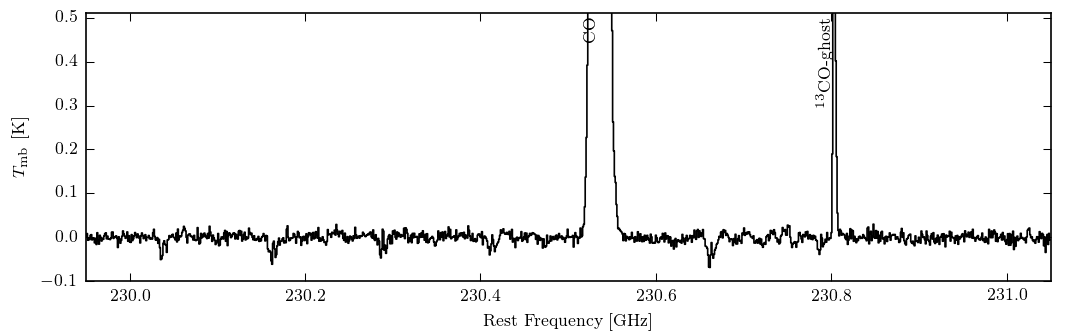} \\
                \includegraphics{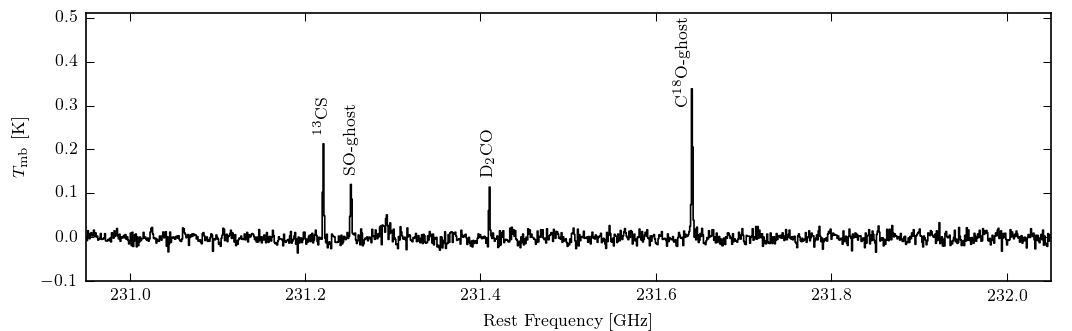} \\
                \end{array}$
                \caption{See Fig.~\ref{fig:largespec1}.}\label{fig:largespec4}
        \end{figure*}
        
        \begin{figure*}[!htb]
                \centering  
                $\begin{array}{c}
                \includegraphics{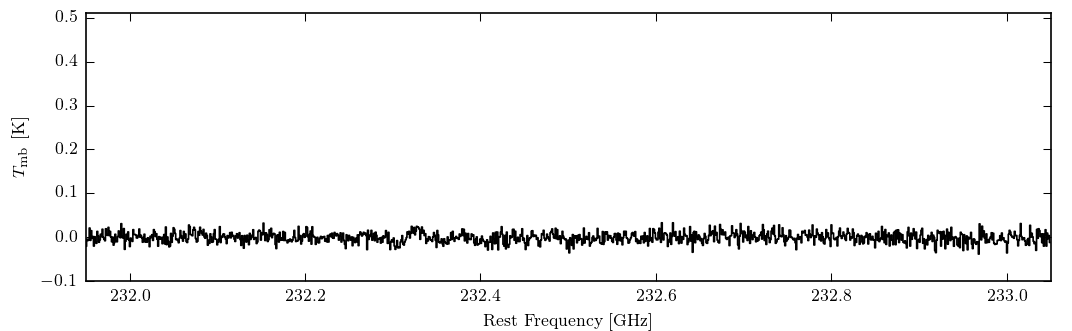} \\
                \includegraphics{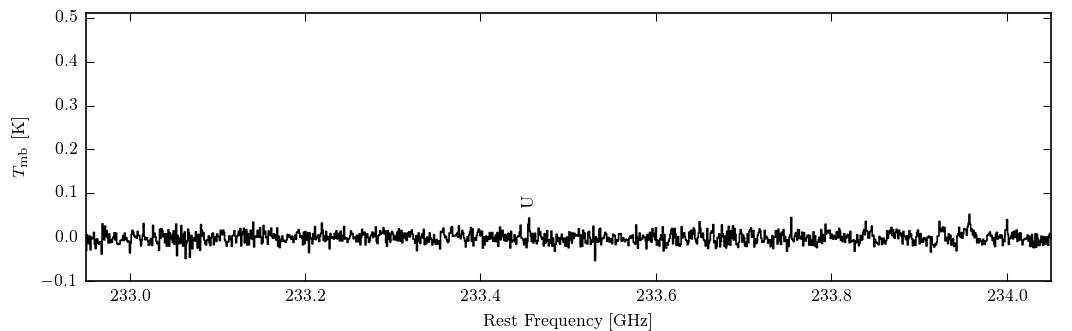} \\
                \includegraphics{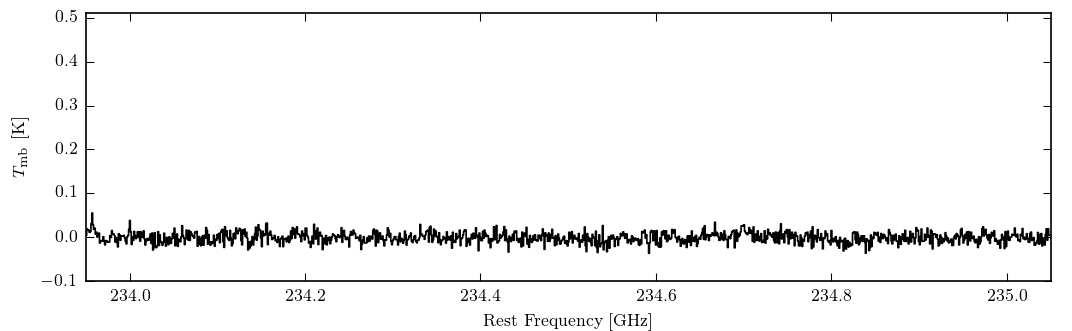} \\
                \includegraphics{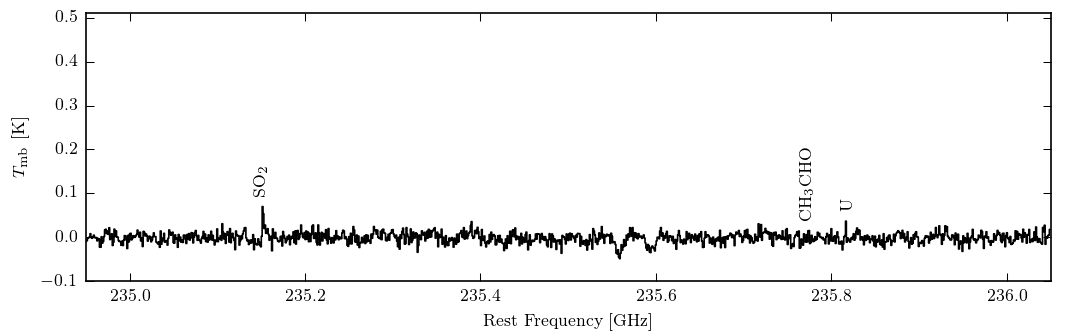} \\
                \end{array}$
                \caption{See Fig.~\ref{fig:largespec1}.}\label{fig:largespec5}
        \end{figure*}
        
        \begin{figure*}[!htb]
                \centering  
                $\begin{array}{c}
                \includegraphics{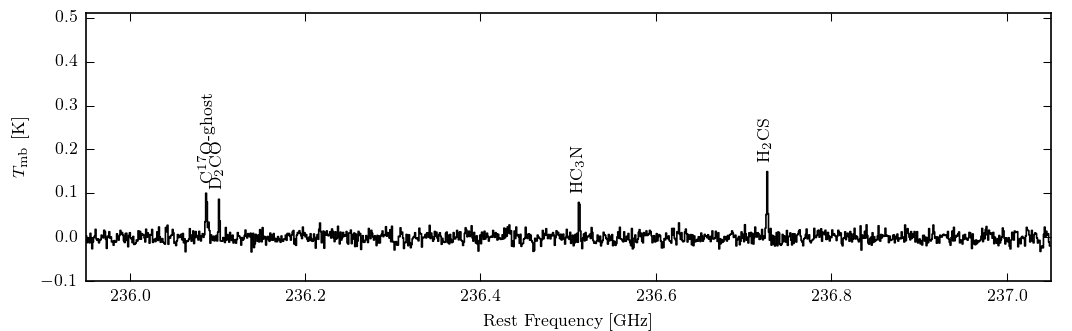} \\
                \includegraphics{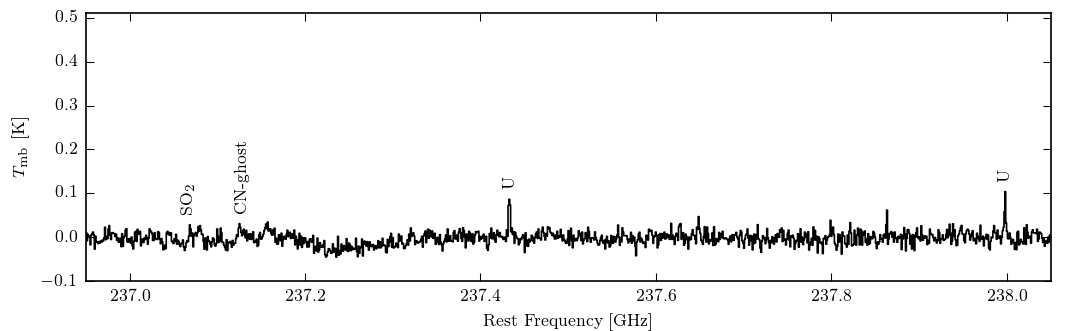} \\
                \includegraphics{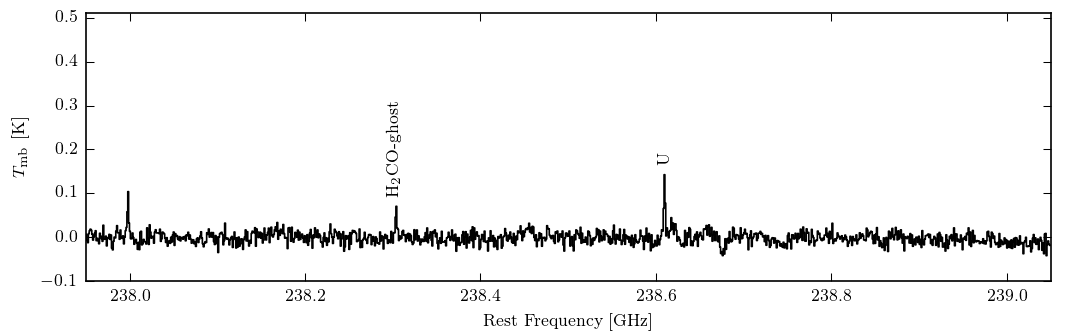} \\
                \includegraphics{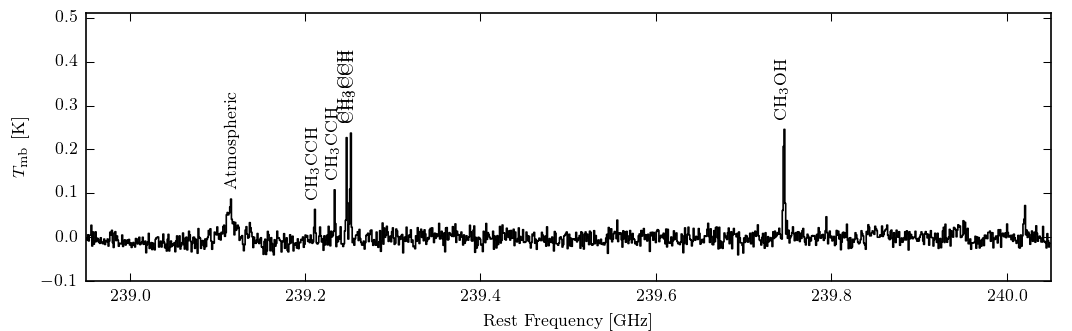} \\
                \end{array}$
                \caption{See Fig.~\ref{fig:largespec1}.}\label{fig:largespec6}
        \end{figure*}
        
        \begin{figure*}[!htb]
                \centering  
                $\begin{array}{c}
                \includegraphics{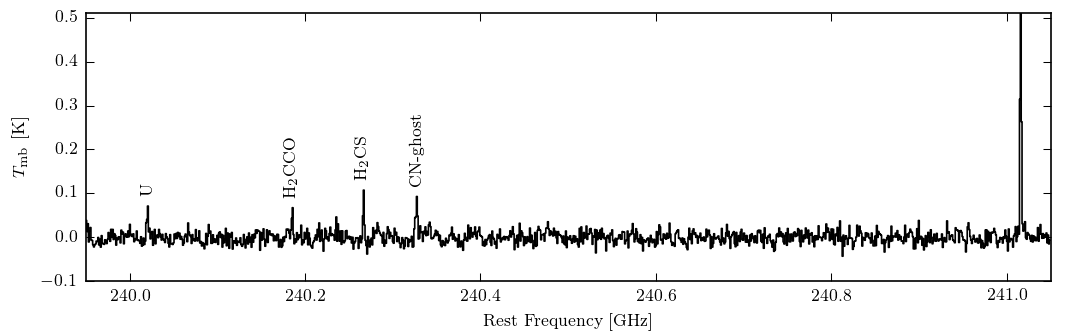} \\
                \includegraphics{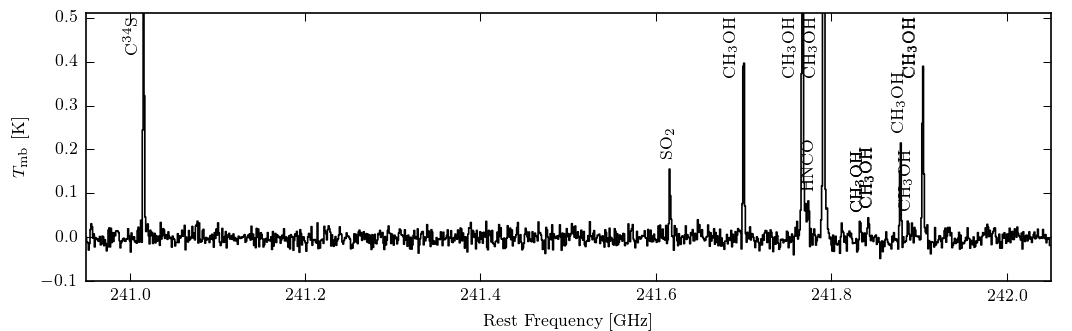} \\
                \includegraphics{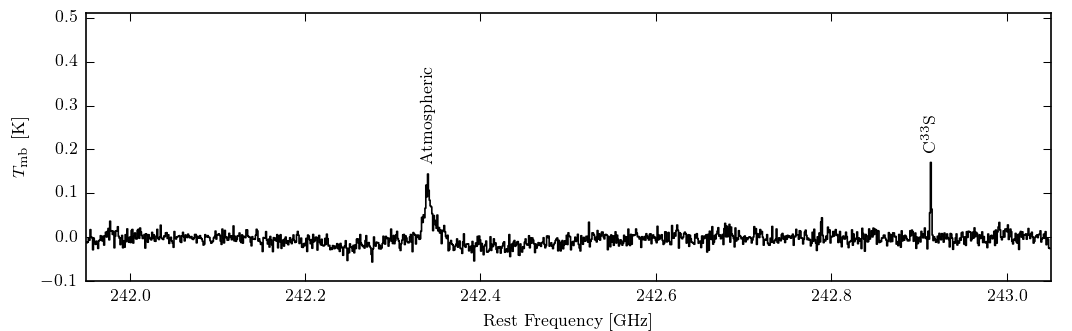} \\
                \includegraphics{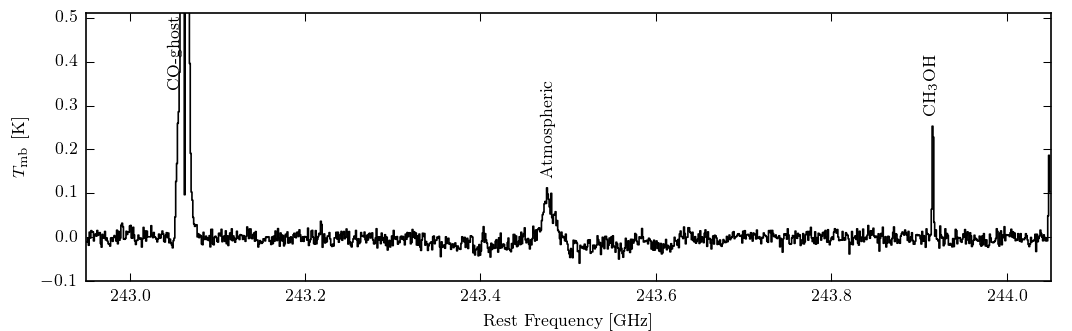} \\
                \end{array}$
                \caption{See Fig.~\ref{fig:largespec1}.}\label{fig:largespec7}
        \end{figure*}
        
        \begin{figure*}[!htb]
                \centering  
                $\begin{array}{c}
                \includegraphics{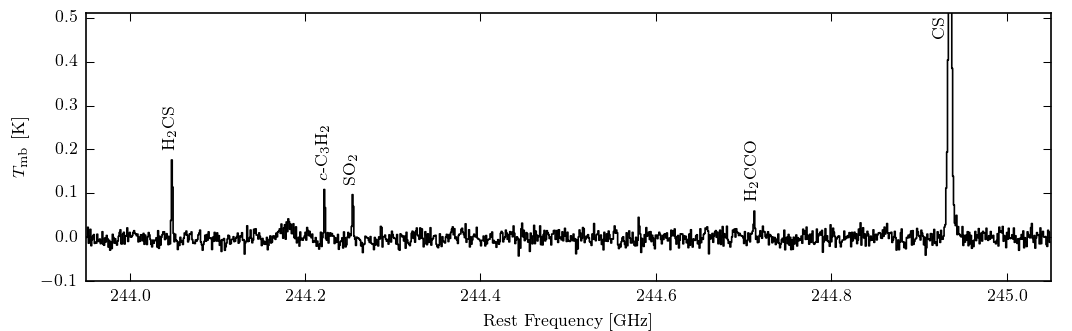} \\
                \includegraphics{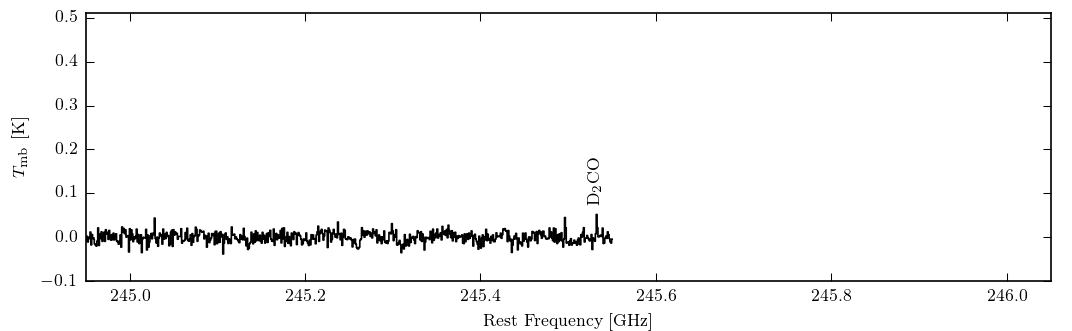} \\
                \end{array}$
                \caption{See Fig.~\ref{fig:largespec1}.}\label{fig:largespec8}
        \end{figure*}
        
        \clearpage
        \section{Complete spectra of the H$_2$CO, CH$_3$OH, and \textit{c}\nobr C$_3$H$_2$ survey}
        \label{app:spectrum_sources}

        \begin{figure*}[!htb]
                \centering  
                $\begin{array}{c}
                \includegraphics{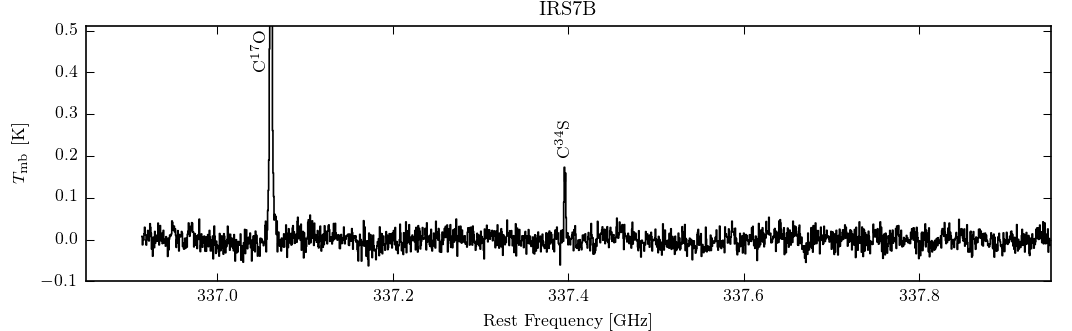} \\
                \includegraphics{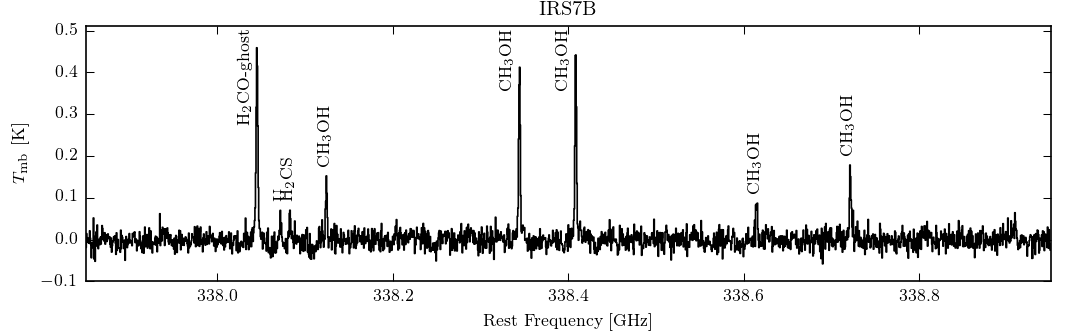} \\
                \includegraphics{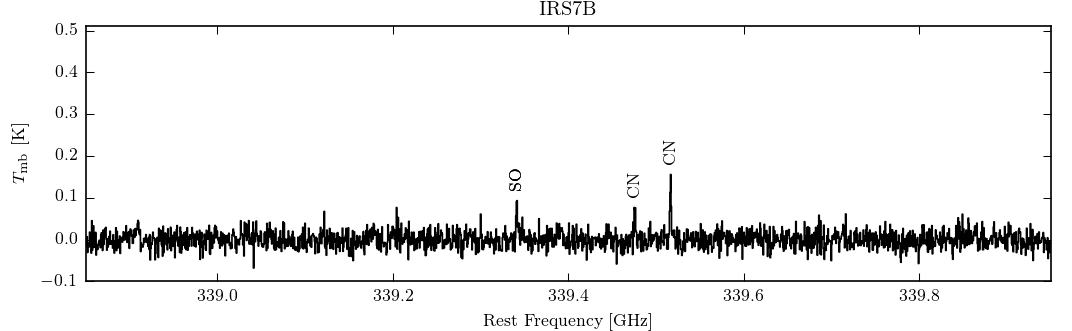} \\
                \includegraphics{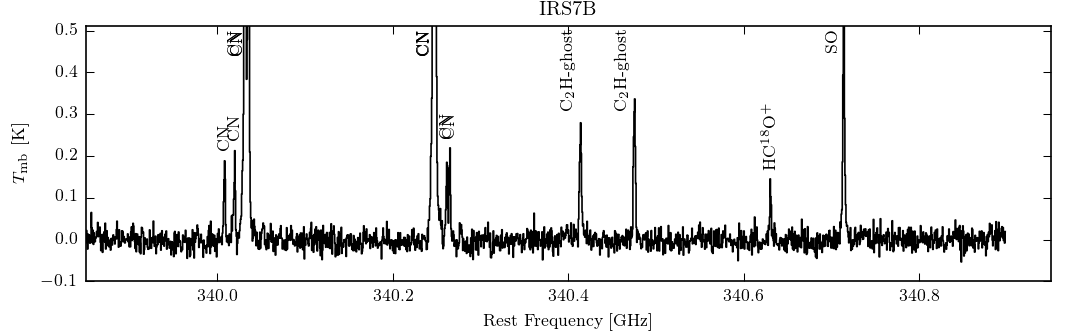} \\
                \end{array}$
                \caption{0.9~mm spectrum of IRS7B, smoothed by a factor 8, corresponding to a channel width of 0.6~\kms.}\label{fig:irs7b_2}
        \end{figure*}
        
        \begin{figure*}[!htb]
                \centering  
                $\begin{array}{c}
                \includegraphics{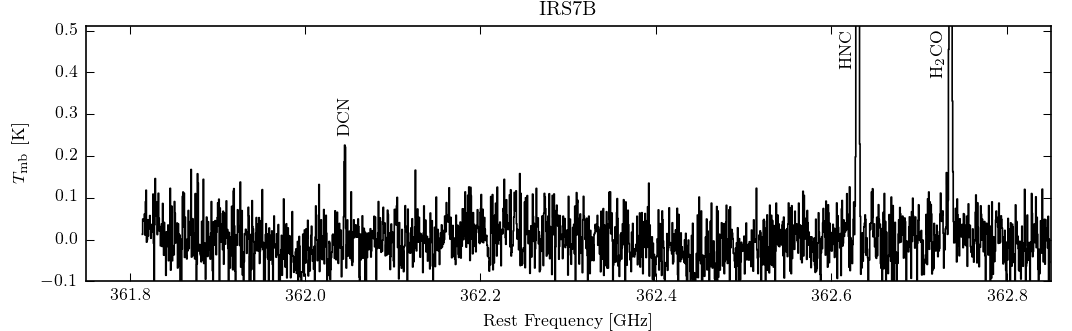} \\
                \includegraphics{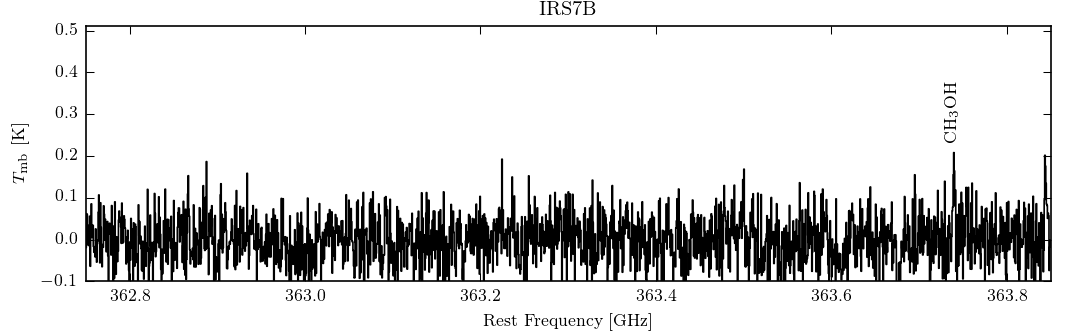} \\
                \includegraphics{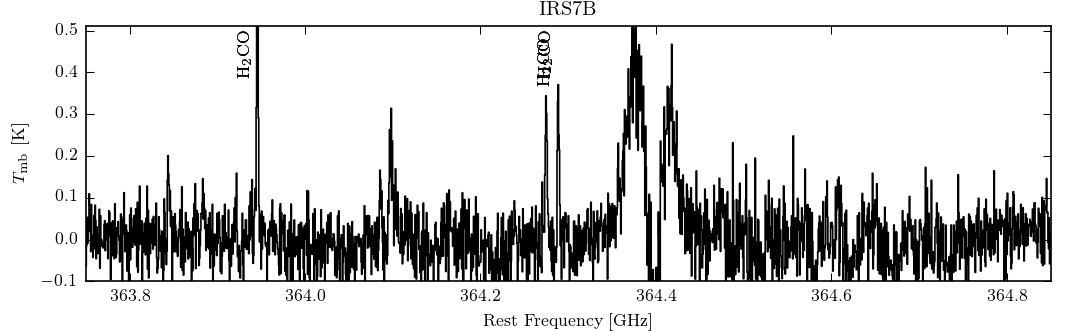} \\
                \includegraphics{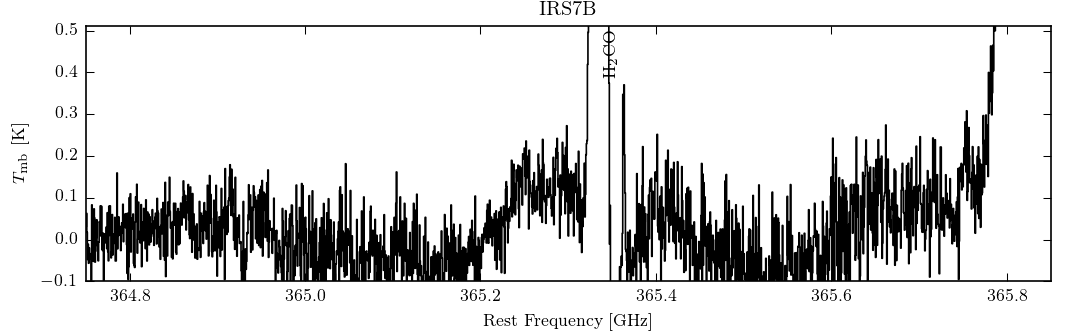} \\
                \end{array}$
                \caption{0.8~mm spectrum of IRS7B, smoothed by a factor 8, corresponding to a channel width of 0.5~\kms. Strong atmospheric lines are visible at 364.1~GHz, 364.4~GHz, and 365.3~GHz.}\label{fig:irs7b_3}
        \end{figure*}
        
        \begin{figure*}[!htb]
                \centering  
                $\begin{array}{c}
                \includegraphics{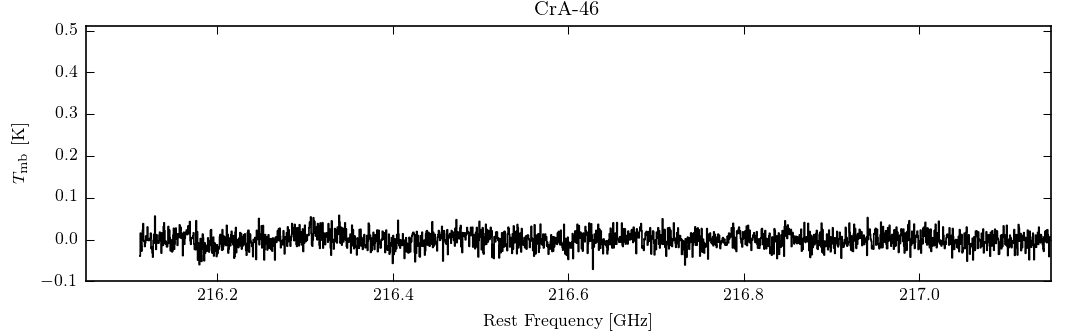} \\
                \includegraphics{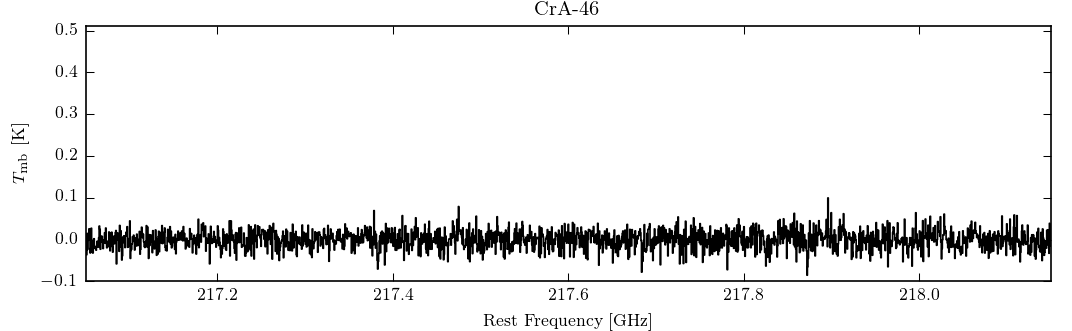} \\
                \includegraphics{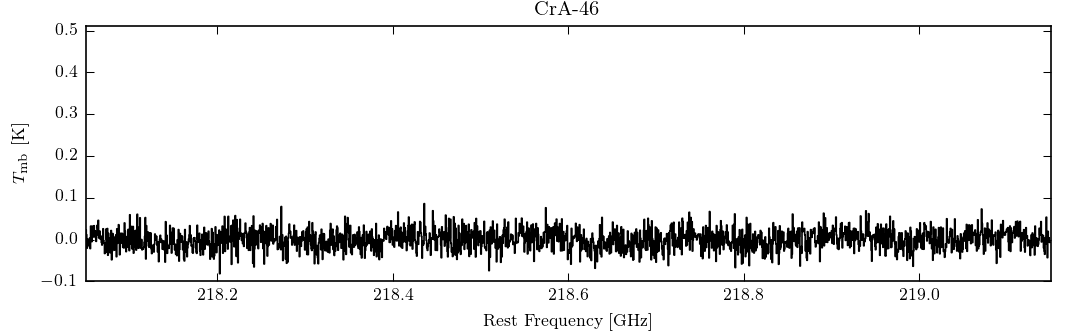} \\
                \includegraphics{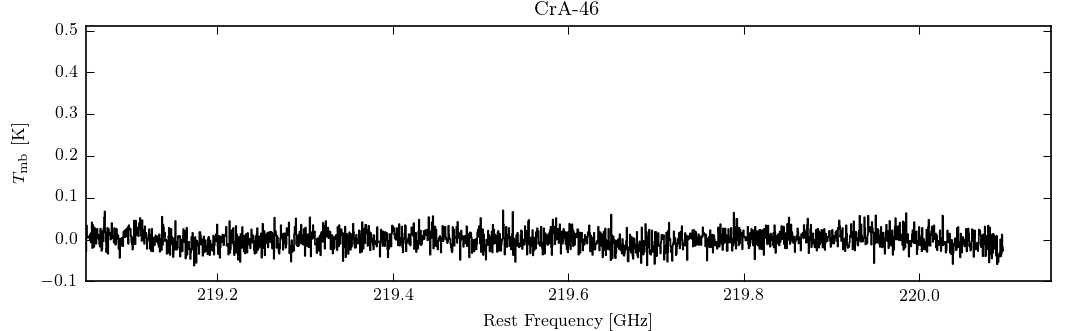} \\
                \end{array}$
                \caption{1.4~mm spectrum of CrA-46, smoothed by a factor 8, corresponding to a channel width of 0.8~\kms.}\label{fig:cra46_1}
        \end{figure*}
        
        \begin{figure*}[!htb]
                \centering  
                $\begin{array}{c}
                \includegraphics{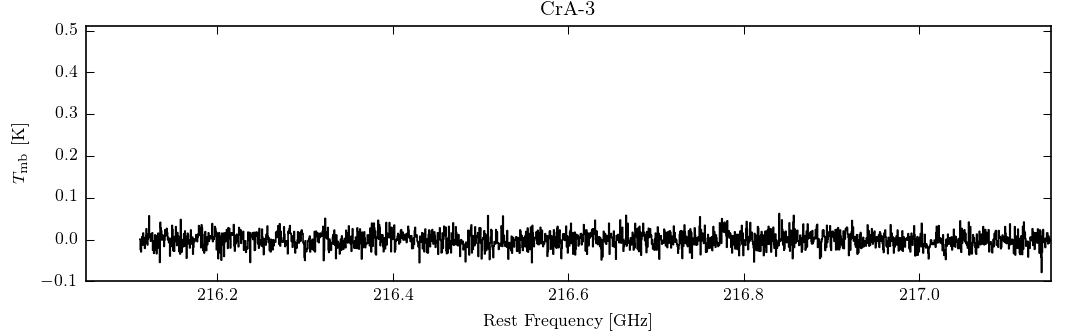} \\
                \includegraphics{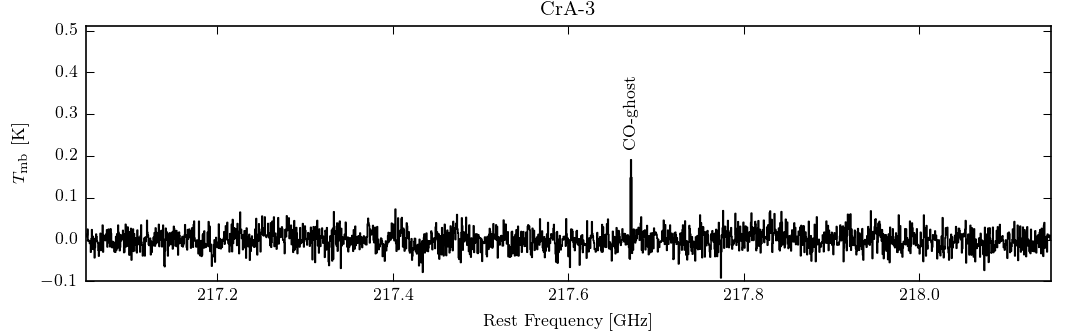} \\
                \includegraphics{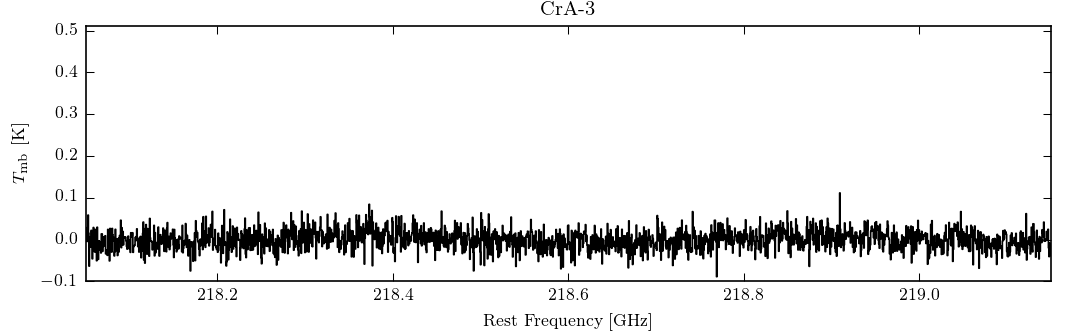} \\
                \includegraphics{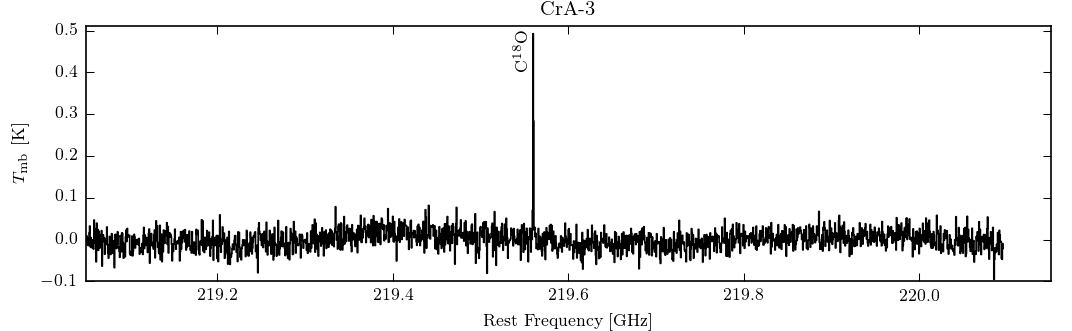} \\
                \end{array}$
                \caption{1.4~mm spectrum of CrA-3, smoothed by a factor 8, corresponding to a channel width of 0.8~\kms.}\label{fig:cra3_1}
        \end{figure*}

        \begin{figure*}[!htb]
        	\centering  
        	$\begin{array}{c}
        	\includegraphics{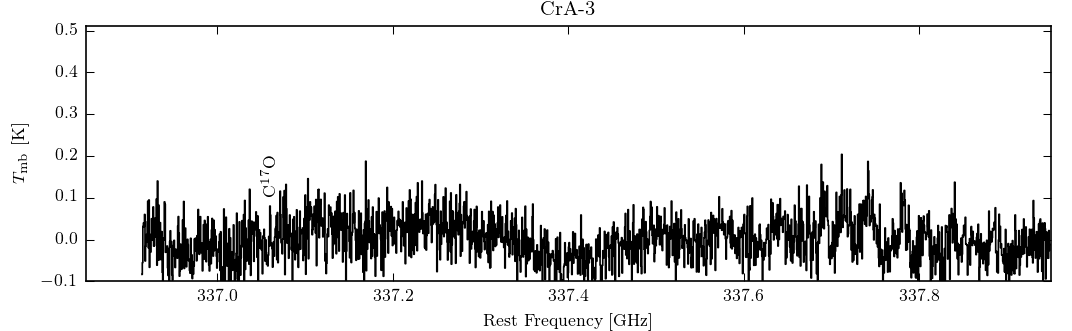} \\
        	\includegraphics{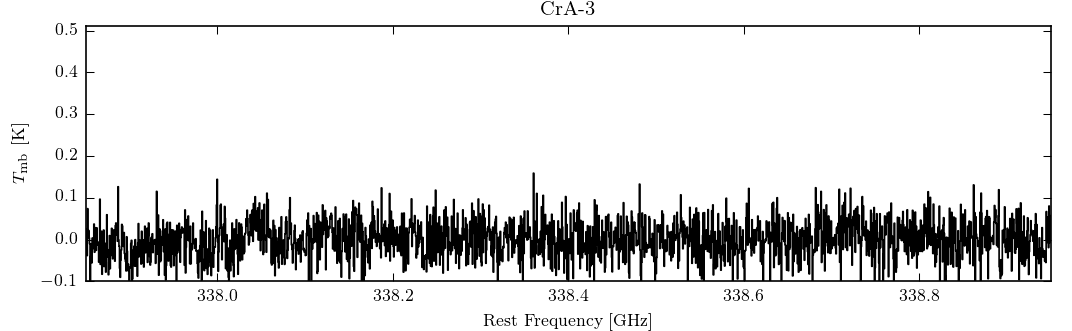} \\
        	\includegraphics{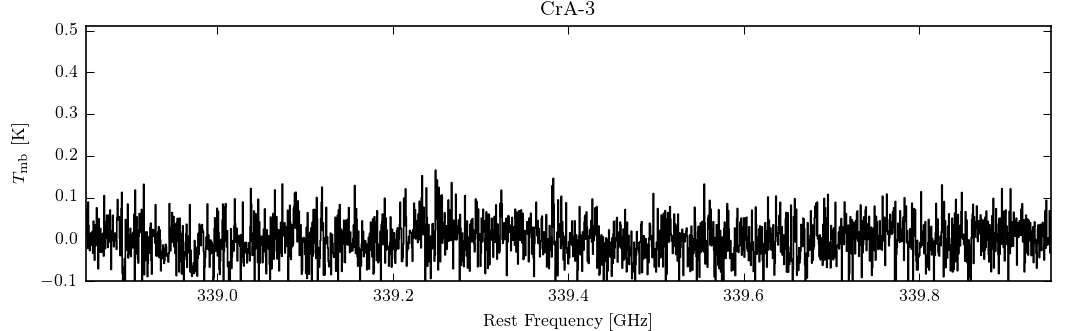} \\
        	\includegraphics{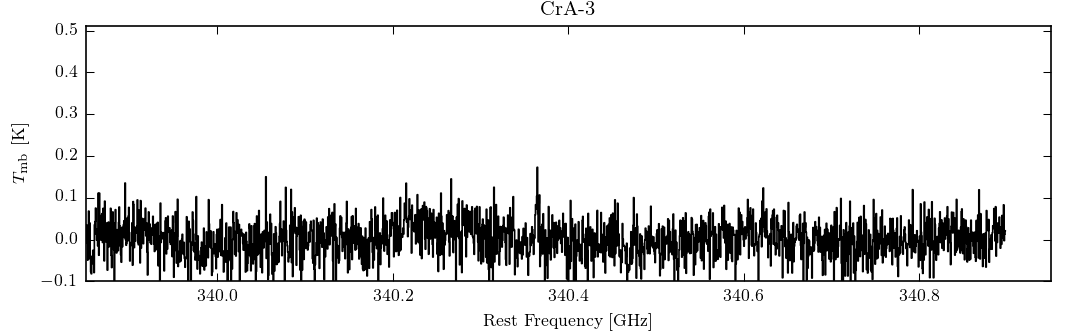} \\
        	\end{array}$
        	\caption{0.9~mm spectrum of CrA-3, smoothed by a factor 8, corresponding to a channel width of 0.6~\kms.}\label{fig:cra3_2}
        \end{figure*}
        
        \begin{figure*}[!htb]
                \centering  
                $\begin{array}{c}
                \includegraphics{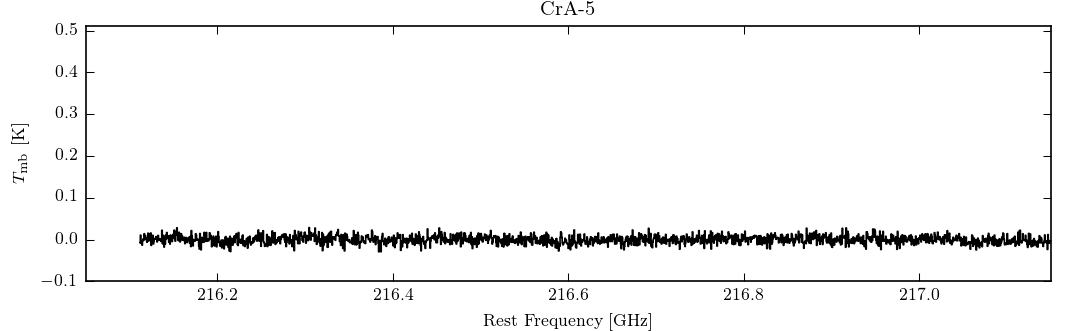} \\
                \includegraphics{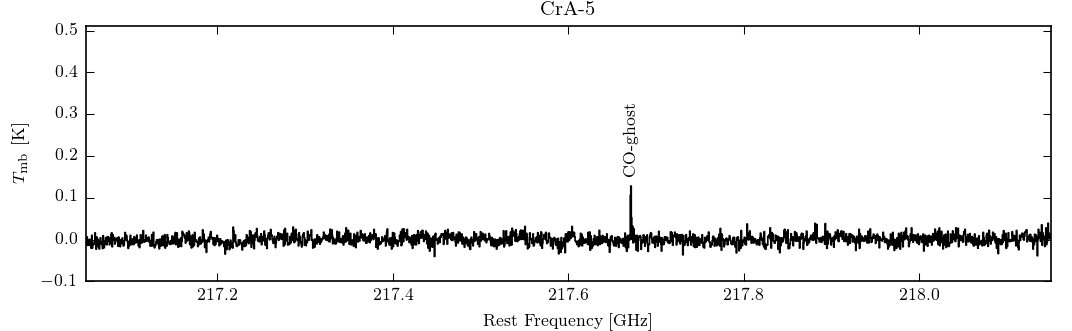} \\
                \includegraphics{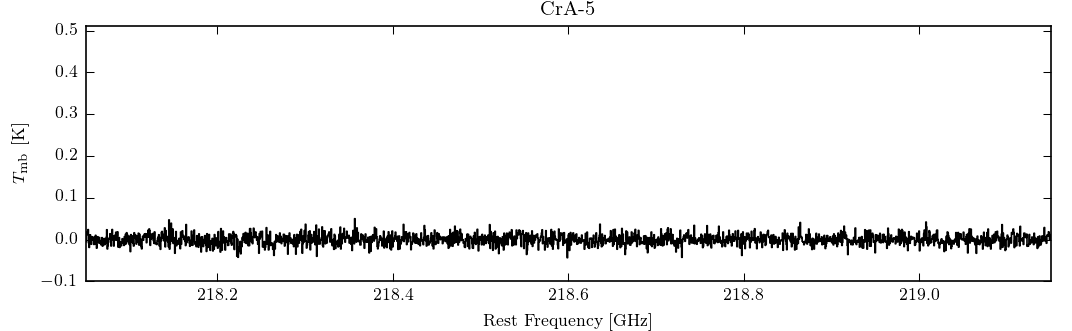} \\
                \includegraphics{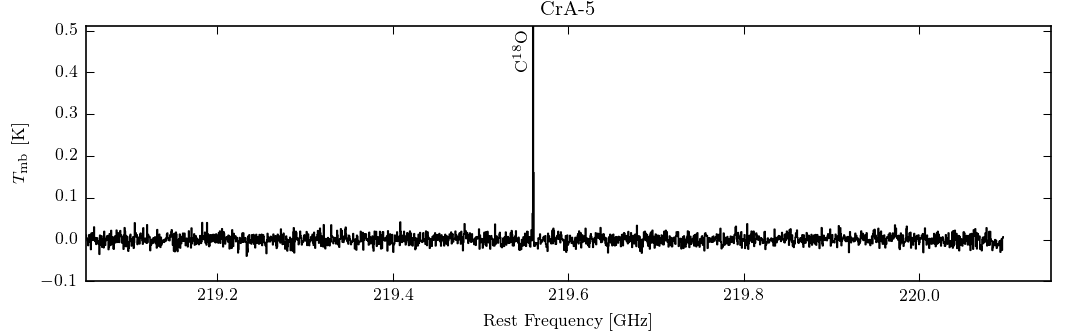} \\
                \end{array}$
                \caption{1.4~mm spectrum of CrA-5, smoothed by a factor 8, corresponding to a channel width of 0.8~\kms.}\label{fig:cra5_1}
        \end{figure*}
        
        \begin{figure*}[!htb]
                \centering  
                $\begin{array}{c}
                \includegraphics{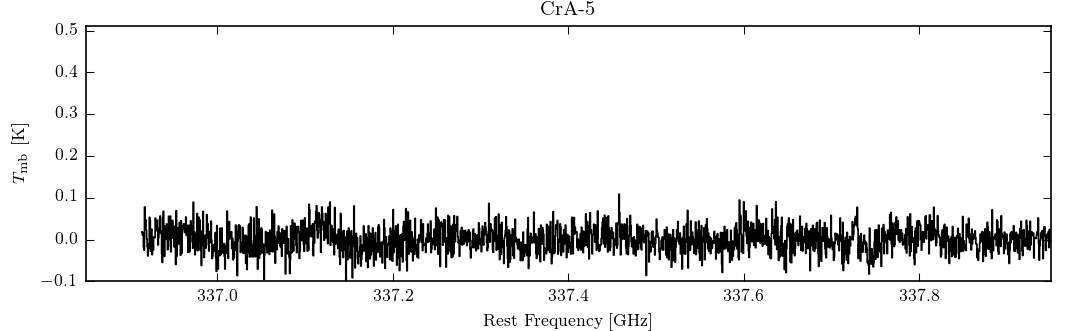} \\
                \includegraphics{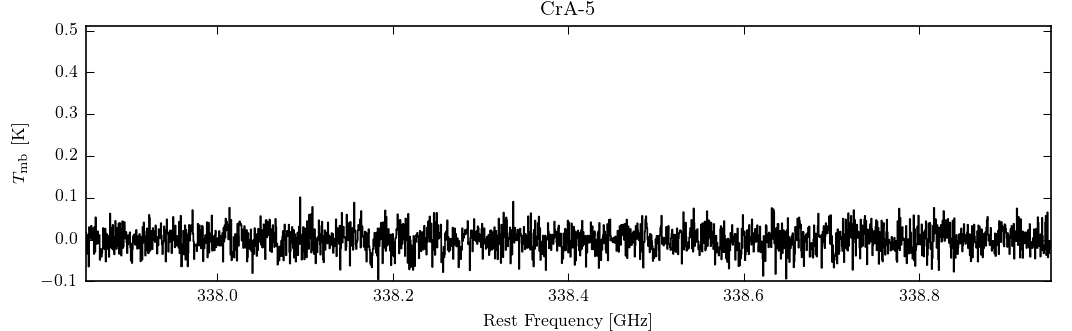} \\
                \includegraphics{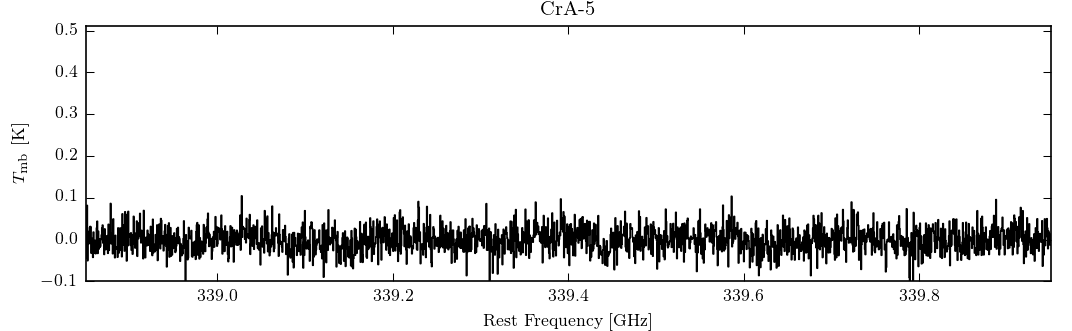} \\
                \includegraphics{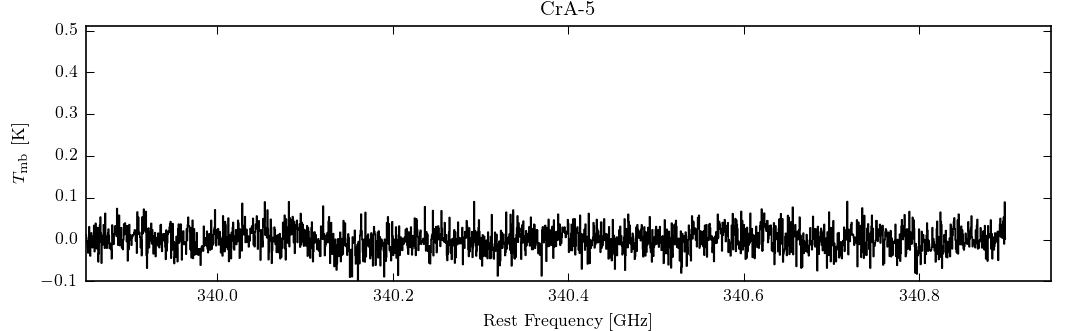} \\
                \end{array}$
                \caption{0.9~mm spectrum of CrA-5, smoothed by a factor 8, corresponding to a channel width of 0.6~\kms.}\label{fig:cra5_2}
        \end{figure*}
        
        \begin{figure*}[!htb]
                \centering  
                $\begin{array}{c}
                \includegraphics{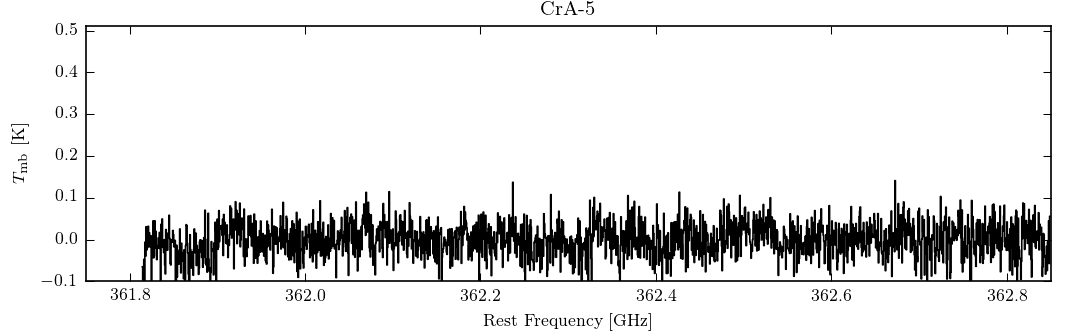} \\
                \includegraphics{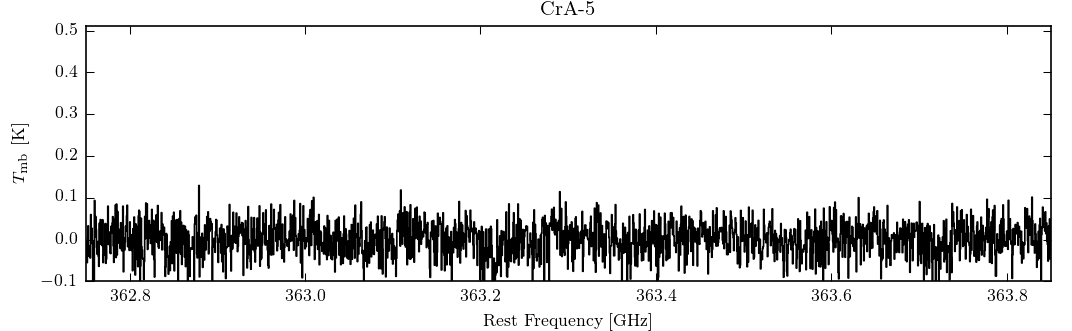} \\
                \includegraphics{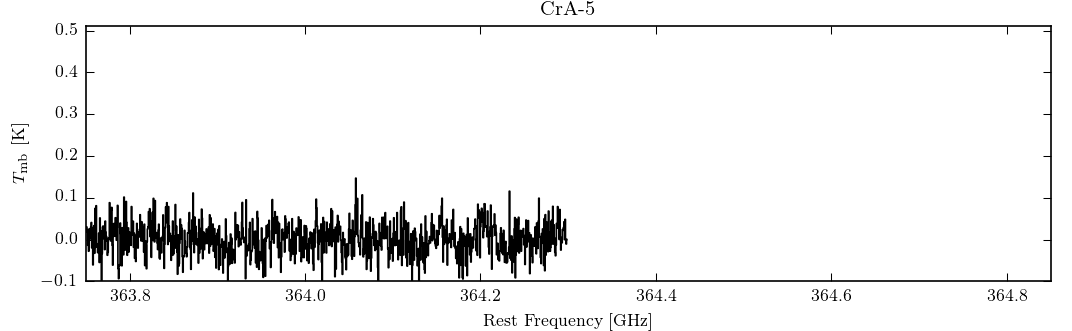} \\
                \end{array}$
                \caption{0.8~mm spectrum of CrA-5, smoothed by a factor 8, corresponding to a channel width of 0.5~\kms.}\label{fig:cra5_3}
        \end{figure*}
        
        \begin{figure*}[!htb]
                \centering  
                $\begin{array}{c}
                \includegraphics{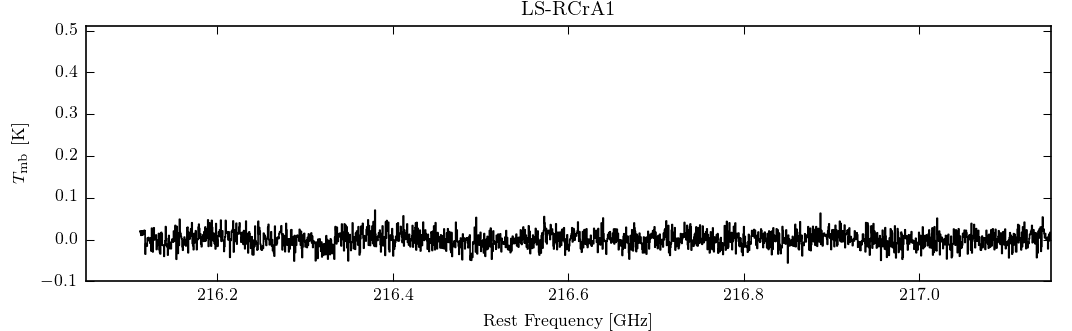} \\
                \includegraphics{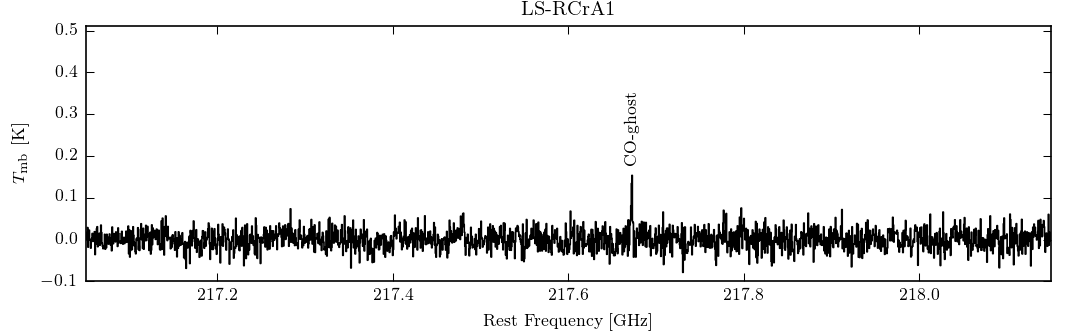} \\
                \includegraphics{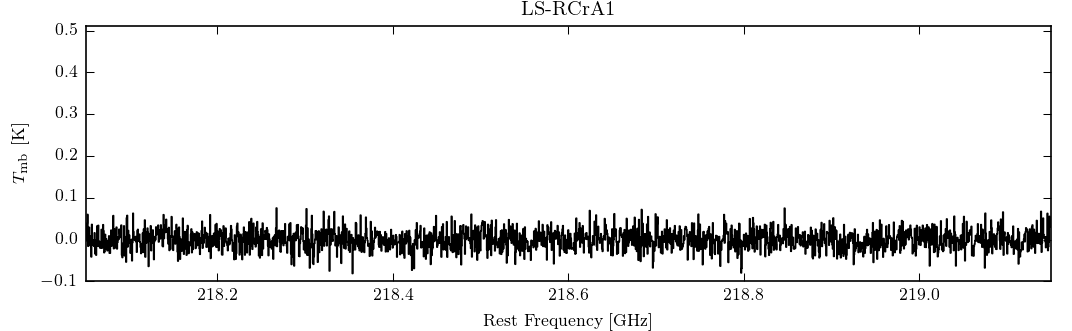} \\
                \includegraphics{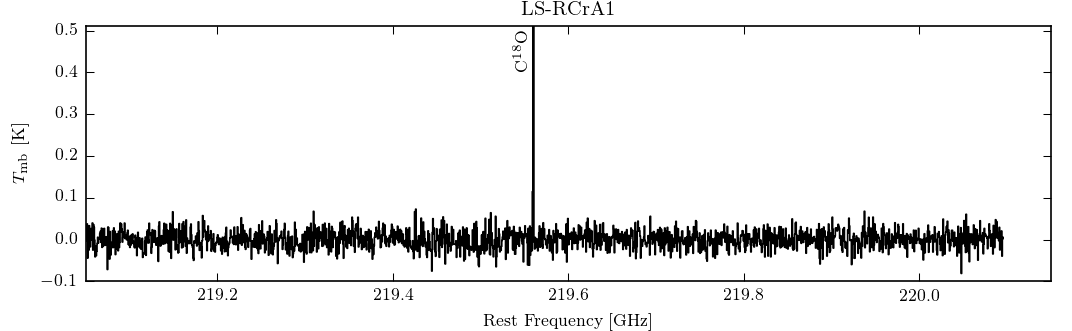} \\
                \end{array}$
                \caption{1.4~mm spectrum of LS-RCrA1, smoothed by a factor 8, corresponding to a channel width of 0.8~\kms.}\label{fig:lsrcra1_1}
        \end{figure*}
        
        \begin{figure*}[!htb]
                \centering  
                $\begin{array}{c}
                \includegraphics{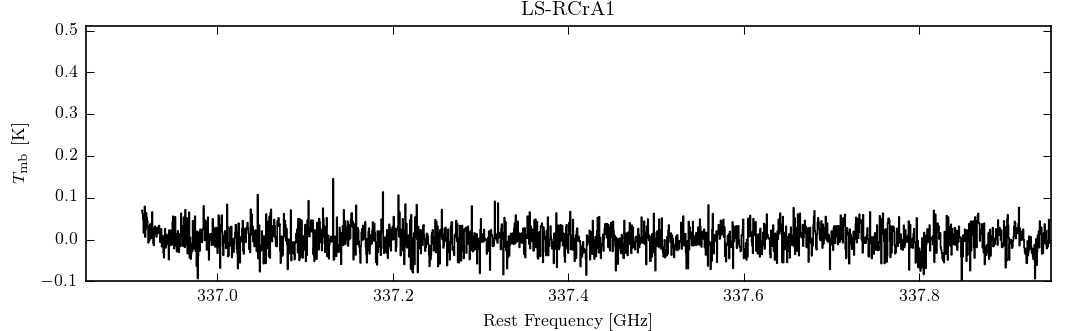} \\
                \includegraphics{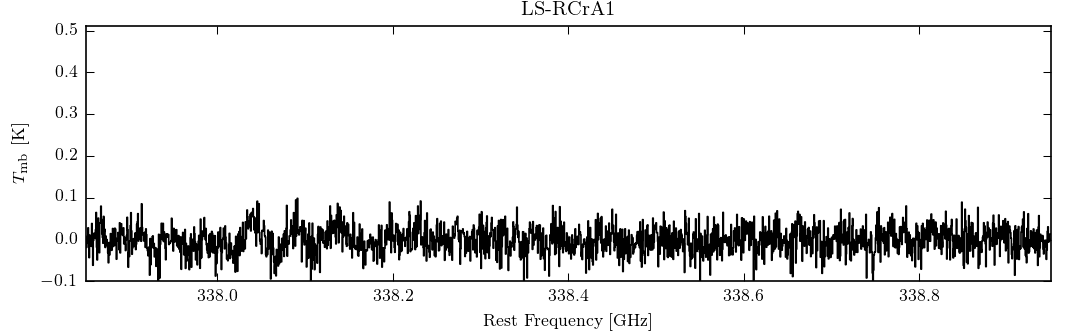} \\
                \includegraphics{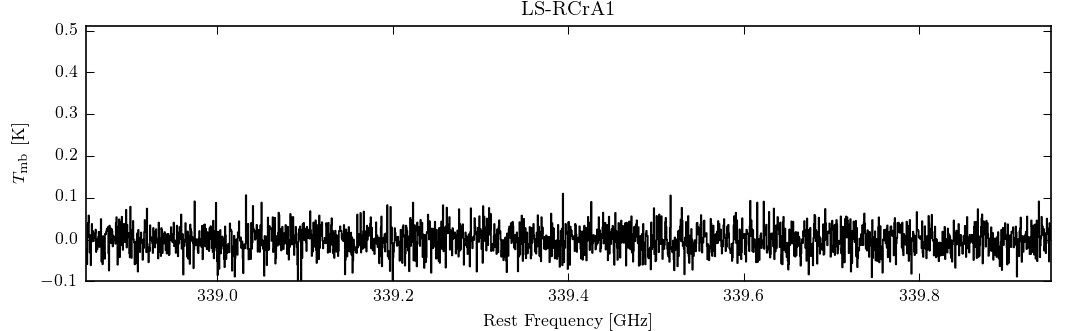} \\
                \includegraphics{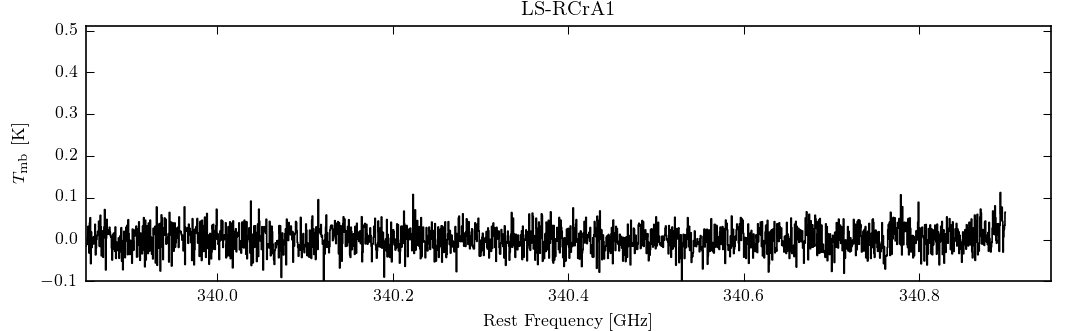} \\
                \end{array}$
                \caption{0.9~mm spectrum of LS-RCrA1, smoothed by a factor 8, corresponding to a channel width of 0.6~\kms.}\label{fig:lsrcra1_2}
        \end{figure*}
        
        \begin{figure*}[!htb]
                \centering  
                $\begin{array}{c}
                \includegraphics{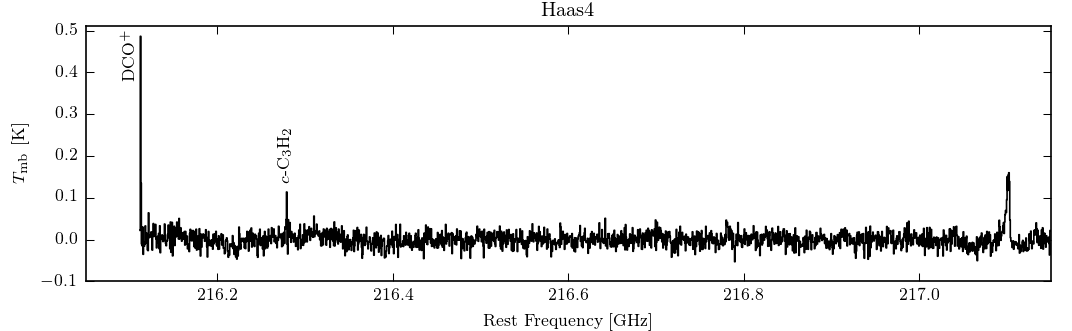} \\
                \includegraphics{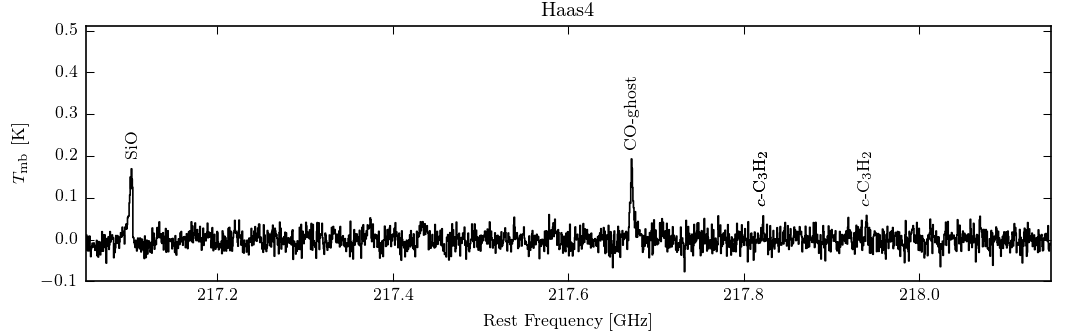} \\
                \includegraphics{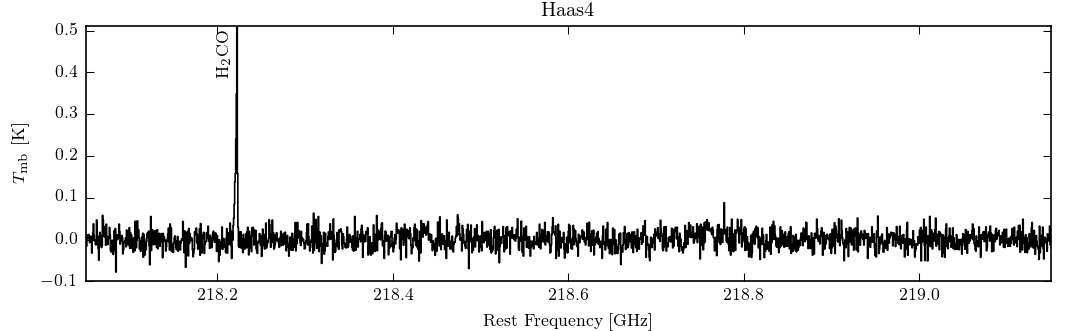} \\
                \includegraphics{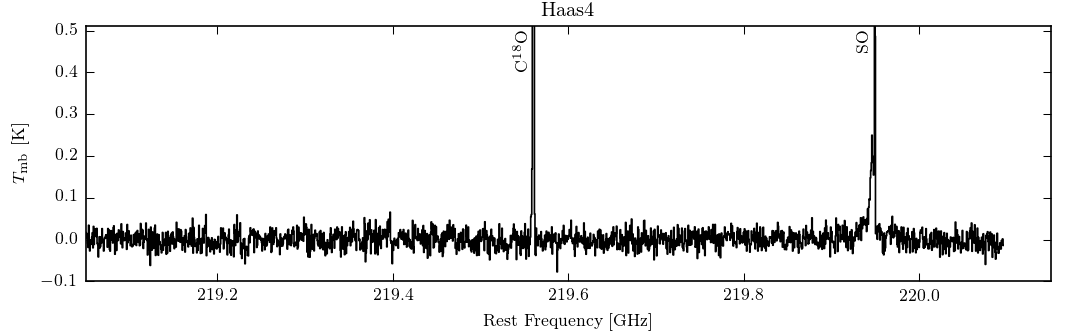} \\
                \end{array}$
                \caption{1.4~mm spectrum of Haas4, smoothed by a factor 8, corresponding to a channel width of 0.8~\kms.}\label{fig:haas4_1}
        \end{figure*}
        
        \begin{figure*}[!htb]
                \centering  
                $\begin{array}{c}
                \includegraphics{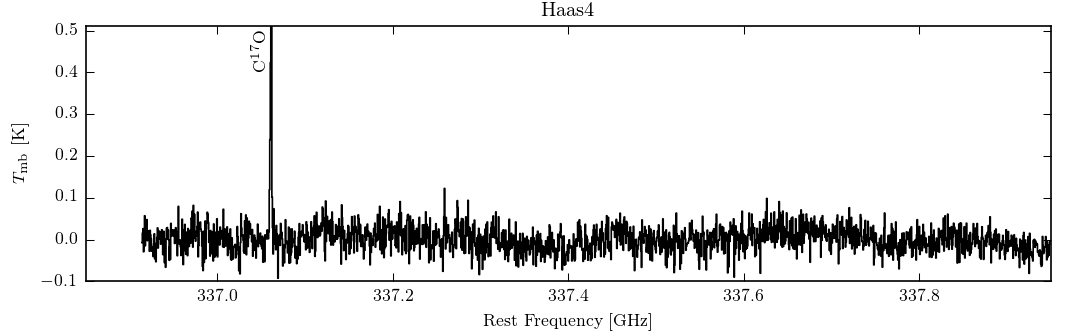} \\
                \includegraphics{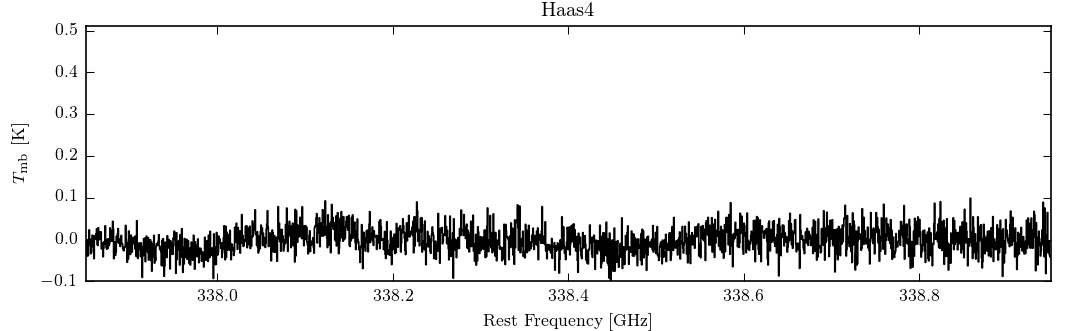} \\
                \includegraphics{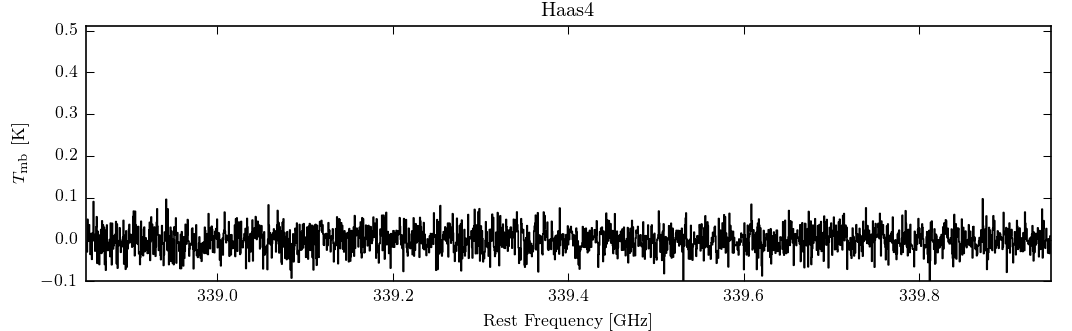} \\
                \includegraphics{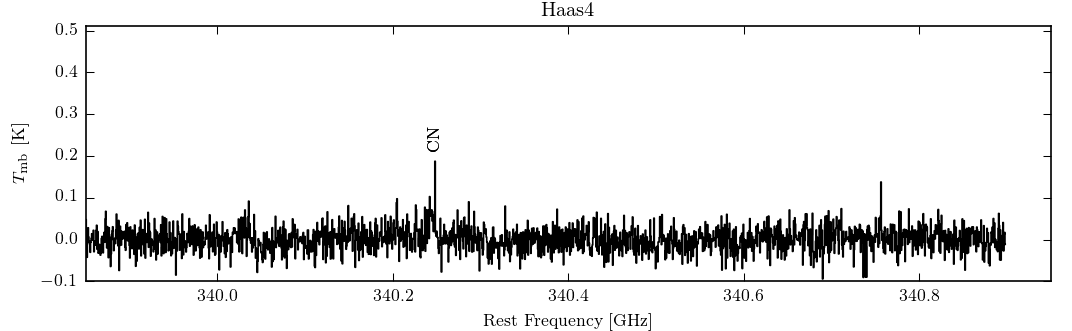} \\
                \end{array}$
                \caption{0.9~mm spectrum of Haas4, smoothed by a factor 8, corresponding to a channel width of 0.6~\kms.}\label{fig:haas4_2}
        \end{figure*}
        
        \begin{figure*}[!htb]
                \centering  
                $\begin{array}{c}
                \includegraphics{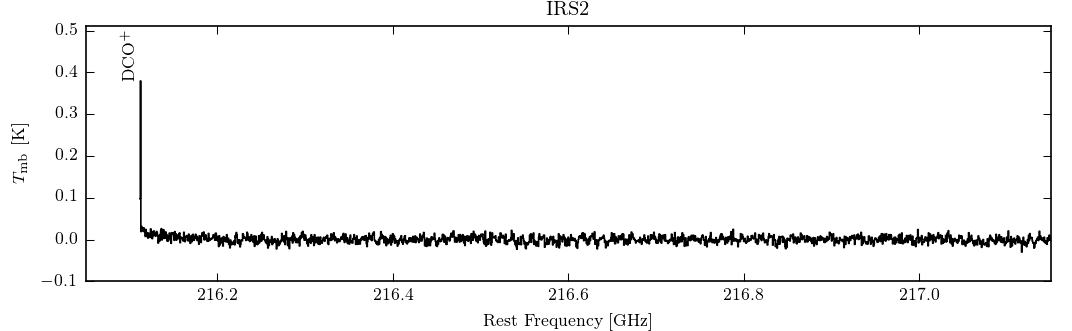} \\
                \includegraphics{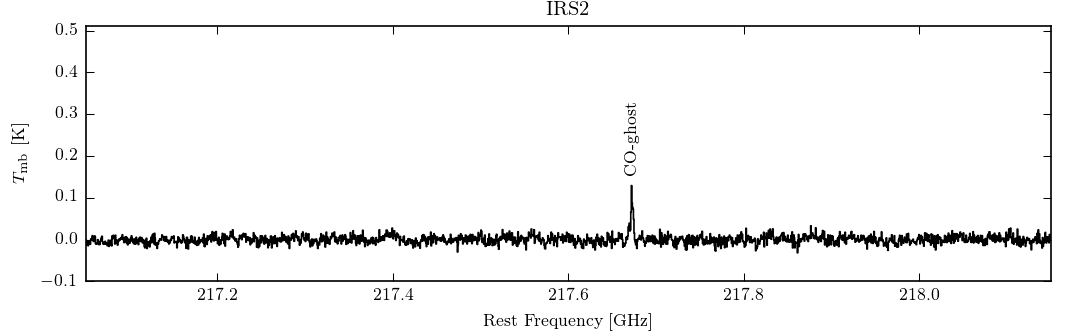} \\
                \includegraphics{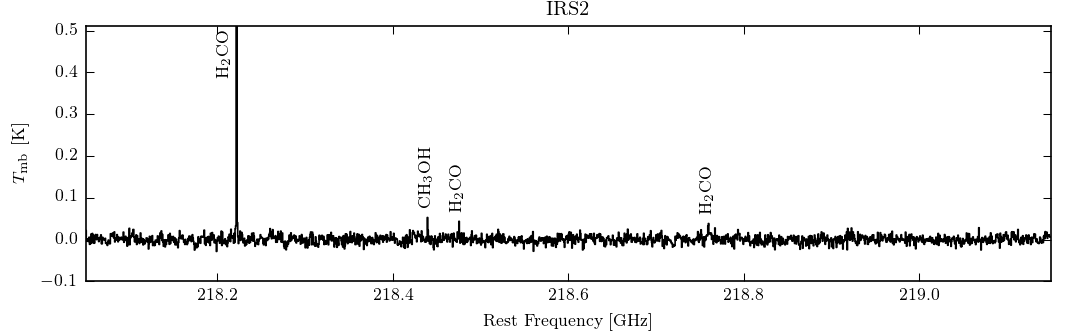} \\
                \includegraphics{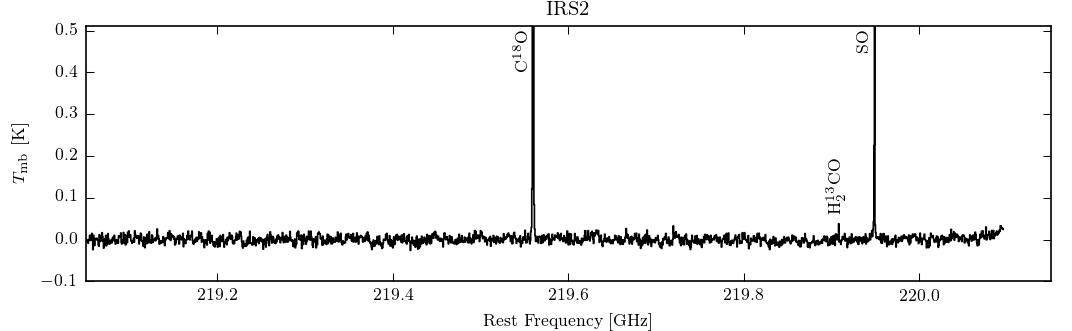} \\
                \end{array}$
                \caption{1.4~mm spectrum of IRS2, smoothed by a factor 8, corresponding to a channel width of 0.8~\kms.}\label{fig:irs2_1}
        \end{figure*}
        
        \begin{figure*}[!htb]
                \centering  
                $\begin{array}{c}
                \includegraphics{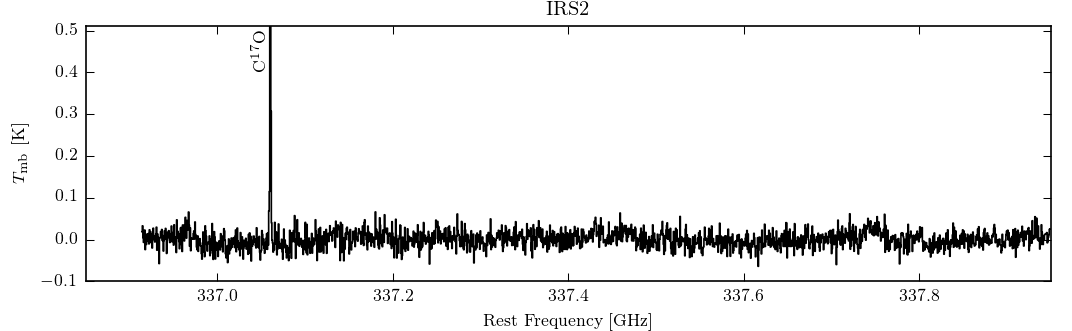} \\
                \includegraphics{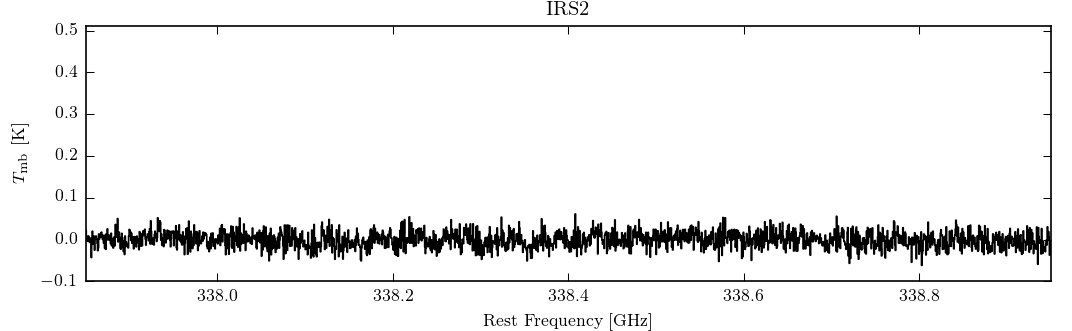} \\
                \includegraphics{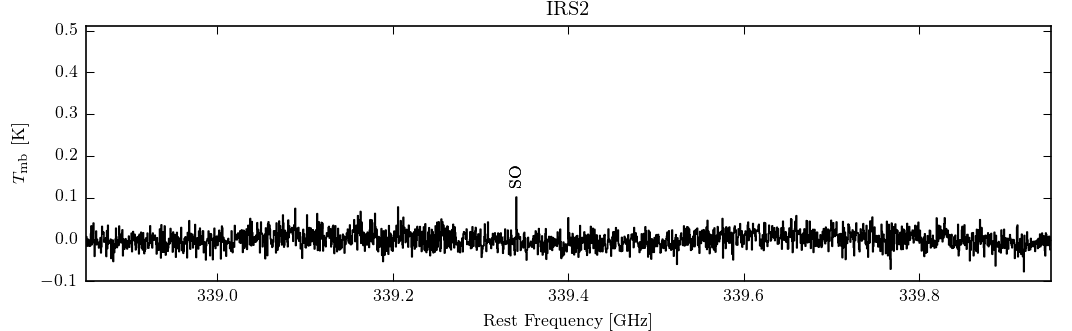} \\
                \includegraphics{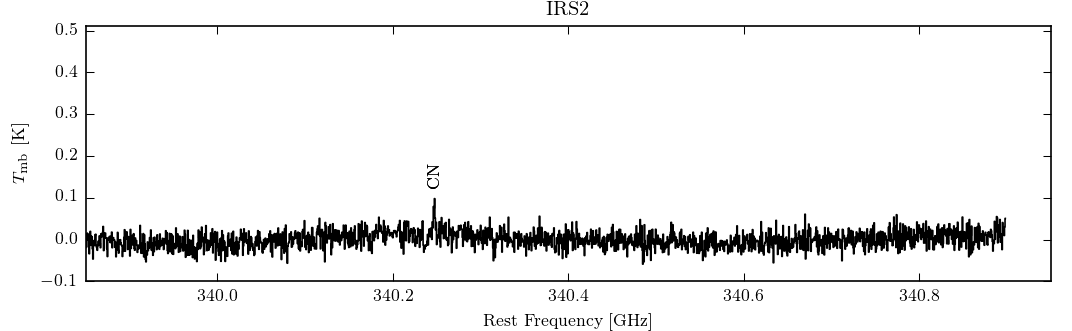} \\
                \end{array}$
                \caption{0.9~mm spectrum of IRS2, smoothed by a factor 8, corresponding to a channel width of 0.6~\kms.}\label{fig:irs2_2}
        \end{figure*}
        
        \begin{figure*}[!htb]
                \centering  
                $\begin{array}{c}
                \includegraphics{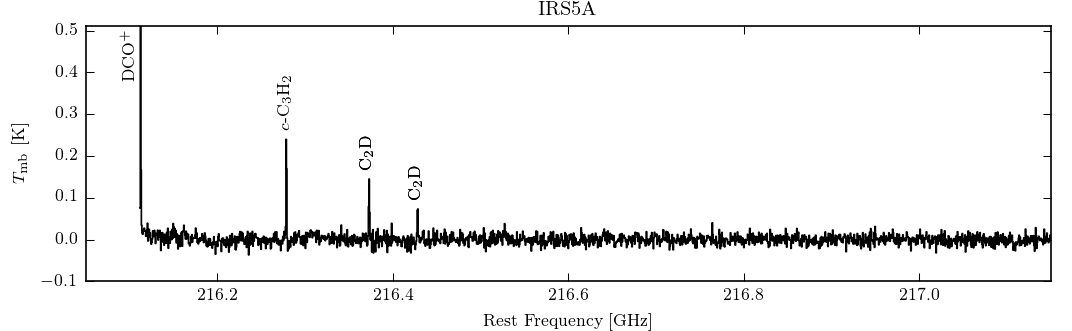} \\
                \includegraphics{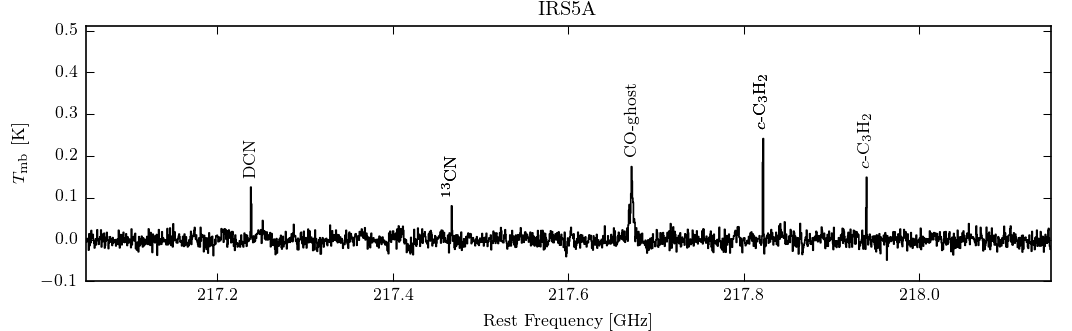} \\
                \includegraphics{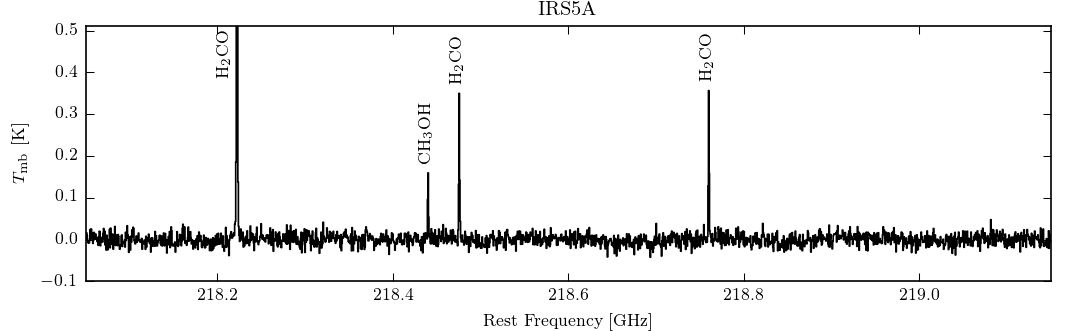} \\
                \includegraphics{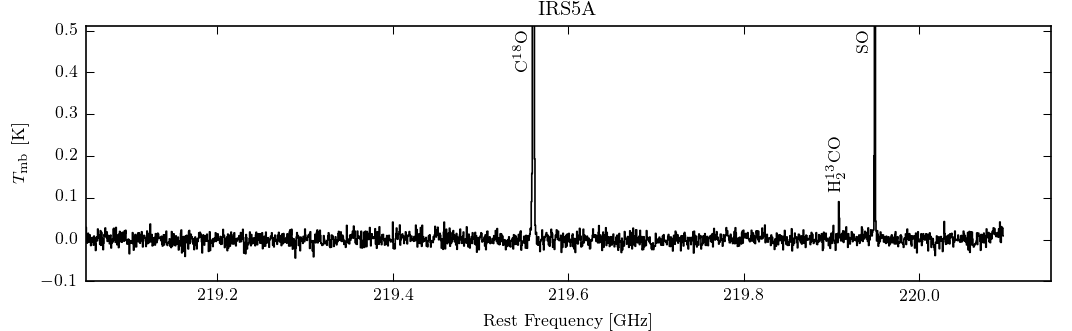} \\
                \end{array}$
                \caption{1.4~mm spectrum of IRS5A, smoothed by a factor 8, corresponding to a channel width of 0.8~\kms.}\label{fig:irs5_1}
        \end{figure*}
        
        \begin{figure*}[!htb]
                \centering  
                $\begin{array}{c}
                \includegraphics{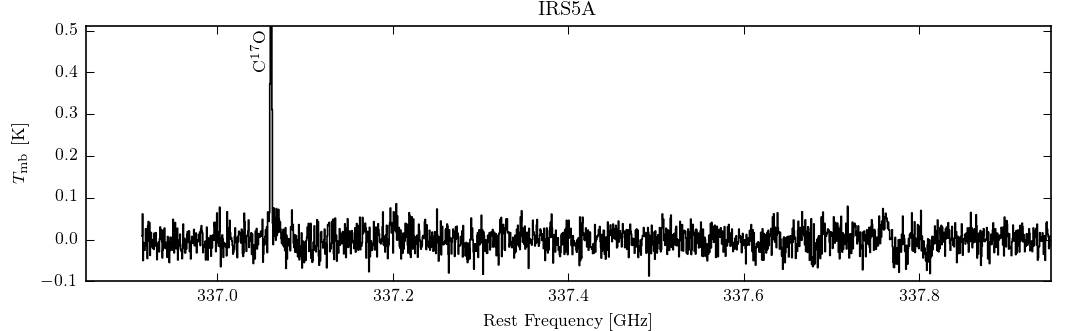} \\
                \includegraphics{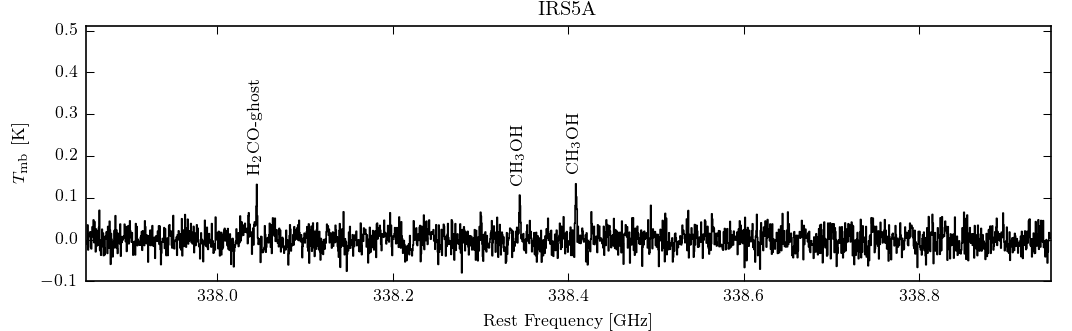} \\
                \includegraphics{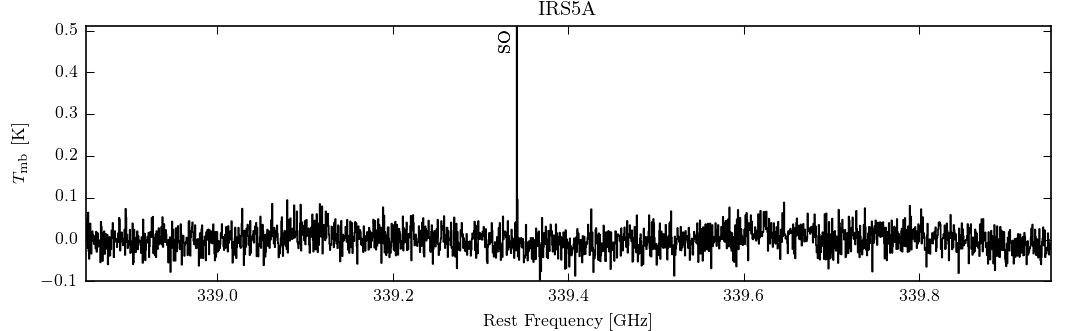} \\
                \includegraphics{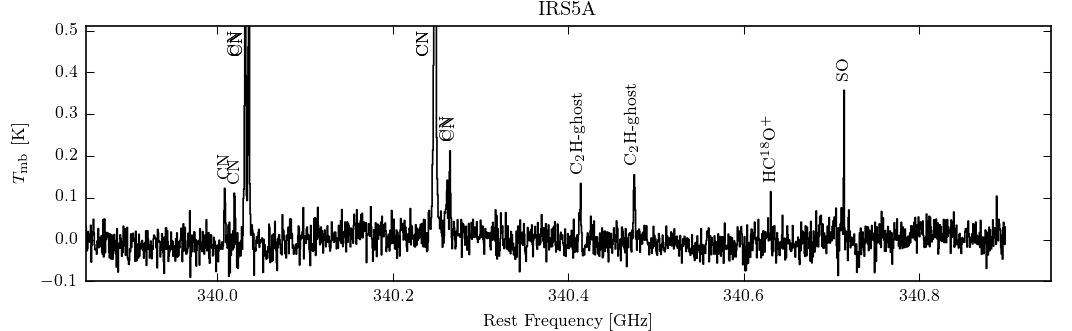} \\
                \end{array}$
                \caption{0.9~mm spectrum of IRS5A, smoothed by a factor 8, corresponding to a channel width of 0.6~\kms.}\label{fig:irs5_2}
        \end{figure*}

        \begin{figure*}[!htb]
                \centering  
                $\begin{array}{c}
                \includegraphics{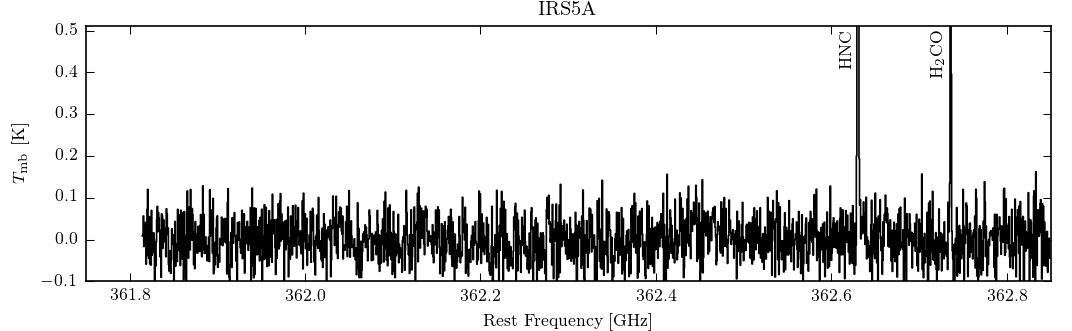} \\
                \includegraphics{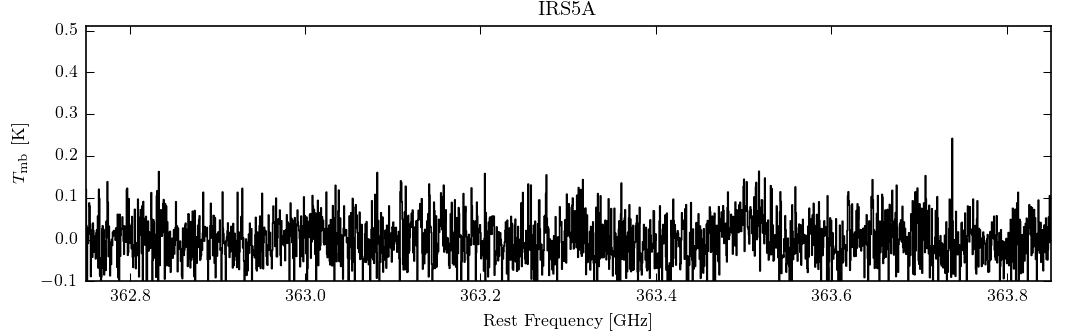} \\
                \includegraphics{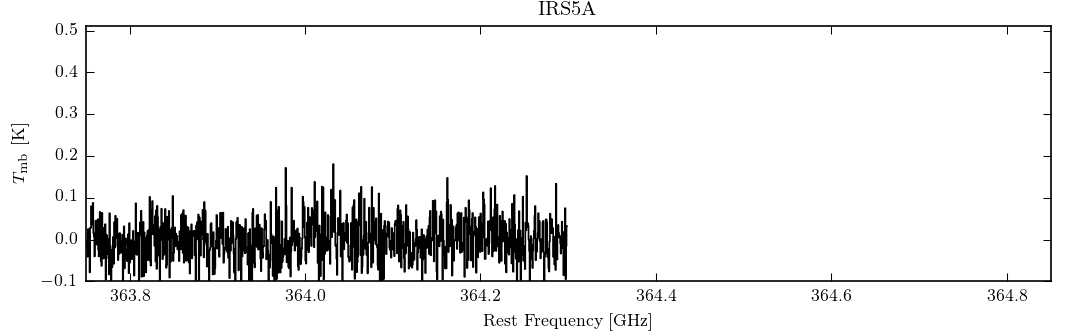} \\
                \end{array}$
                \caption{0.8~mm spectrum of IRS5A, smoothed by a factor 8, corresponding to a channel width of 0.5~\kms.}\label{fig:irs5_3}
        \end{figure*}

        \begin{figure*}[!htb]
                \centering  
                $\begin{array}{c}
                \includegraphics{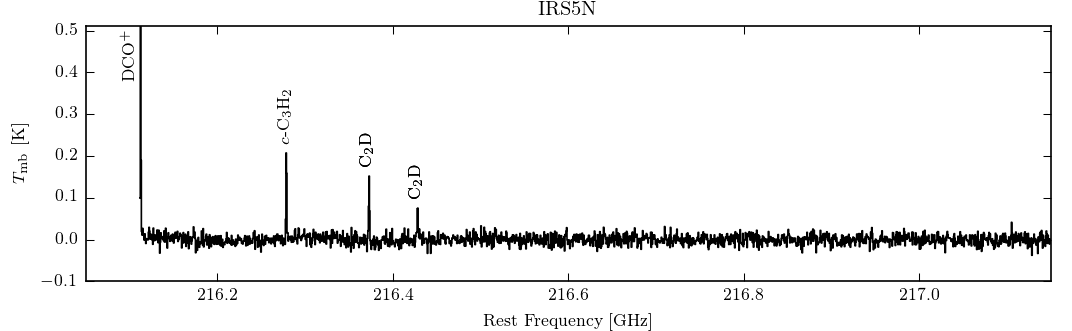} \\
                \includegraphics{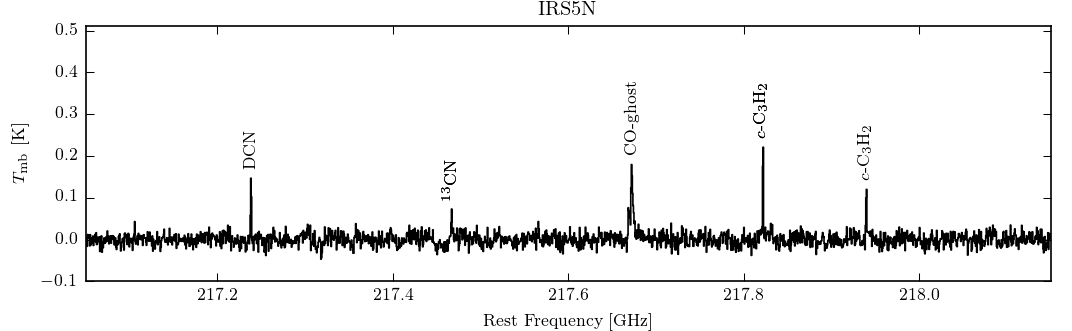} \\
                \includegraphics{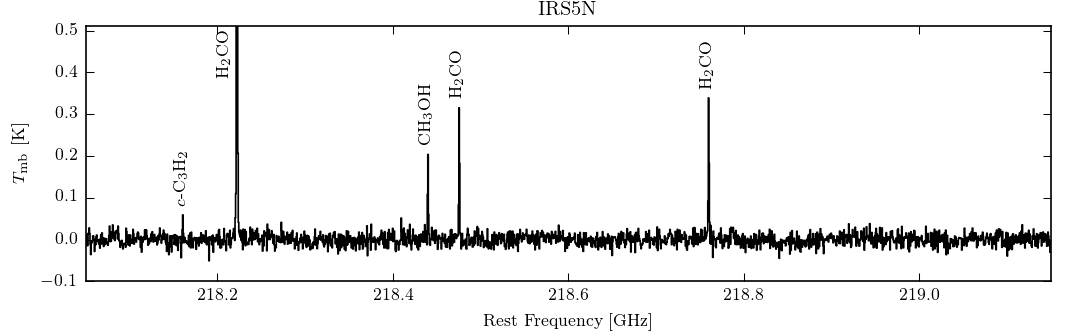} \\
                \includegraphics{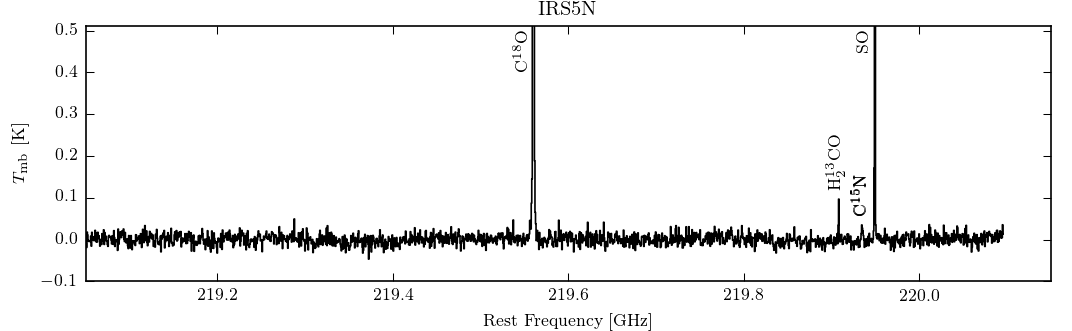} \\
                \end{array}$
                \caption{1.4~mm spectrum of IRS5N, smoothed by a factor 8, corresponding to a channel width of 0.8~\kms.}\label{fig:irs5n_1}
        \end{figure*}
        
        \begin{figure*}[!htb]
                \centering  
                $\begin{array}{c}
                \includegraphics{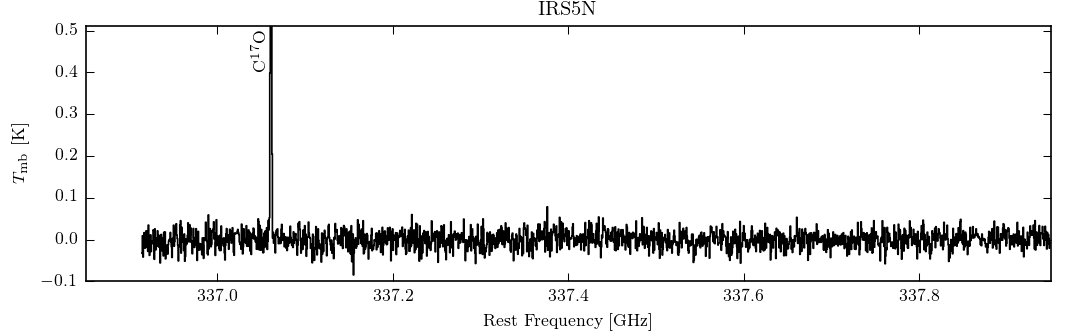} \\
                \includegraphics{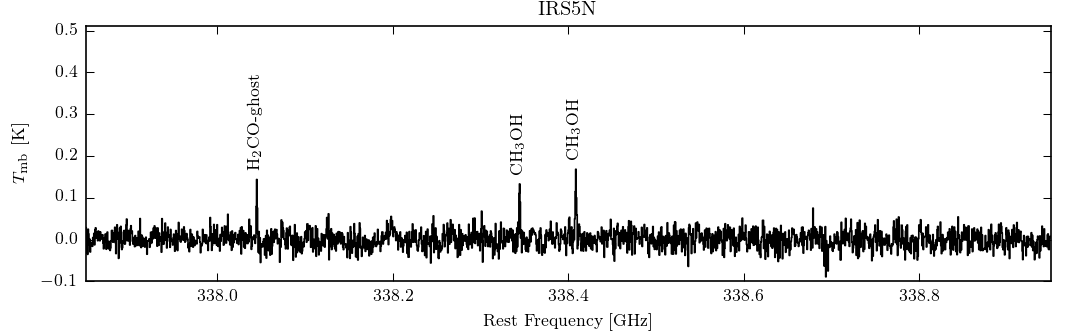} \\
                \includegraphics{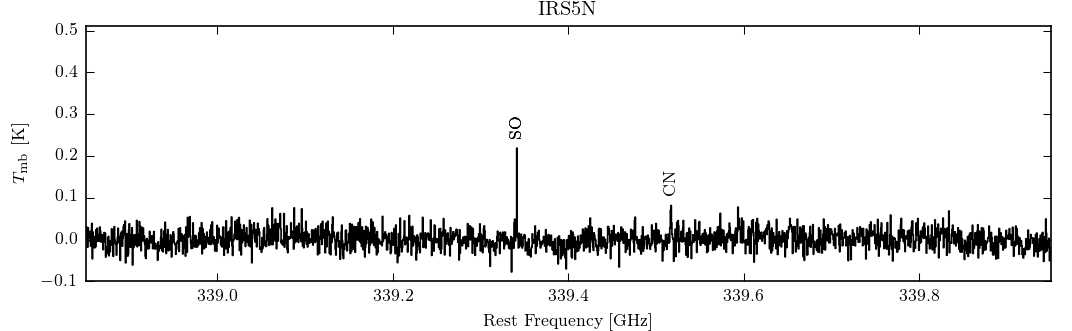} \\
                \includegraphics{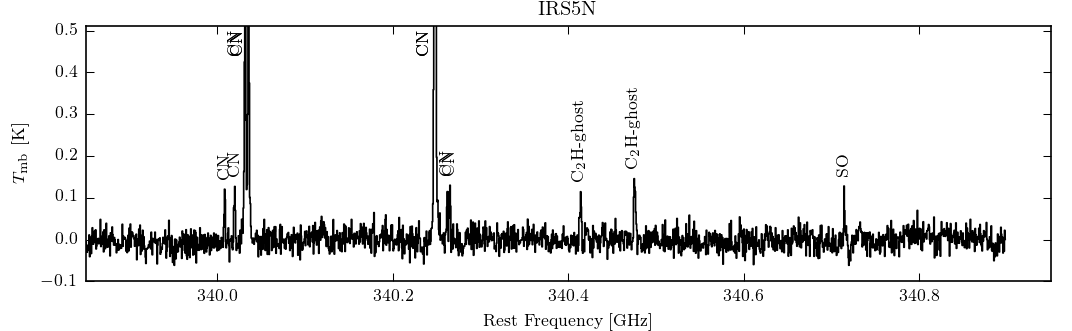} \\
                \end{array}$
                \caption{0.9~mm spectrum of IRS5N, smoothed by a factor 8, corresponding to a channel width of 0.6~\kms.}\label{fig:irs5n_2}
        \end{figure*}

        \begin{figure*}[!htb]
                \centering  
                $\begin{array}{c}
                \includegraphics{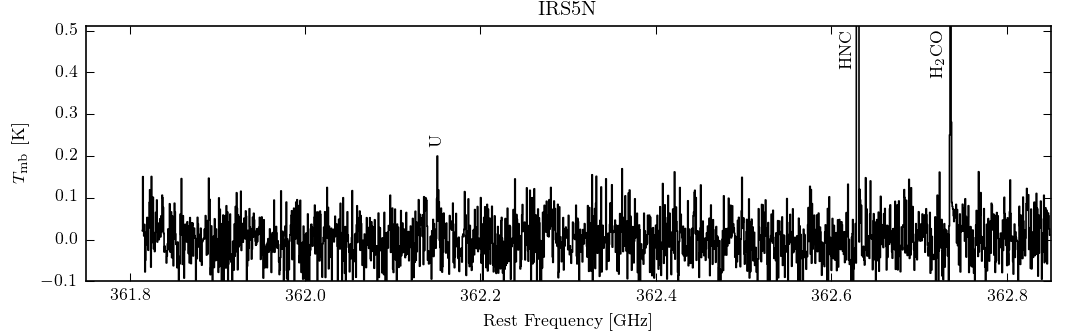} \\
                \includegraphics{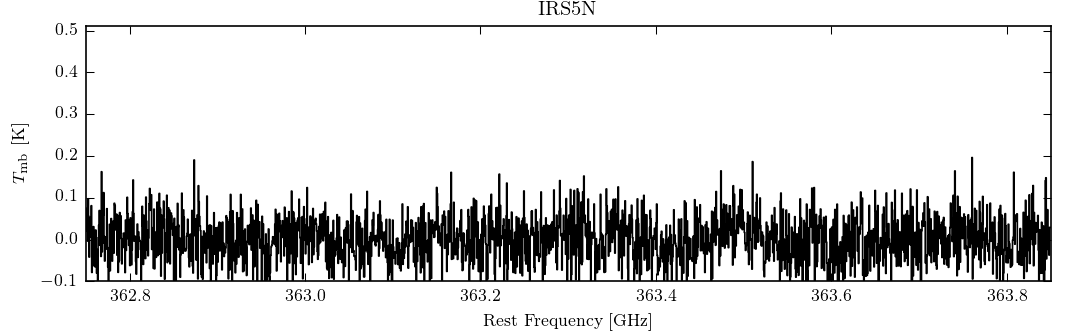} \\
                \includegraphics{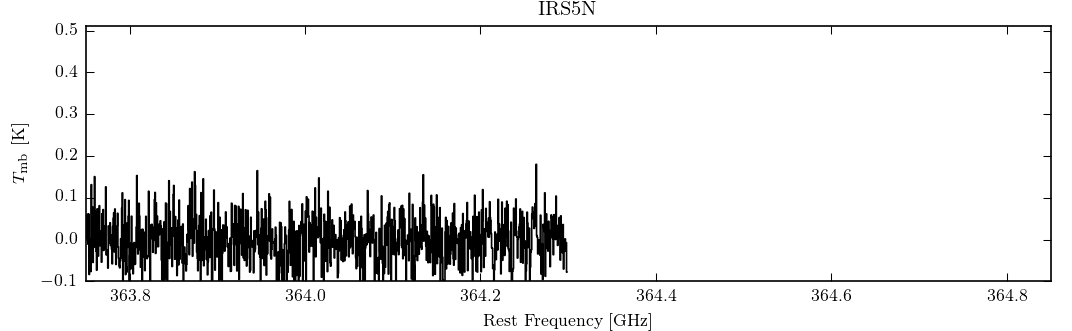} \\
                \end{array}$
                \caption{0.8~mm spectrum of IRS5N, smoothed by a factor 8, corresponding to a channel width of 0.5~\kms.}\label{fig:irs5n_3}
        \end{figure*}

        \begin{figure*}[!htb]
                \centering  
                $\begin{array}{c}
                \includegraphics{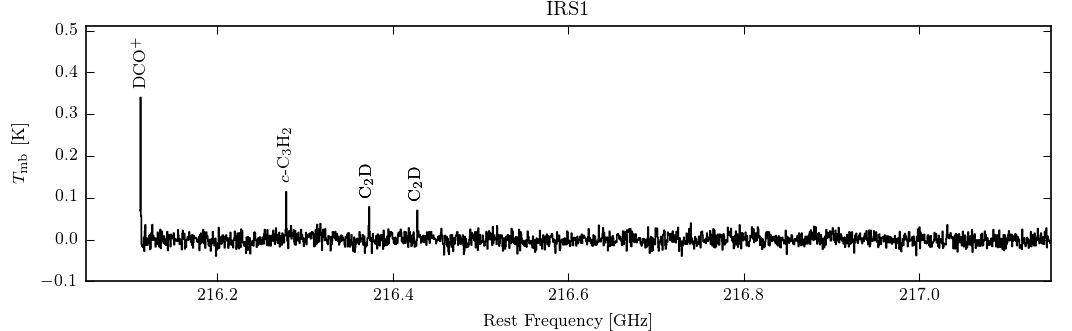} \\
                \includegraphics{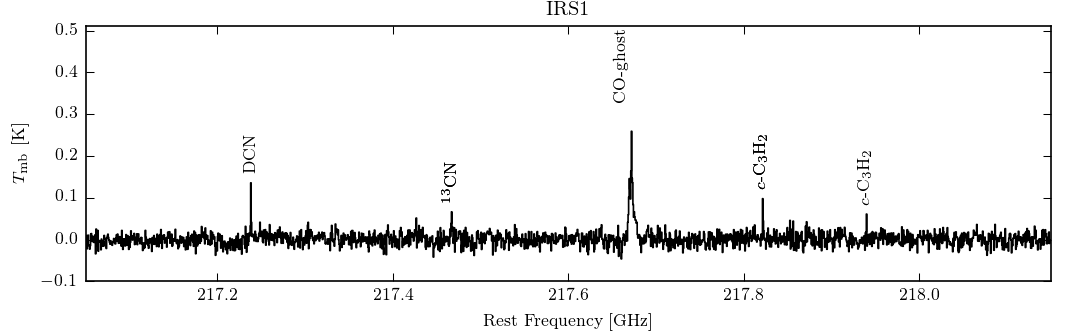} \\
                \includegraphics{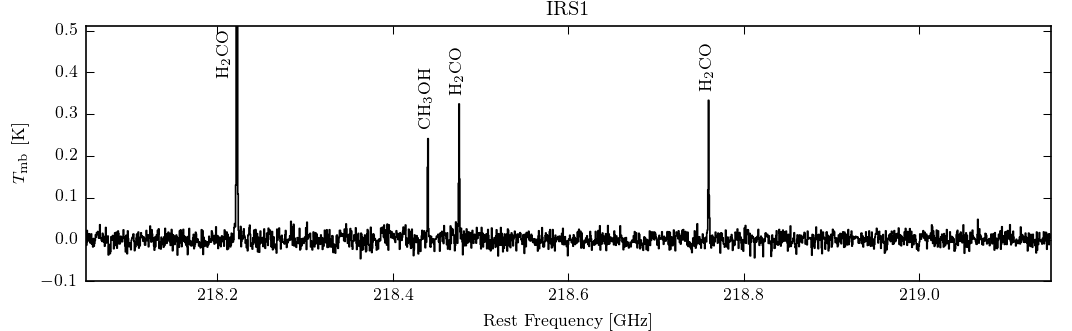} \\
                \includegraphics{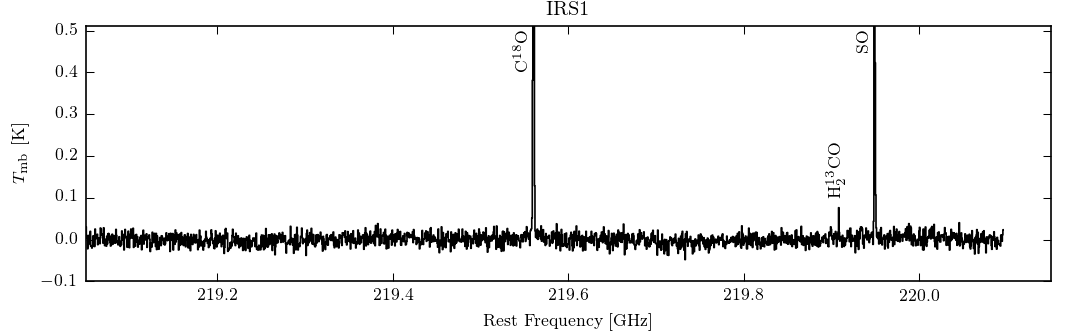} \\
                \end{array}$
                \caption{1.4~mm spectrum of IRS1, smoothed by a factor 8, corresponding to a channel width of 0.8~\kms.}\label{fig:irs1_1}
        \end{figure*}
        
        \begin{figure*}[!htb]
                \centering  
                $\begin{array}{c}
                \includegraphics{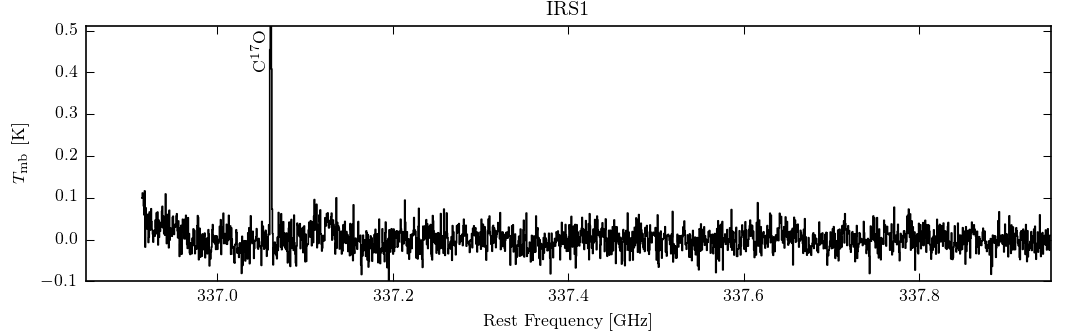} \\
                \includegraphics{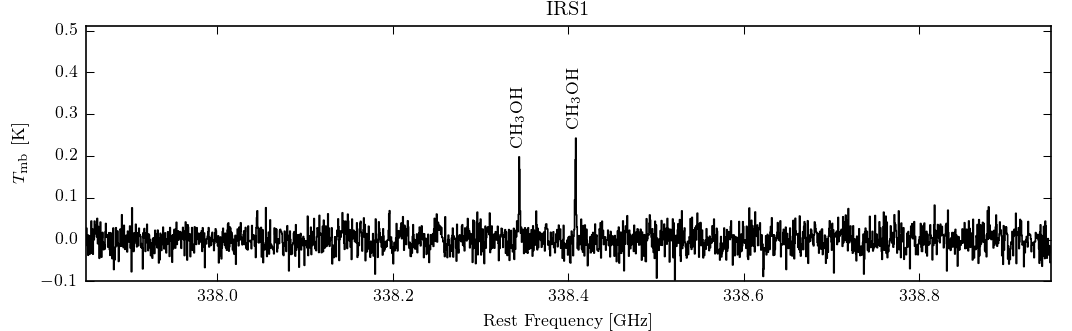} \\
                \includegraphics{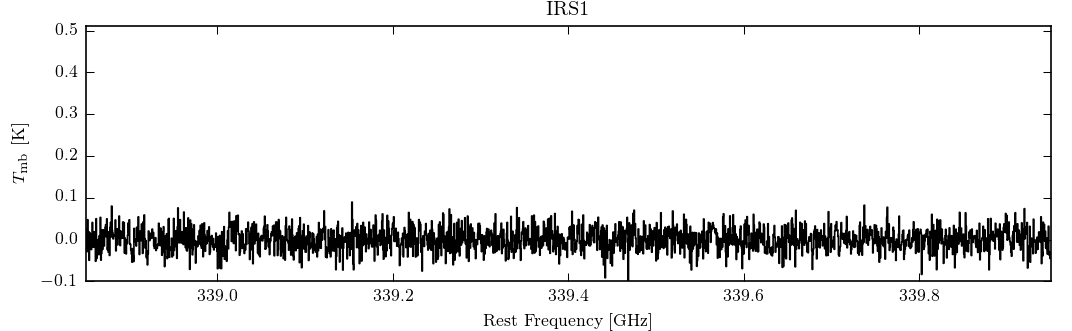} \\
                \includegraphics{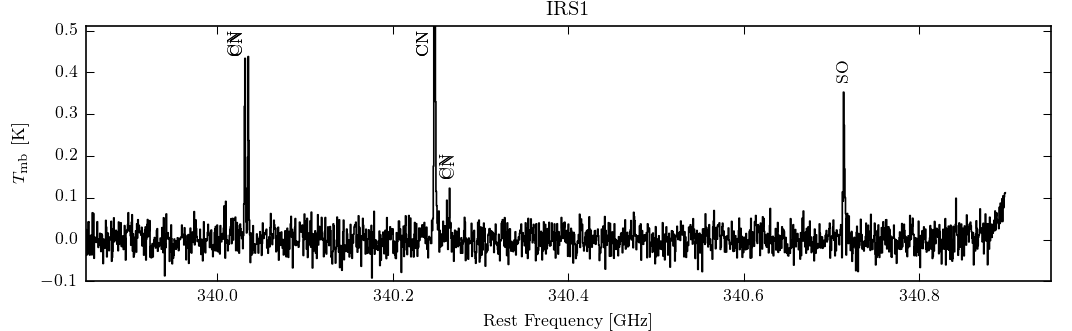} \\
                \end{array}$
                \caption{0.9~mm spectrum of IRS1, smoothed by a factor 8, corresponding to a channel width of 0.6~\kms.}\label{fig:irs1_2}
        \end{figure*}
        
        \begin{figure*}[!htb]
                \centering  
                $\begin{array}{c}
                \includegraphics{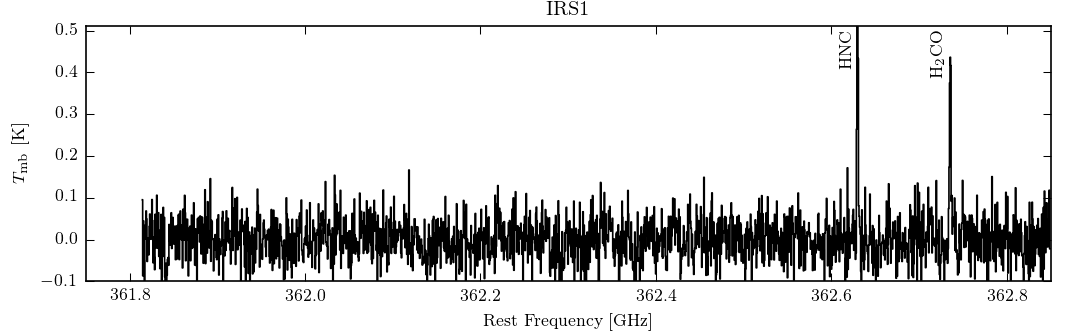} \\
                \includegraphics{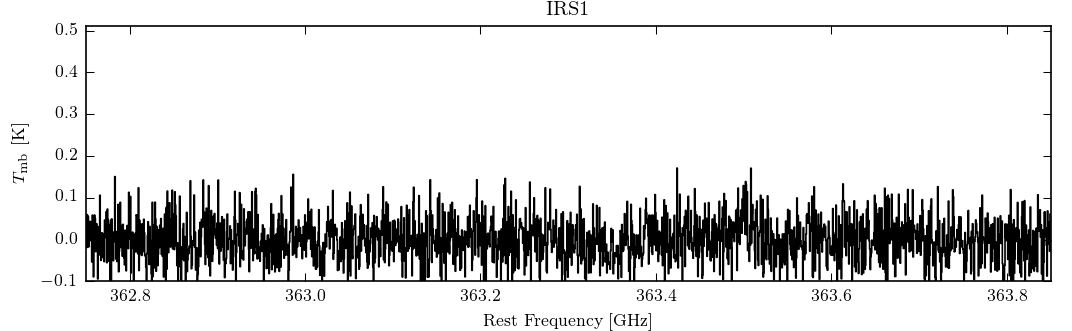} \\
                \includegraphics{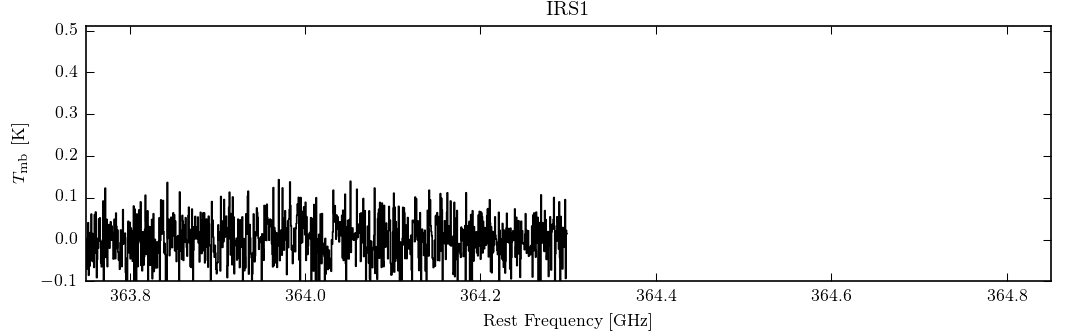} \\
                \end{array}$
                \caption{0.8~mm spectrum of IRS1, smoothed by a factor 8, corresponding to a channel width of 0.5~\kms.}\label{fig:irs1_3}
        \end{figure*}
        
        \begin{figure*}[!htb]
                \centering  
                $\begin{array}{c}
                \includegraphics{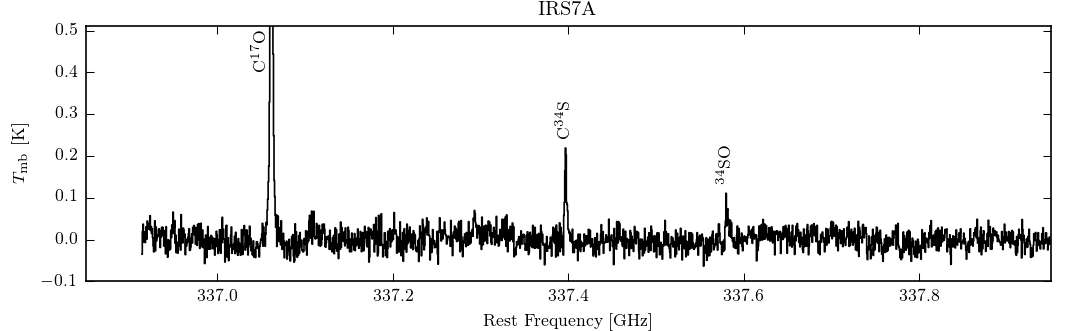} \\
                \includegraphics{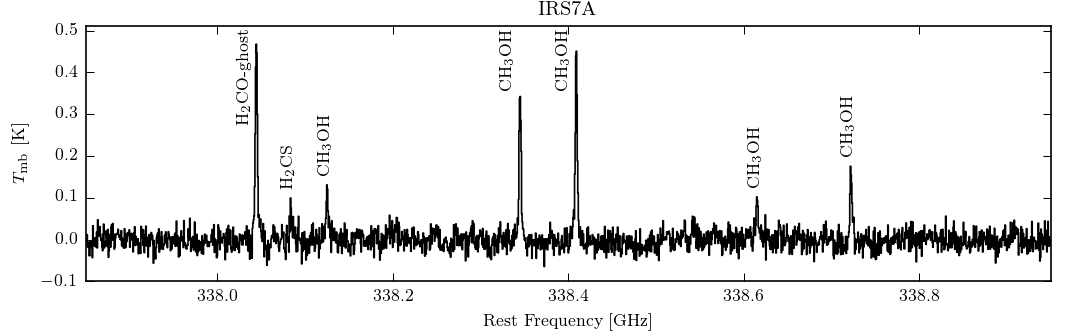} \\
                \includegraphics{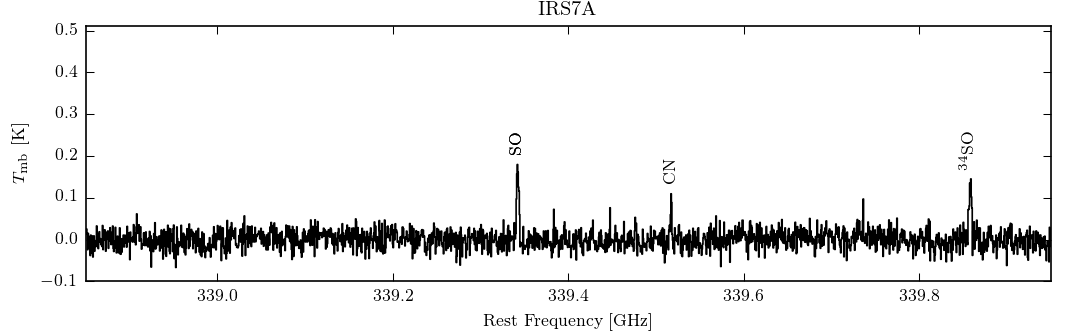} \\
                \includegraphics{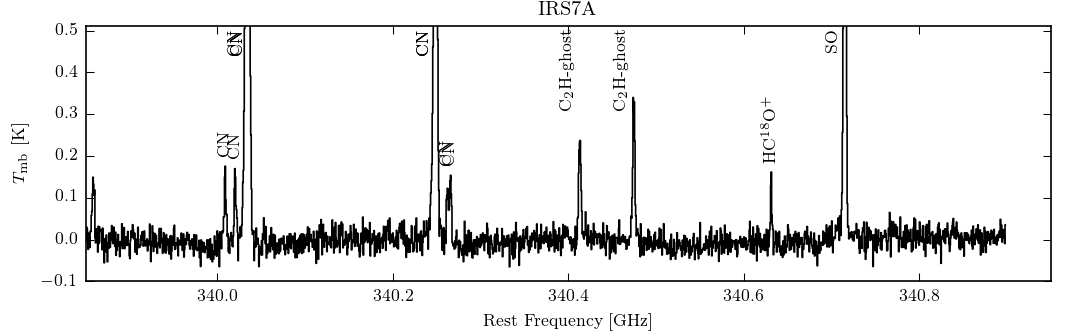} \\
                \end{array}$
                \caption{0.9~mm spectrum of IRS7A, smoothed by a factor 8, corresponding to a channel width of 0.6~\kms.}\label{fig:irs7a_2}
        \end{figure*}
        
        \begin{figure*}[!htb]
                \centering  
                $\begin{array}{c}
                \includegraphics{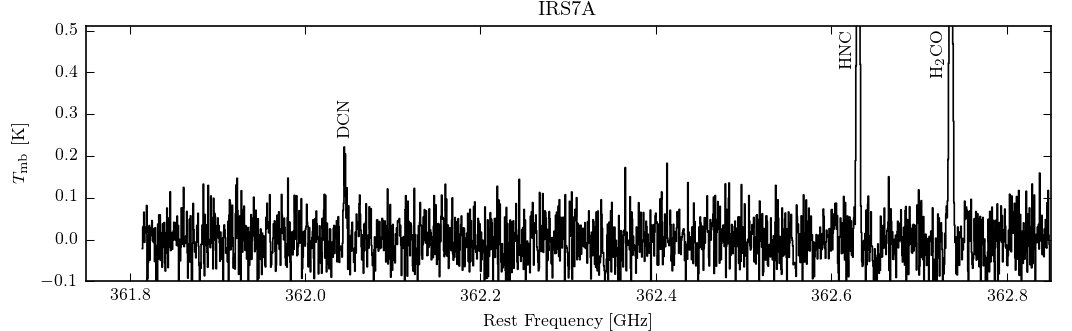} \\
                \includegraphics{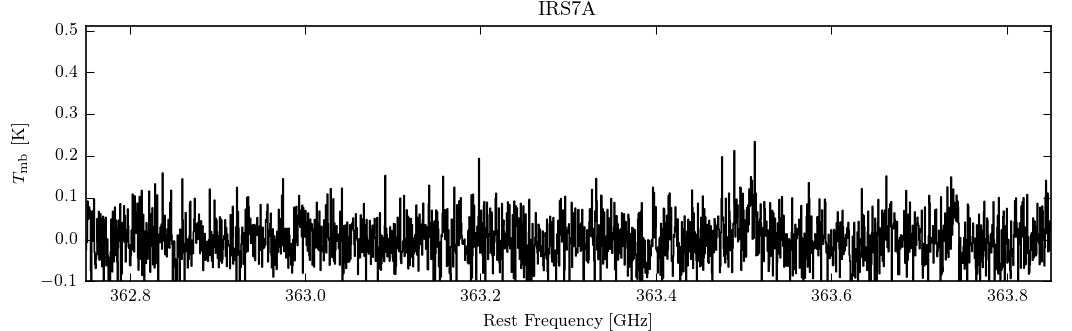} \\
                \includegraphics{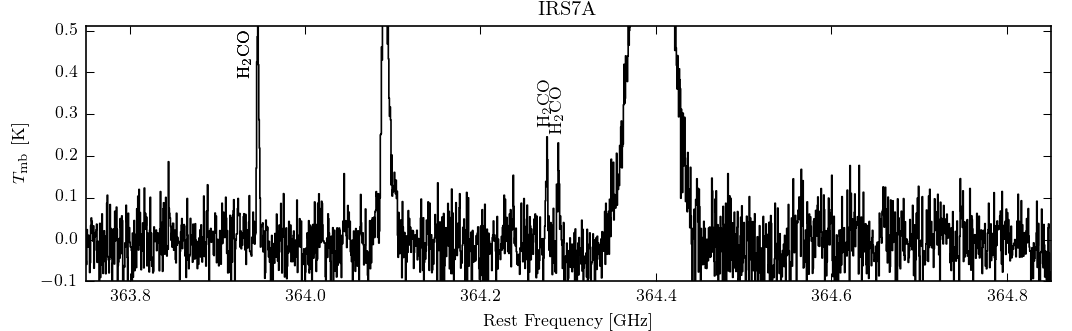} \\
                \includegraphics{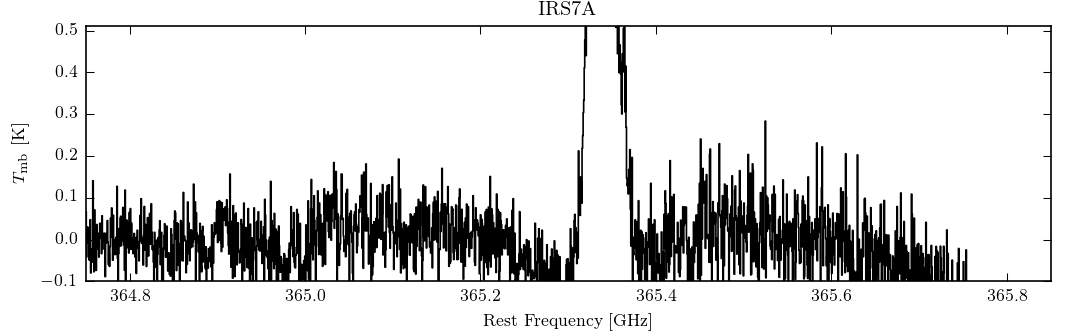} \\
                \end{array}$
                \caption{0.8~mm spectrum of IRS7A, smoothed by a factor 8, corresponding to a channel width of 0.5~\kms. Strong atmospheric lines are visible at 364.1~GHz, 364.4~GHz, and 365.3~GHz.}\label{fig:irs7a_3}
        \end{figure*}
        
        \begin{figure*}[!htb]
                \centering  
                $\begin{array}{c}
                \includegraphics{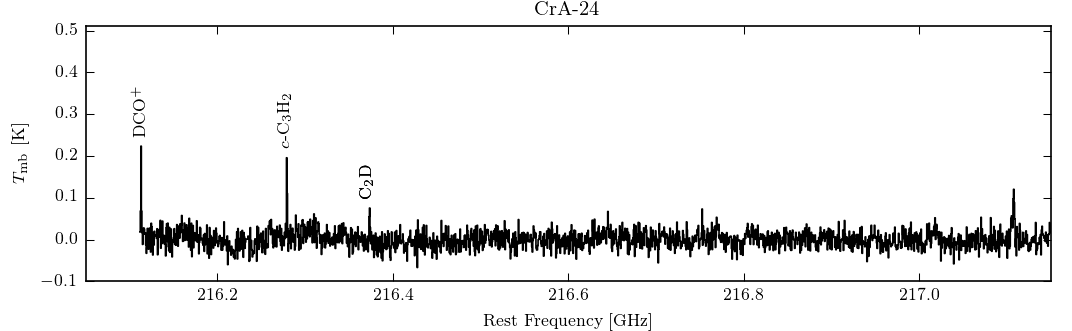} \\
                \includegraphics{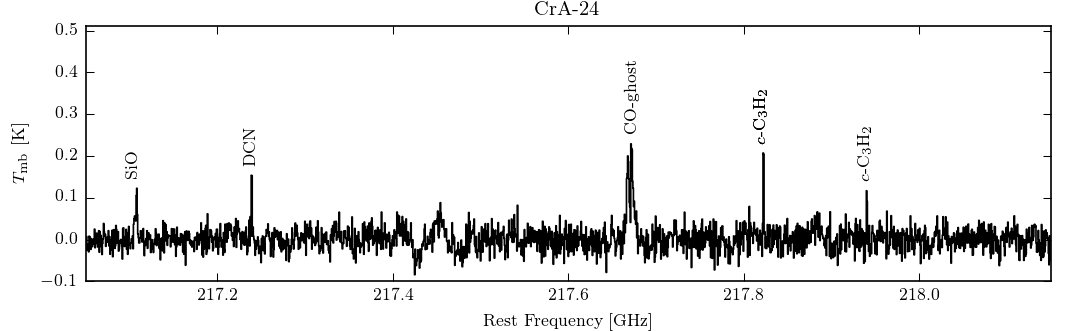} \\
                \includegraphics{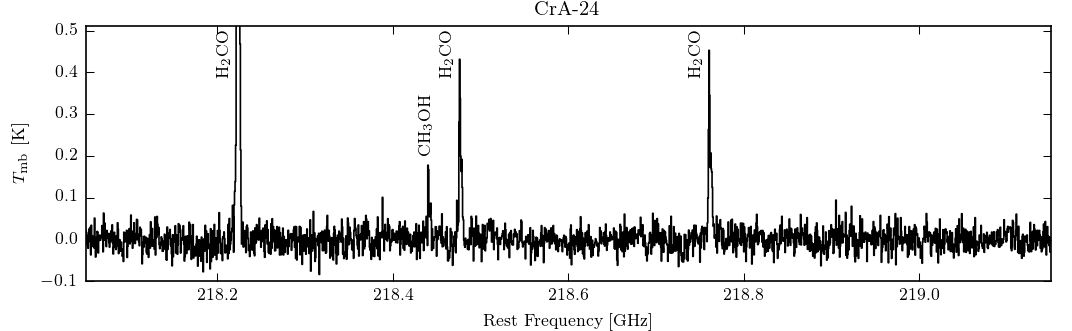} \\
                \includegraphics{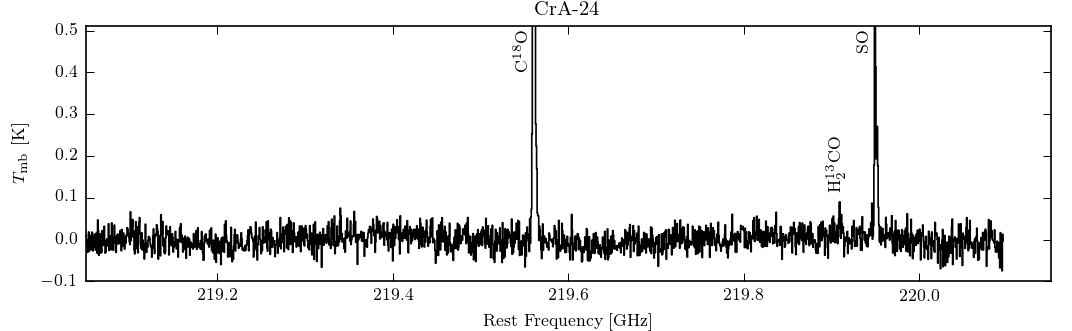} \\
                \end{array}$
                \caption{1.4~mm spectrum of CrA-24, smoothed by a factor 8, corresponding to a channel width of 0.8~\kms. Atmospheric line at 217.7~GHz.}\label{fig:cra24_1}
        \end{figure*}
        
        \begin{figure*}[!htb]
                \centering  
                $\begin{array}{c}
                \includegraphics{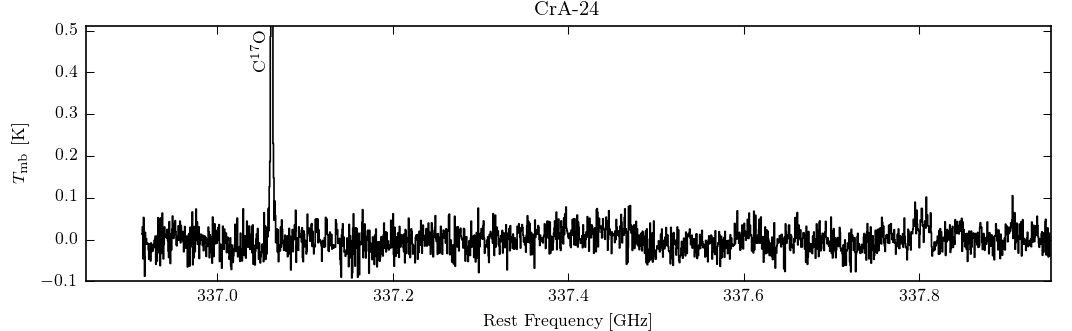} \\
                \includegraphics{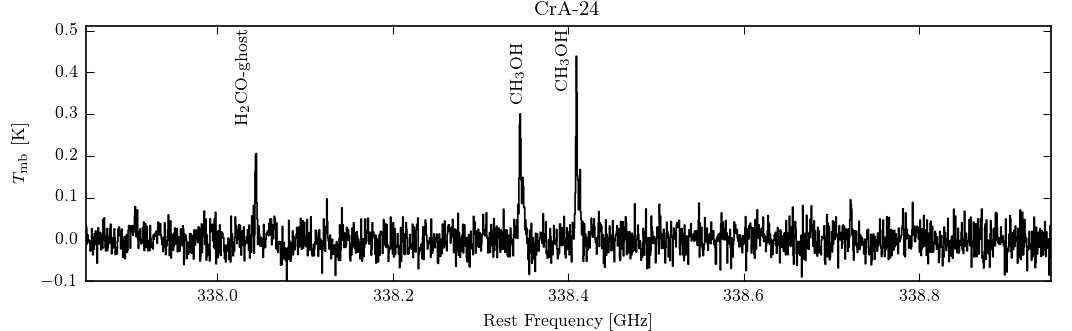} \\
                \includegraphics{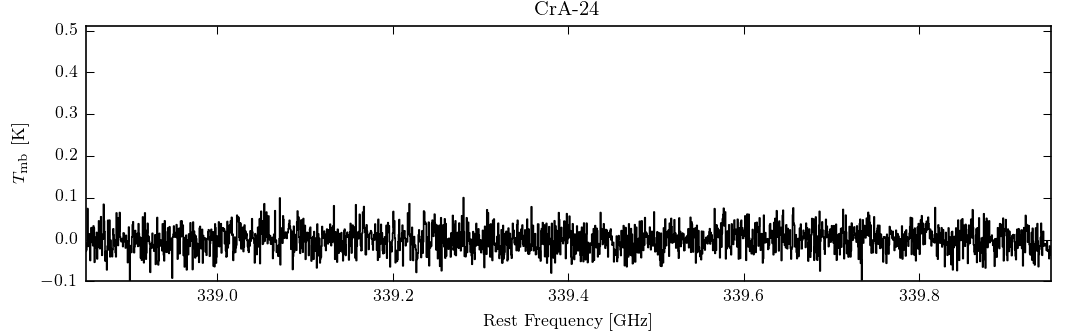} \\
                \includegraphics{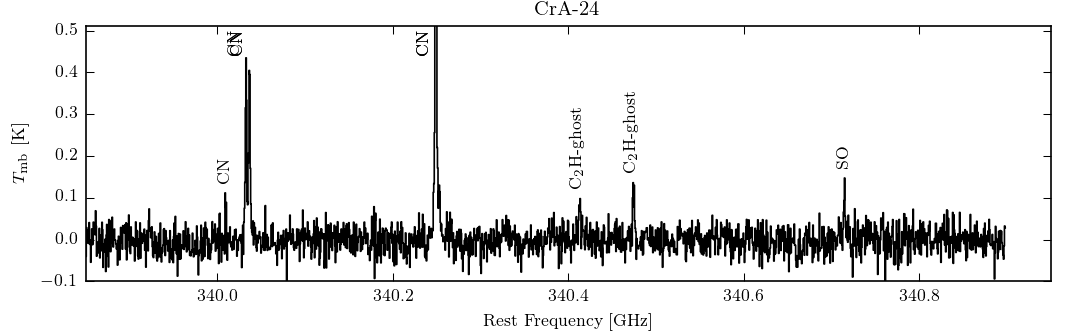} \\
                \end{array}$
                \caption{0.9~mm spectrum of CrA-24, smoothed by a factor 8, corresponding to a channel width of 0.6~\kms.}\label{fig:cra24_2}
        \end{figure*}
        
        \begin{figure*}[!htb]
                \centering  
                $\begin{array}{c}
                \includegraphics{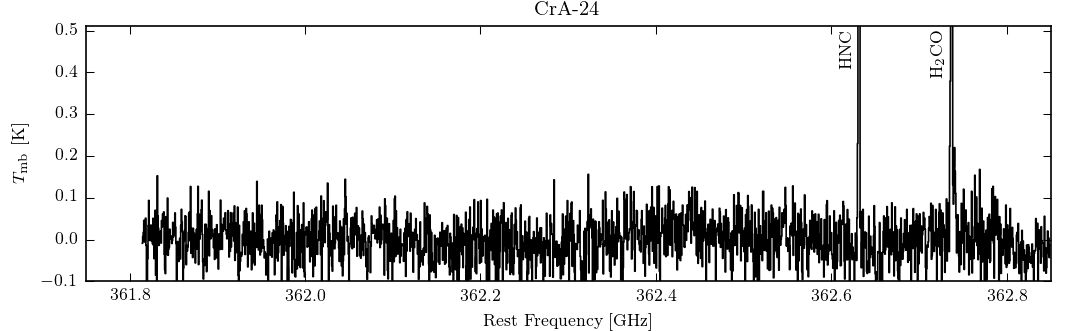} \\
                \includegraphics{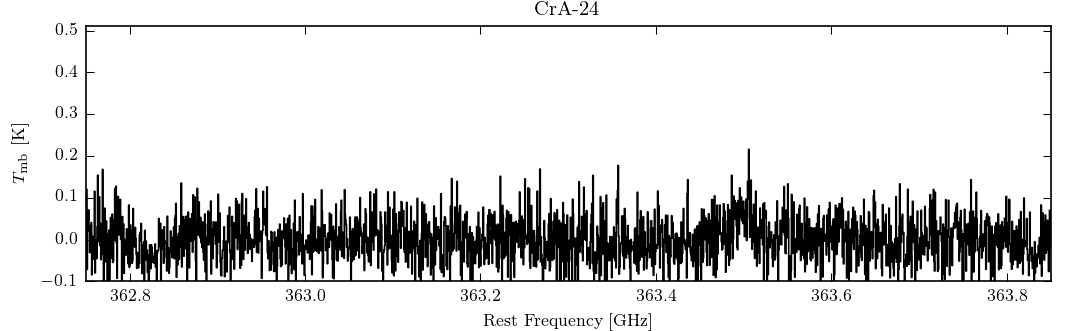} \\
                \includegraphics{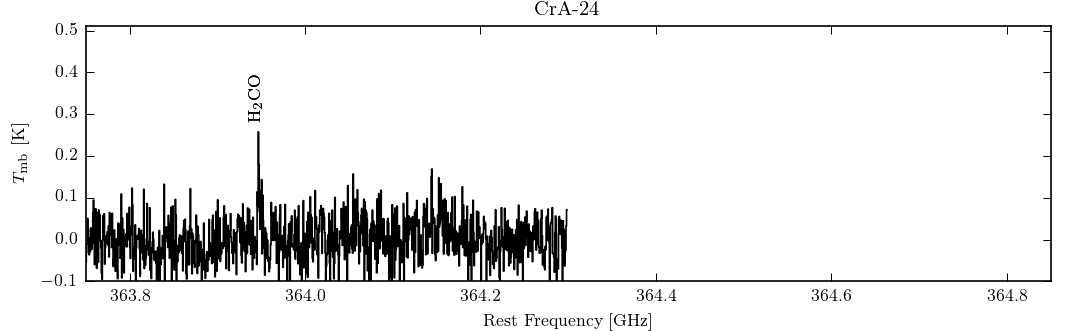} \\
                \end{array}$
                \caption{0.8~mm spectrum of CrA-24, smoothed by a factor 8, corresponding to a channel width of 0.5~\kms.}\label{fig:cra24_3}
        \end{figure*}
        
        \begin{figure*}[!htb]
                \centering  
                $\begin{array}{c}
                \includegraphics{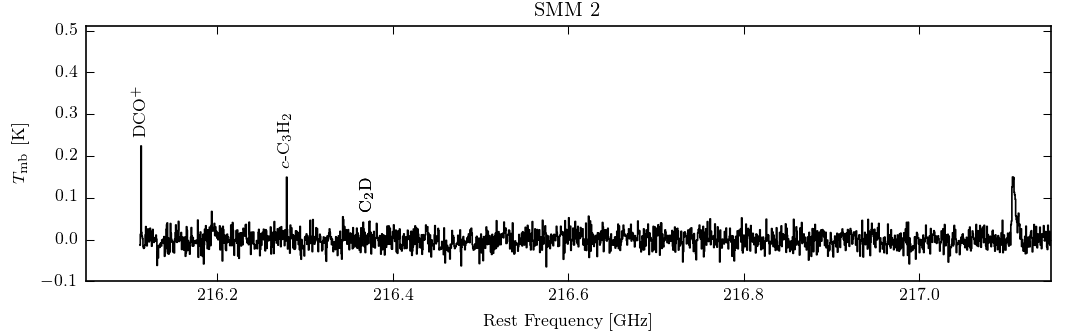} \\
                \includegraphics{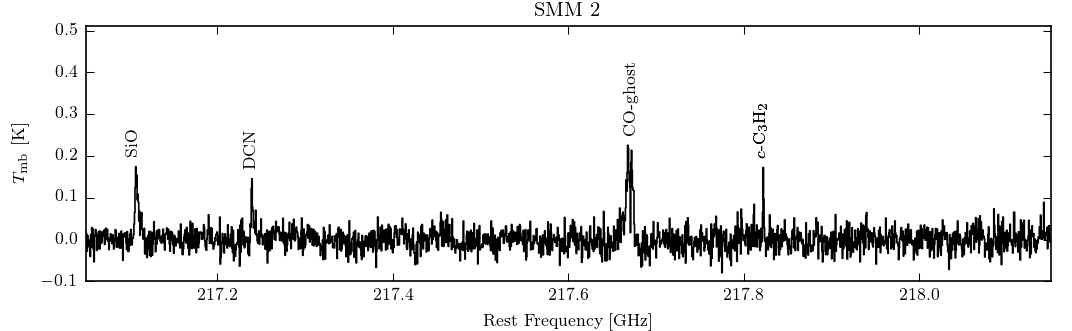} \\
                \includegraphics{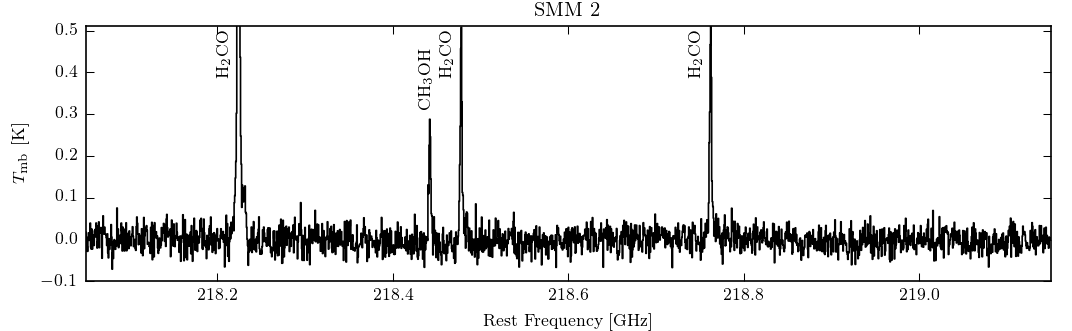} \\
                \includegraphics{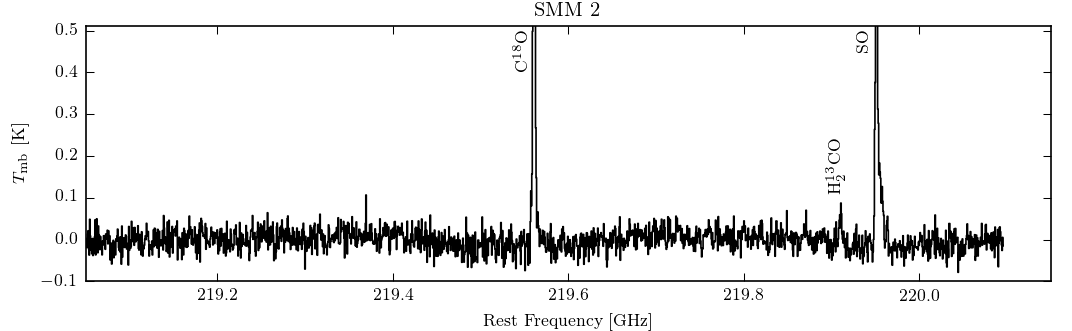} \\
                \end{array}$
                \caption{1.4~mm spectrum of SMM~2, smoothed by a factor 8, corresponding to a channel width of 0.8~\kms.}\label{fig:smm2_1}
        \end{figure*}
        
        \begin{figure*}[!htb]
                \centering  
                $\begin{array}{c}
                \includegraphics{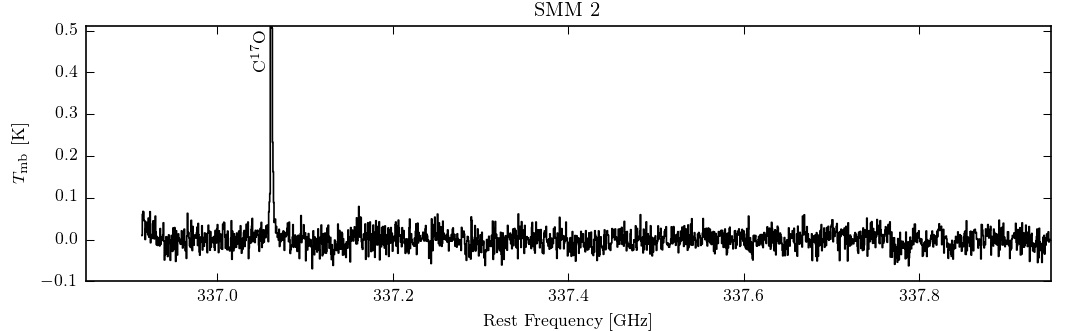} \\
                \includegraphics{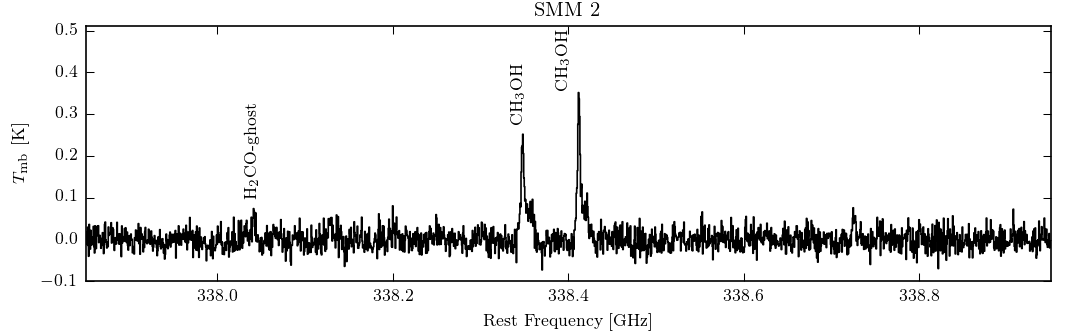} \\
                \includegraphics{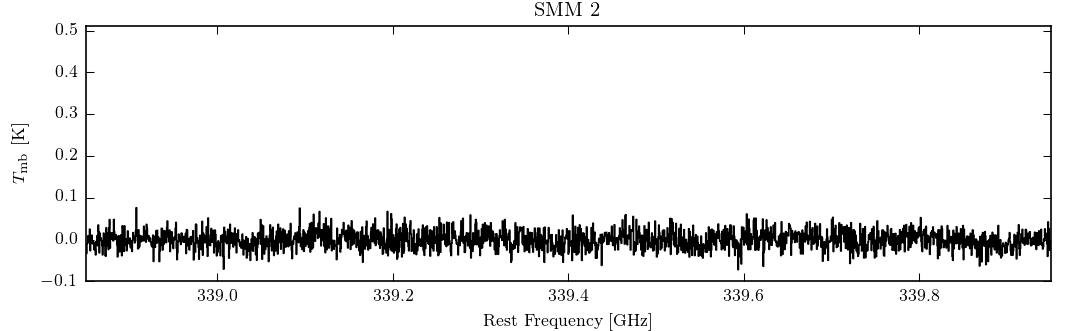} \\
                \includegraphics{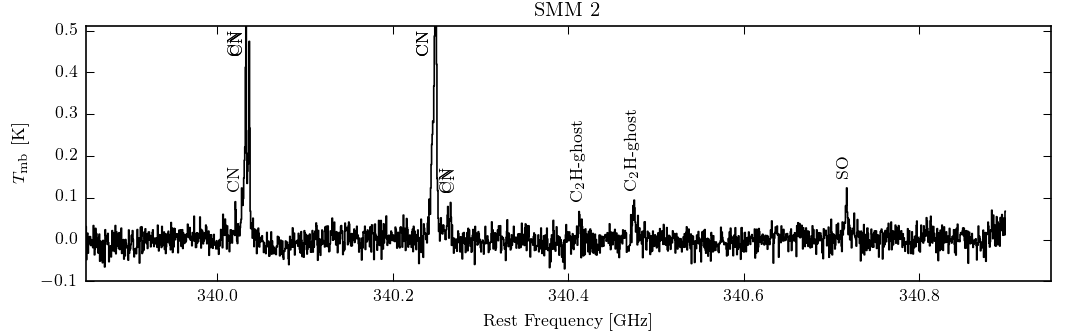} \\
                \end{array}$
                \caption{0.9~mm spectrum of SMM~2, smoothed by a factor 8, corresponding to a channel width of 0.6~\kms.}\label{fig:smm2_2}
        \end{figure*}
        
        \begin{figure*}[!htb]
                \centering  
                $\begin{array}{c}
                \includegraphics{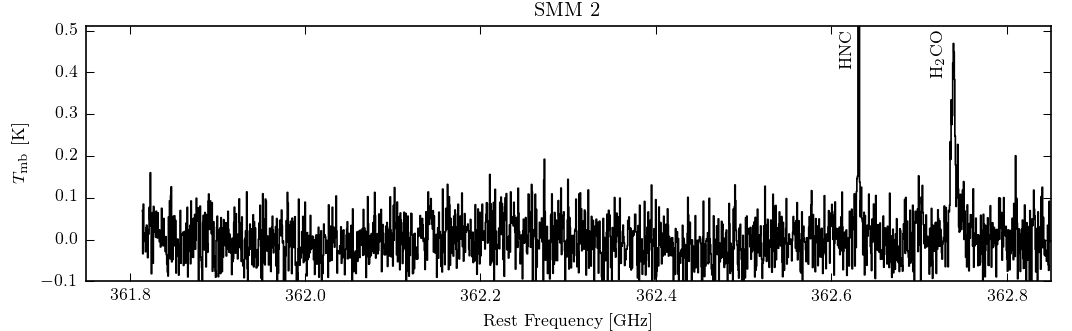} \\
                \includegraphics{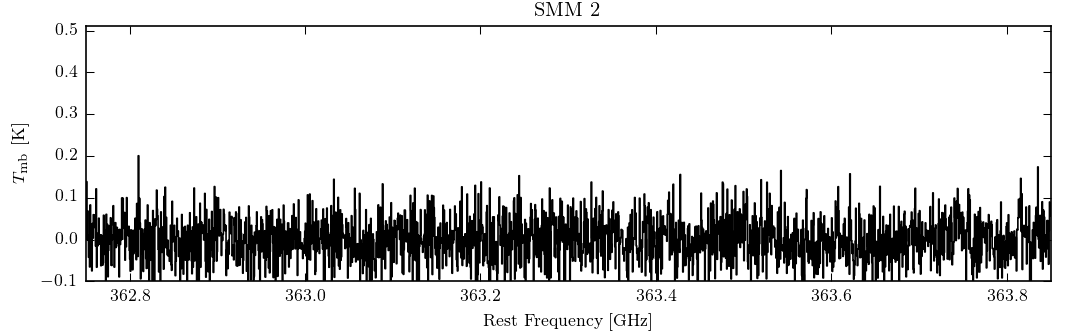} \\
                \includegraphics{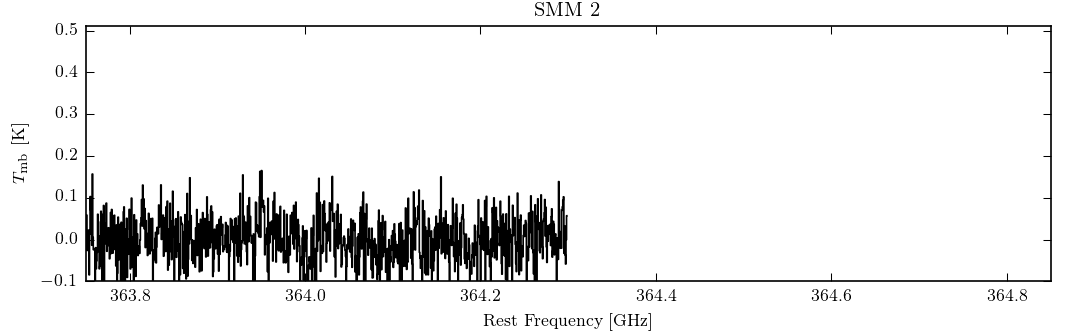} \\
                \end{array}$
                \caption{0.8~mm spectrum of SMM~2, smoothed by a factor 8, corresponding to a channel width of 0.5~\kms.}\label{fig:smm2_3}
        \end{figure*}
        
        \begin{figure*}[!htb]
                \centering  
                $\begin{array}{c}
                \includegraphics{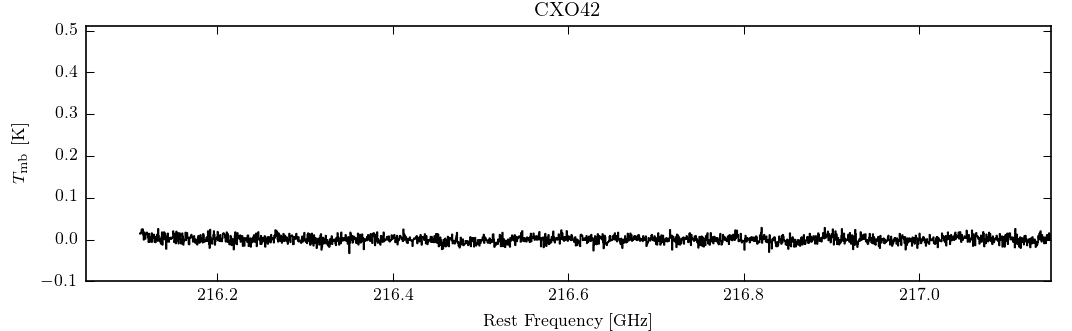} \\
                \includegraphics{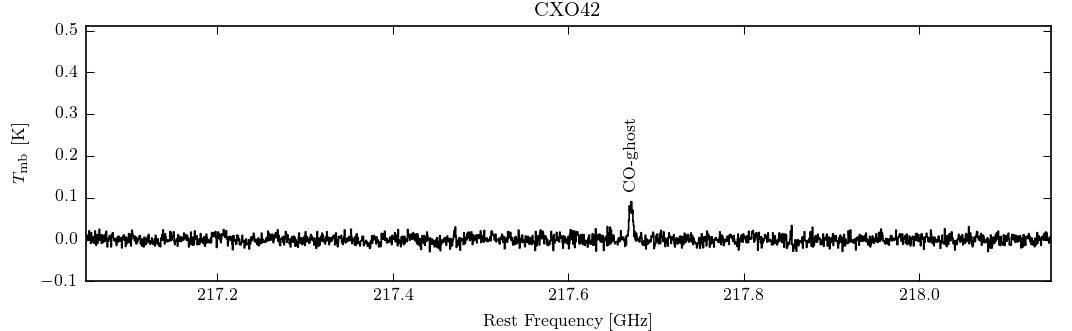} \\
                \includegraphics{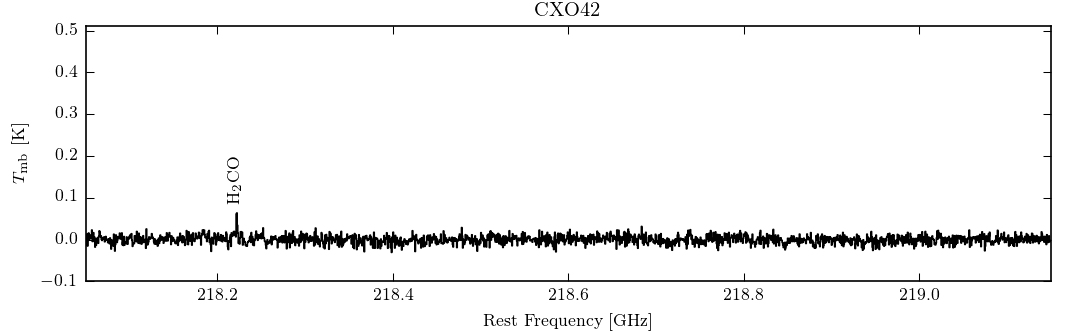} \\
                \includegraphics{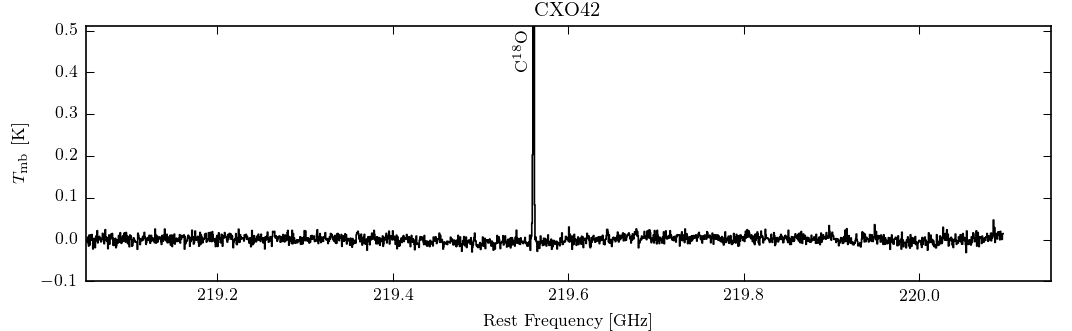} \\
                \end{array}$
                \caption{1.4~mm spectrum of CXO42, smoothed by a factor 8, corresponding to a channel width of 0.8~\kms.}\label{fig:cxo42_1}
        \end{figure*}
        
        \begin{figure*}[!htb]
                \centering  
                $\begin{array}{c}
                \includegraphics{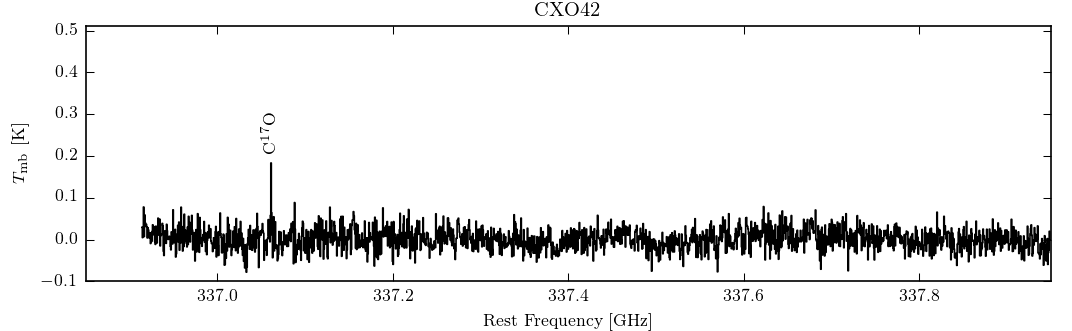} \\
                \includegraphics{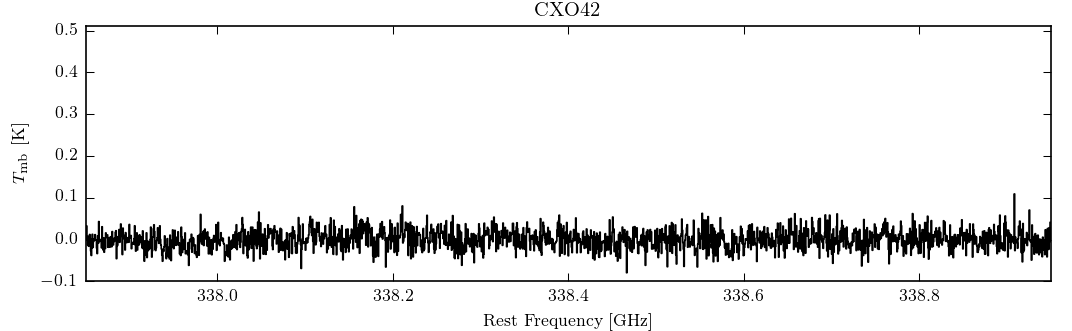} \\
                \includegraphics{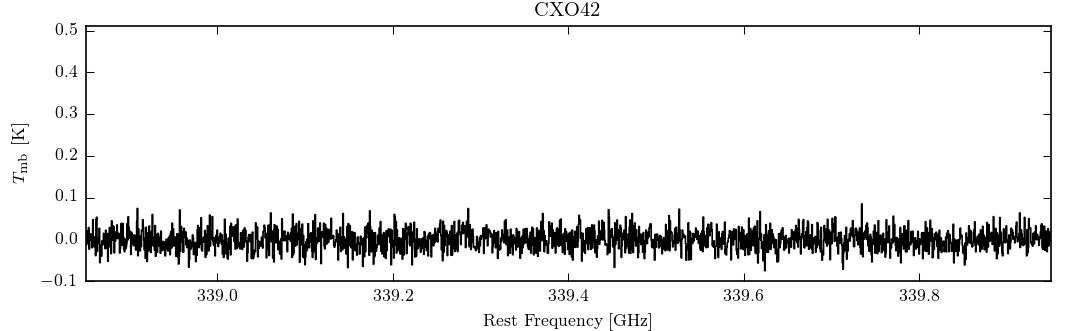} \\
                \includegraphics{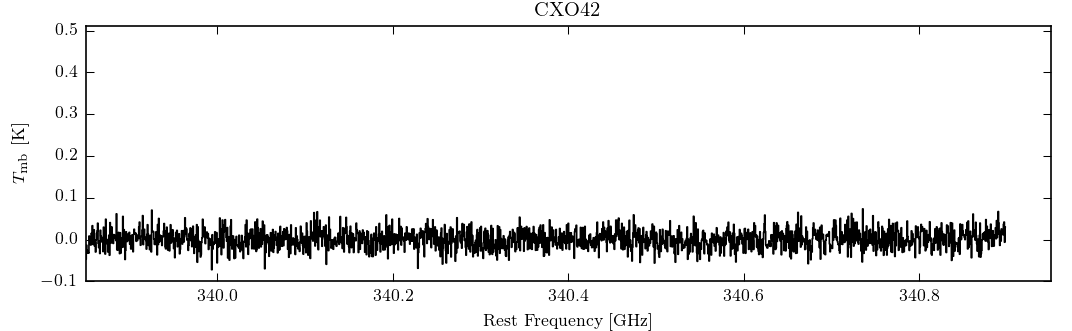} \\
                \end{array}$
                \caption{0.9~mm spectrum of CXO42, smoothed by a factor 8, corresponding to a channel width of 0.6~\kms.}\label{fig:cxo42_2}
        \end{figure*}

        \begin{figure*}[!htb]
                \centering  
                $\begin{array}{c}
                \includegraphics{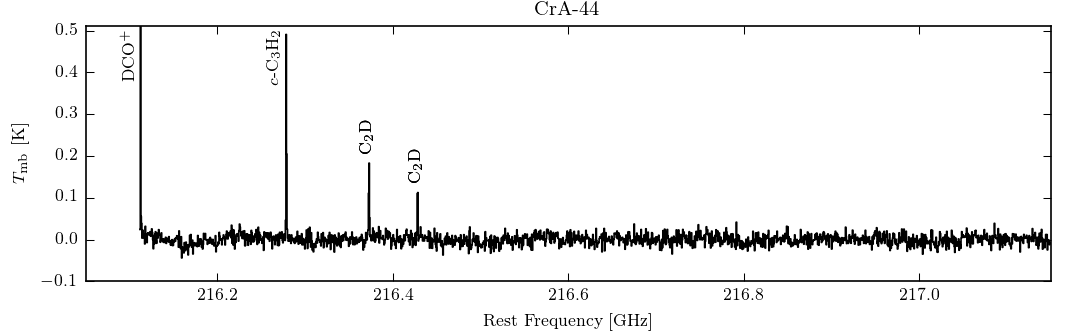} \\
                \includegraphics{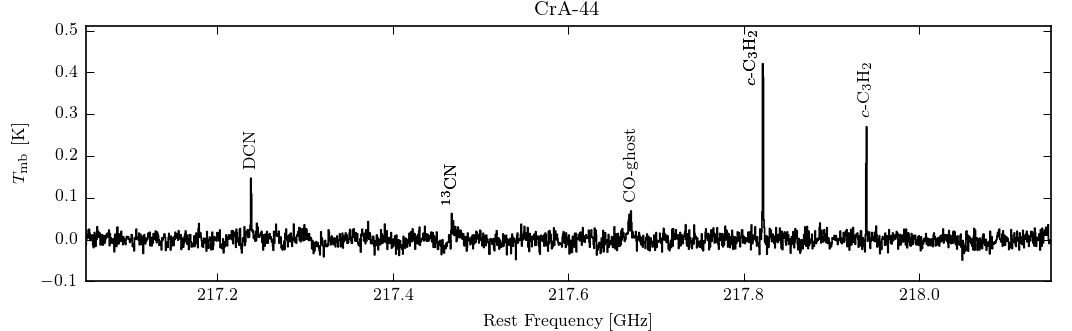} \\
                \includegraphics{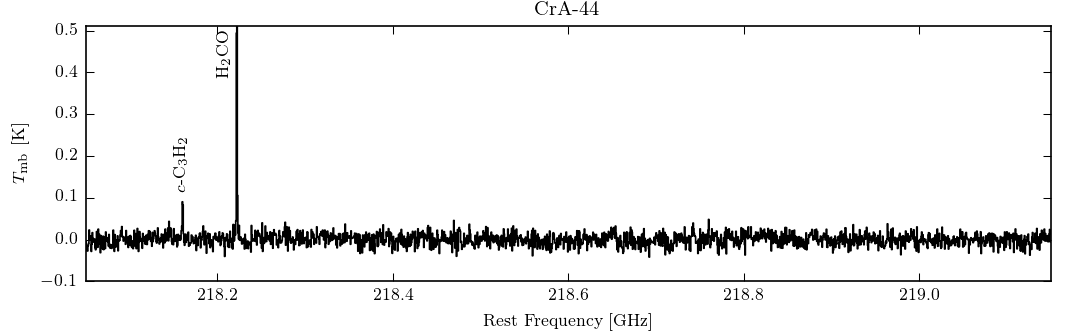} \\
                \includegraphics{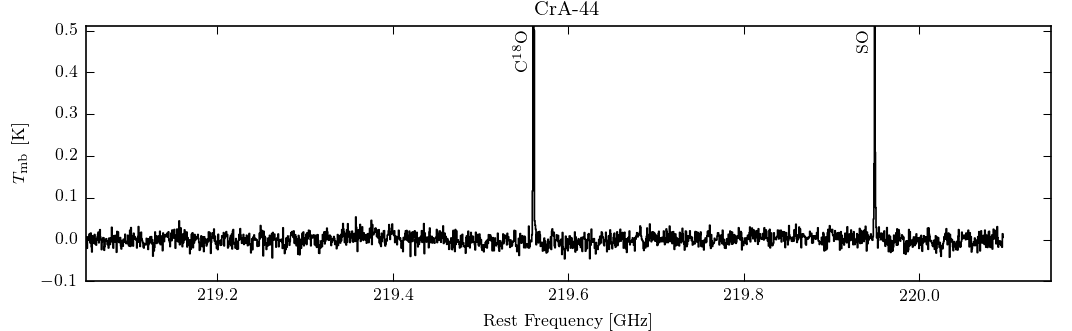} \\
                \end{array}$
                \caption{1.4~mm spectrum of CrA-44, smoothed by a factor 8, corresponding to a channel width of 0.8~\kms.}\label{fig:cra44_1}
        \end{figure*}
        
        \begin{figure*}[!htb]
                \centering  
                $\begin{array}{c}
                \includegraphics{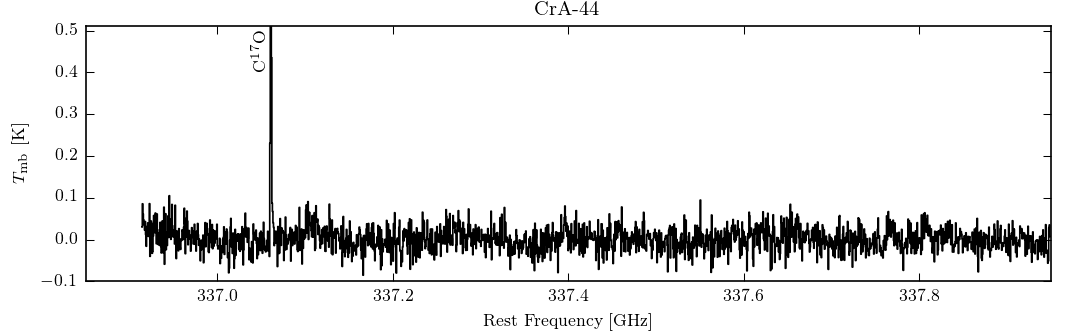} \\
                \includegraphics{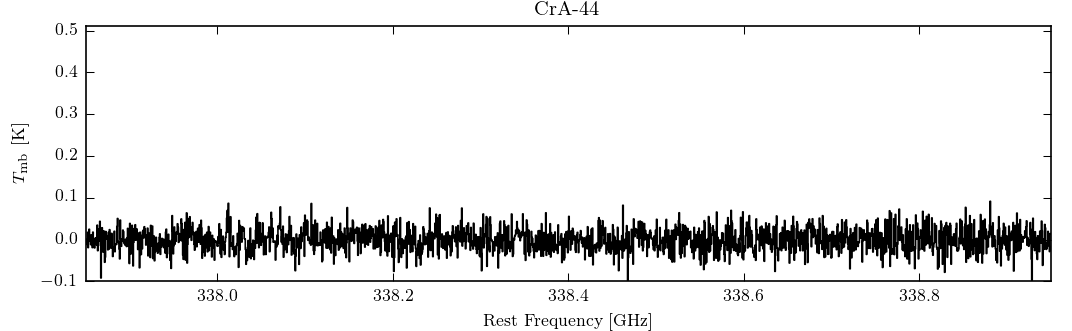} \\
                \includegraphics{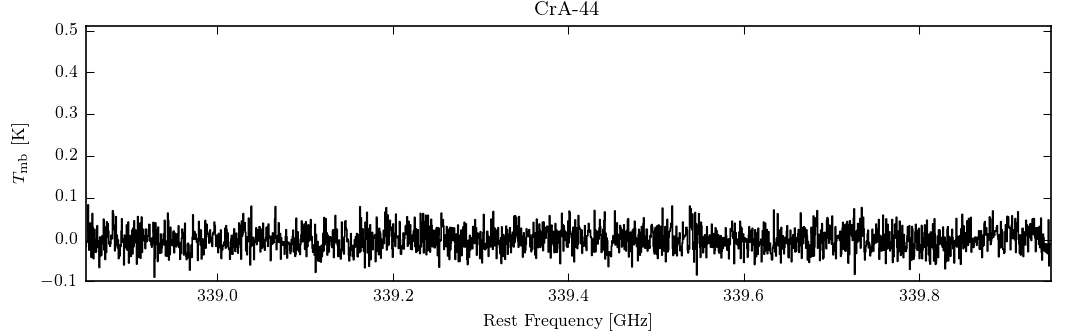} \\
                \includegraphics{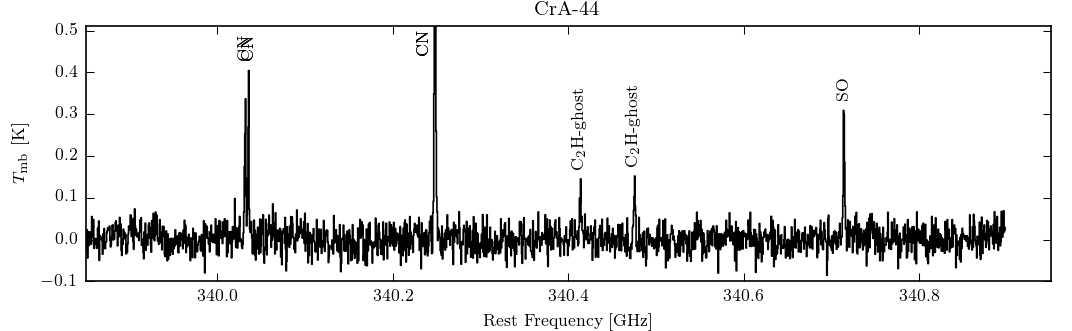} \\
                \end{array}$
                \caption{0.9~mm spectrum of CrA-44, smoothed by a factor 8, corresponding to a channel width of 0.6~\kms.}\label{fig:cra44_2}
        \end{figure*}
        
        \begin{figure*}[!htb]
                \centering  
                $\begin{array}{c}
                \includegraphics{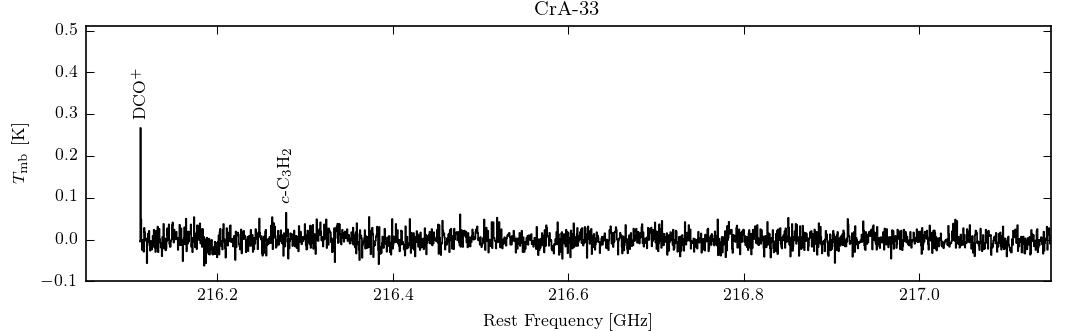} \\
                \includegraphics{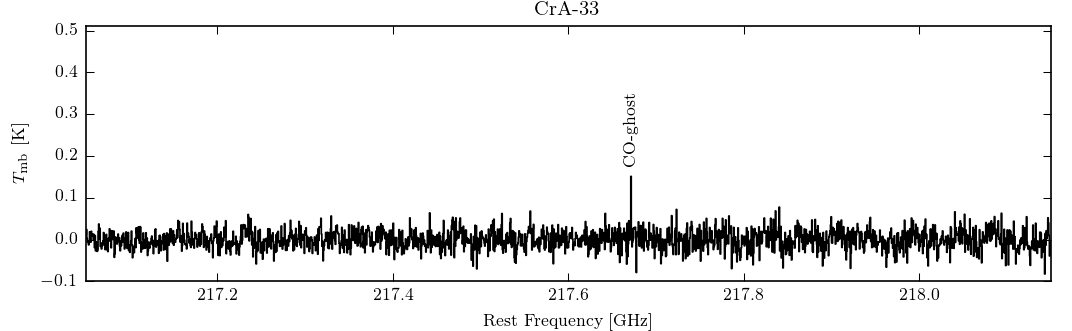} \\
                \includegraphics{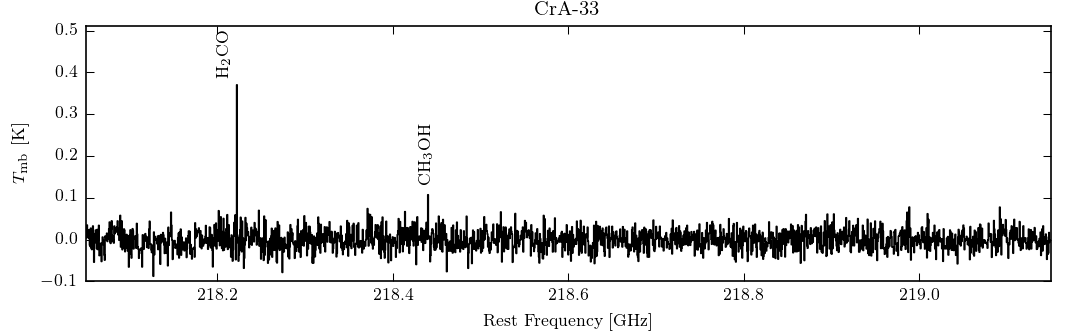} \\
                \includegraphics{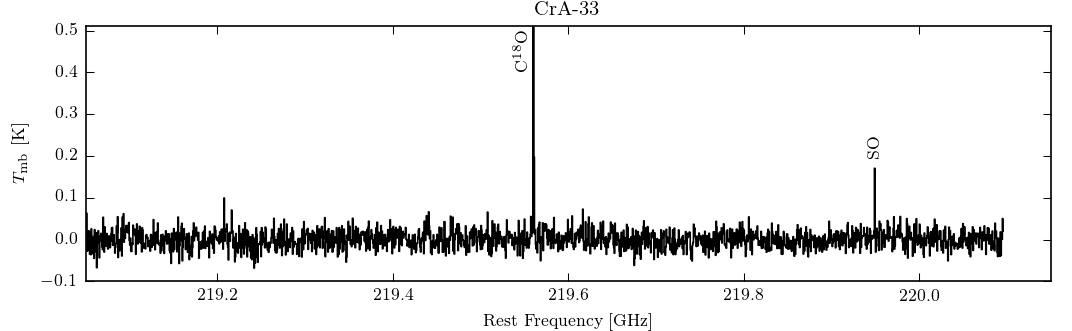} \\
                \end{array}$
                \caption{1.4~mm spectrum of CrA-33, smoothed by a factor 8, corresponding to a channel width of 0.8~\kms.}\label{fig:cra33_1}
        \end{figure*}
        
        \clearpage
        \begin{figure*}[!htb]
                \centering  
                $\begin{array}{c}
                \includegraphics{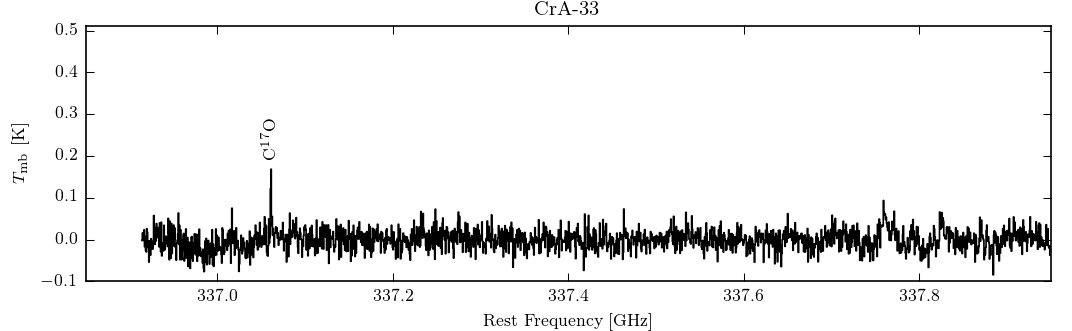} \\
                \includegraphics{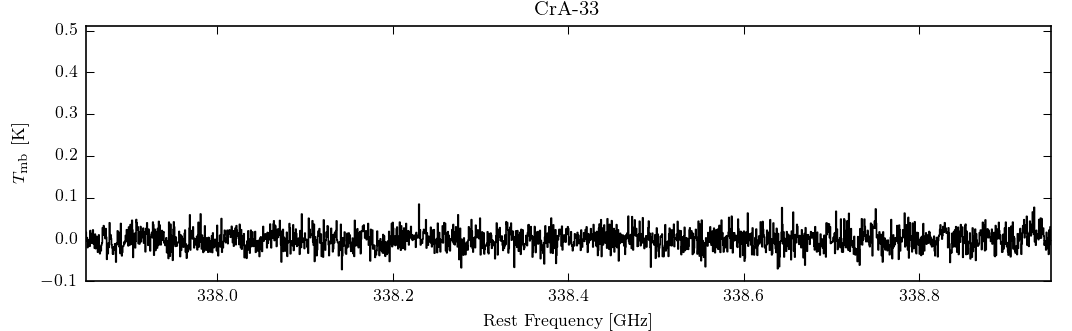} \\
                \includegraphics{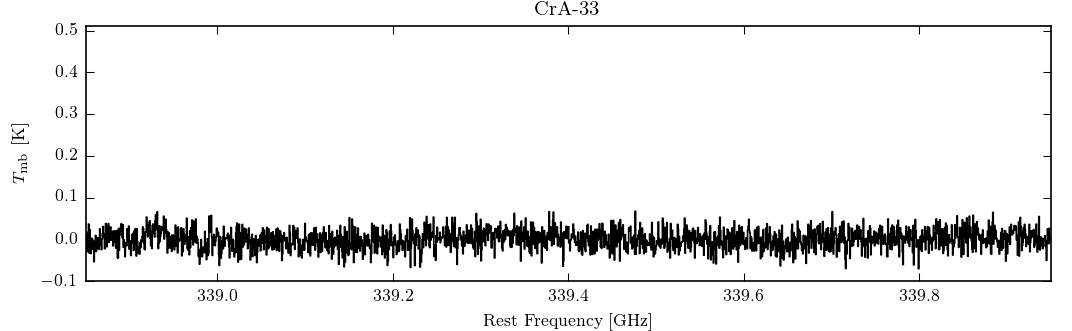} \\
                \includegraphics{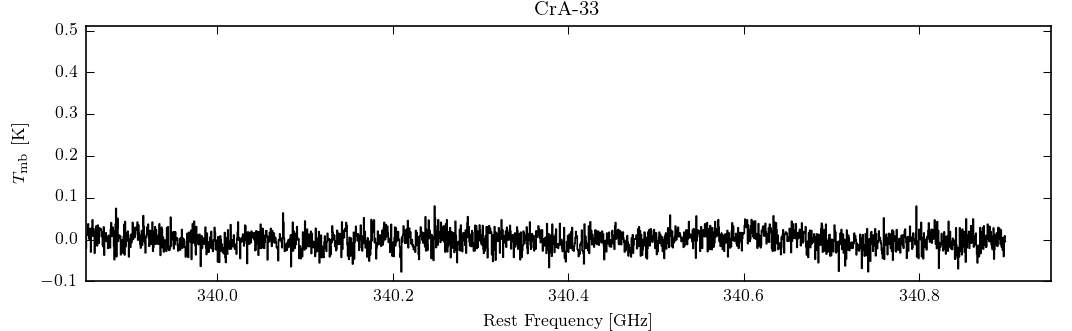} \\
                \end{array}$
                \caption{0.9~mm spectrum of CrA-33, smoothed by a factor 8, corresponding to a channel width of 0.6~\kms.}\label{fig:cra33_2}
        \end{figure*}
        
        \clearpage
        \begin{figure*}[!htb]
                \centering  
                $\begin{array}{c}
                \includegraphics{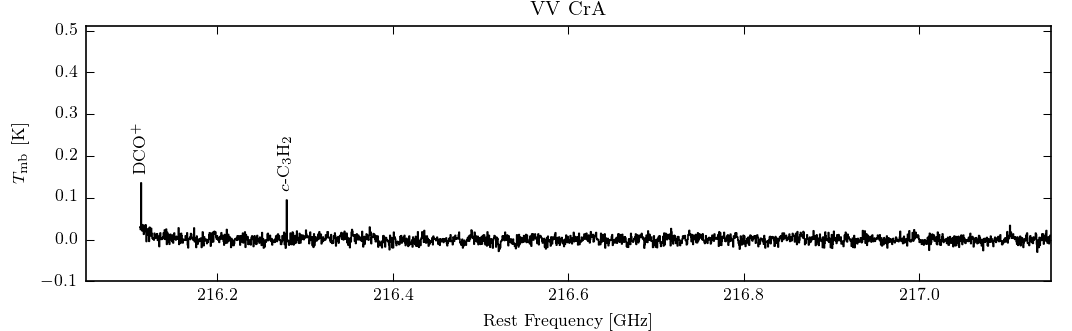} \\
                \includegraphics{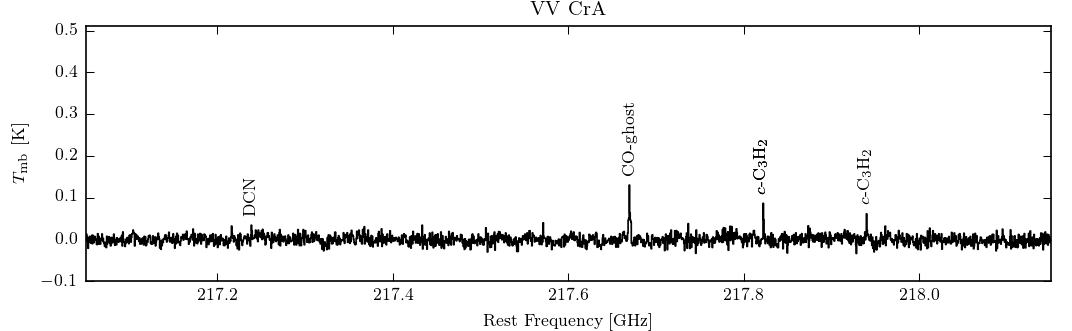} \\
                \includegraphics{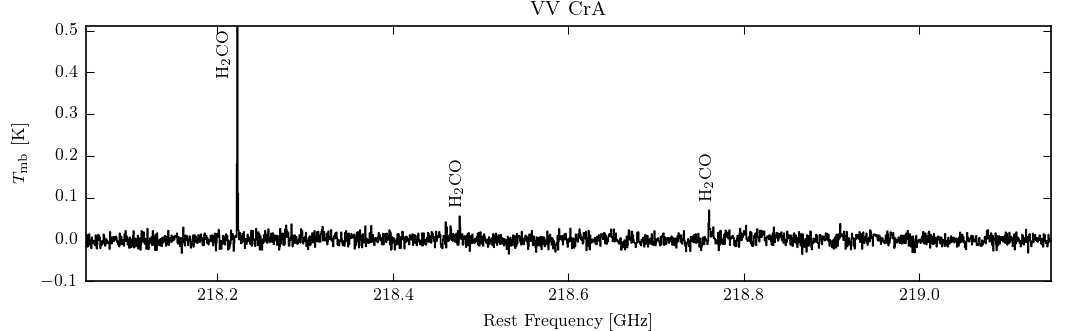} \\
                \includegraphics{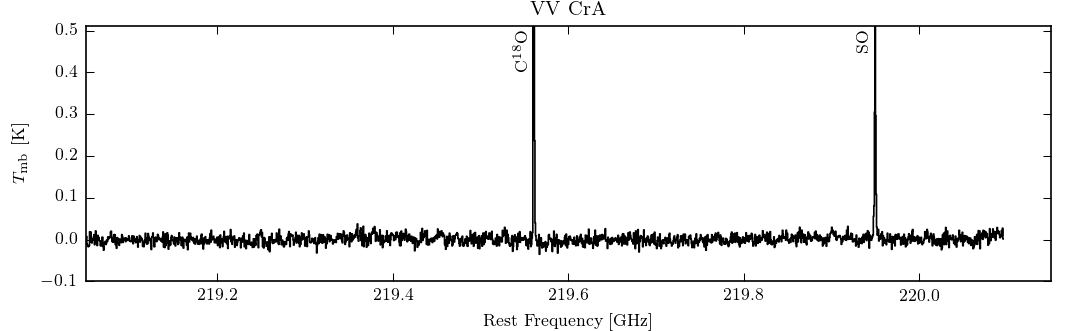} \\
                \end{array}$
                \caption{1.4~mm spectrum of VV~CrA, smoothed by a factor 8, corresponding to a channel width of 0.8~\kms.}\label{fig:vvcra_1}
        \end{figure*}
        
        \begin{figure*}[!htb]
                \centering  
                $\begin{array}{c}
                \includegraphics{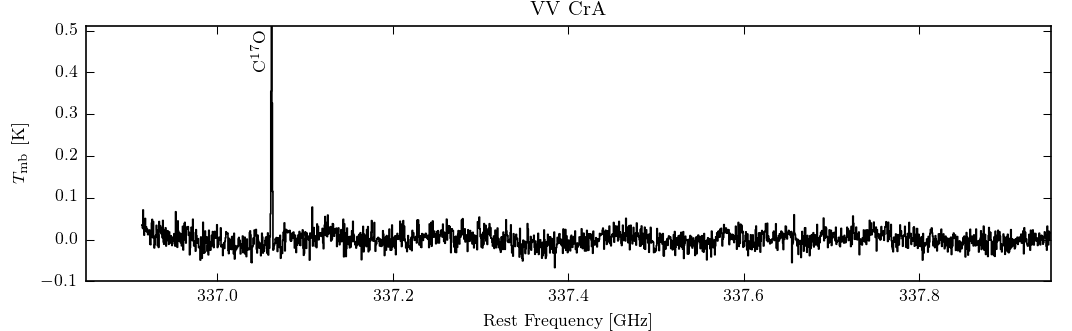} \\
                \includegraphics{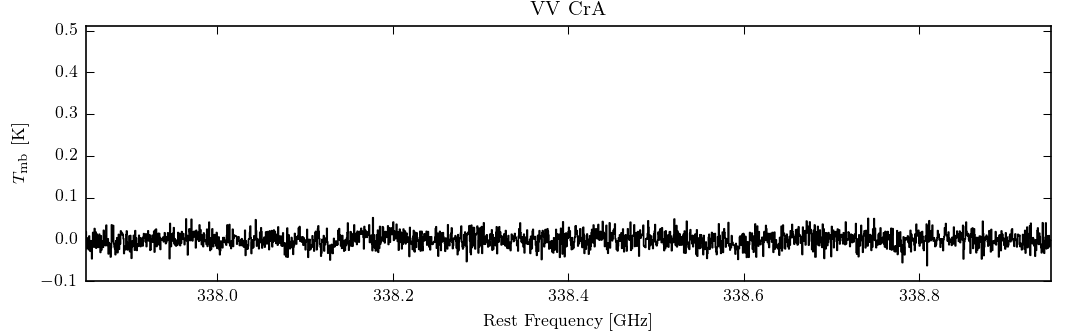} \\
                \includegraphics{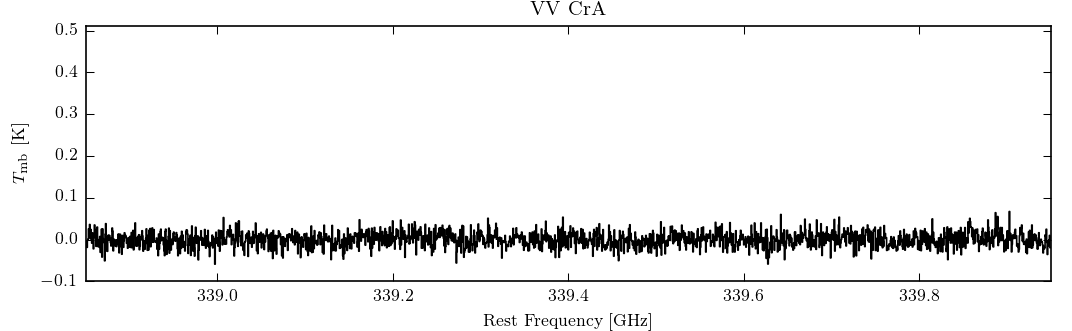} \\
                \includegraphics{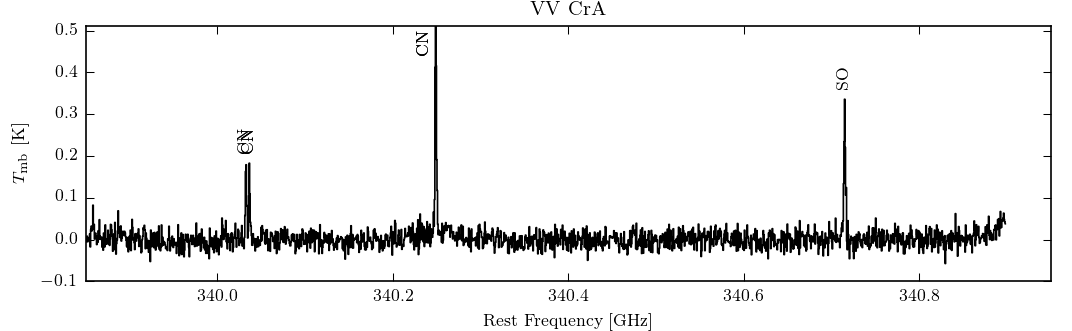} \\
                \end{array}$
                \caption{0.9~mm spectrum of VV~CrA, smoothed by a factor 8, corresponding to a channel width of 0.6~\kms.}\label{fig:vvcra_2}
        \end{figure*}
        \begin{figure*}[!htb]
                \centering  
                $\begin{array}{c}
                \includegraphics{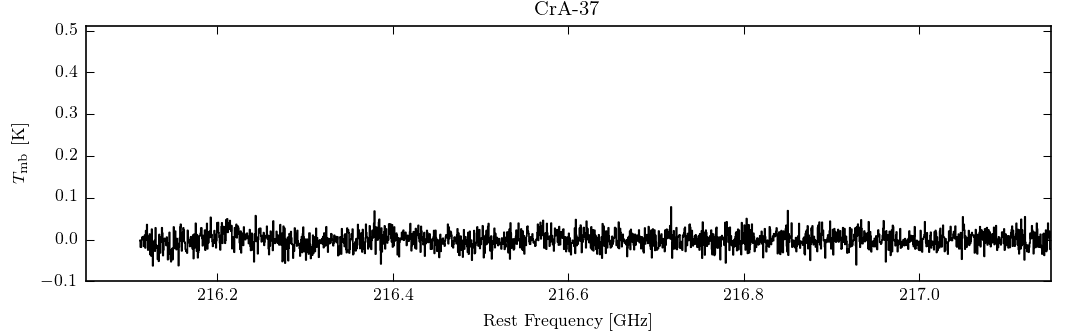} \\
                \includegraphics{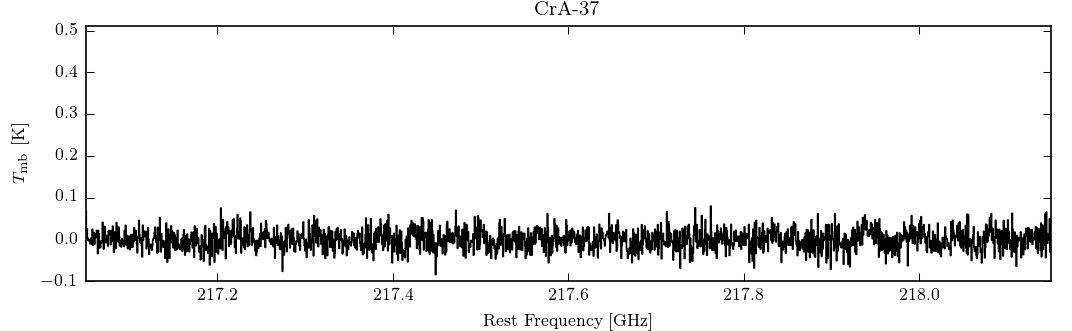} \\
                \includegraphics{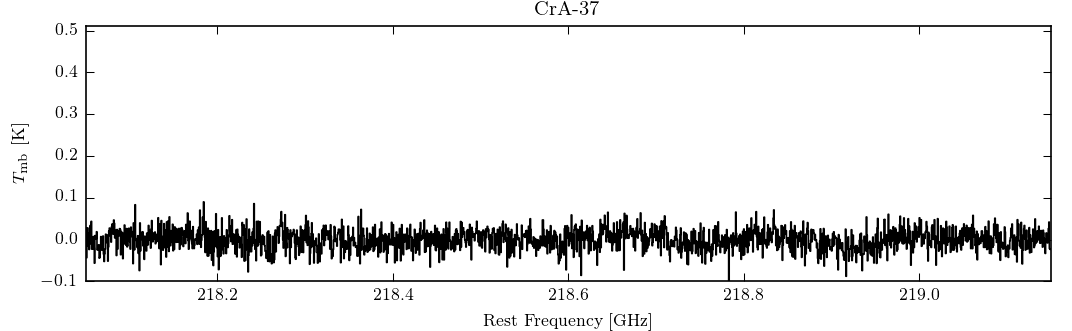} \\
                \includegraphics{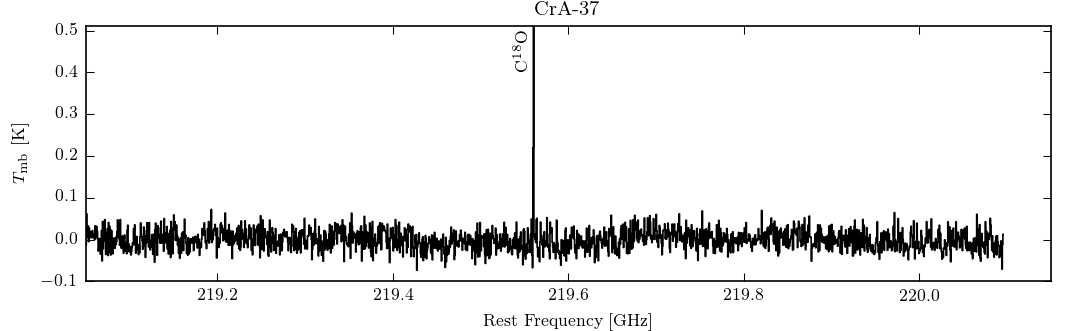} \\
                \end{array}$
                \caption{1.4~mm spectrum of CrA-37, smoothed by a factor 8, corresponding to a channel width of 0.8~\kms.}\label{fig:cra37_1}
        \end{figure*}
        
        \begin{figure*}[!htb]
                \centering  
                $\begin{array}{c}
                \includegraphics{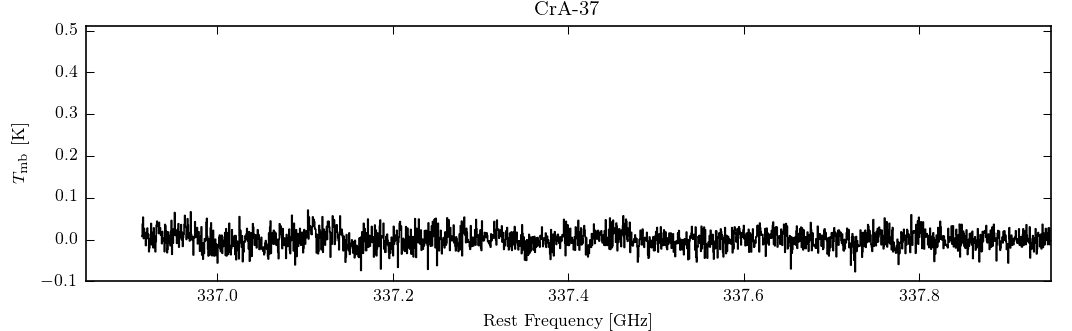} \\
                \includegraphics{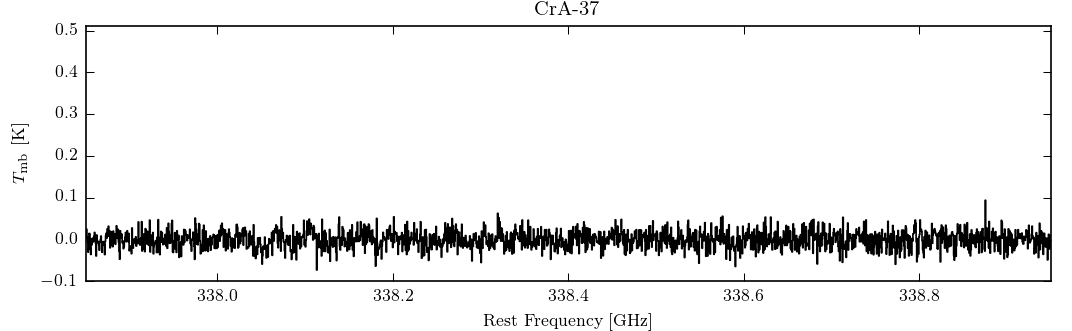} \\
                \includegraphics{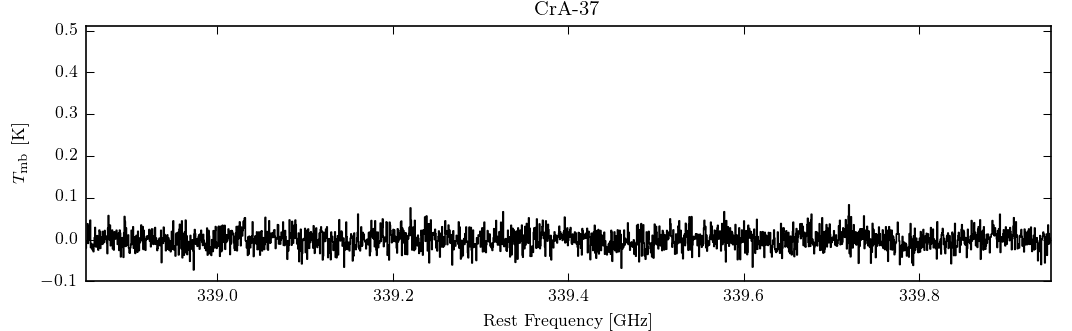} \\
                \includegraphics{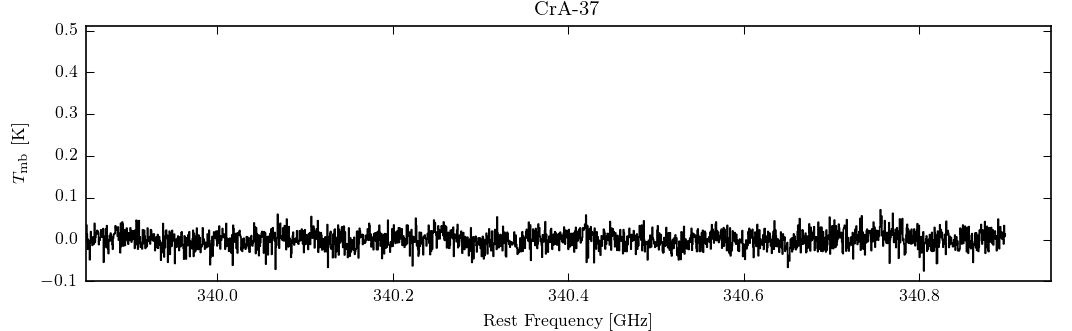} \\
                \end{array}$
                \caption{0.9~mm spectrum of CrA-37, smoothed by a factor 8, corresponding to a channel width of 0.6~\kms.}\label{fig:cra37_2}
        \end{figure*}

\clearpage
\twocolumn
\section{Unidentified spectral lines}
\label{app:ulines}

We here present our attempts to identify all U lines detected towards IRS7B: The faint line at 229.7760~GHz (assuming $v_\mathrm{LSR} = 5.7$~km~s$^{-1}$) could be the CH$_3$CHO line at 229.7750~GHz, but that would give this line a $v_{\mathrm{LSR}}=4.5$~km~s$^{-1}$, which is 1.2~km~s$^{-1}$ lower than the median LSR velocity and 0.5~km~s$^{-1}$ lower than any other line, including the more certain CH$_3$CHO lines (also if taking the JPL uncertainty of 50~kHz on the measured laboratory line frequency into account). The line at 225.160~GHz coincides with a CH$_3$SH line, but this is not the strongest line expected from this species in the studied frequency range. The 237.998~GHz line is consistent with a \textit{c}\nobr H$^{13}$CCCH line, but if this is a true detection, the \textit{c}\nobr H$^{13}$CCCH abundance would be of the same order as the \textit{c}\nobr C$_3$H$_2$ abundance, which is very unlikely. The remaining U lines do not align with any spectral lines found in Splatalogue of species expected to be present at this rms level. Two of the detected lines agree with known U lines at 223.756~GHz and 233.456~GHz. As a result of incomplete image-band rejection, a few so-called ghost lines appear at the image frequencies of the strongest spectral lines (CO isotopologue lines and strong H$_2$CO, CH$_3$OH, and CN lines). We did not tabulate these ghost lines, but they are labelled in the spectra in Appendix~\ref{app:spectrum_irs7b}. For all identified and unidentified lines, the image frequencies were investigated to exclude the possibility that any reported line would be such a ghost line.

\section{Non-LTE models of H$_2$CO emission}
\label{app:h2co}

\citet{mangum93} showed that ratios of H$_2$CO transitions involving the same $J_\mathrm{u}$-level but from different $K$-ladders (such as the $3_{03}\rightarrow2_{02}/3_{22}\rightarrow2_{21}$ and $5_{05}\rightarrow4_{04}/5_{24}\rightarrow4_{23}$ ratios) are excellent tracers of the kinetic temperature of the gas because they only operate through collisional excitation. However, at densities $n\lesssim10^8$~cm$^{-3}$, the different $J_\mathrm{u}$-levels are not fully thermalised, which means that ratios of transitions involving different $J_\mathrm{u}$-levels (such as $3_{03}\rightarrow2_{02}/5_{05}\rightarrow4_{04}$) are sensitive not only to the temperature, but also to the molecular density $n(\mathrm{H}_2)$. 

\begin{figure*}[!htb]
                \centering  
                $\begin{array}{cc}
                \includegraphics{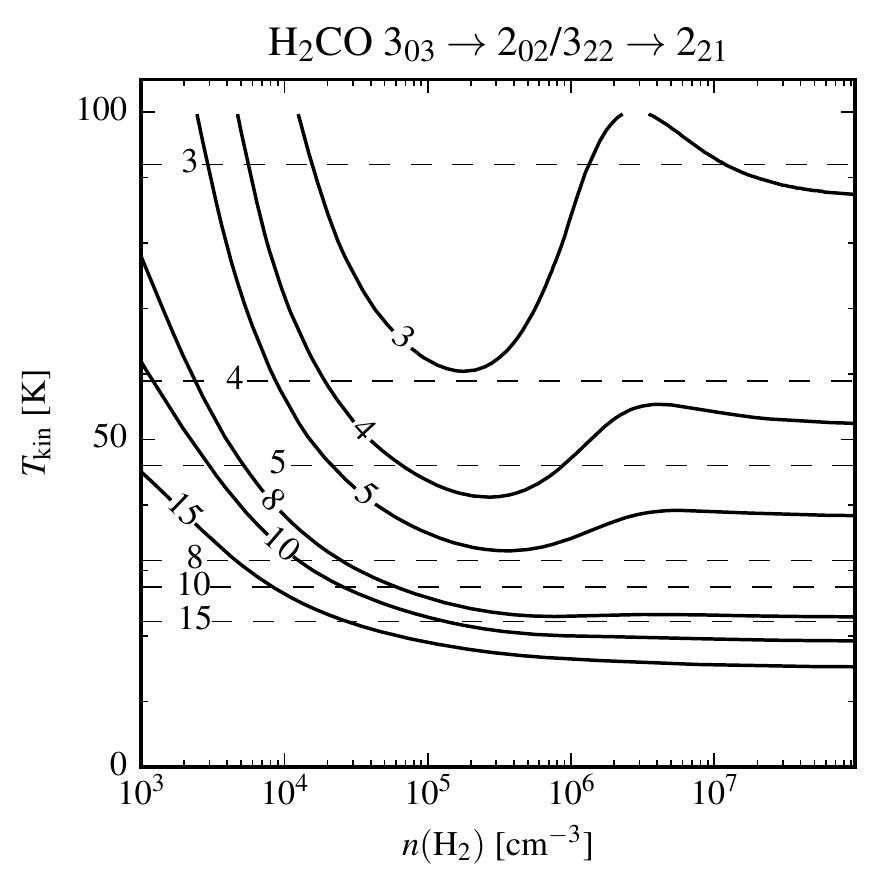} &
                \includegraphics{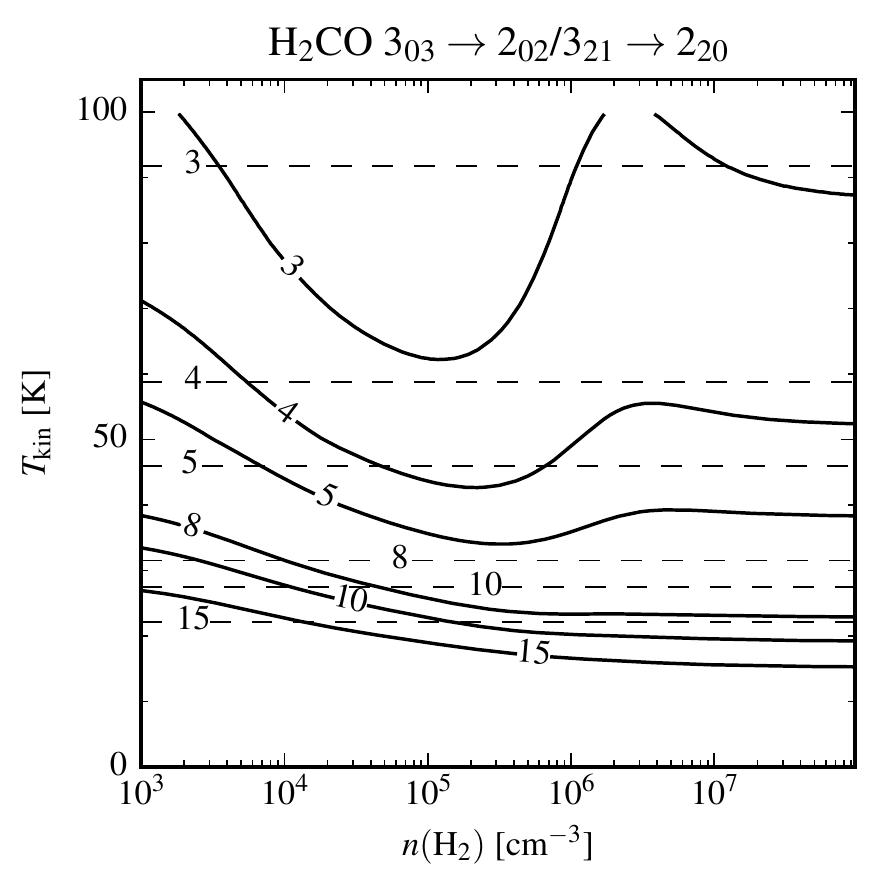} \\
                \includegraphics{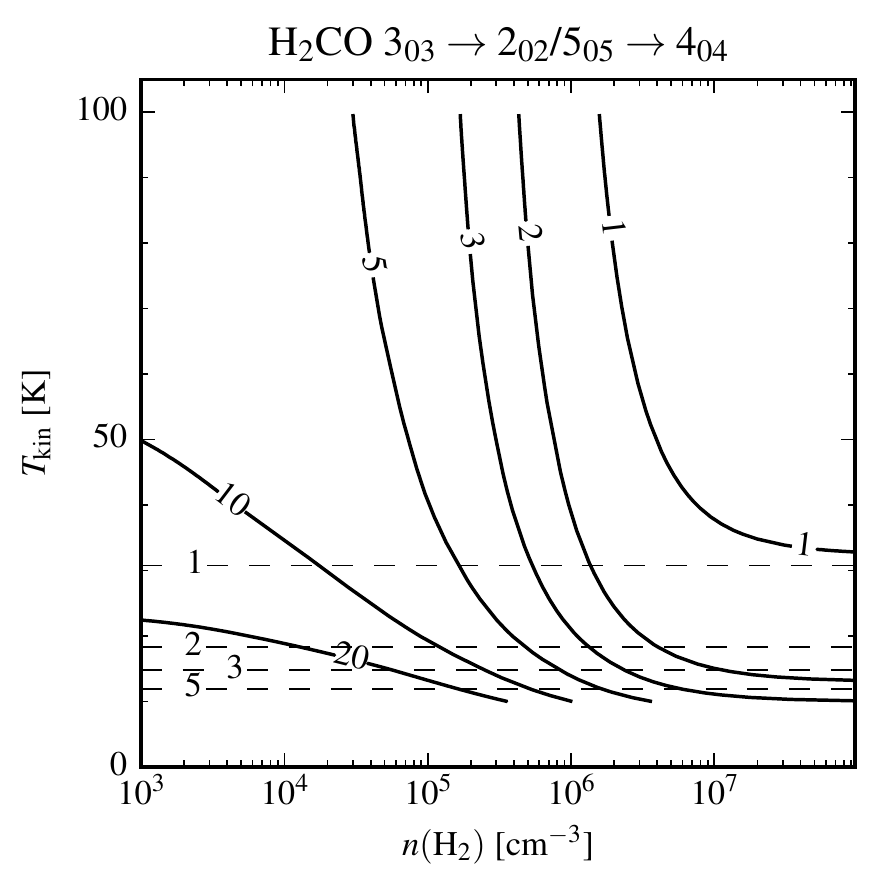} &
                \includegraphics{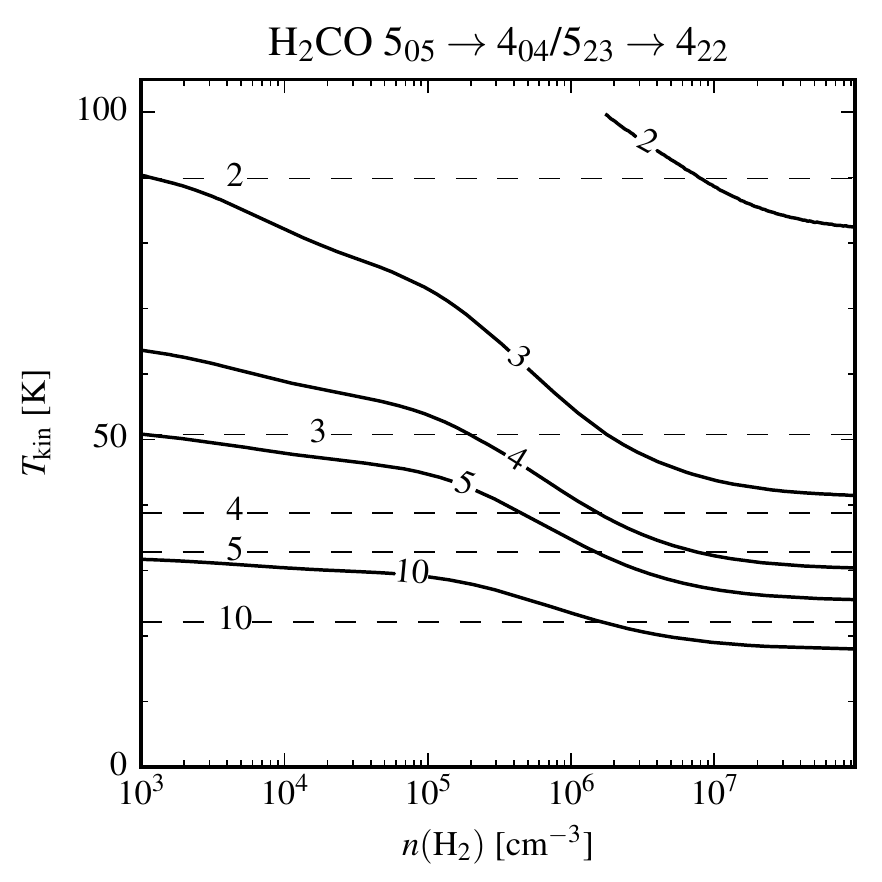} \\
                \end{array}$
                \caption{Ratios of H$_2$CO line intensities as modelled by RADEX (solid) and assuming LTE (dashed) with a column density $N=10^{14}$~cm$^{-2}$ and line widths of 2~km~s$^{-1}$.}\label{fig:radex}
        \end{figure*}

\citet{mangum93} used LVG models to derive a method for extracting temperature, density, and column density from H$_2$CO line observations. \citet{jansen_phd} performed RADEX modelling of H$_2$CO line ratios to show which H$_2$CO line ratios can be used to trace the kinetic temperature at rather low column densities, investigating certain line ratios at a p-H$_2$CO column density of $N=10^{12}$~cm$^{-2}$. In Fig.~\ref{fig:radex} we show such plots at a p-H$_2$CO column density $N=10^{14}$~cm$^{-2}$, which agrees better with the properties of the sources in this study. The $3_{03}\rightarrow2_{02}/5_{05}\rightarrow4_{04}$ ratio clearly is a particularly poor temperature probe, while the $3_{03}\rightarrow2_{02}/3_{22}\rightarrow2_{21}$, $3_{03}\rightarrow2_{02}/3_{21}\rightarrow2_{20}$, and $5_{05}\rightarrow4_{04}/5_{23}\rightarrow4_{22}$ ratios probe the temperature well at $T\lesssim50$~K and $n\gtrsim10^5$~cm$^{-2}$, also at this relatively high column density. We used the p-H$_2$CO molecular data file from the LAMDA database \citep{lamda}, which uses the collisional rates for p\nobr H$_2$CO-H$_2$ from \citet{wiesenfeld13}. If we instead use the older \citet{green91} p\nobr H$_2$CO-He collisional rates corrected to collisions with H$_2$ by correctional factors for pressure broadening and relative collision velocities the estimated $n(\mathrm{H}_2)$ values become $\sim50\%$ higher.

\begin{figure}[!htb]
                \centering  
                \includegraphics{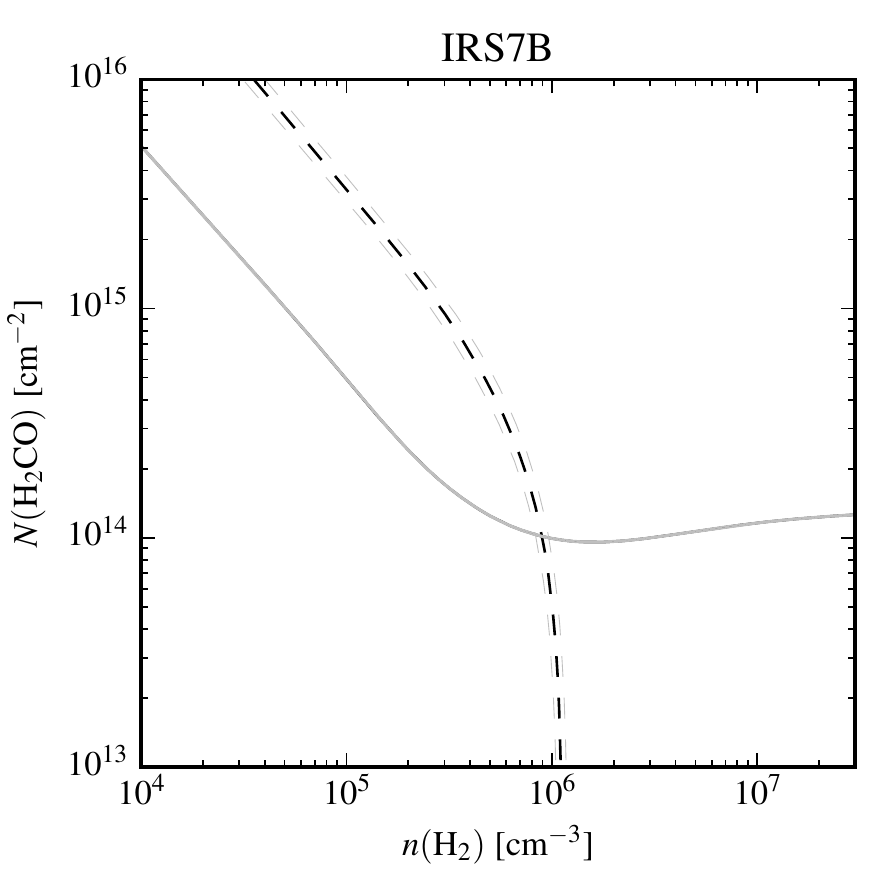} \\
                \caption{RADEX fit for H$_2$CO in IRS7B, given $T=40$~K (from the rotational diagram fit of the full IRS7B survey). The solid line shows the measured $3_{03}\rightarrow2_{02}$ line intensity, and the dashed line is the measured $3_{03}\rightarrow2_{02}/5_{05}\rightarrow4_{04}$ line ratio. $3\sigma$ errors are shown as grey lines, but can barely be distinguished from the measured values. When we estimate uncertainties, the errors on the fitted temperatures are also taken into account.}\label{fig:irs7b_radex}
        \end{figure}

To acquire the H$_2$CO temperature $T$, column density $N$, and the H$_2$ density $n$ towards the sources where we have measurements of both $J_\mathrm{u}=3$ and $J_\mathrm{u}=5$ transitions, we used the following recipe developed from the method of \citet{mangum93}:

\begin{enumerate}
        \item Measure the rotational temperatures of the $J_\mathrm{u}=3$ and $J_\mathrm{u}=5$ transitions separately and assume that the kinetic temperature is the weighted average of these two temperatures. If only the $J_\mathrm{u}=3$ transitions can be measured, we use this as the kinetic temperature (and conversely, we use only the $J_\mathrm{u}=5$ temperature in IRS7A, since the $J_\mathrm{u}=3$ transitions were not covered). 
        \item Use the non-LTE radiative transfer code RADEX \citep{vandertak07} to calculate model line strengths of the $3_{03}\rightarrow2_{02}$ and $5_{05}\rightarrow4_{04}$ lines for the kinetic temperature and a grid of $n$ and $N$ (where $N$ is the total H$_2$CO column density assuming an ortho-to-para ratio of 1.6).
        \item On this $(n,N)$ model grid, plot lines corresponding to the measured $3_{03}\rightarrow2_{02}/5_{05}\rightarrow4_{04}$ line ratio and the measured $3_{03}\rightarrow2_{02}$ line strength. The values of $n$ and $N$ are determined from the intersection of these lines (see example in Fig.~\ref{fig:irs7b_radex}).
\end{enumerate}

\end{appendix}

\end{document}